\documentclass[twocolumn,numbering,showpacs]{revtex4-1}

\usepackage{graphicx}
\usepackage{color}
\usepackage{latexsym}
\usepackage{amsmath,amssymb}        
\usepackage{hyperref}
\usepackage{mathrsfs}
\begin{document}

% -----> TITLE 

\title{Relaxation of structures resulting from head-on mergers of ultralight bosonic dark matter cores}

% ----->   AUTHORS   <-----

\author{Ana A. Avilez${}^{1}$ and F. S. Guzm\'an${}^{2}$}
\affiliation{${}^{1}$ Facultad de Ciencias F\'isico-Matem\'aticas, Benem\'erita Universidad Aut\'onoma de Puebla, 
Avenida San Claudio y 18 Sur, Colonia San Manuel, Edificio FM1-101B,
Ciudad Universitaria, C.P. 72570, Puebla, Puebla, M\'exico.
\\
		${}^{2}$ Laboratorio de Inteligencia Artificial y Superc\'omputo,
	      Instituto de F\'{\i}sica y Matem\'{a}ticas, Universidad
              Michoacana de San Nicol\'as de Hidalgo. Edificio C-3, Cd.
              Universitaria, 58040 Morelia, Michoac\'{a}n,
              M\'{e}xico.
              }
% --->   DATE

\date{\today}

% -----> ABSTRACT

\begin{abstract}
In this work we study some features of head-on mergers of equilibrium solutions of the Gross-Pitaevskii-Poisson system that rules the dynamics of the ultralight bosonic dark matter model. The importance of equilibrium solutions is that they play the role of halo cores in structure formation simulations. We consider a given range of initial conditions in order to  sample the parameter space in terms of mass ratio and head-on momentum. In each case we analyze the relaxation process induced by gravitational cooling in the high and low momentum regimes and estimate the relaxation time scales in each case. We detect a low frequency mode in the whole parameters space and it was found that the resulting configuration oscillates under this mode with amplitude that depends on the mass ratio and head-on momentum. In some cases the resulting configuration oscillates with changes in density of two orders of magnitude and with a matter distribution that is far from isotropic. These results could contribute to the collection of possible mass distributions considered in the reconstruction of mass profiles obtained in structure formation simulations.
\end{abstract}

% ----->   PACS

\pacs{keywords: dark matter -- Bose condensates}
%07.05.Tp Computer modeling and simulation
%07.05.Mh Neural networks, fuzzy logic, artificial intelligence
%05.45.Tp Time series analysis
%04.30.-w Gravitational waves

% ----->   MAKETITLE   <-----

\maketitle

% ----->     INTRODUCTION     <-----
% --------------------------------------------
\section{Introduction}
\label{sec:introduction}

Lately ultralight bosonic dark matter has gained great interest  since such type  of dark matter candidate gives rise to reasonable predictions
at galactic, cluster of galaxies and cosmic scales \cite{MatosUrena2000,Hu2000,BohmerHarko2007,MarshFerreira2010,Hui2016}.
In contrast to the standard cold dark matter model, in this model the formation of substructure at small scales is suppressed and therefore the missing-satellite problem is avoided. Additionally, given the smallness of the mass of the boson, its characteristic wave-length is of order of kpc, as a consequence all particles making up a macroscopic configuration lay in the ground state within a Bose-Einstein condensate of critical temperature in the TeV scale \cite{Magana2009}. The evolution of structures is then modeled by the mean field approximation of the condensate  represented by a wave function that obeys the Gross-Pitaevskii equation along with the Poisson equation (GPP system) describing the self-gravity of the system. At local scales, phenomenologically this is interesting since core-like density profiles arise naturally without feedback from baryons \cite{Schive2014,ChavanisNew}.  From structure formation numerical simulations, it has been concluded that this type of dark matter forms the same structures as CDM at cosmological scales, and consequently this model predictions are compatible with observations of large scale structure \cite{Schwabe:2016,Veltmaat:2018,Woo:2008}.

An interesting feature to look at is the relaxation process of structures formed due to mergers of smaller substructures. It has been illustrated that the wave-like nature of this fuzzy dark matter implies oscillatory and interference phenomena due to the nature of the GPP equations \cite{Mocz2017,Schive2014,Schive:2014hza,Schwabe:2016}. From large scale simulations it has been inferred that the structures have a core-halo density profile. It is claimed in fact, that the core shows a density profile similar to that of equilibrium solutions of the Gross-Pitaevskii-Poisson system \cite{Schive2014,Schwabe:2016}. Nevertheless, as we show in this paper, the dynamics of the interaction among structures has some cases that might affect the stability and dynamics of the relaxation process and thus restrict the core-tail density profile of structures.

Various works in the last years have found that the results of  mergers tend toward a virialized configurations that show  a stationary density profile with a core-like central shape and a NFW-tail at external radii arisen from the interference during the merger \cite{Schive2014,Schive:2014hza,Schwabe:2016,Mocz2017}. In \cite{Mocz2017,Du2017} the spatial structure of this core-tail density profile was characterized by deriving a core-to-halo relation. Additionally, the oscillatory nature of this interaction among structures was linked to further phenomena in galactic dynamics. Particularly, it is claimed that some signatures 
of the interference patterns in the tail can arise as turbulence and vorticity. More recently, the existence of nonvirialized overdense wavelets and core oscillations that may provide further tests to dynamical aspects of this kind of dark matter in galactic dynamics was pointed out \cite{Mocz2018,Marsh2018}.

The results of the present work  and \cite{GuzmanAvilez2018} place various new questions and open issues over the table regarding  the dynamics and relaxation of structures resulting from mergers. Specifically,  we find that mergers arising from head-on encounters of axial-symmetric configurations virialize,
 depending on the final size of the structure and the mass of the boson, in time-scales that run from tenths of gigayears to the age of the Universe. Thus, there is not a final configuration with a steady density profile. At most, a core-tail structure of density arises as a time-average profile \cite{GuzmanAvilez2018}. Furthermore, in this work we find that solutions to the GPP equations for the head-on merger of structures are complex and enclose 
rich dynamical aspects that have been left behind in previous studies.

In this paper we present the  relaxation process of systems resulting from the head-on merger of two cores. This is a much simpler system than those arising in structure formation simulations, where many structures interact and clump together to form structures with a classifiable density profile. However simple as it is, the merger of two equilibrium configurations illustrates how complicated the relaxation process is. It also illustrates in some cases, how the configurations resulting from mergers might show tremendous dynamical behavior that might prevent the coexistence of dark and luminous matter as found for specific situations in \cite{GonzalezGuzman2016}.
The analysis of relaxation processes of localized clumps in structure formation simulations based on the GPP system of equations, is restricted by computational power, since the resolution required to resolve binary mergers is unpractical. Thus, our study of binary head-on mergers can use 2D simulations with axial symmetry.
This allows one to monitor the evolution of the system during a given time window and follow the relaxation process. It also allows one to be systematic in the sense that we can study various scenarios involving different head-on momentum and mass ratios.

The paper is organized as follows. In Section \ref{sec:methods} we describe the methods used for the solution of the GPP system and the set up of the initial conditions for the merger analysis. In Section \ref{sec:results} we present our results. Finally in Section \ref{sec:final} we describe the conclusions of our analysis.

% -------------------------------------------
% ----->     Section     <-----
% ----------------bla---------------------------
\section{Solution of the Head-on merger}
\label{sec:methods}

% ---------->.    Subsection    <---------
\subsection{Evolution}

We solve the GPP system of equations which can be written in code units as

\begin{eqnarray}
i\partial_t \Psi &=& -\frac{1}{2}\nabla^2 + V \Psi\nonumber \\
\nabla^2 V &=& |\Psi|^2,
\label{eq:gpp}
\end{eqnarray}

\noindent which drives the dynamics of a Bose Condensate. Here, $\Psi$ represents the wave function of the system ground state and $|\Psi|^2$ is interpreted as the macroscopic density of the condensate and $V$ is the gravitational potential sourced by the condensate itself. Notice that we do not include the self-interaction term and consider only the free field regime know as the Fuzzy dark matter case, which is the one analyzed in 3D simulations \cite{Schwabe:2016,Schive2014,Schive:2014hza}.

System (\ref{eq:gpp}) is solved as a constrained initial value problem for $\Psi$, in a finite domain, with boundary conditions, given initial data for $\Psi$ consistent with $V$. Since the system is not expected to be stationary, Poisson equation acts as a constraint that has to be fulfilled during the evolution.

We solve this problem numerically in 3D with axial symmetry following the recipe  in \citep{BernalGuzman2006a,BernalGuzman2006b}. Basically, we define the problem on a two dimensional numerical domain in  cylindrical coordinates $r\in[0,r_{max}] \times z\in[z_{min},z_{max}]$, and solve it using a second order finite differences approximation on a uniform grid, integrated in time using the method of lines. We solve Poisson equation for $V$ with a Successive Over Relaxation (SOR) method with optimal acceleration parameter, at initial time and during the evolution when required at the intermediate steps of the time integration.

In order to simulate a system with no influence from the exterior,  we implemented a sponge by adding an imaginary potential starting from a zone beyond that of physical interest and until the boundary of the numerical domain $V\rightarrow V+V_{im}$, which acts as a sink of particles \citep{GuzmanRivera2010}. Specifically, we implement this sponge outside a sphere of radius $r_{sponge}$ in a region outside of the region of the mergers studied. In order to absorb as many modes as possible, we choose $V_{im}$ to have a smooth profile of the form $V_{im}=-\frac{V_0}{2}[2+\tanh (r-r_{sponge})/\delta - \tanh(r_{sponge}/\delta)]$, where $\delta$ is the width of a transition region centered at $r=r_{sponge}$ between the physical zone where $V_{im}=0$ and the zone of the sponge.

% ---------->.    Subsection    <---------
\subsection{Initial conditions}

Initial conditions correspond to the head-on collision of two equilibrium solutions of the GPP system and is set as follows. Equilibrium configurations are stationary solutions with spherical symmetry, constructed under the assumption of spherical symmetry and harmonic time dependence of the wave function $\Psi=e^{-i\omega t}\psi(r)$, which defines a Sturm-Liouville problem with eigenvalue $\omega$ (see e.g. \cite{GuzmanUrena2004} for a complete description).
For the binary system one needs to construct a wave function consisting on the superposition of the wave functions of the two equilibrium solutions and the method to define the initial wave function is as follows. First, the two equilibrium configurations that will collide are placed along the $z$ axis at a given distance from the coordinate origin. Second, we add a finite head-on momentum $p_z$ to each configuration in order to consider various scenarios starting from the free-fall case. Third, we also want to collide configurations with different mass ratios. We aim to sweep a wide range of these two parameters in order to obtain a complete catalog of initial conditions which can be mapped to those of any possible real system given the scale symmetry
of the GPP system as we shall explain in short.

The details of this initial set up are the following. One does not need to construct equilibrium configuration solutions for different mass values by solving the resulting Sturm-Liouville problem, instead we exploit the scale invariance of the GPP system of equations \cite{GuzmanUrena2004}. This invariance indicates that under the transformation 
$t=\lambda^2 \hat{t}$, 
$x=\lambda \hat{x}$,
$\Psi = \hat{\Psi}/\lambda^2$, 
$V = \hat{V}/\lambda^2$, where $x$ represents any spatial coordinate, and $\lambda$ is any number, the system of equations (\ref{eq:gpp}) remains invariant. Consequently, solving the GPP system for a given configuration means one has solved the system for all the equilibrium configurations. This invariance is also reflected in derived quantities like density and mass
$\rho=\lambda^4 \hat{\rho}$, 
$M=\lambda\hat{M}$ that are helpful in the construction of different mass configurations.

In practice the workhorse equilibrium configuration is that with the central value of the wave function  $ \psi(r=0)=1$ that we will call $\psi(r)_1$ and has mass we call $M_1$. With this configuration one can construct another configuration with a different mass $M_{\lambda}$ by choosing a value for $\lambda$ in the scaling relations above; consequently one would have $M_{\lambda}=\lambda M_1$. In practice one only needs to use the wave function associated to the workhorse configuration and rescale it with the expression $\psi(r)_{\lambda} = \lambda^2 \psi(r)_1$.

Therefore, in order to collide two unequal mass configurations we choose the first one to be the workhorse configuration described by $\psi(r)_1$ and a scaled configuration described by $\psi(r)_{\lambda}$ for a given $\lambda$. Notice that according to the scaling relations,   the mass ratio between the configuration with mass $M_{\lambda} = \lambda M_{1}$ is precisely $MR=M_{\lambda}/ M_1 =\lambda$.
By virtue of this scaling property, it is possible to sweep the parameter space for binary mergers of configurations with arbitrary mass, by simply rescaling the workhorse configuration.

In order to monitor the evolution of the merger, it is desirable to set the center of mass at the center of the numerical domain. %, which in our case coincides with the center of coordinates $(r=0,z=0)$. 
For this we take the convention of $0<\lambda<1$ in all cases, so that $M_{\lambda} < M_1$ always. We choose the configuration with mass $M_1$ to be represented by the wave function $\psi_L$ ($L$ for left),  centered initially at a point $(0,-z_0\lambda)$ with $z_0>0$. The configuration with mass $M_{\lambda}$ will be represented by the wave function $\psi_R$ ($R$ for right), and will be centered at the position $(0,z_0)$. The location of the two configurations defined in this manner implies the center of mass lies at the coordinate origin.

Now, we add momentum along the head-on direction to each of the blobs. For this,  we define a momentum for the left configuration $p_z^{L}$ and a momentum for the configuration on the right $p_z^{R}=p_z^{L} / \lambda$, which maintain the center of mass at the coordinate origin. 
The head-on momentum is applied by redefining $\psi_L\rightarrow e^{i{\bf p_z \cdot x}}\psi_L$ and $\psi_R \rightarrow e^{-i{\bf p_z \cdot x}}\psi_R$. Finally the wave function at initial time is $\psi_L + \psi_R$. Our simulations are parametrized  by $p_z^{L}$ that we simply will call $p_z$ in the reminder of the paper.

\begin{figure}
\includegraphics[width= 7cm]{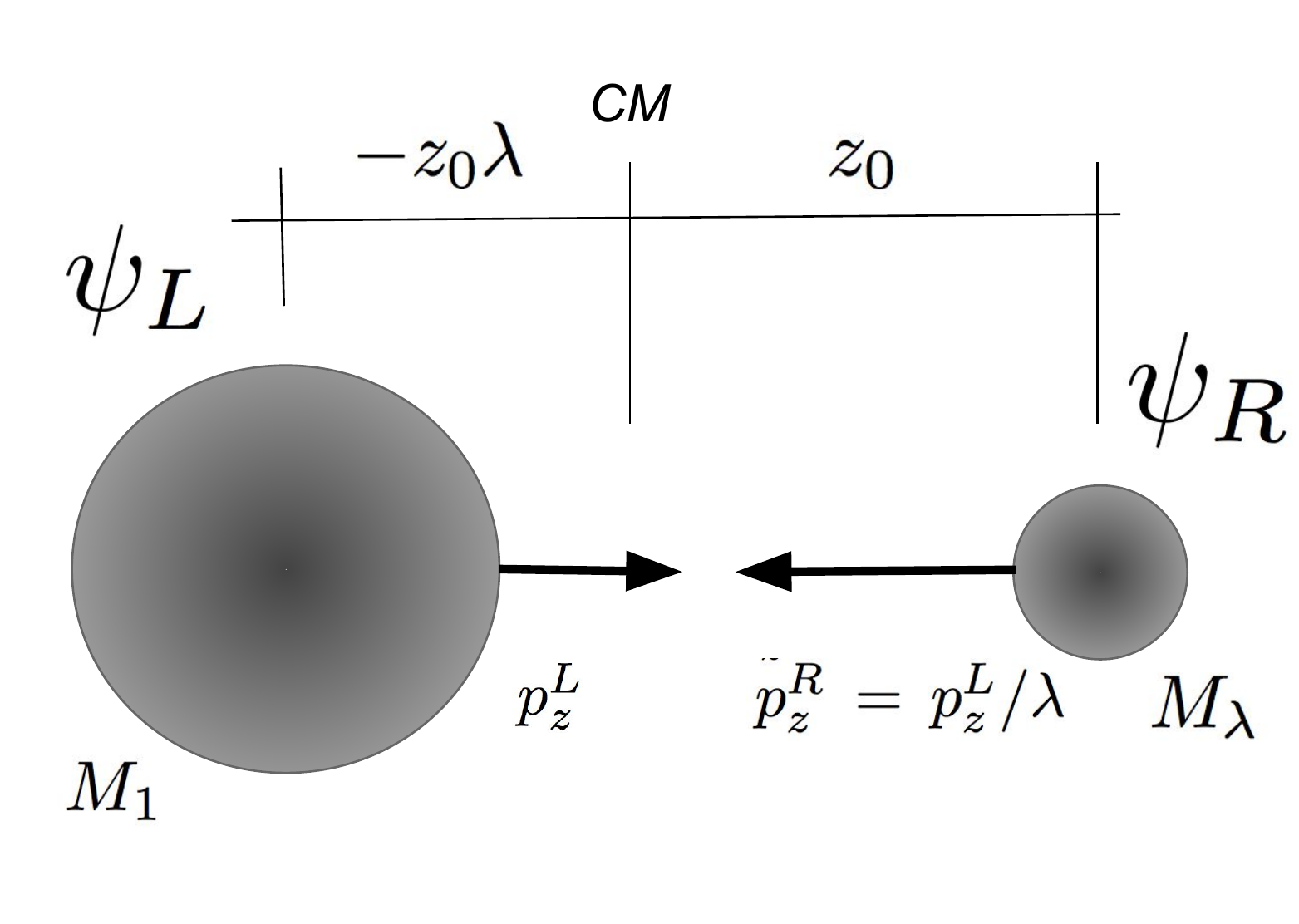}
\caption{Illustration of the initial conditions. The circles represent each one of the configurations. In all the cases $M_1>M_{\lambda}$, $p_z^{R}=p_z^{L} / \lambda> p_z^{L}$ and $\lambda z_0< z_0$.}
\label{fig:initial_conditions}
\end{figure}

In Figure \ref{fig:initial_conditions} we show a scheme of the initial conditions, where  $M_1>M_{\lambda}$ as assumed in all the cases. The center of mass is then located closer to the configuration with mass $M_1$ since $\lambda = MR <1$. The head-on momentum as defined above $p_z^{R}=p_z^{L} / \lambda$, is bigger for the configuration of smaller mass $M_{\lambda}$. An important consideration is that we change the value of $MR$ by keeping the value of $M_1$ fixed, whereas we change the value of $\lambda$ to obtain the mass of the lighter configuration $M_{\lambda}$, its momentum $p_{z}^{R}$ and the distance to the center of mass, all at once.

%%%%%%%%%%%%%%%%%%%%%%%%%%%%%%%%%%%%%%%%%%%%%%%%
%%%%%%%%%%%%%.   SECTION      %%%%%%%%%%%%%%%%%%%%%%%%
%%%%%%%%%%%%%%%%%%%%%%%%%%%%%%%%%%%%%%%%%%%%%%%%
%%%%%%%%%%%%%%%%%%%%%%%%%%%%%%%%%%%%%%%%%%%%%%%%

\section{Results}
\label{sec:results}

\begin{figure*}
\centering
\includegraphics[width= 3.25cm]{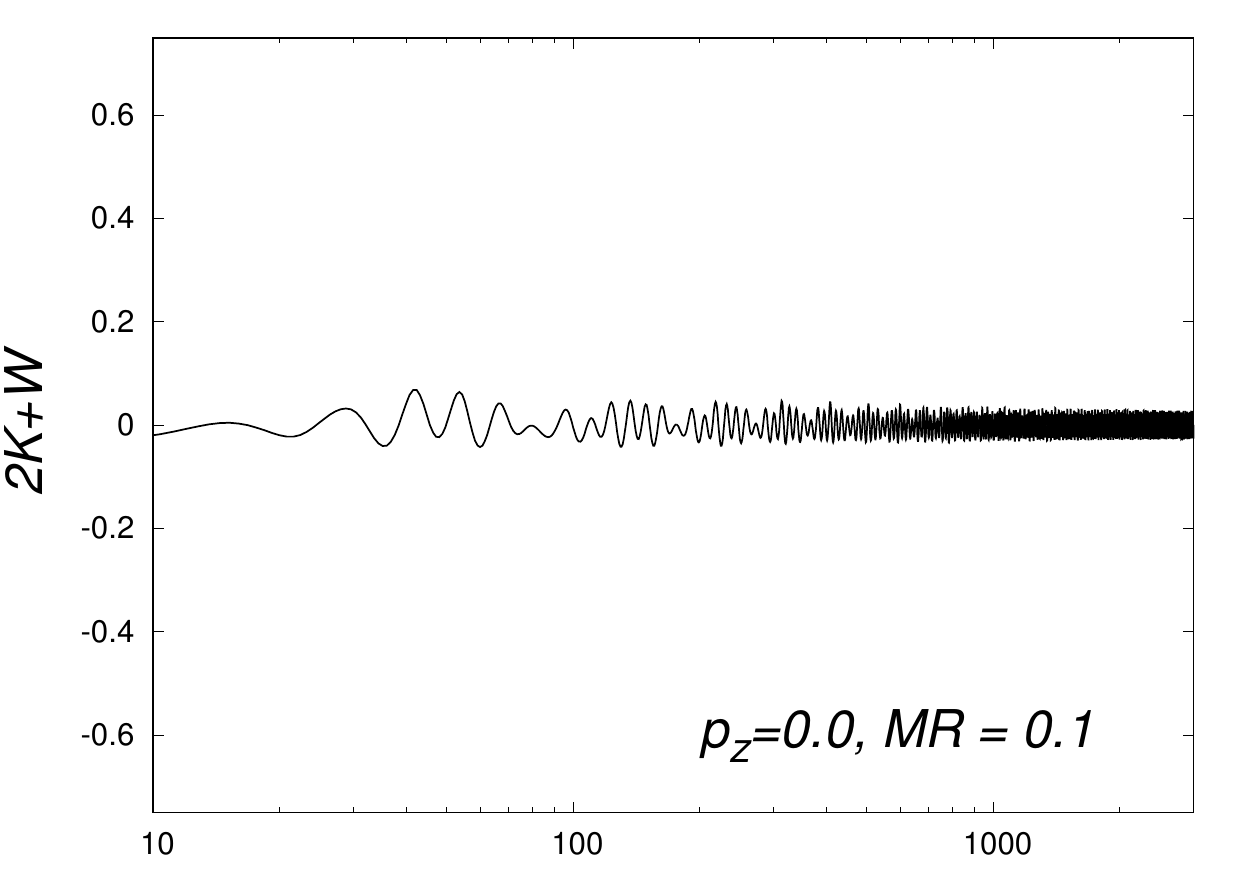}
\includegraphics[width= 3.25cm]{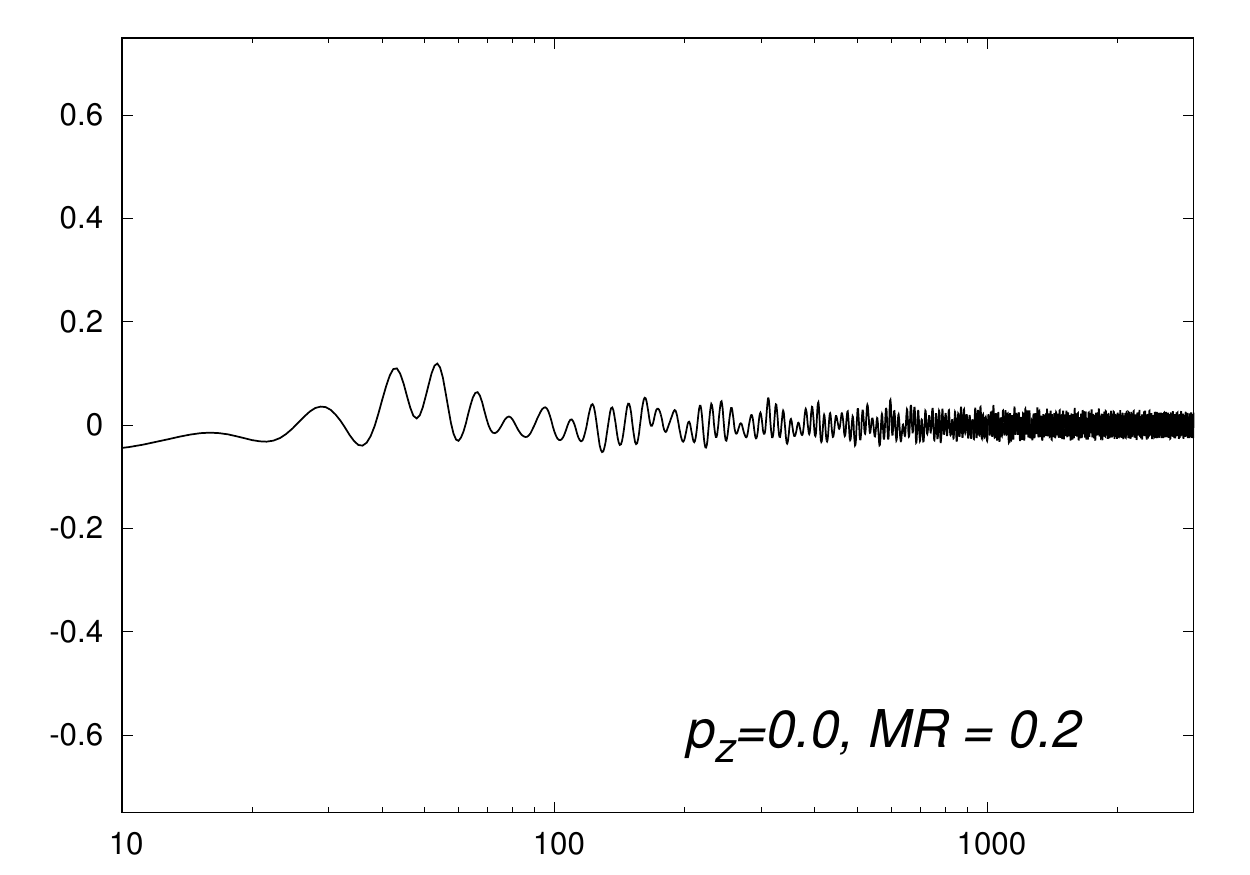}
\includegraphics[width= 3.25cm]{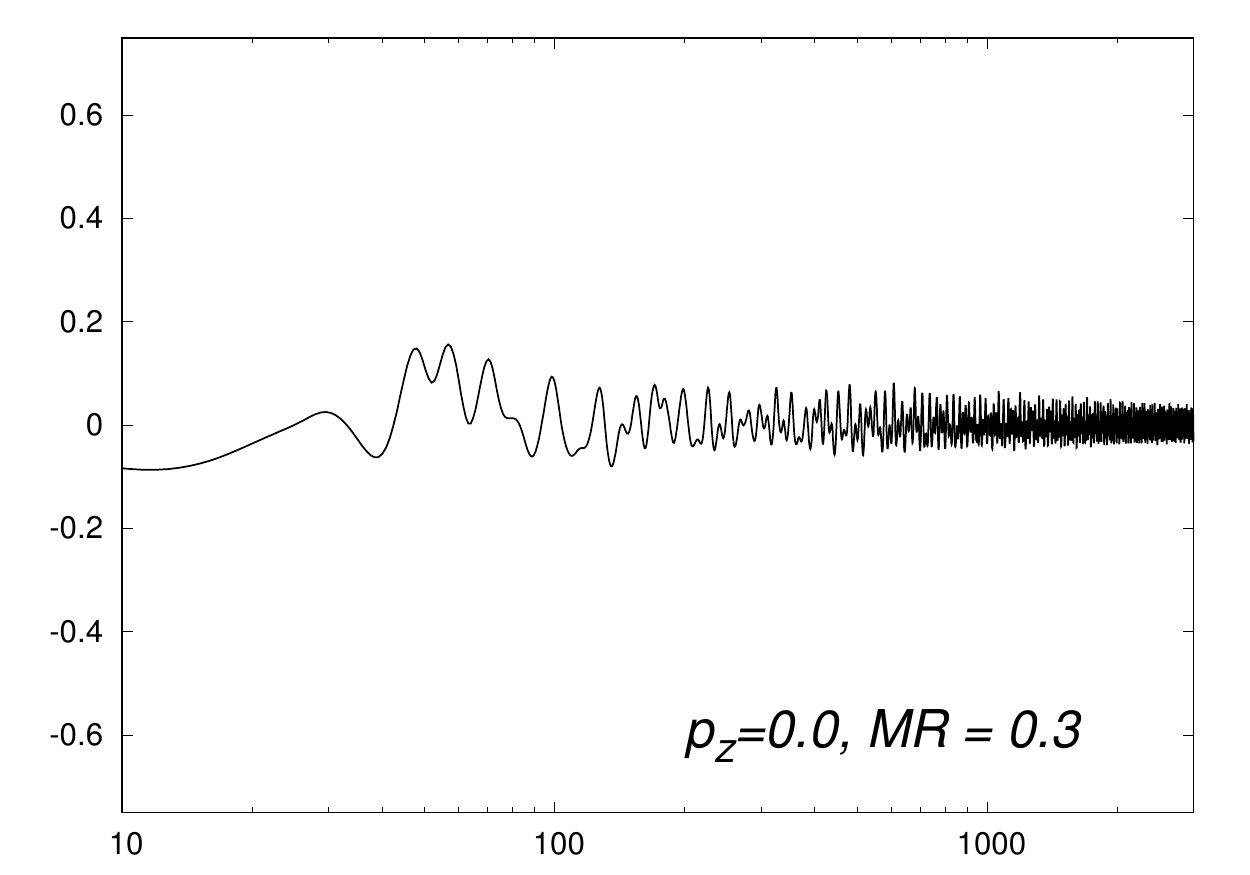}
\includegraphics[width= 3.25cm]{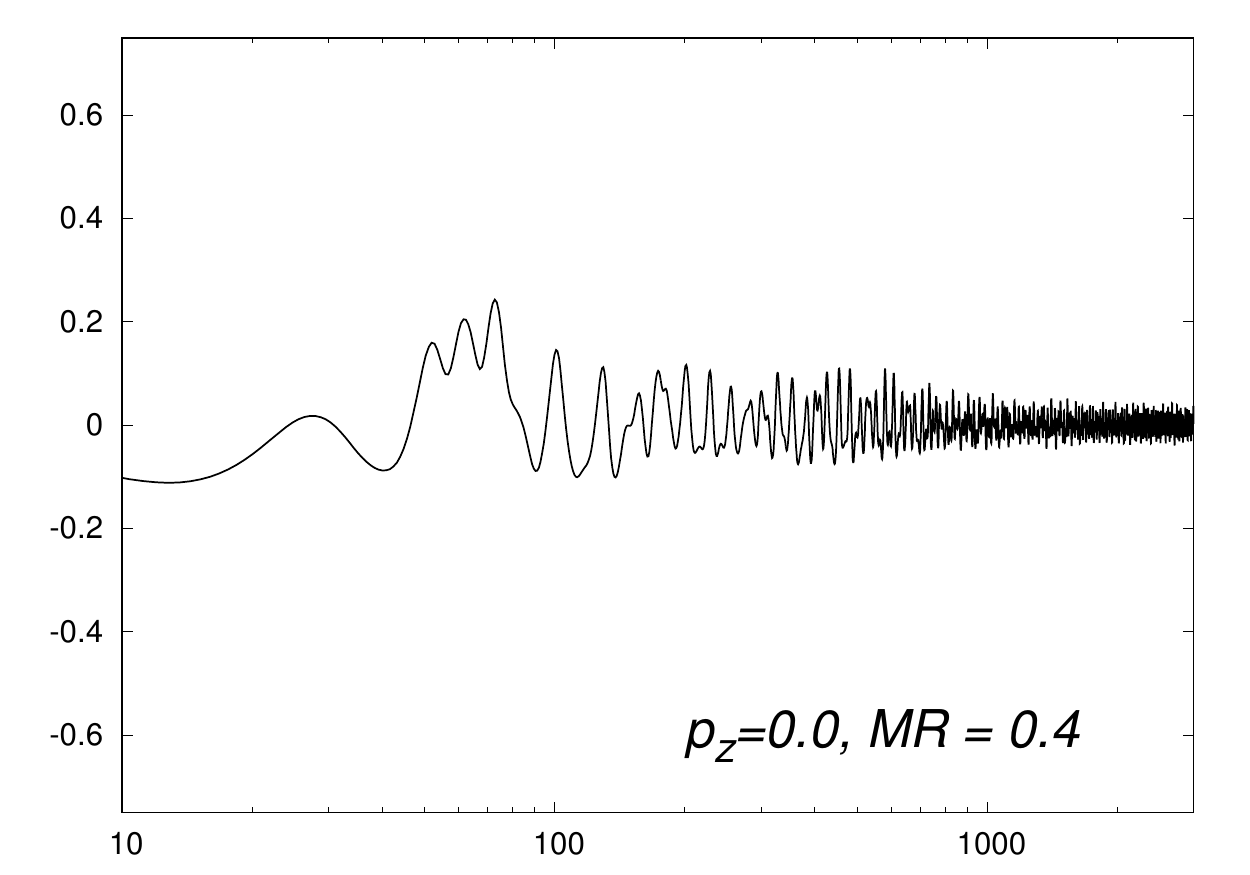}
\includegraphics[width= 3.25cm]{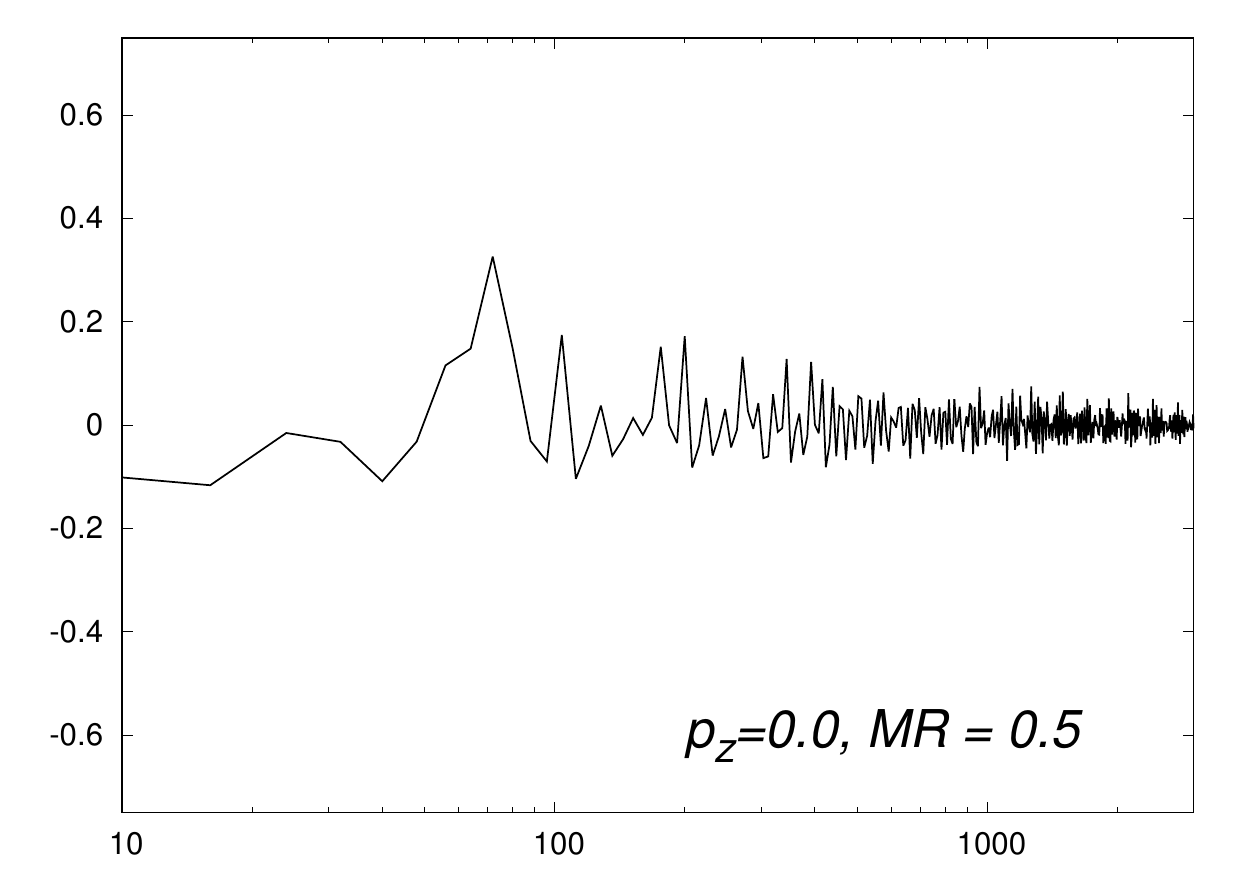}
\includegraphics[width= 3.25cm]{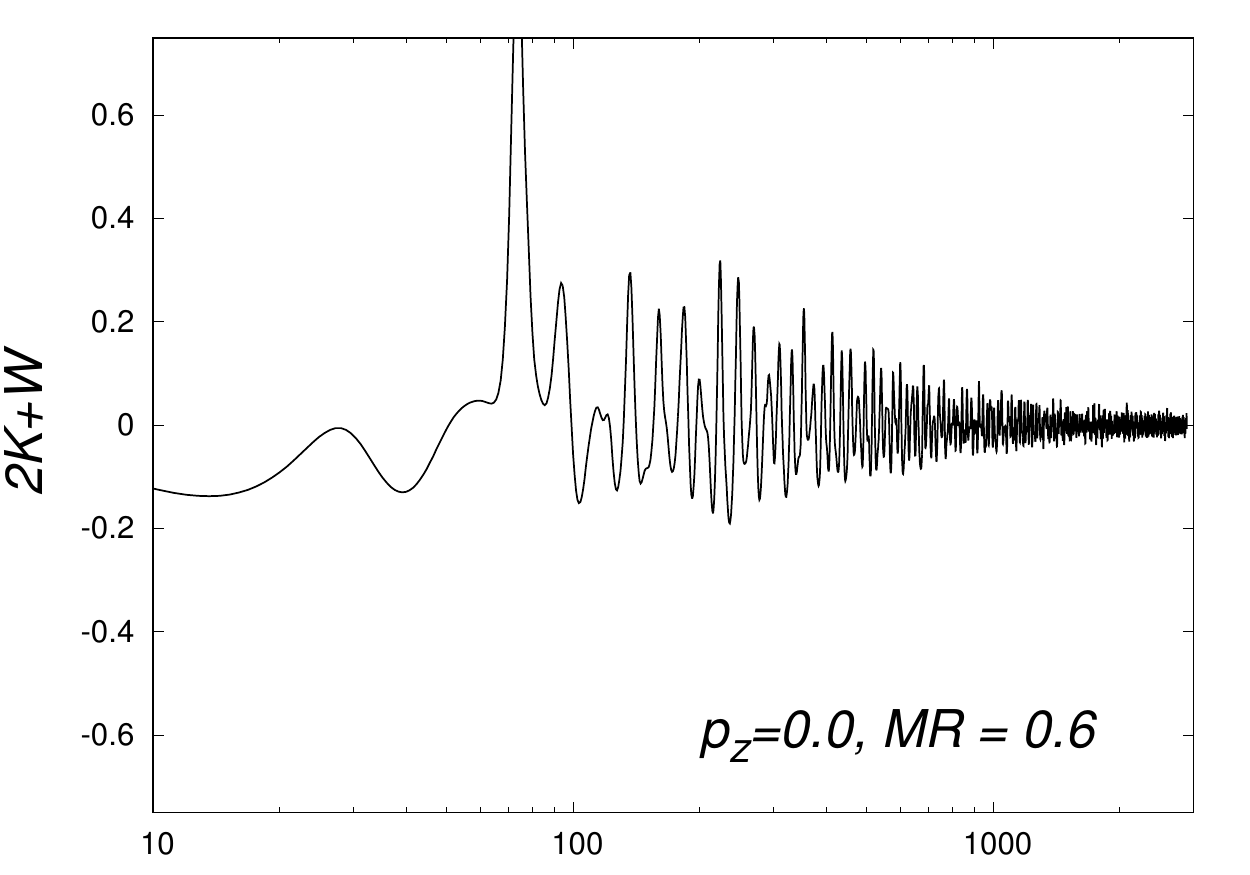}
\includegraphics[width= 3.25cm]{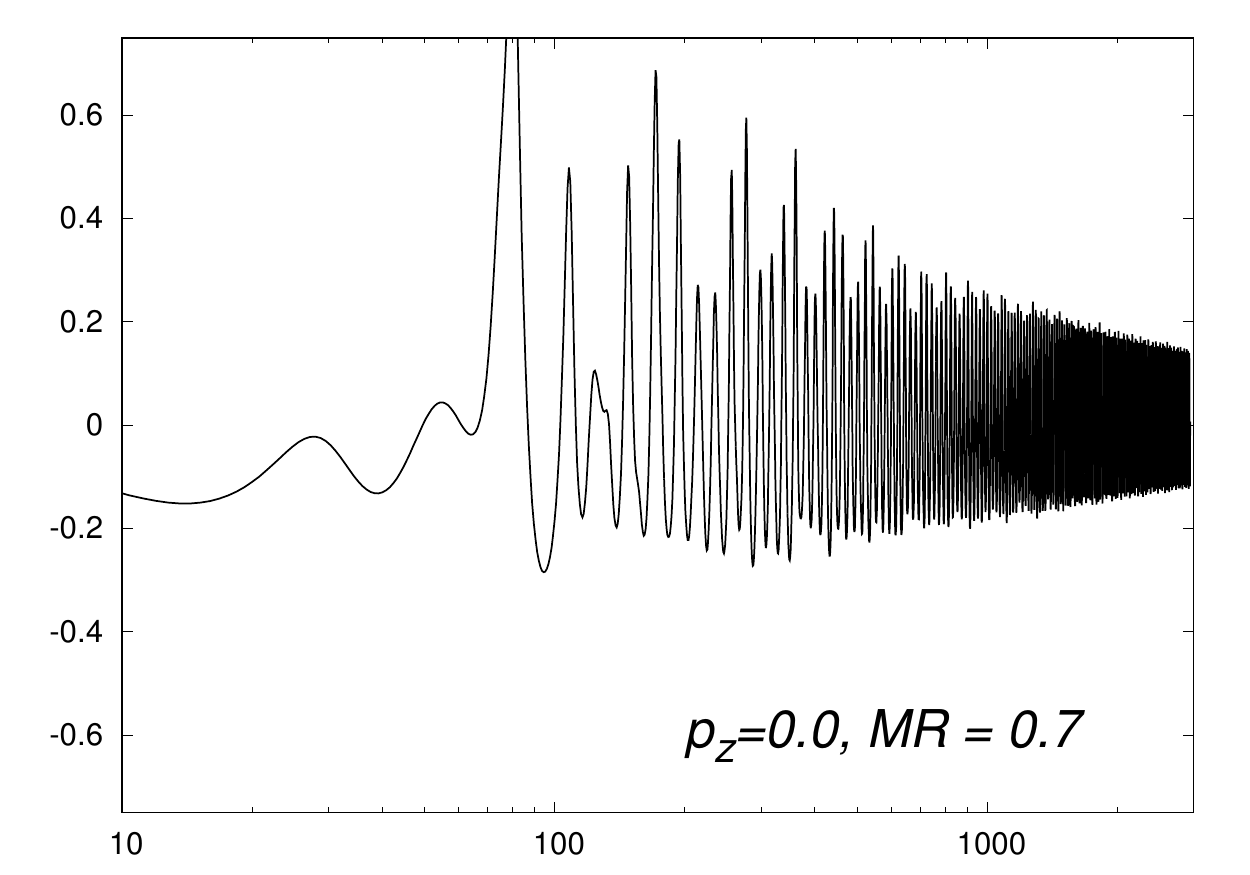}
\includegraphics[width= 3.25cm]{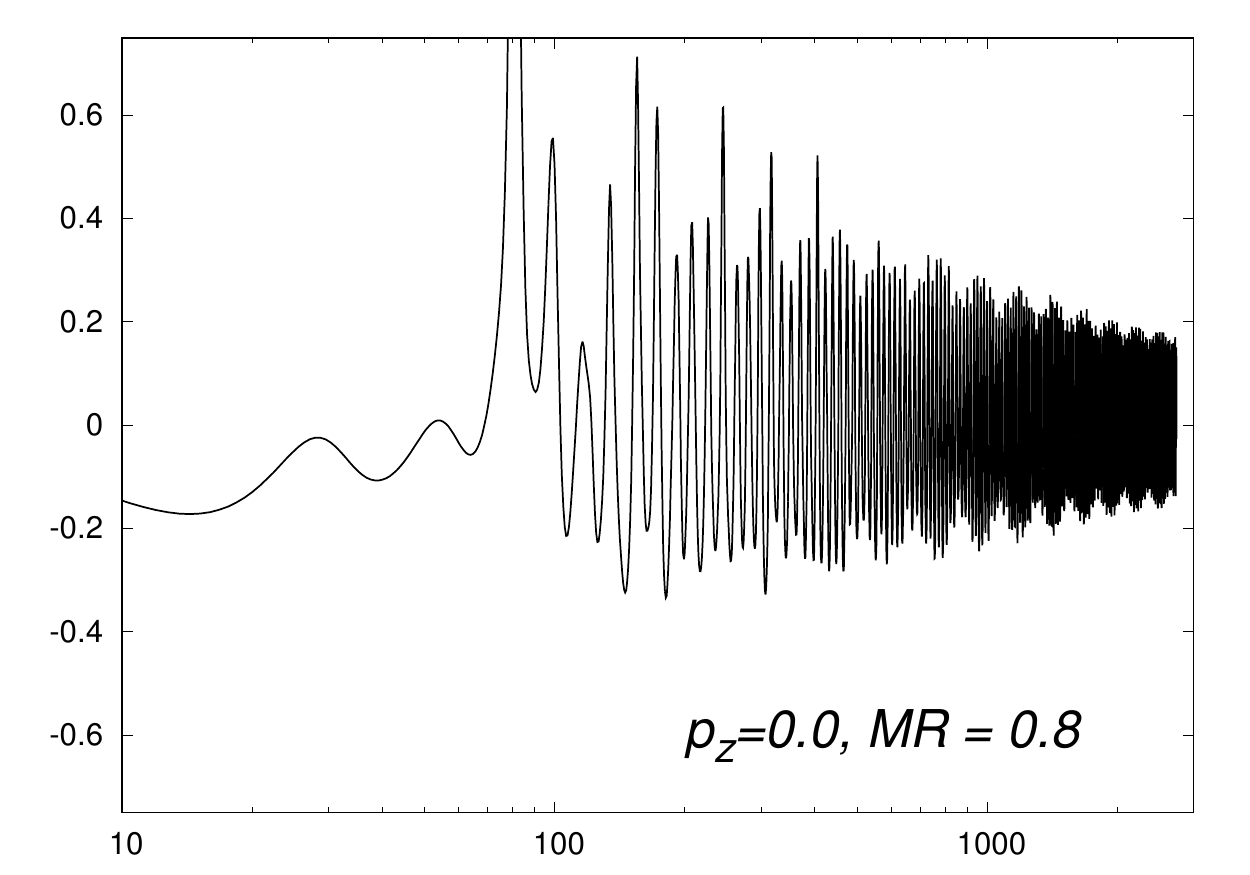}
\includegraphics[width= 3.25cm]{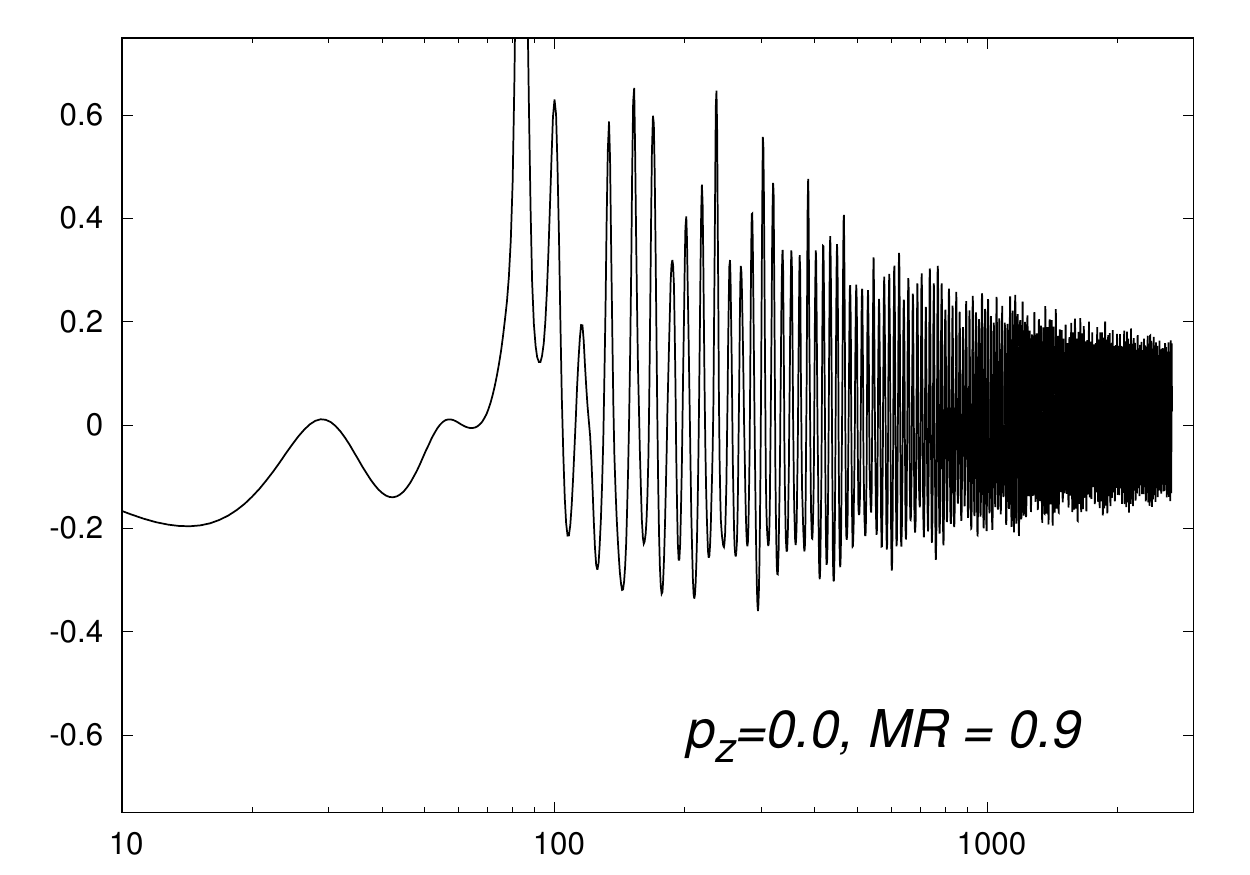}
\includegraphics[width= 3.25cm]{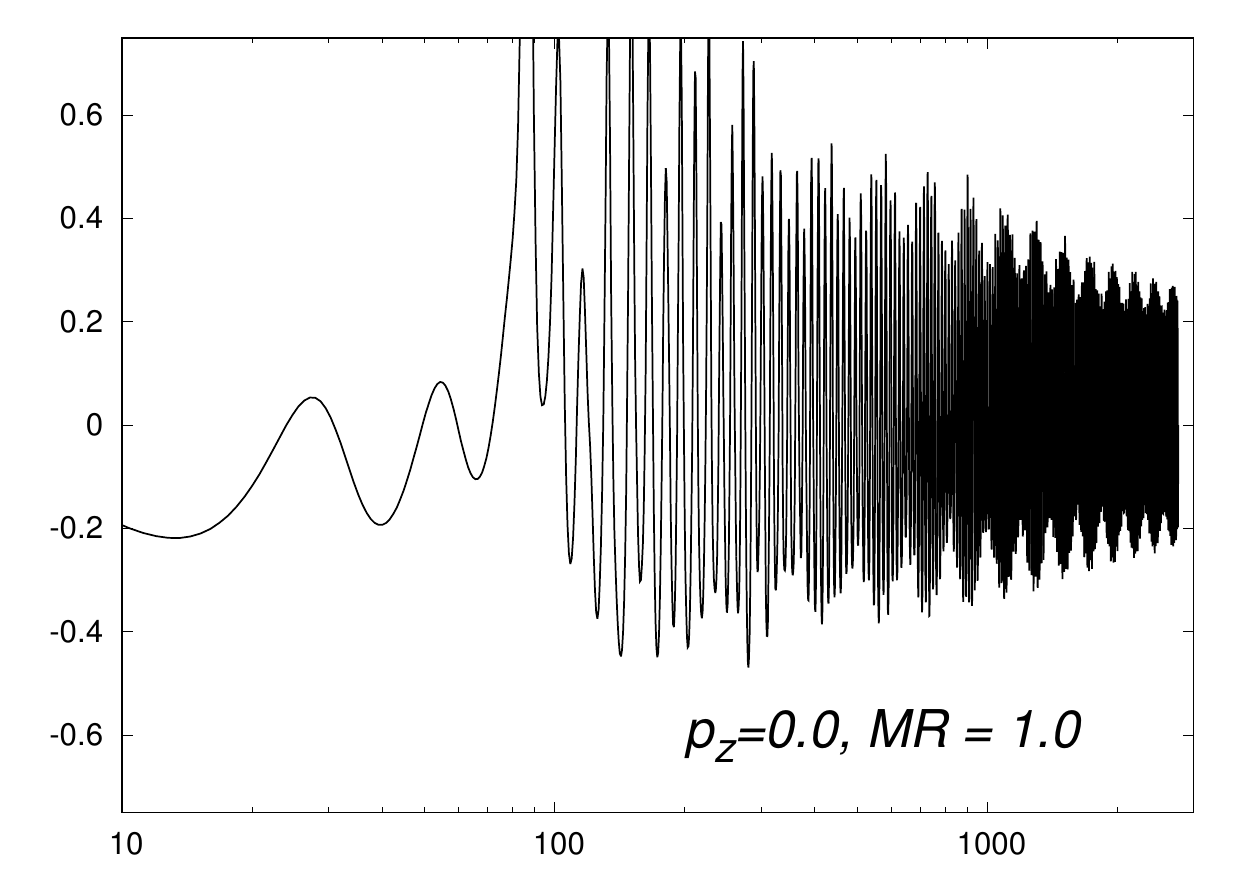}
\includegraphics[width= 3.25cm]{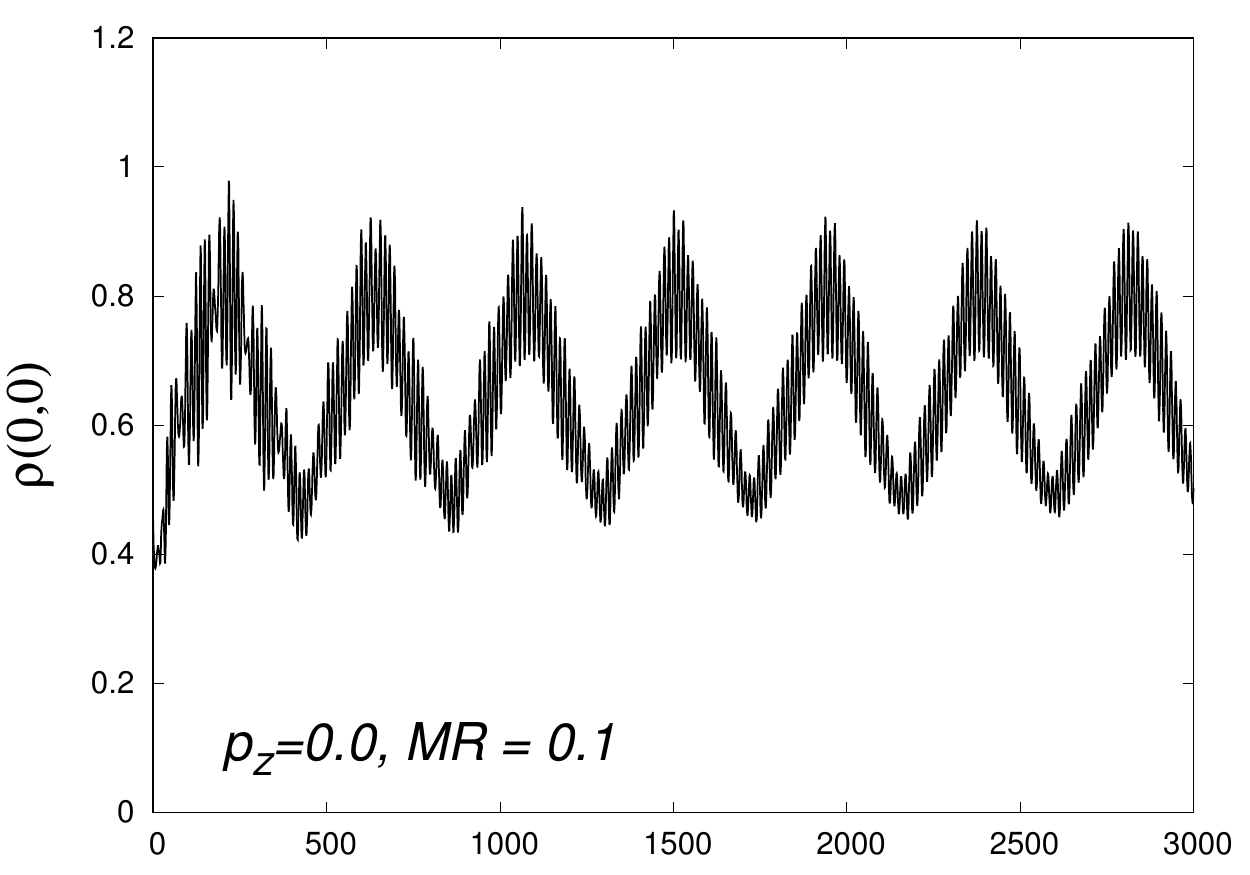}
\includegraphics[width= 3.25cm]{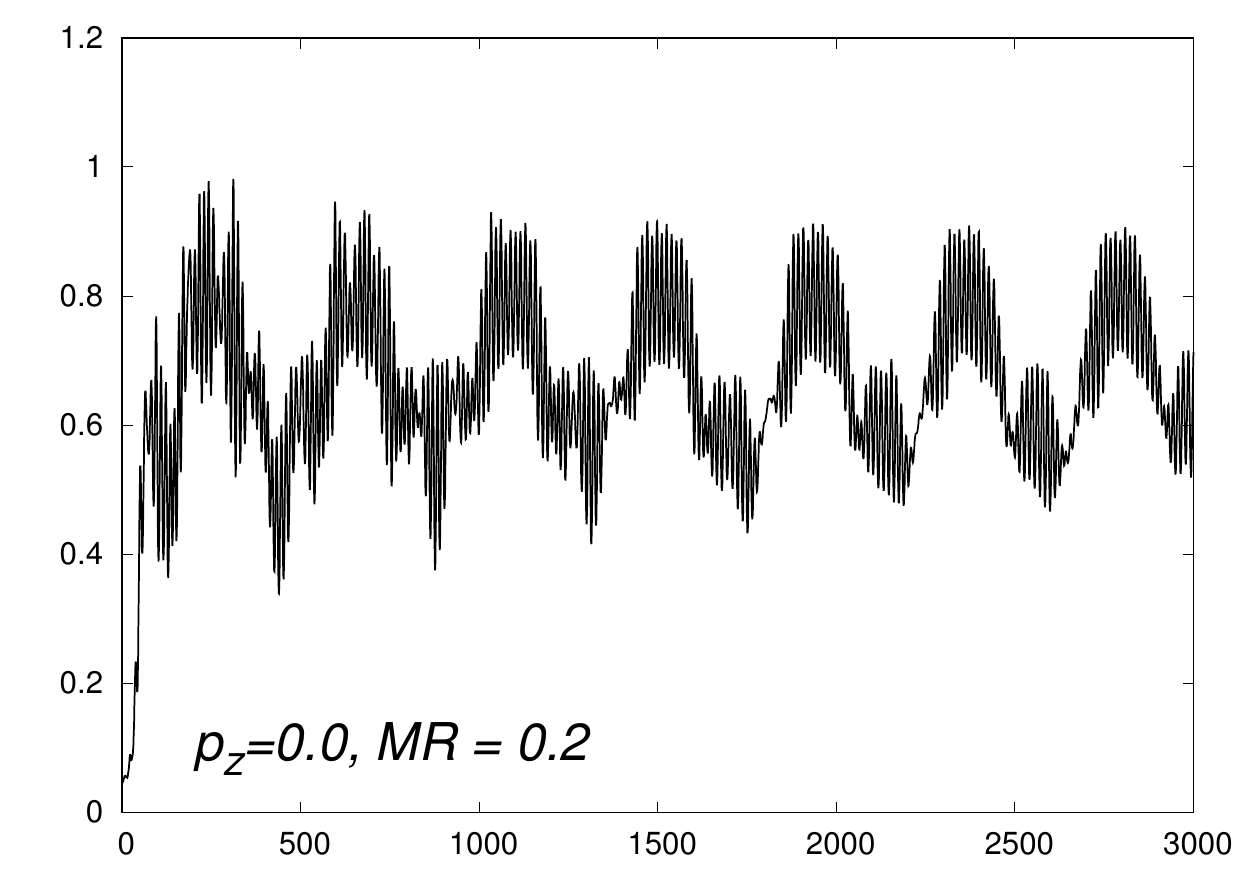}
\includegraphics[width= 3.25cm]{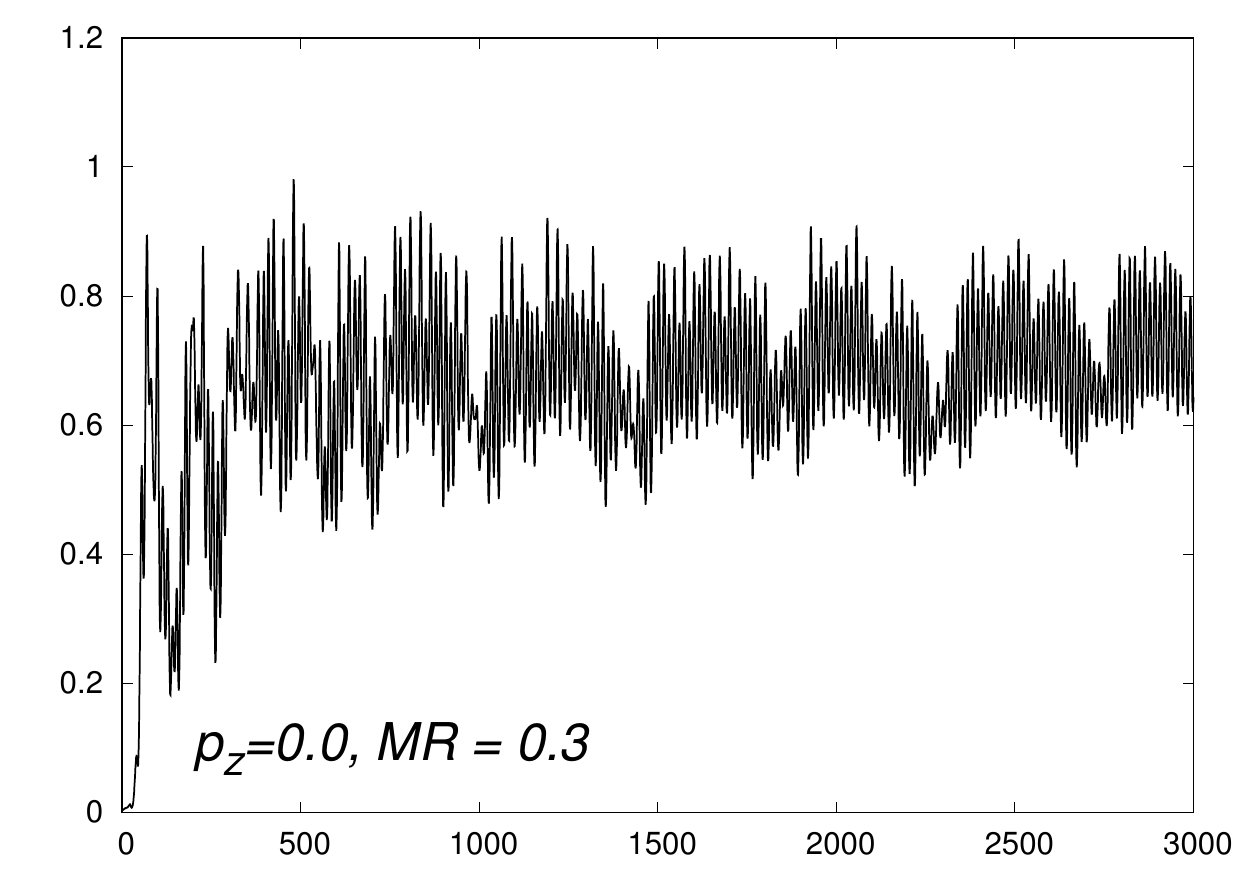}
\includegraphics[width= 3.25cm]{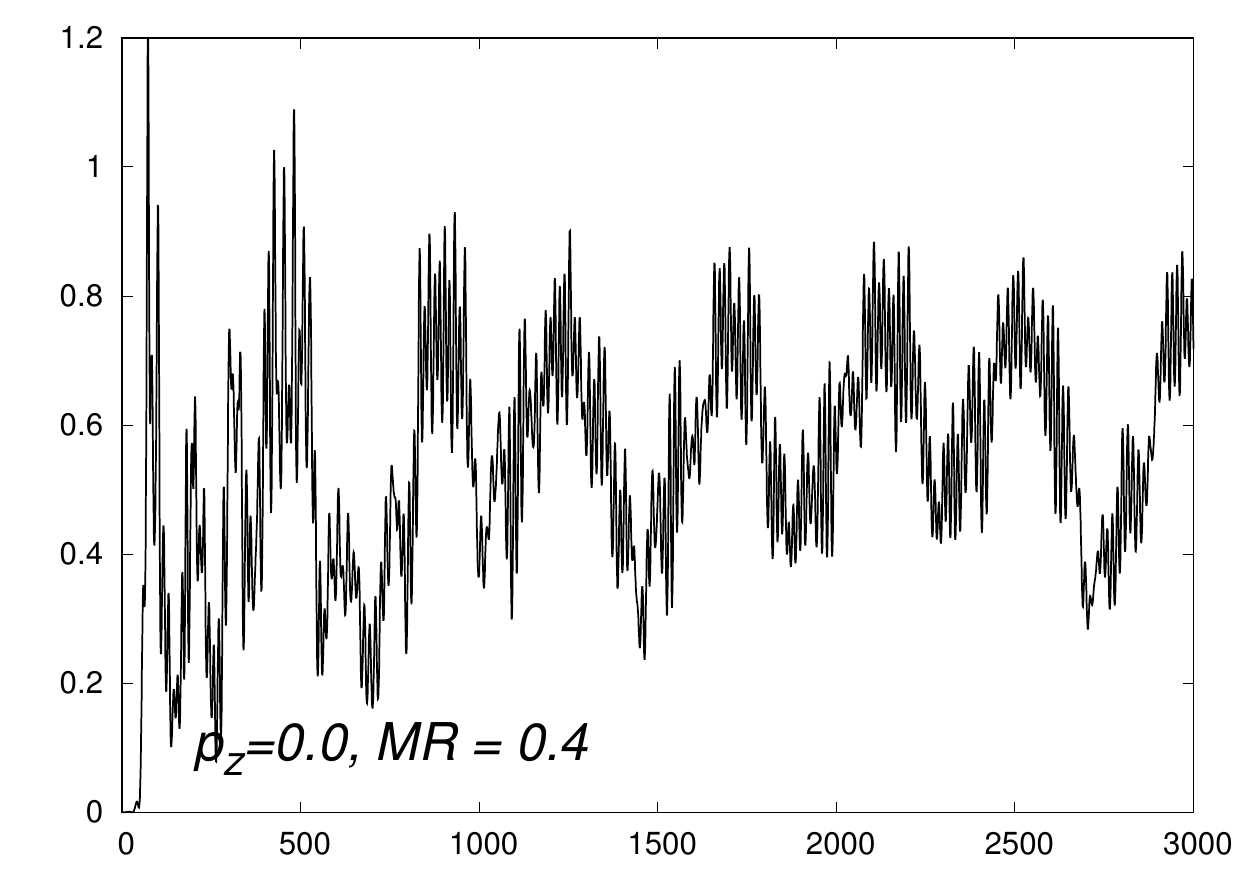}
\includegraphics[width= 3.25cm]{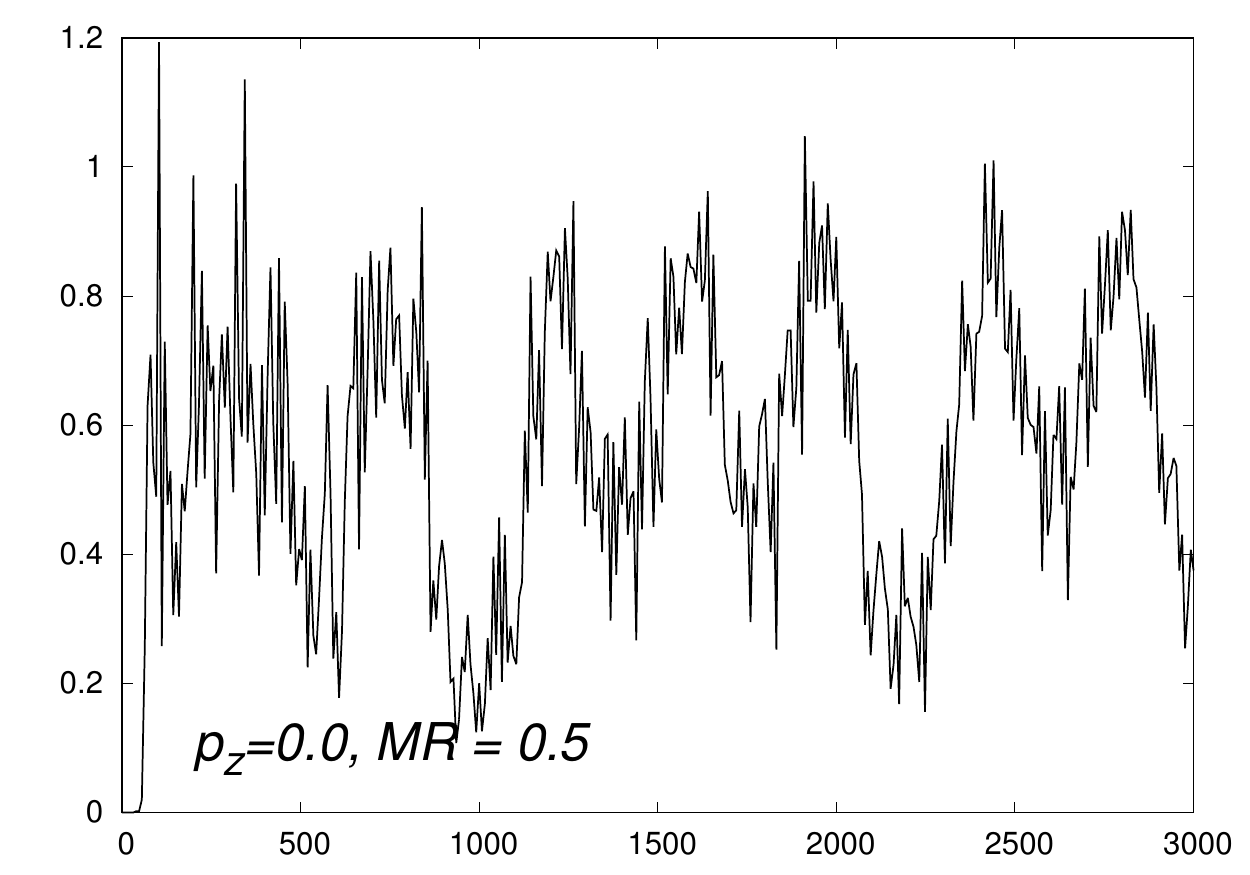}
\includegraphics[width= 3.25cm]{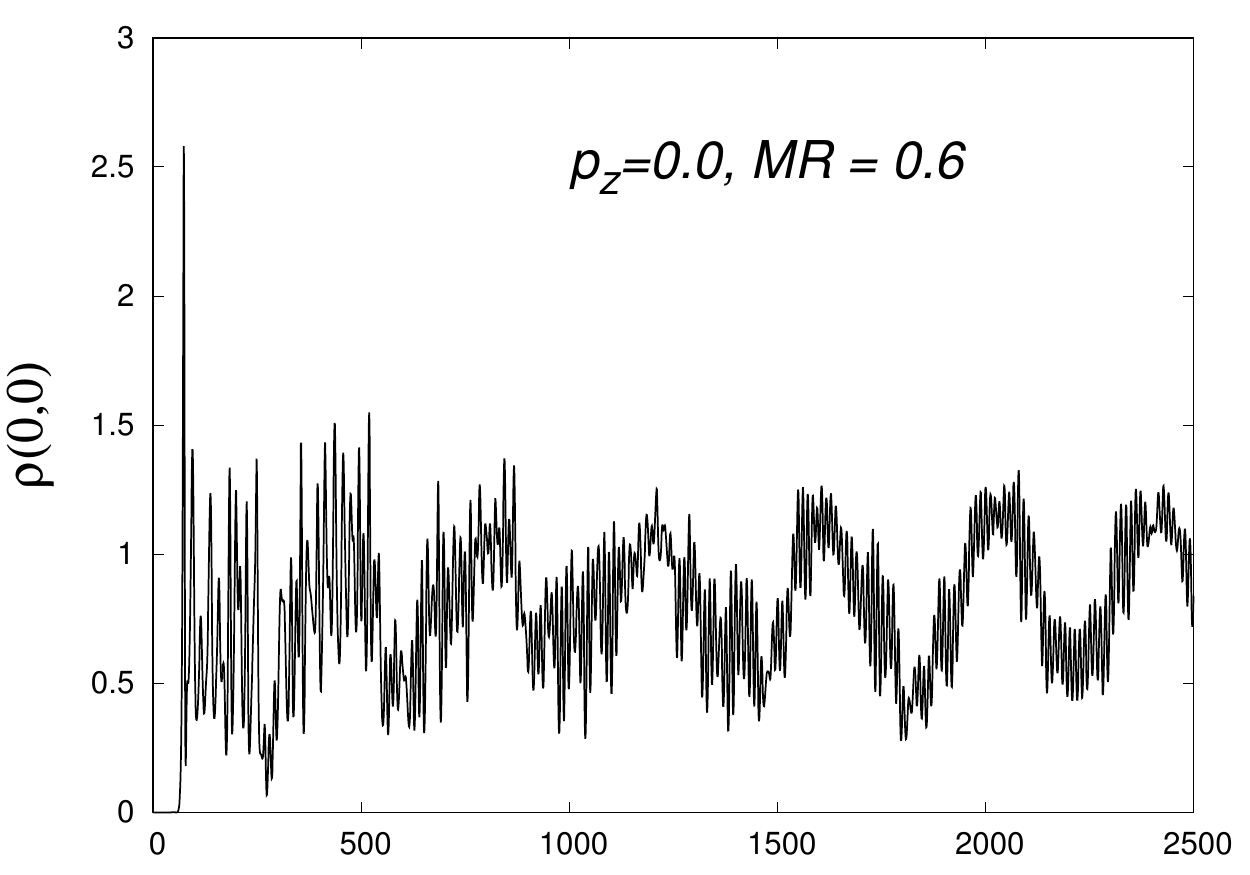}
\includegraphics[width= 3.25cm]{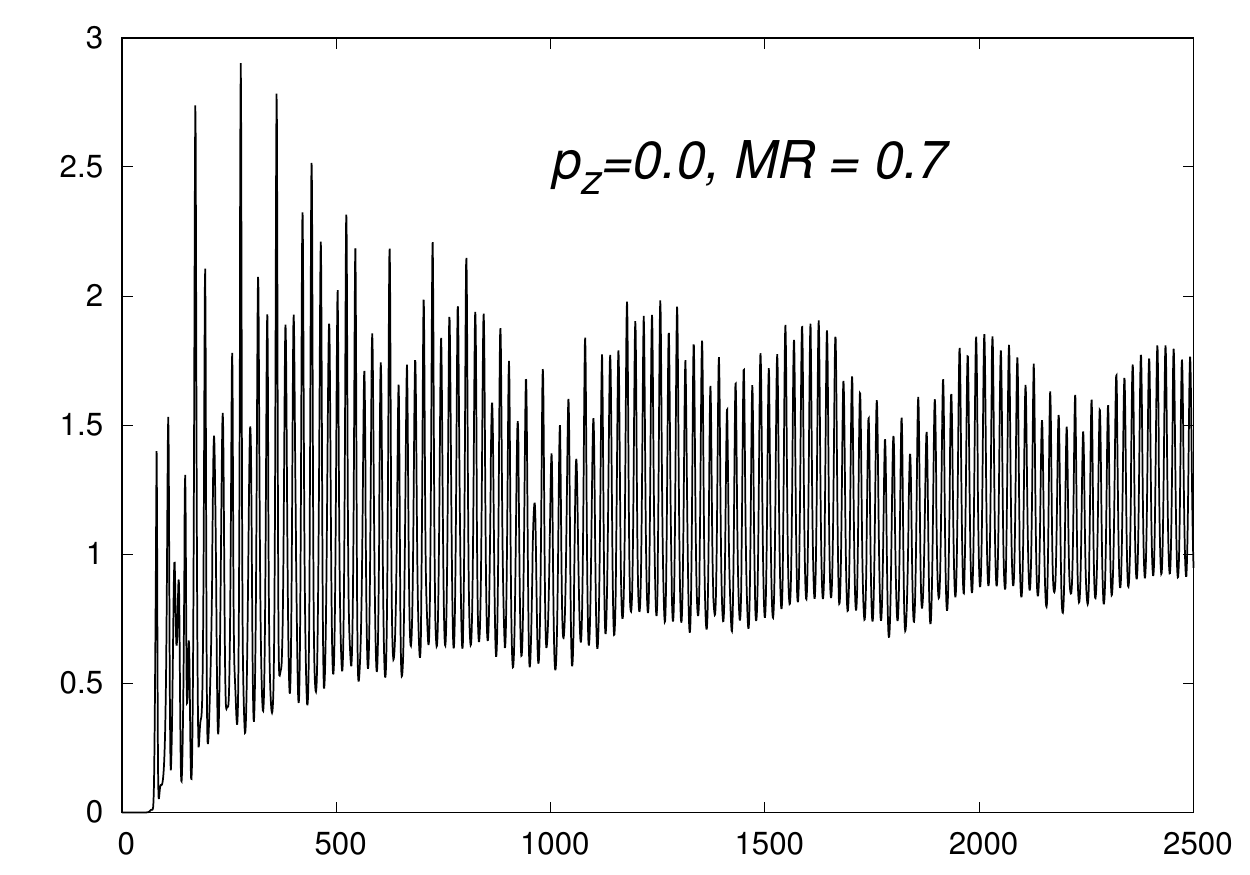}
\includegraphics[width= 3.25cm]{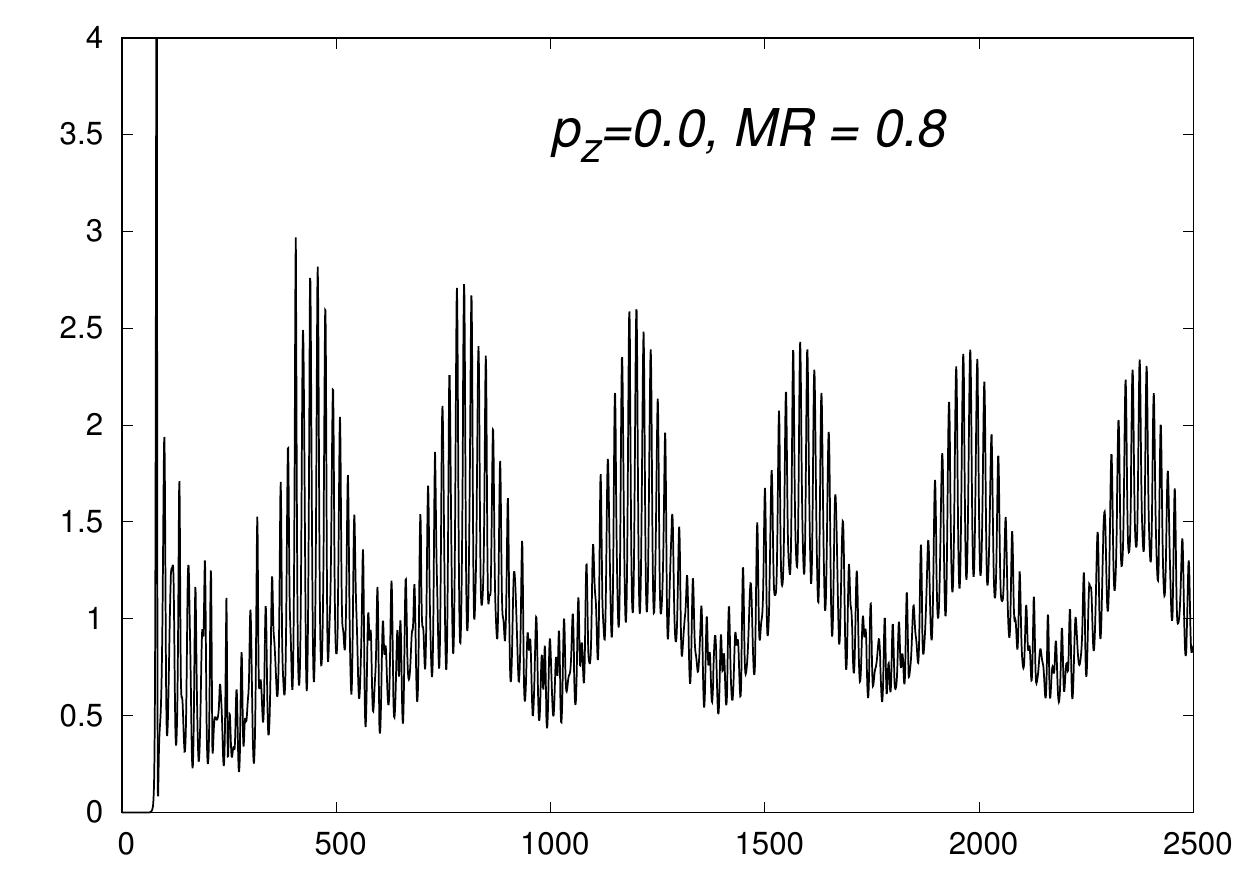}
\includegraphics[width= 3.25cm]{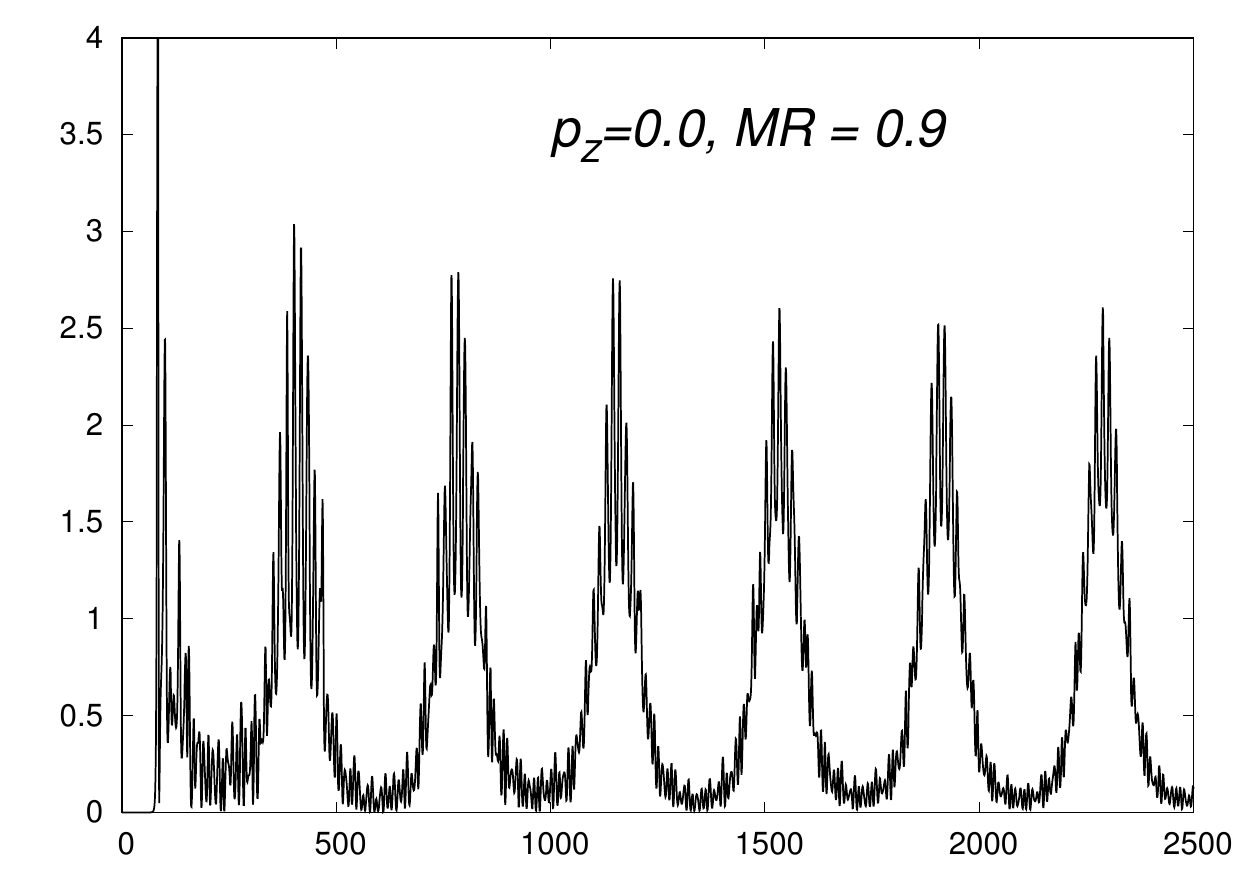}
\includegraphics[width= 3.25cm]{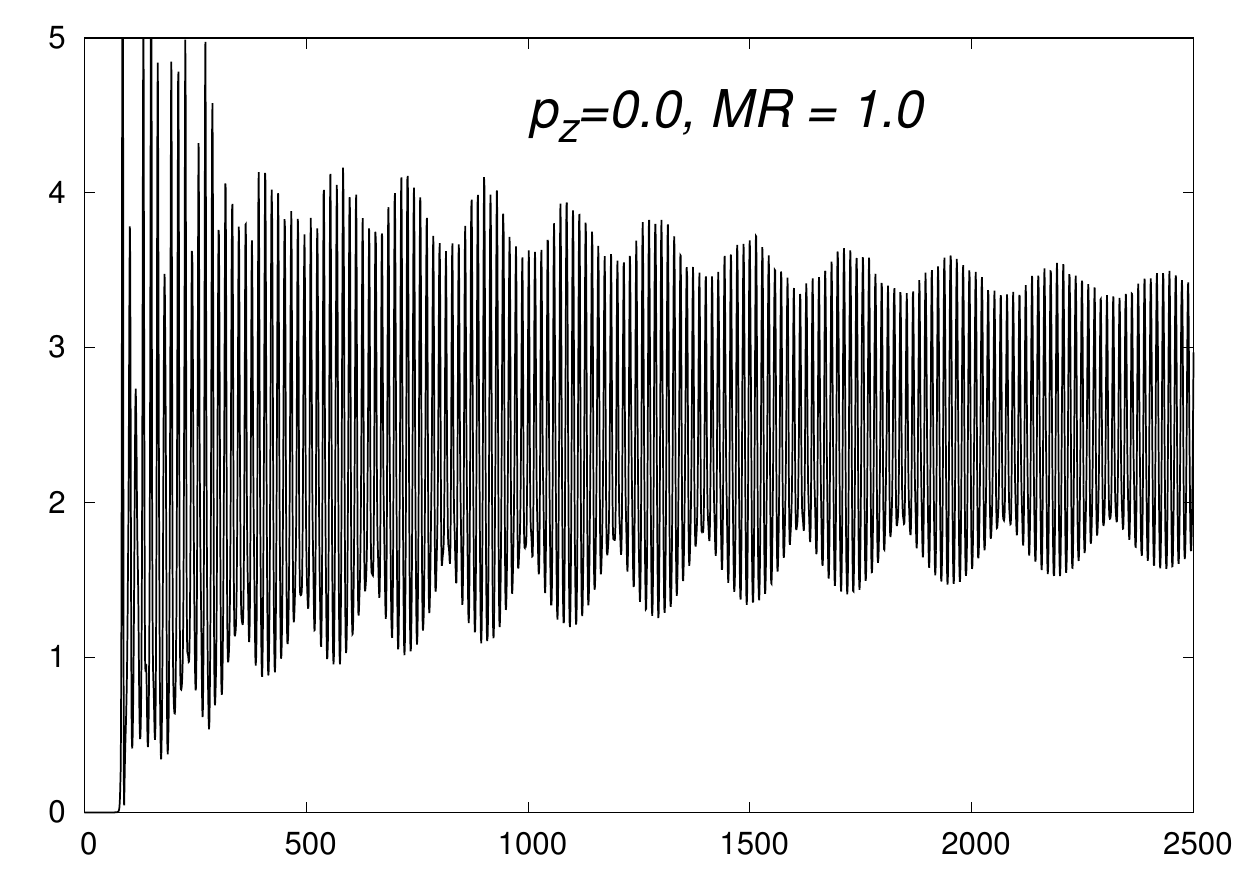}
\includegraphics[width= 3.25cm]{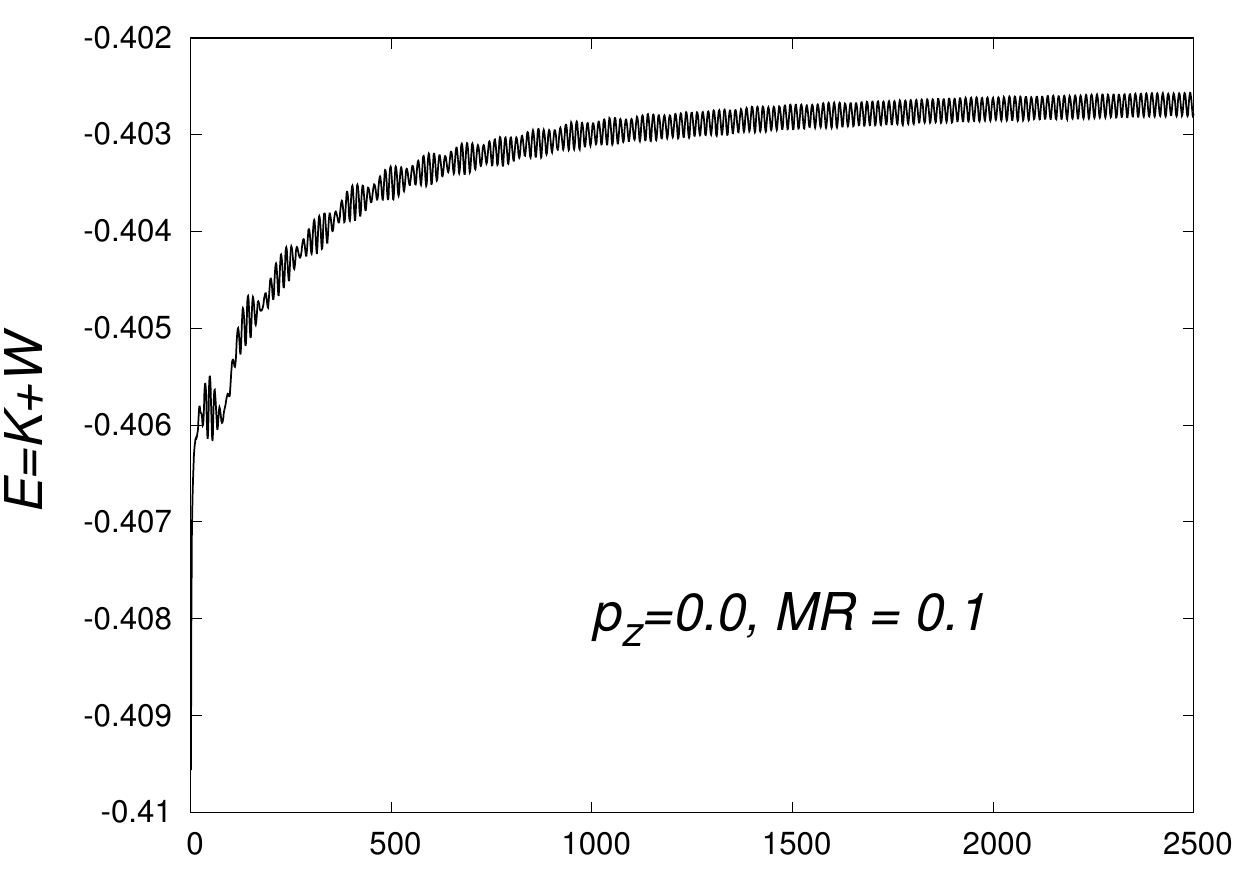}
\includegraphics[width= 3.25cm]{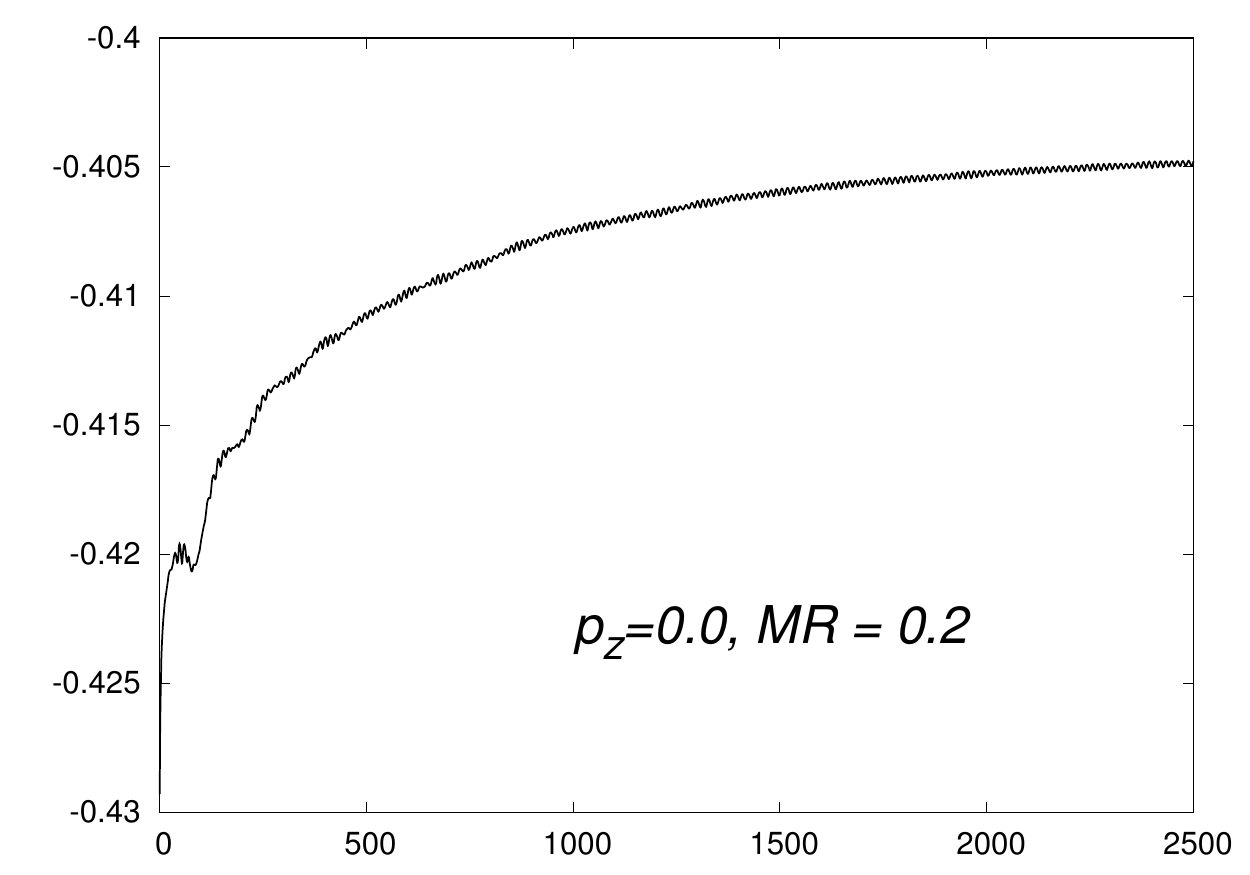}
\includegraphics[width= 3.25cm]{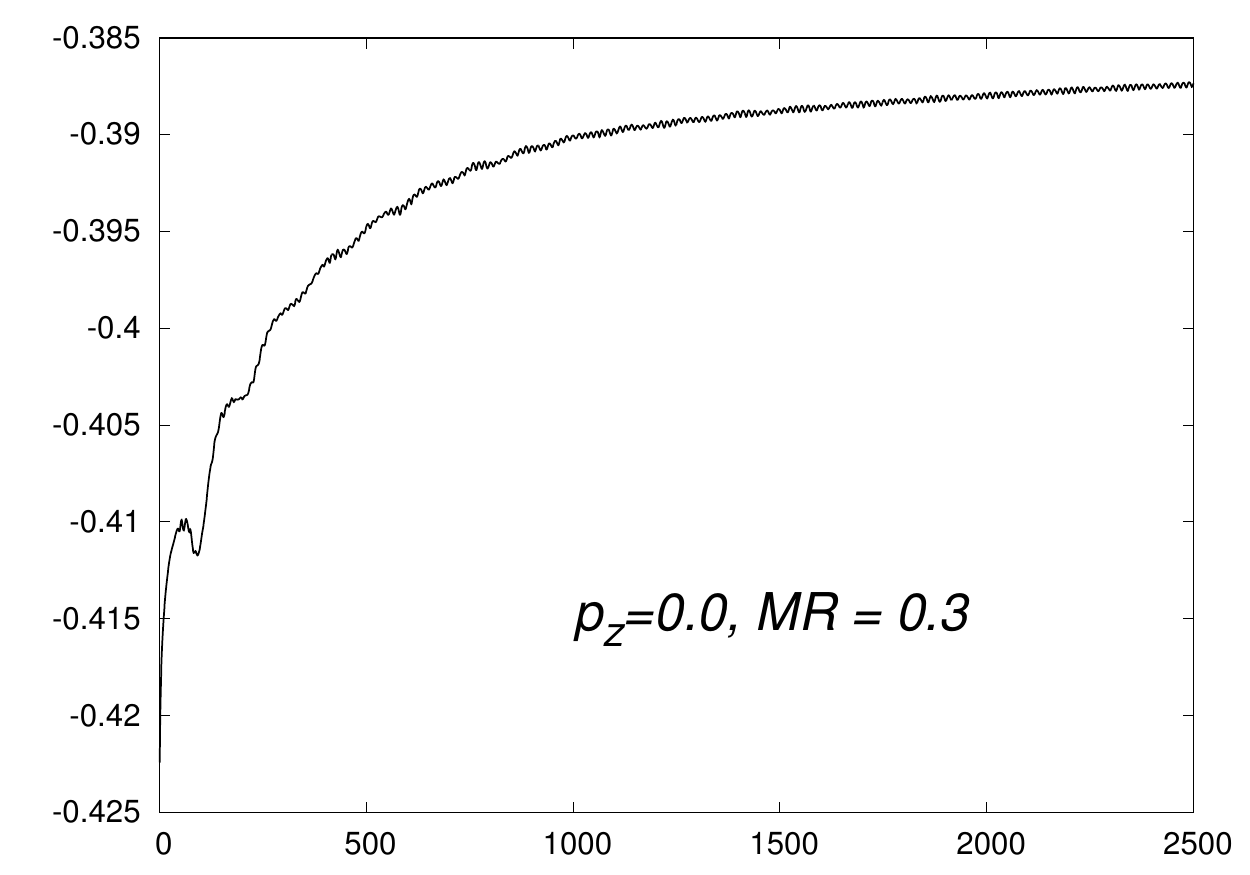}
\includegraphics[width= 3.25cm]{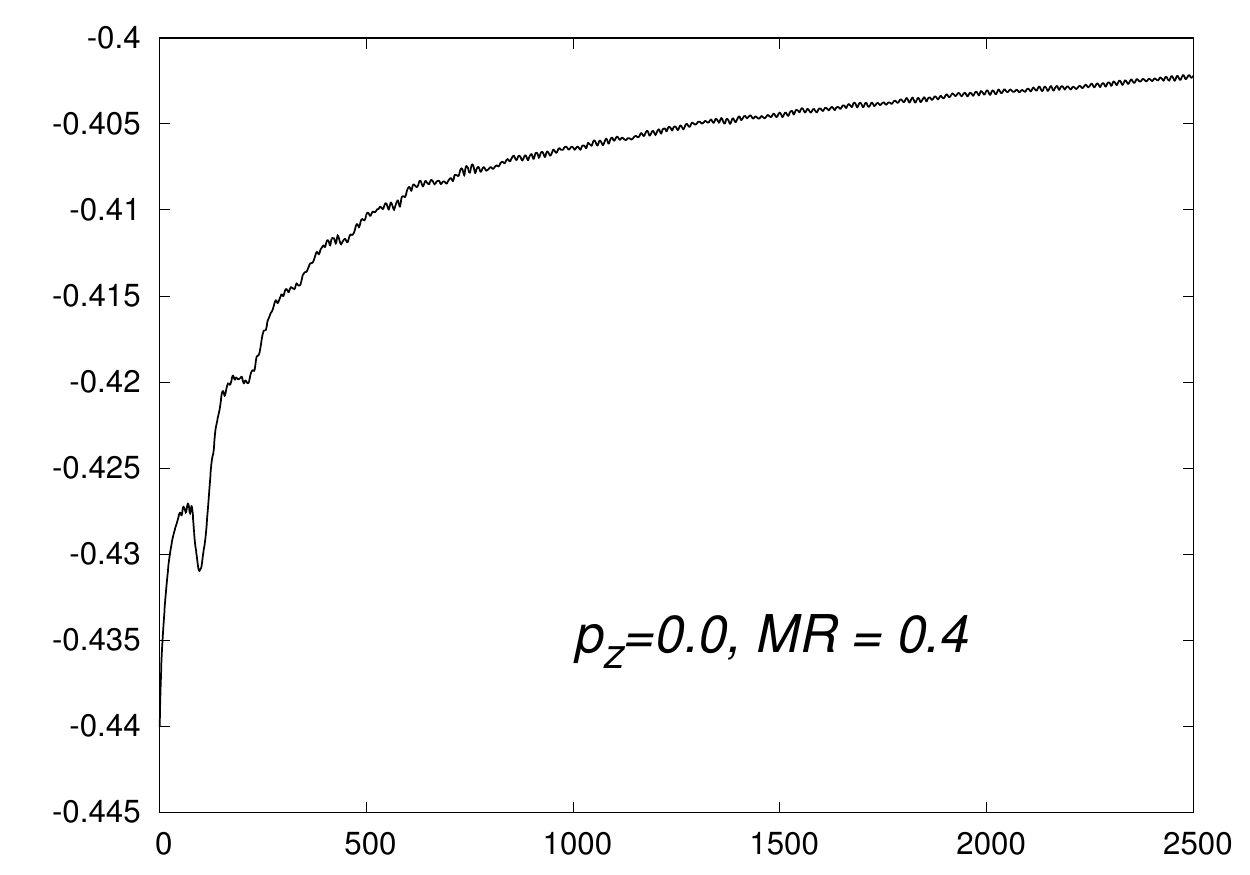}
\includegraphics[width= 3.25cm]{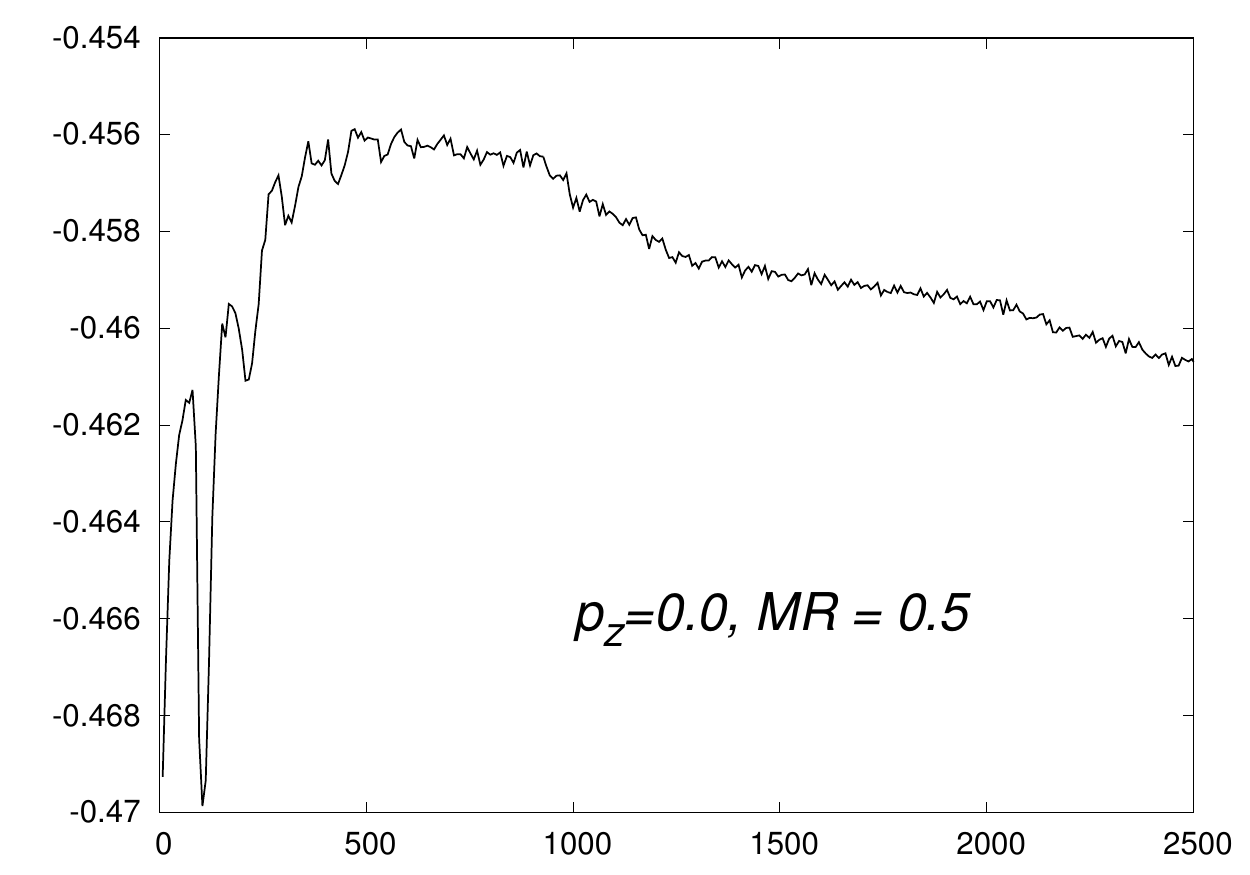}
\includegraphics[width= 3.25cm]{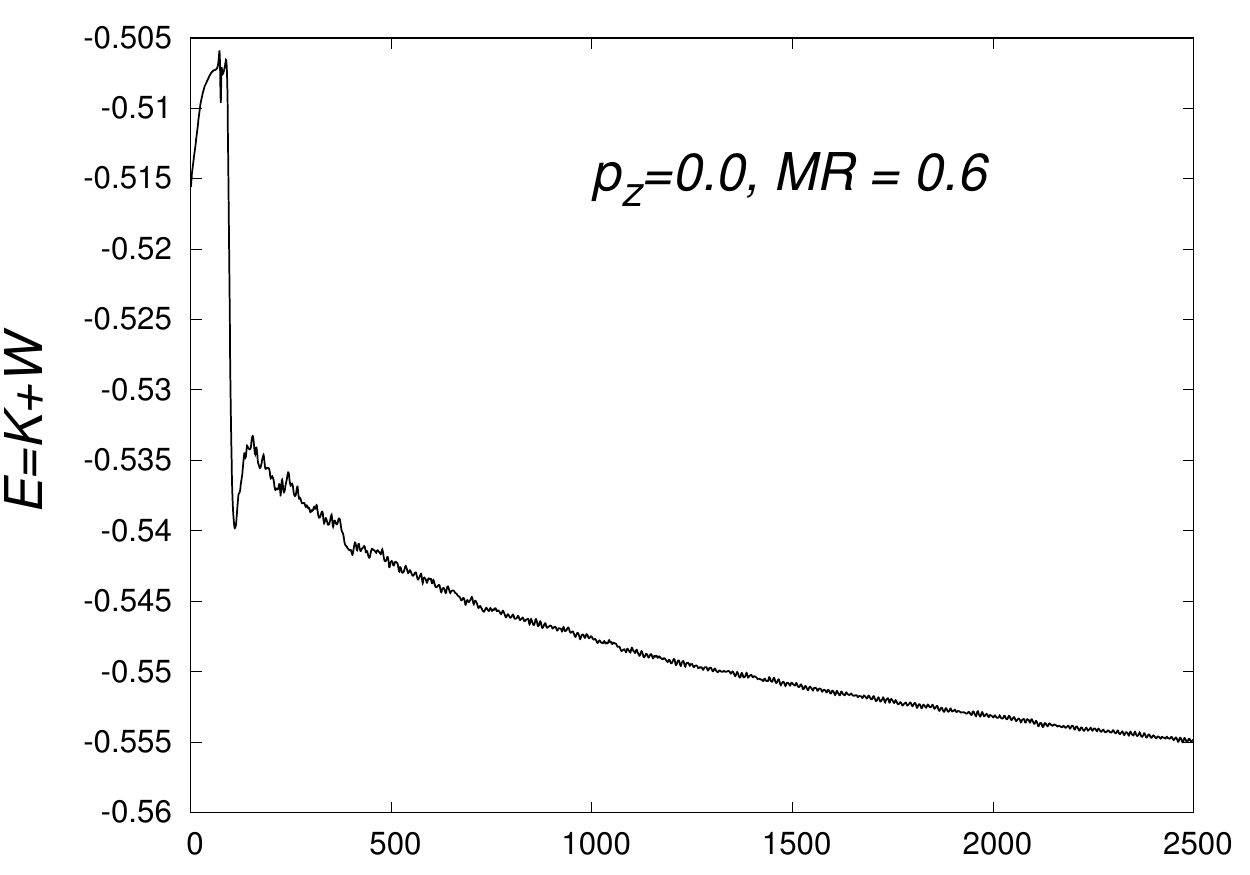}
\includegraphics[width= 3.25cm]{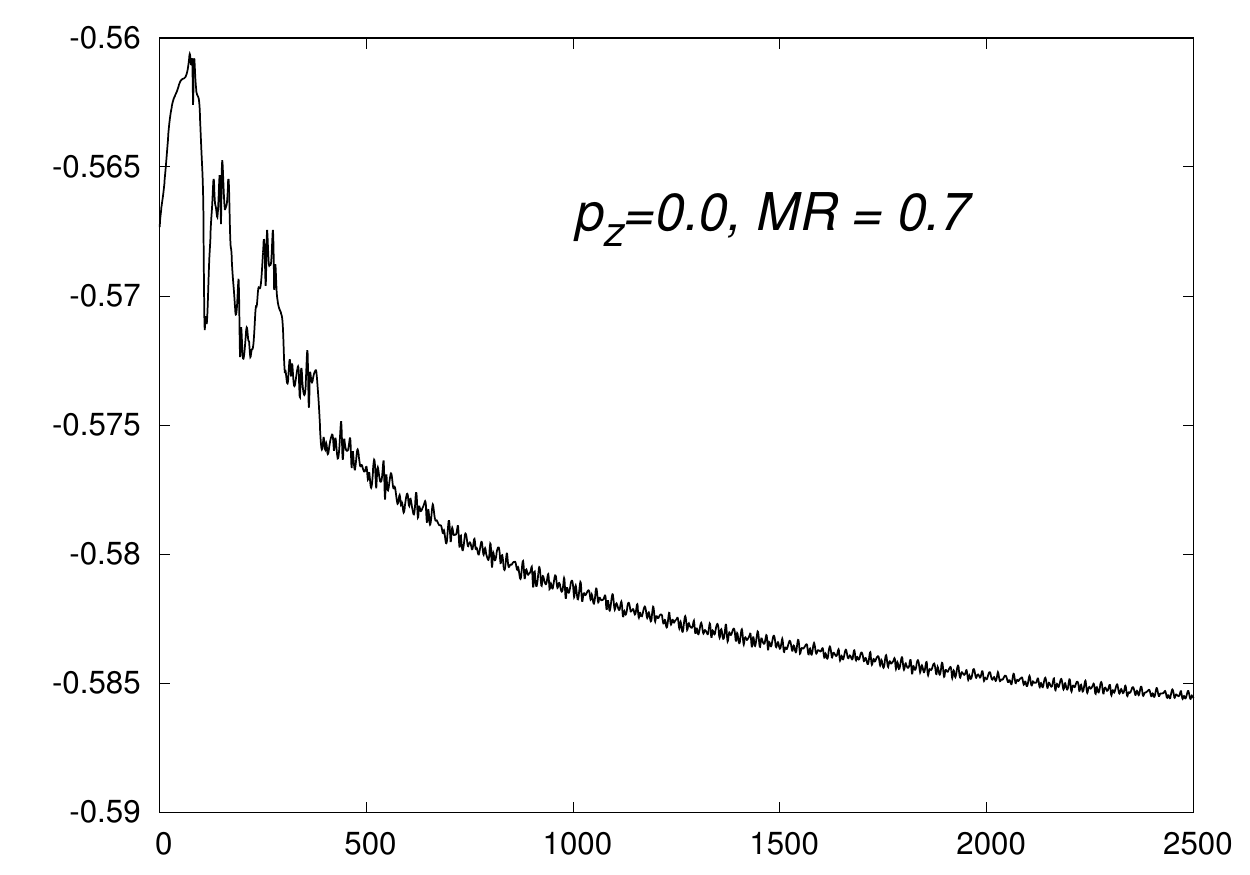}
\includegraphics[width= 3.25cm]{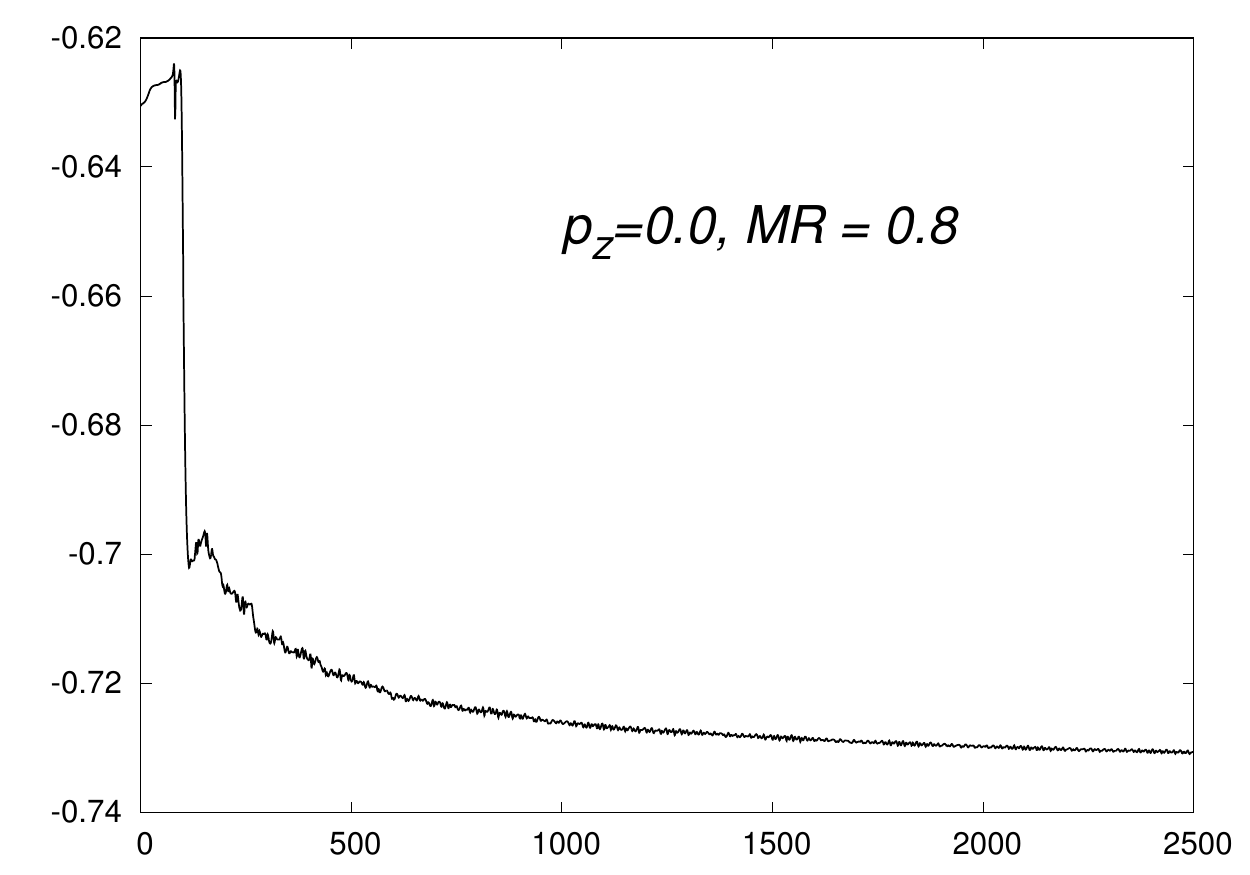}
\includegraphics[width= 3.25cm]{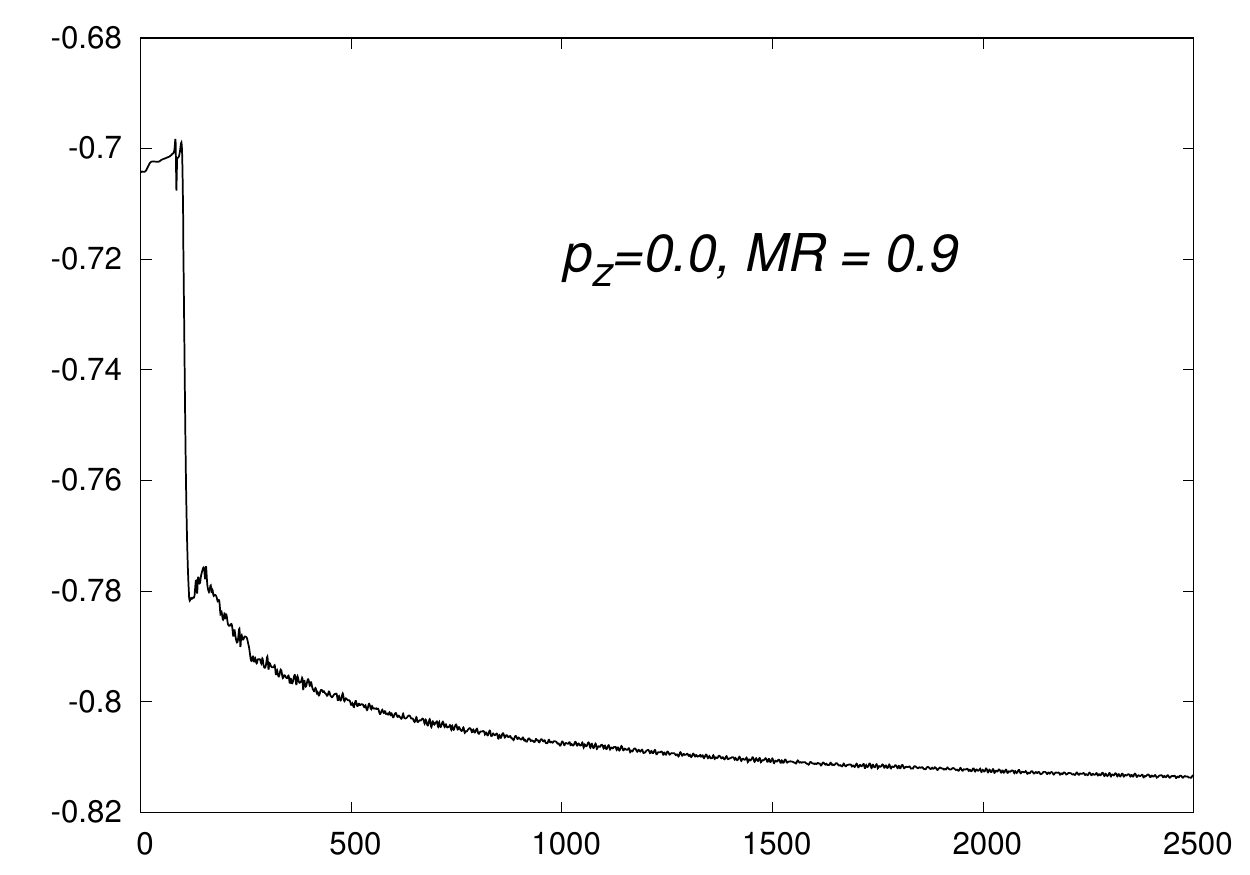}
\includegraphics[width= 3.25cm]{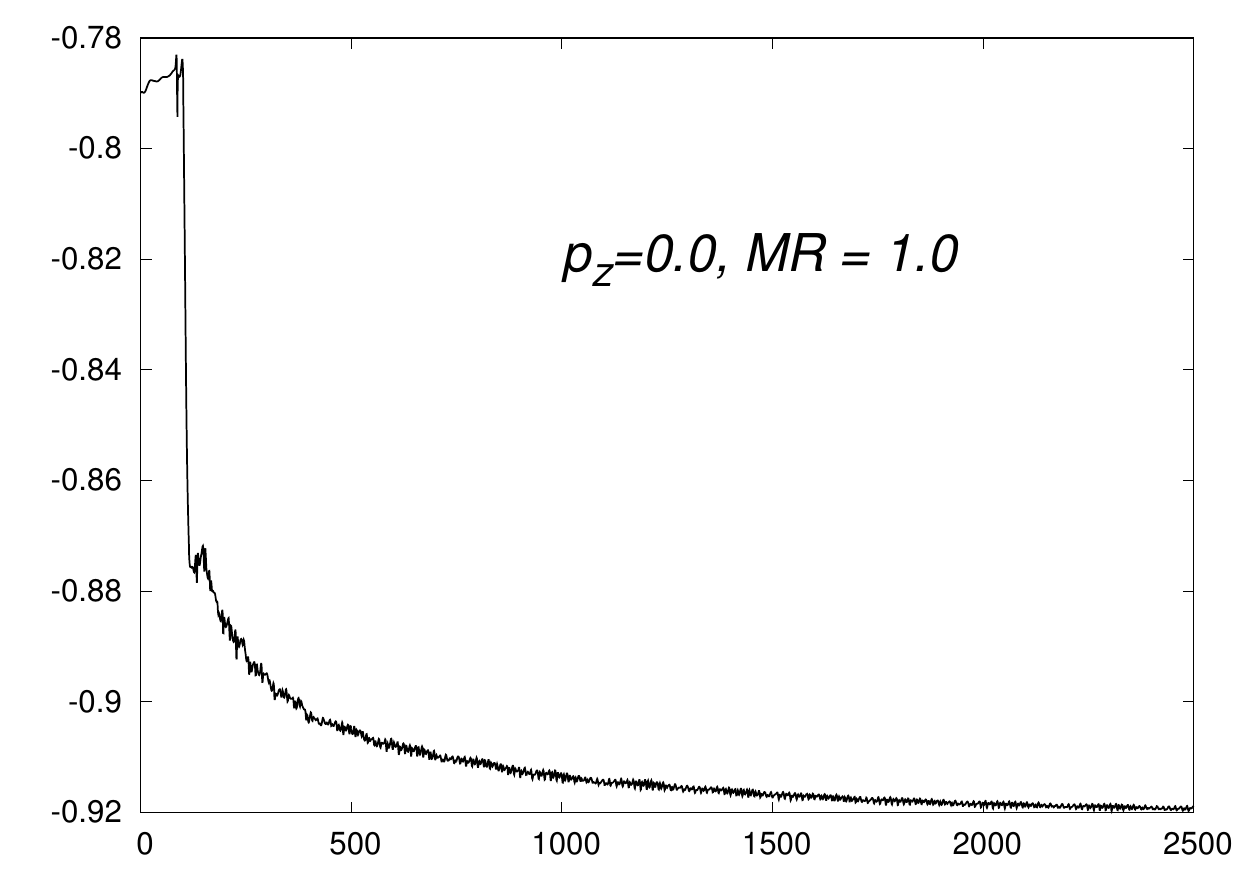}
\caption{Diagnostics of the relaxation process for the case $p_z=0.0$ and  the various mass ratios $MR=0.1,...,1.0$. We show the quantity $2K+W$, the central density where the final structure centers $\rho(0,0)$ and the total energy $E=K+W$ as function of time $t$.}
\label{fig:coolingpz0_0}
\end{figure*}

\begin{figure*}
\centering
\includegraphics[width= 3.25cm]{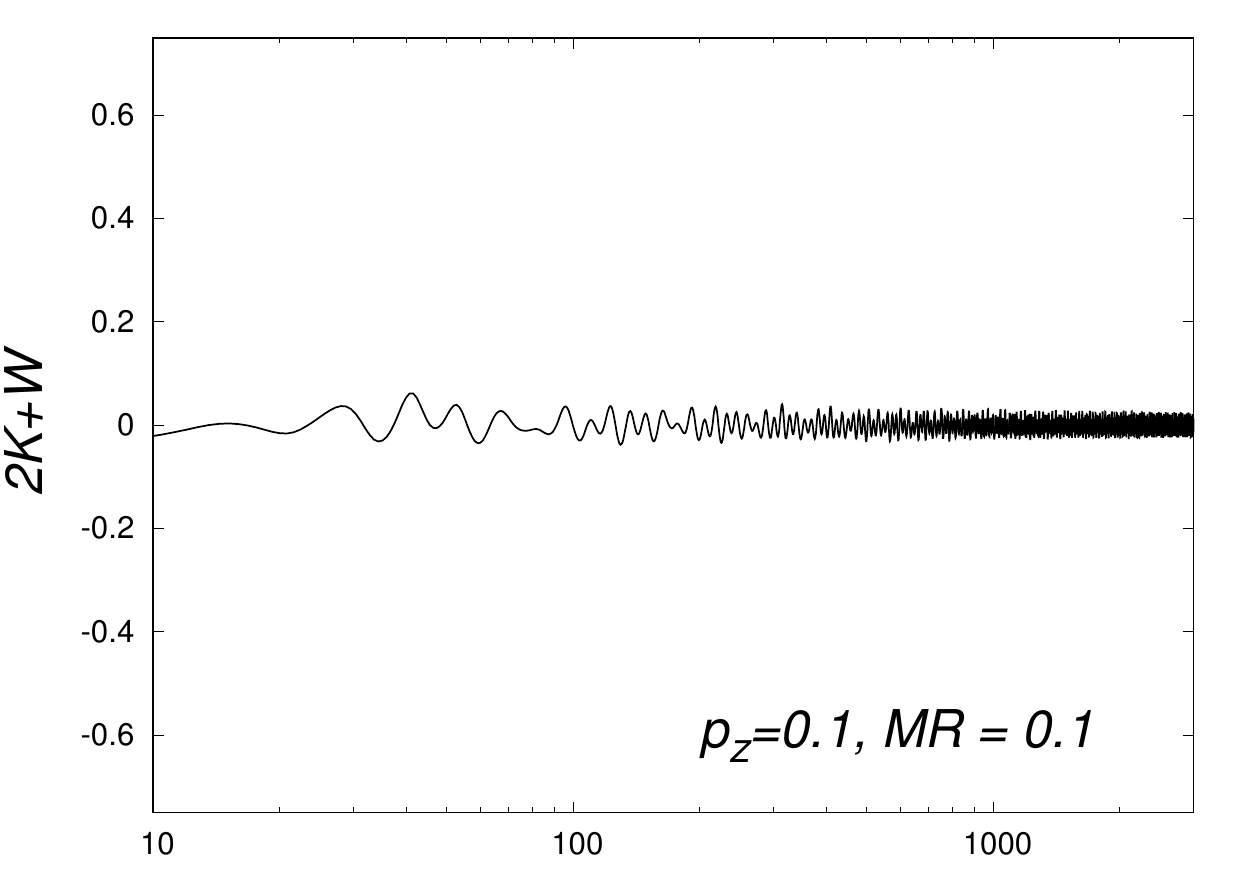}
\includegraphics[width= 3.25cm]{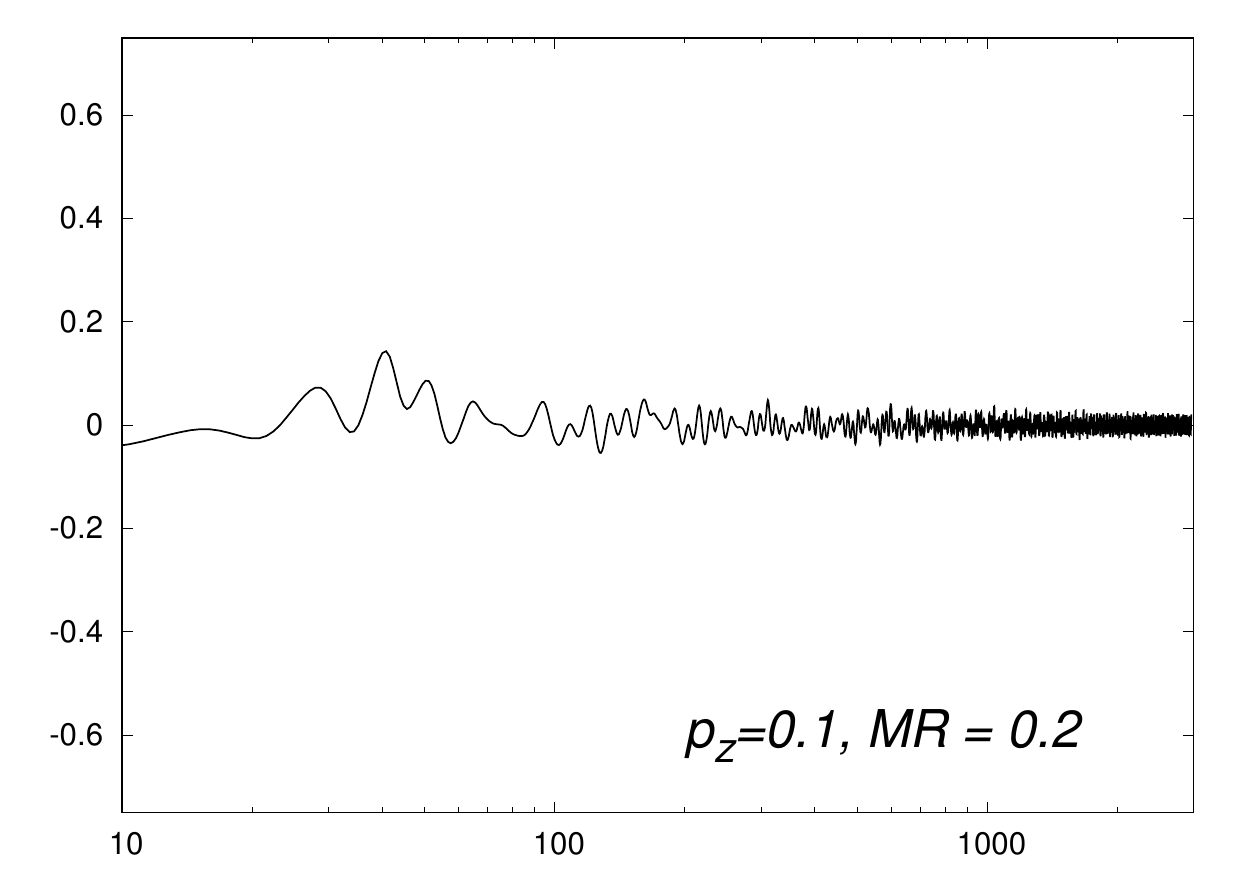}
\includegraphics[width= 3.25cm]{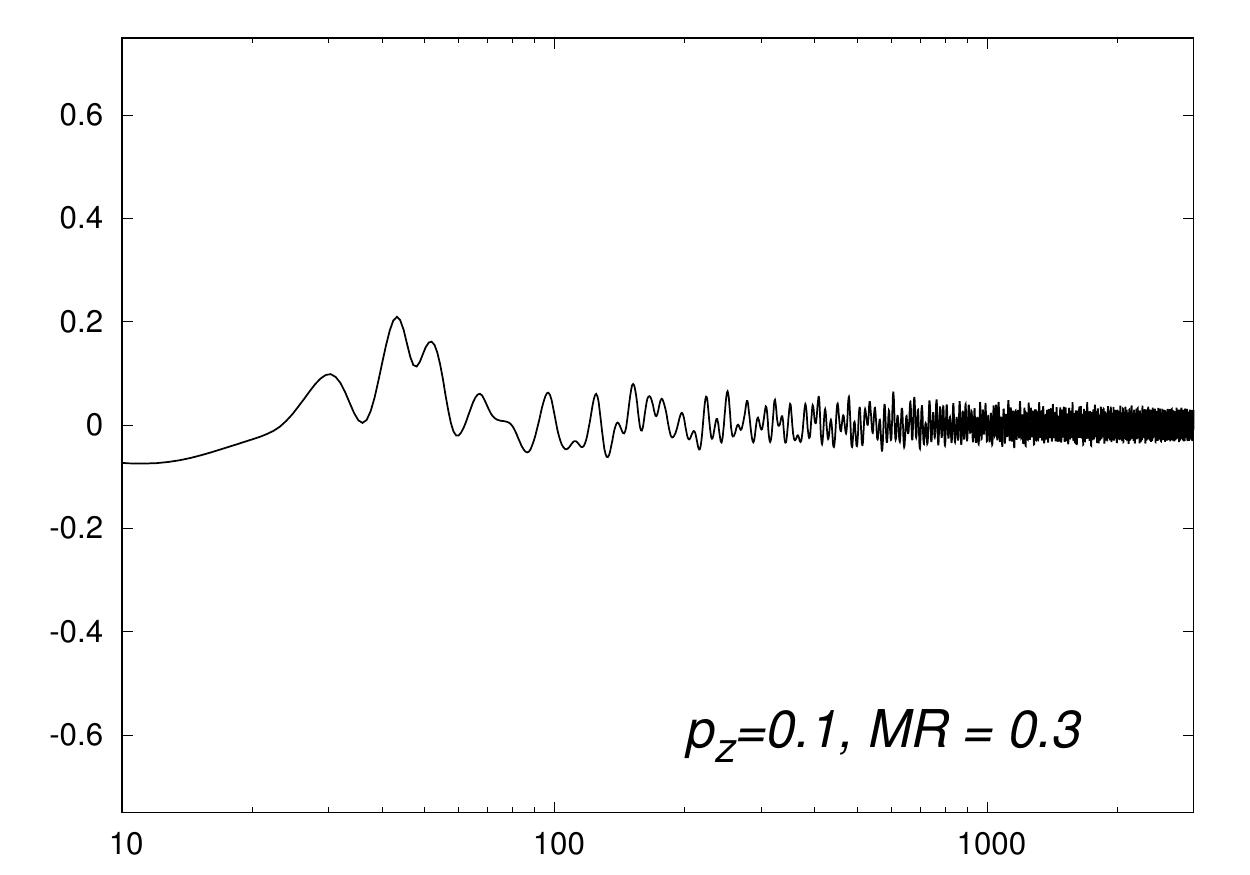}
\includegraphics[width= 3.25cm]{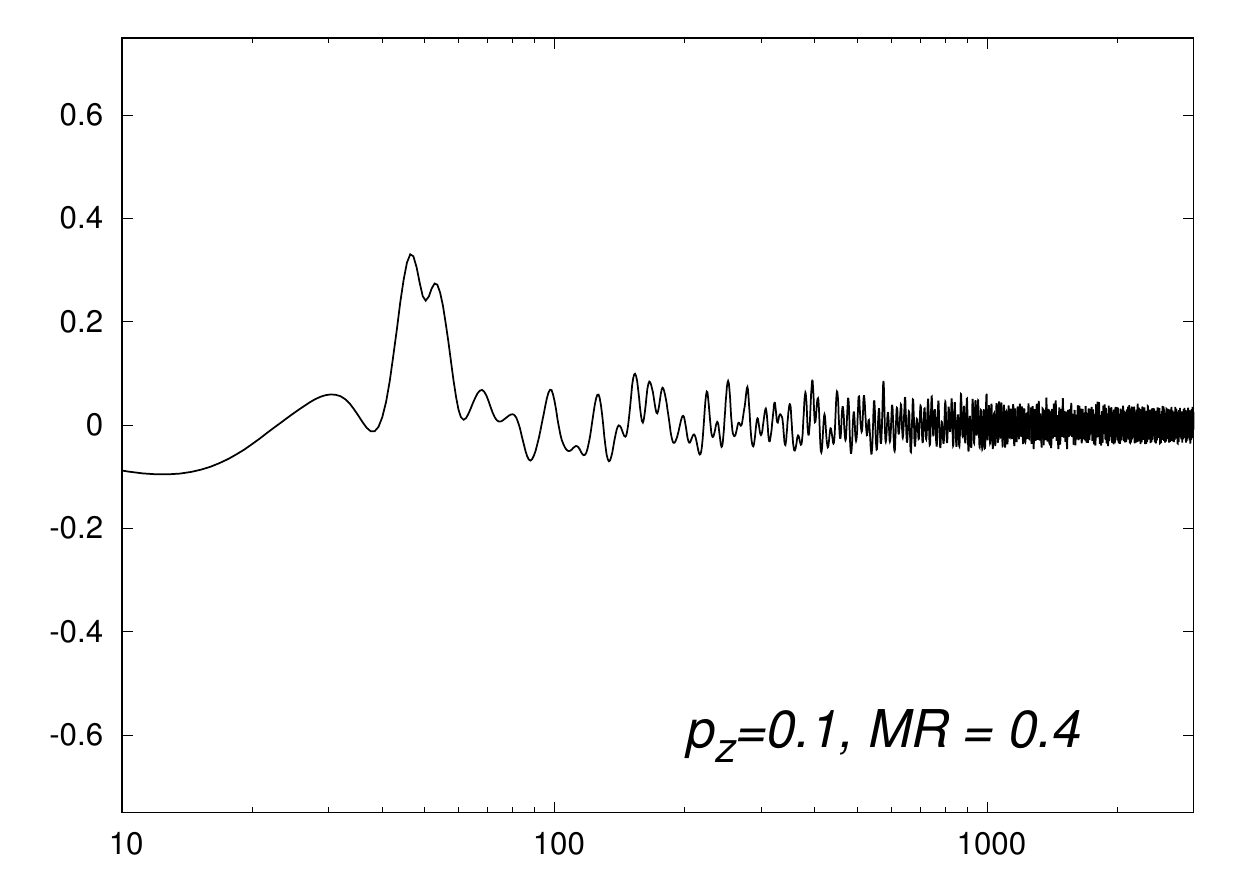}
\includegraphics[width= 3.25cm]{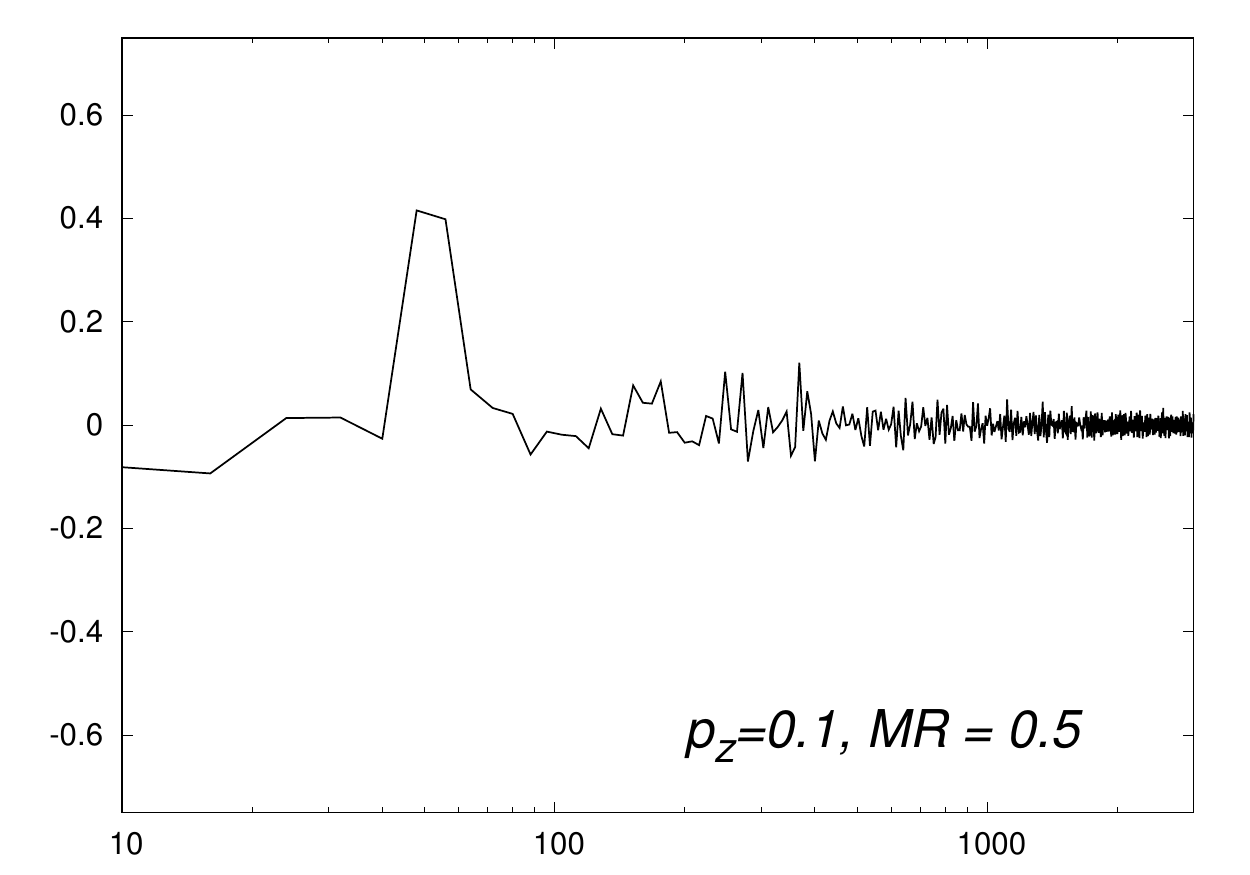}
\includegraphics[width= 3.25cm]{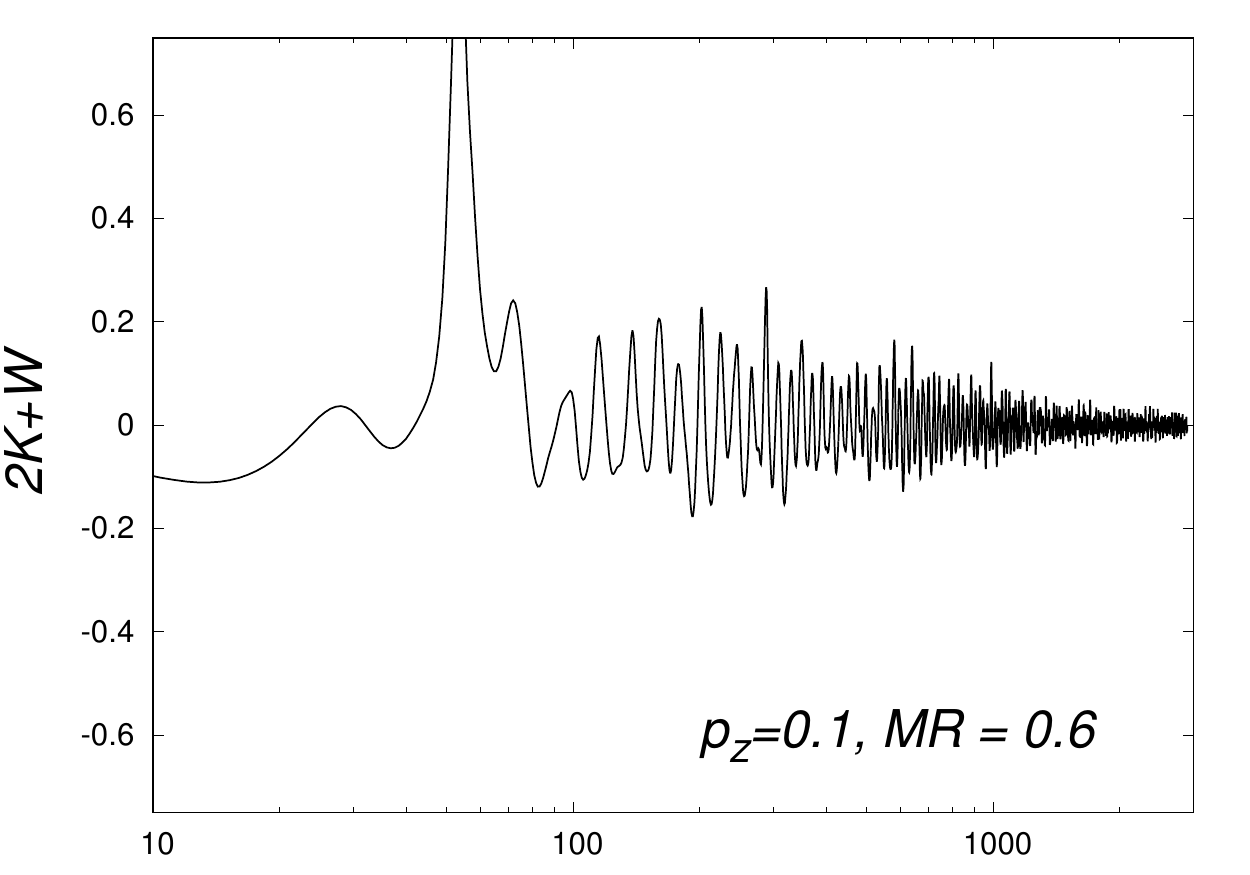}
\includegraphics[width= 3.25cm]{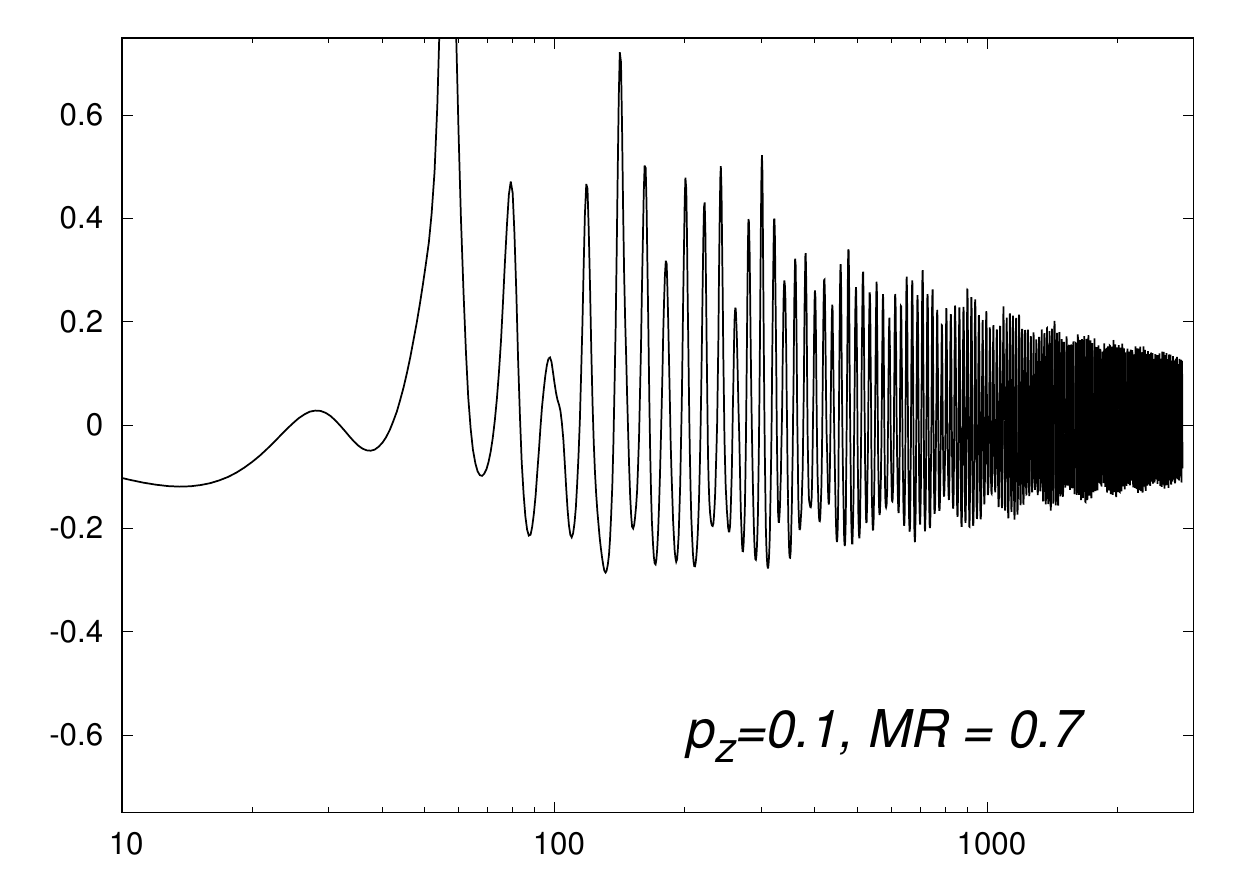}
\includegraphics[width= 3.25cm]{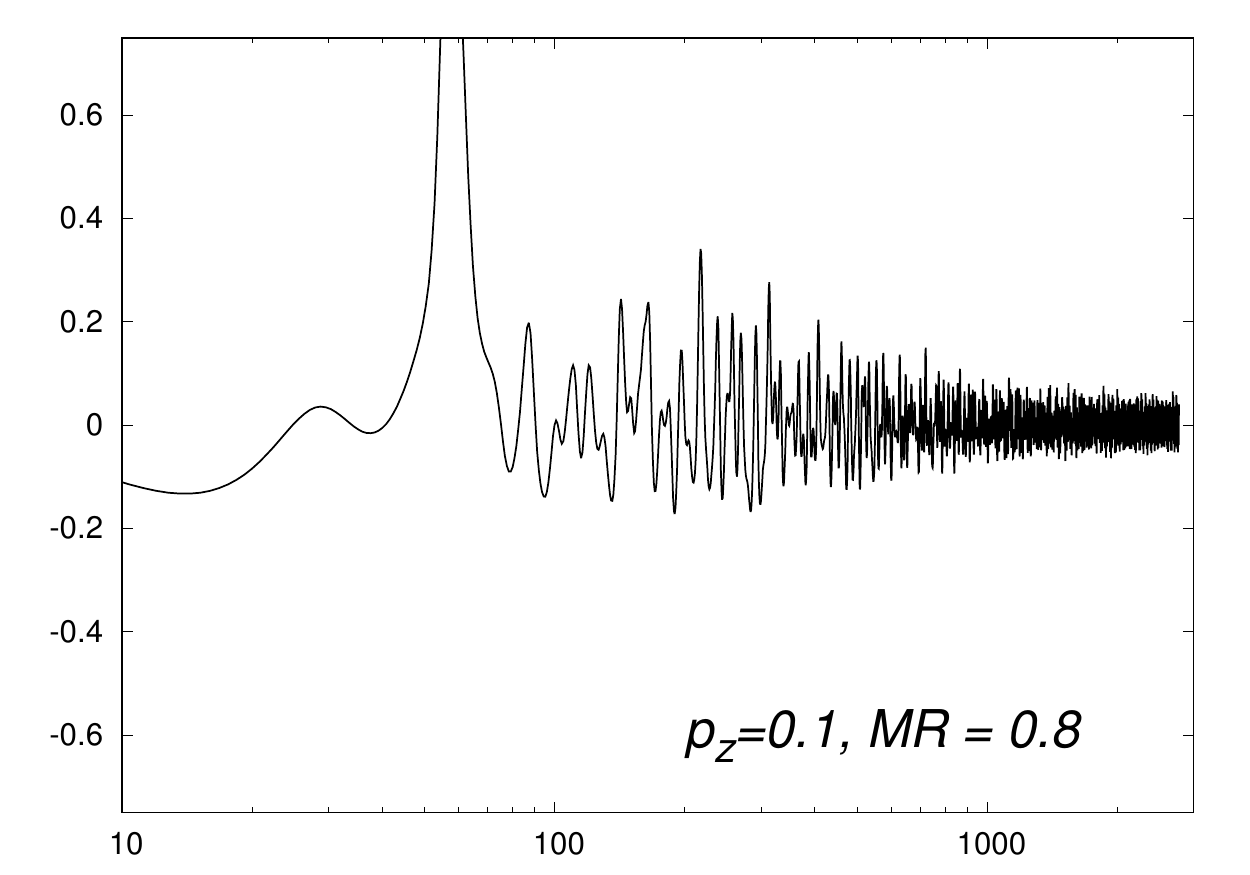}
\includegraphics[width= 3.25cm]{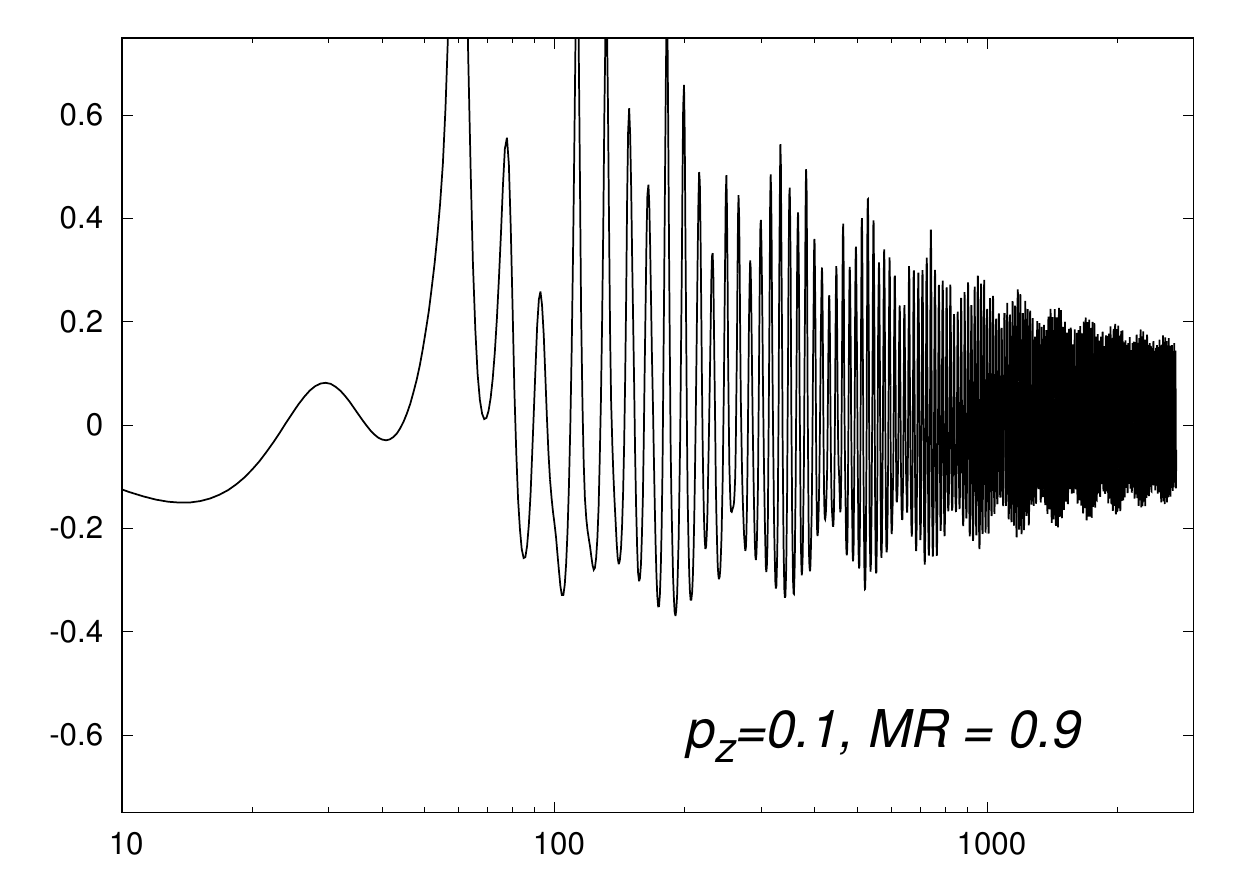}
\includegraphics[width= 3.25cm]{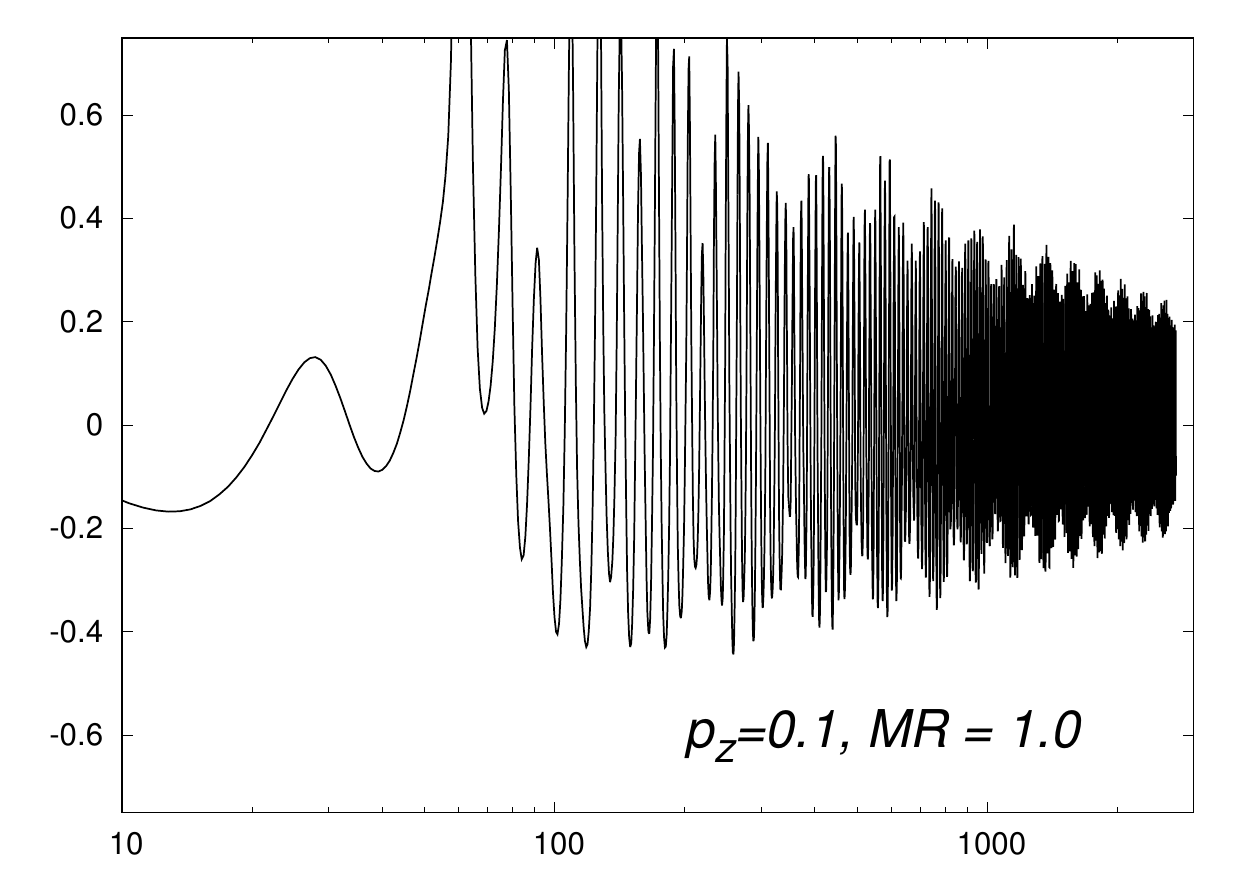}
\includegraphics[width= 3.25cm]{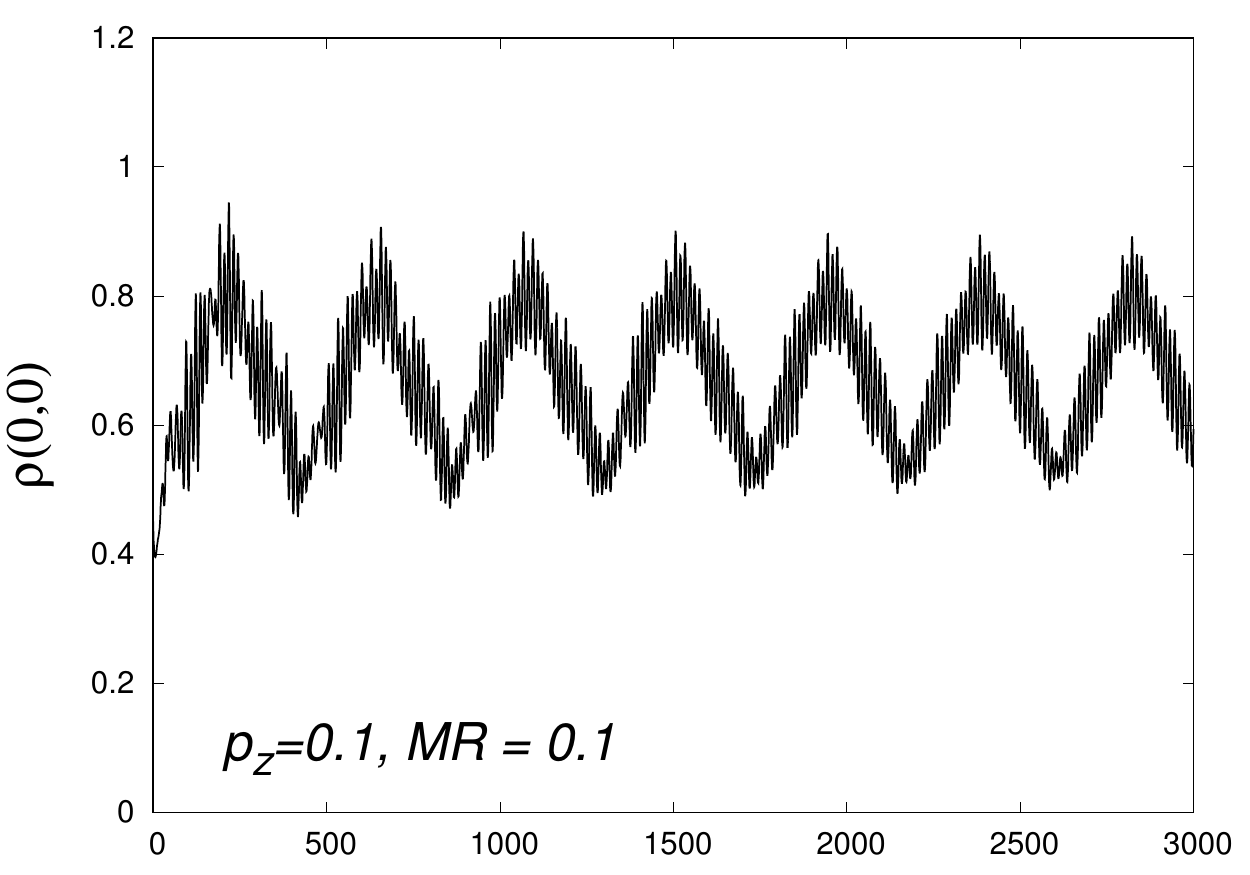}
\includegraphics[width= 3.25cm]{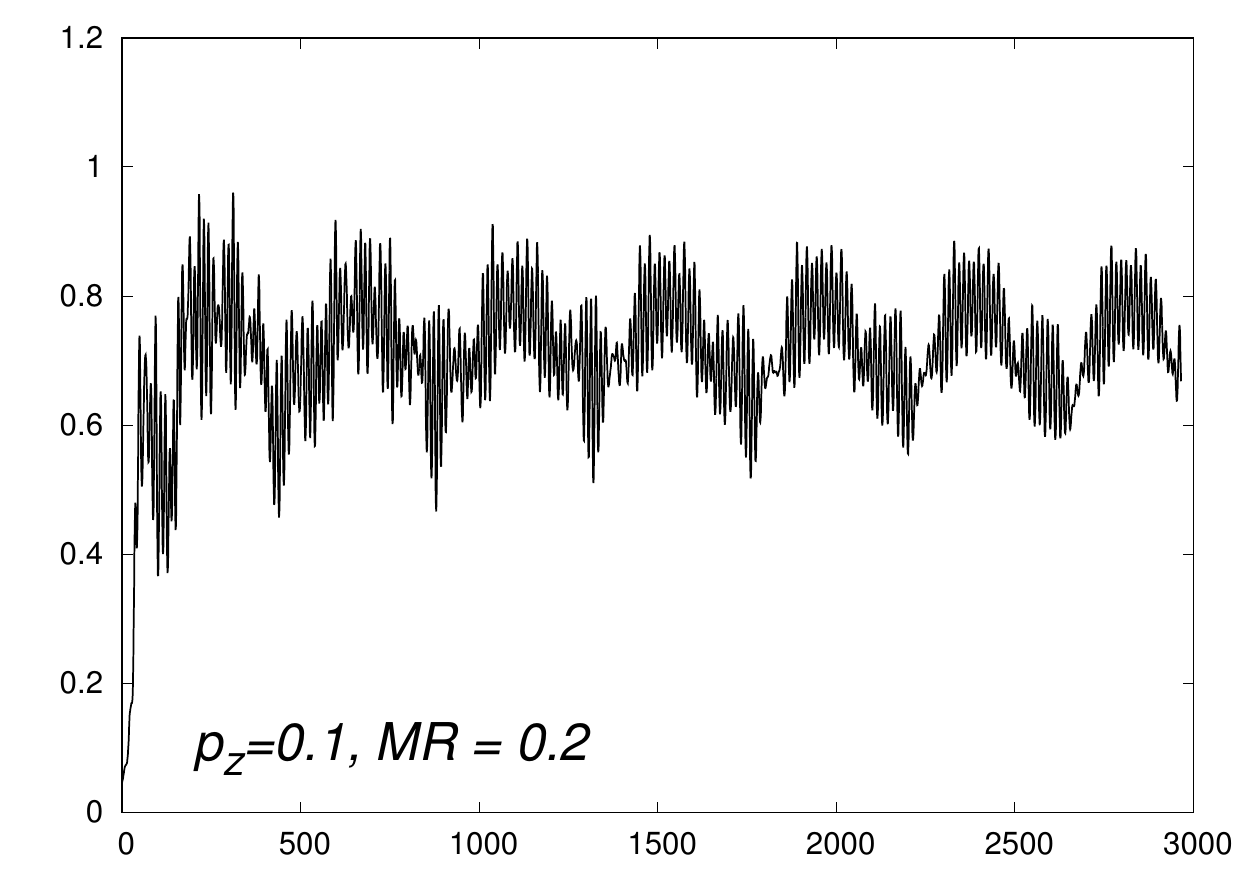}
\includegraphics[width= 3.25cm]{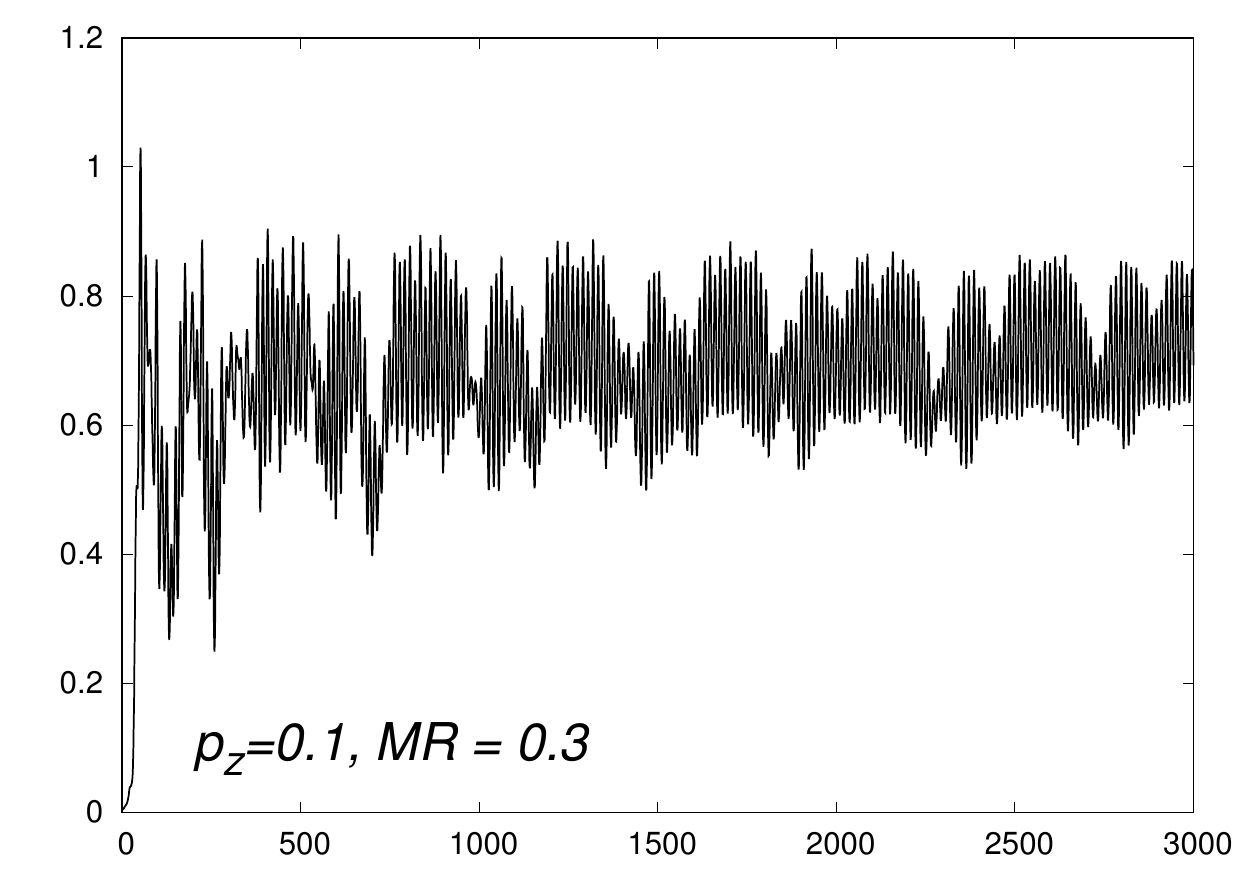}
\includegraphics[width= 3.25cm]{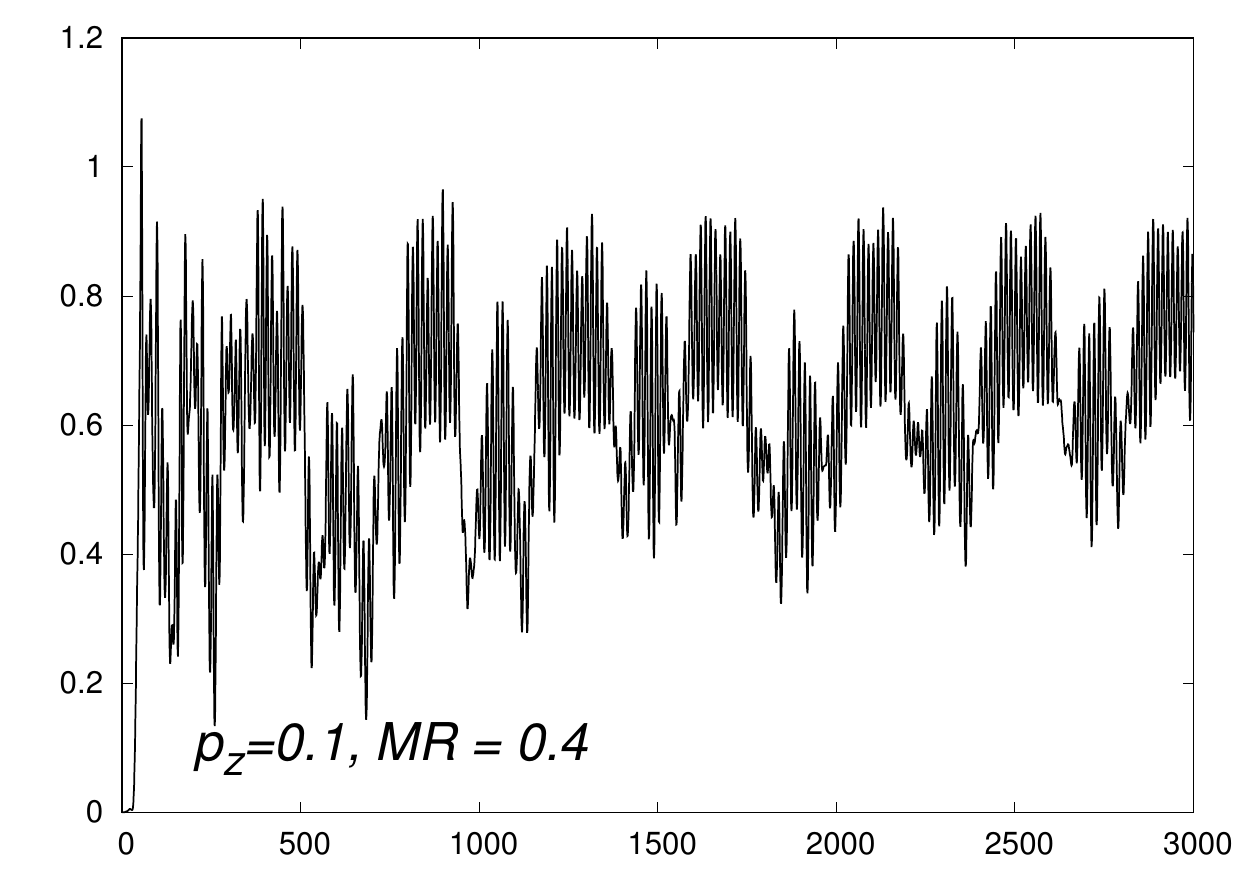}
\includegraphics[width= 3.25cm]{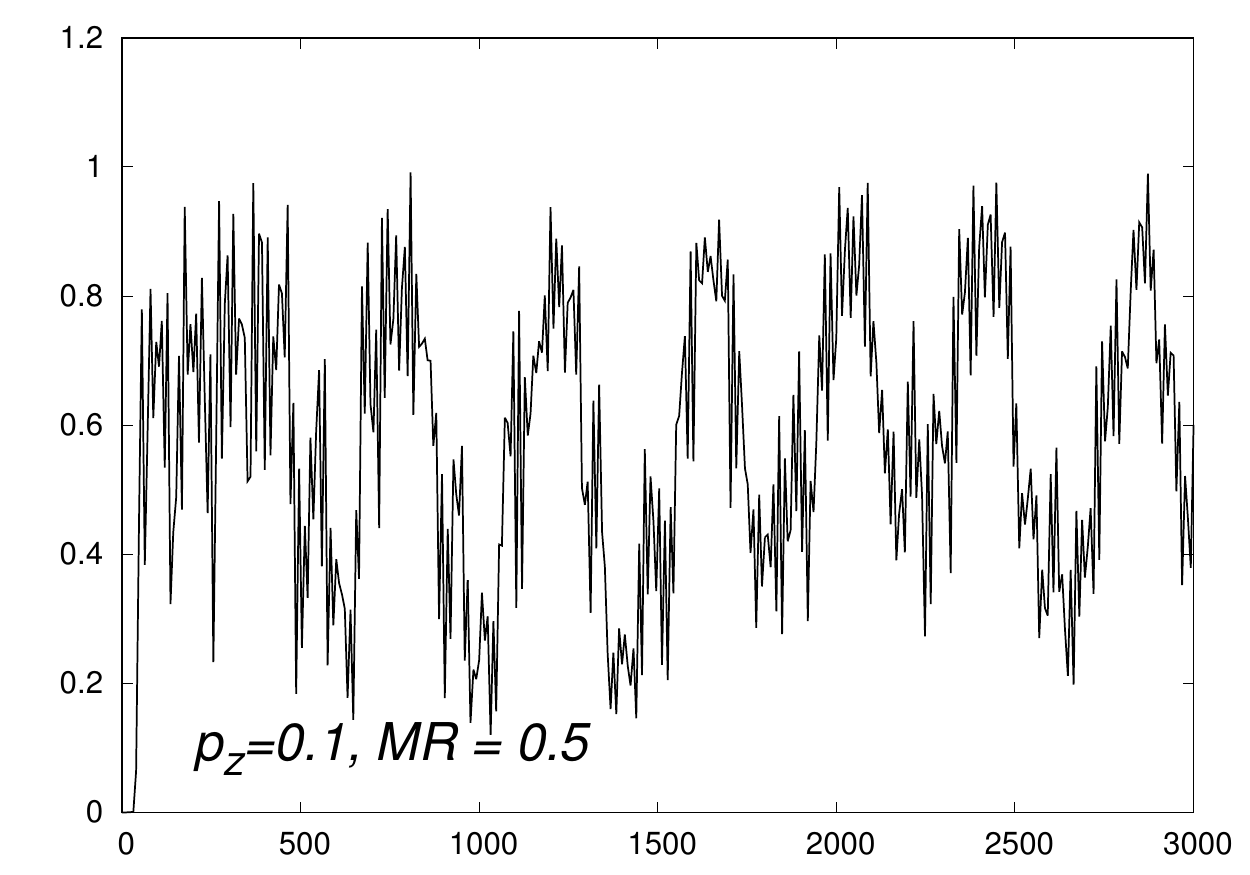}
\includegraphics[width= 3.25cm]{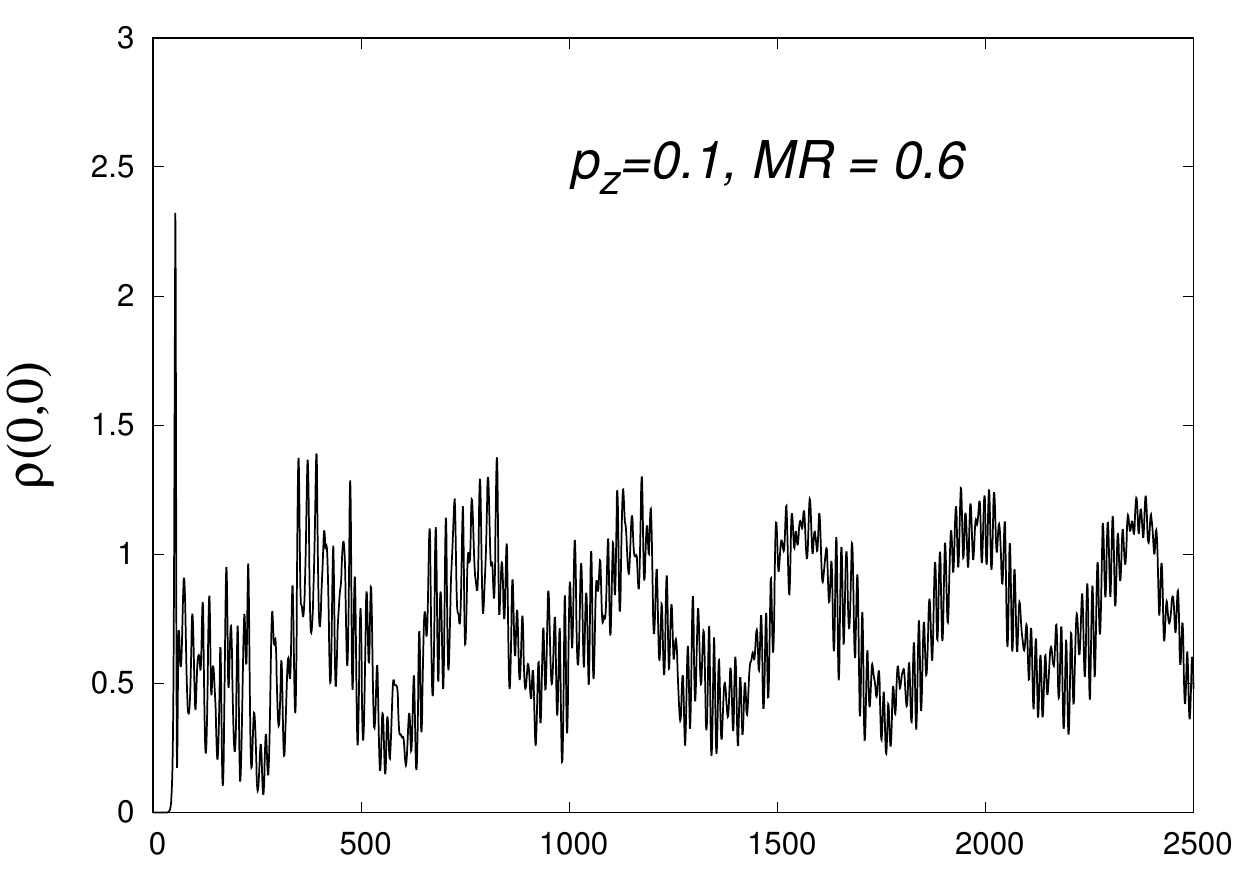}
\includegraphics[width= 3.25cm]{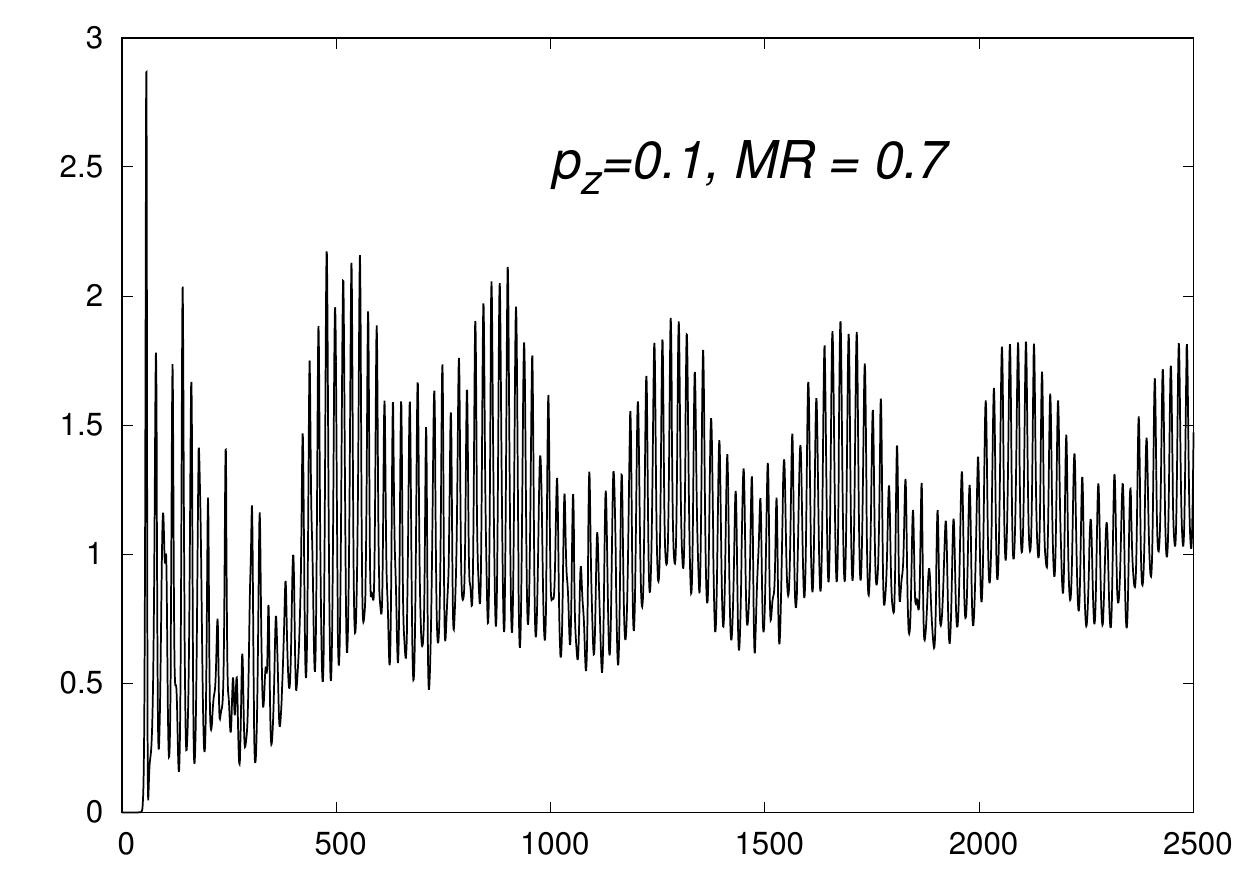}
\includegraphics[width= 3.25cm]{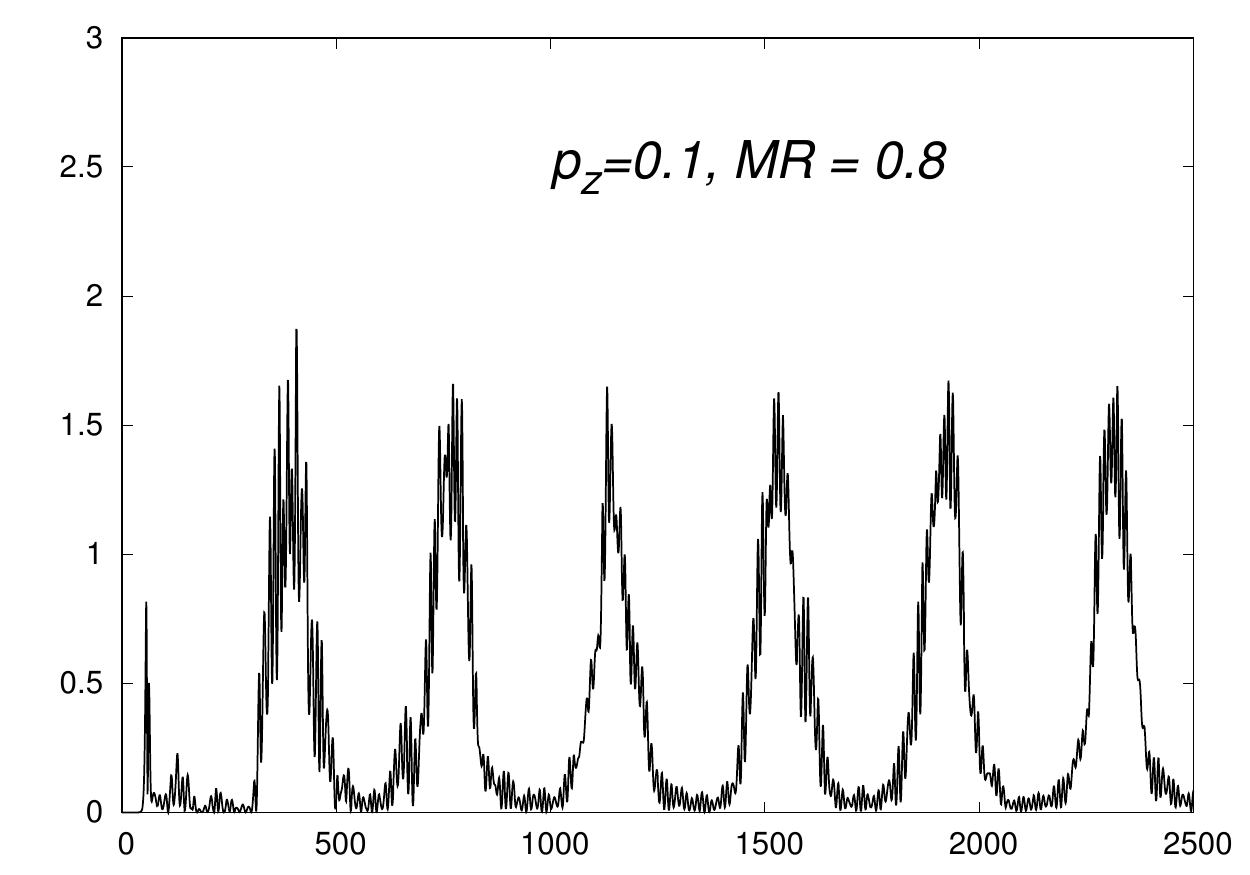}
\includegraphics[width= 3.25cm]{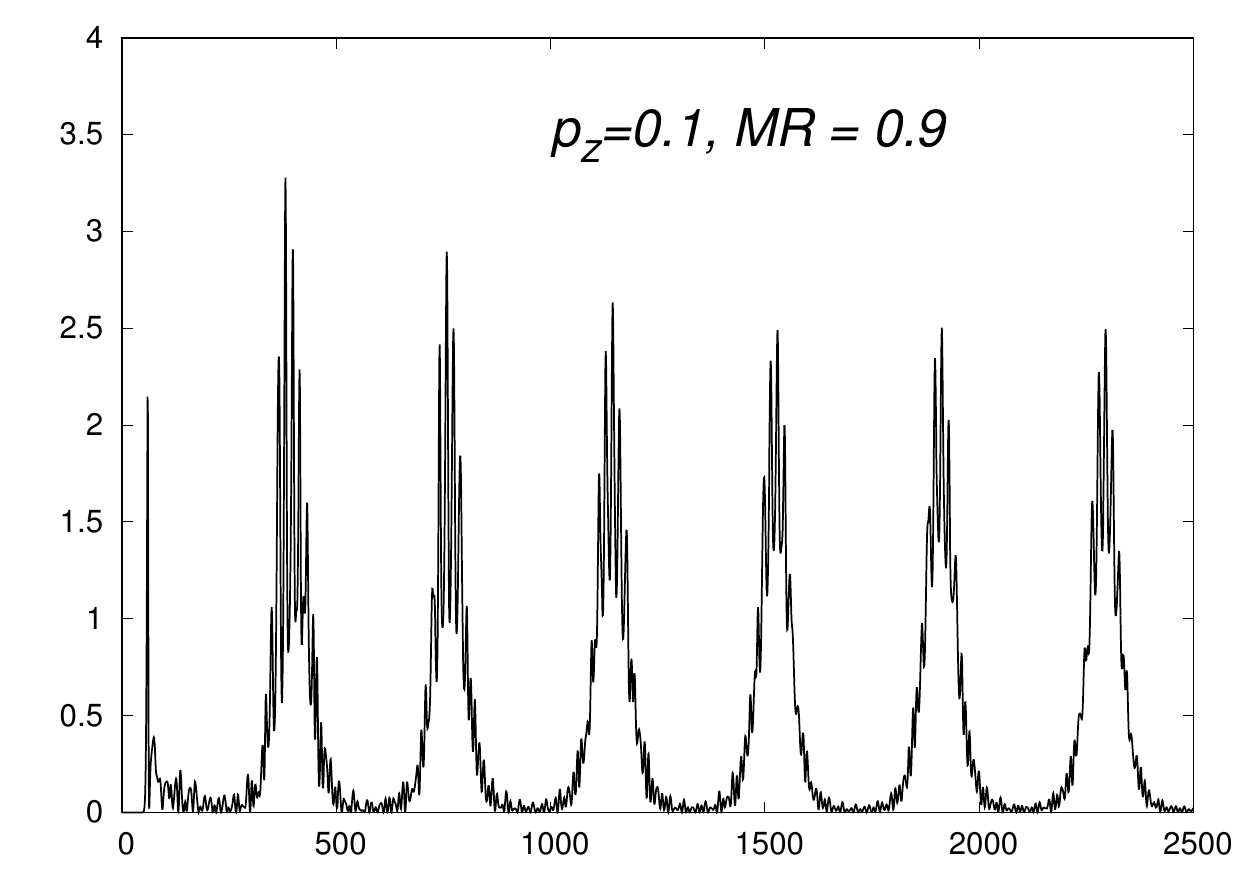}
\includegraphics[width= 3.25cm]{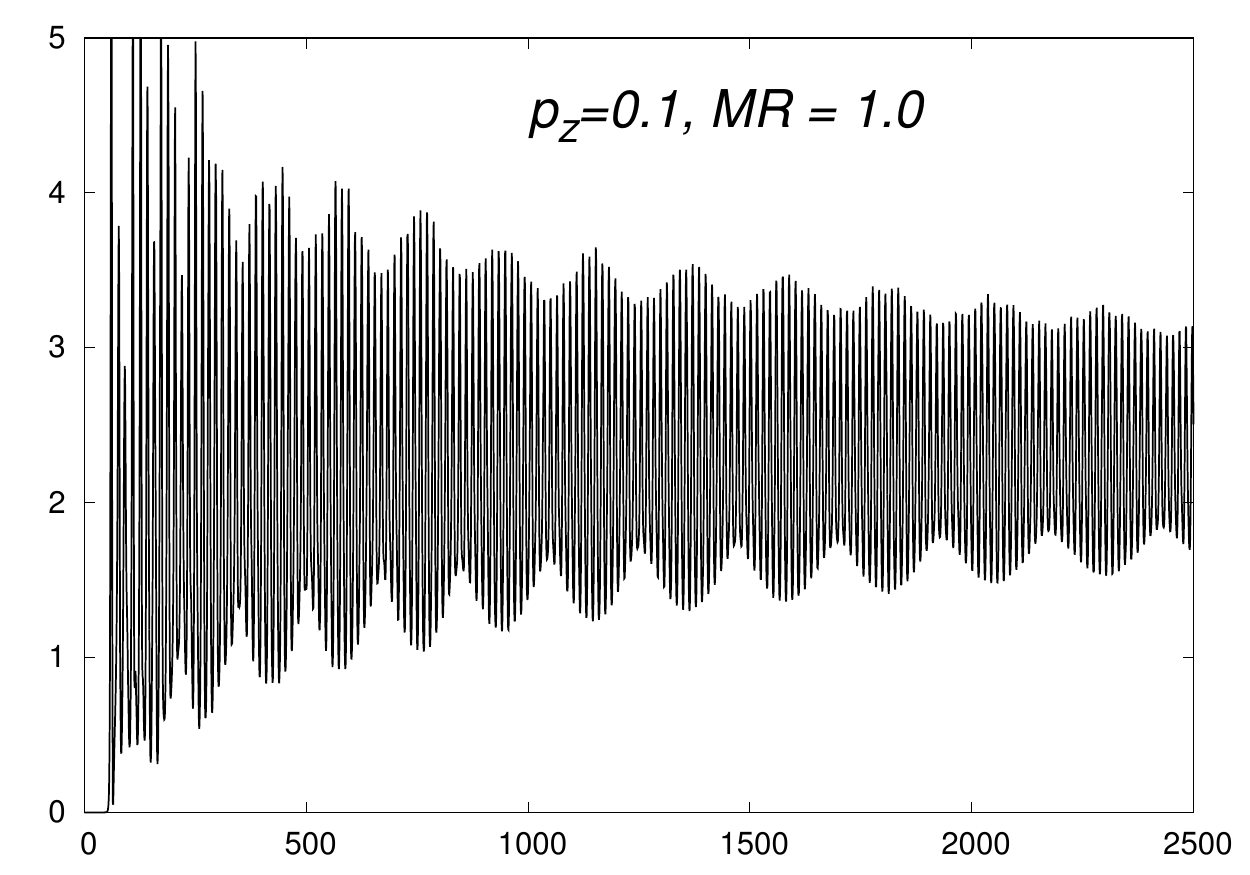}
\includegraphics[width= 3.25cm]{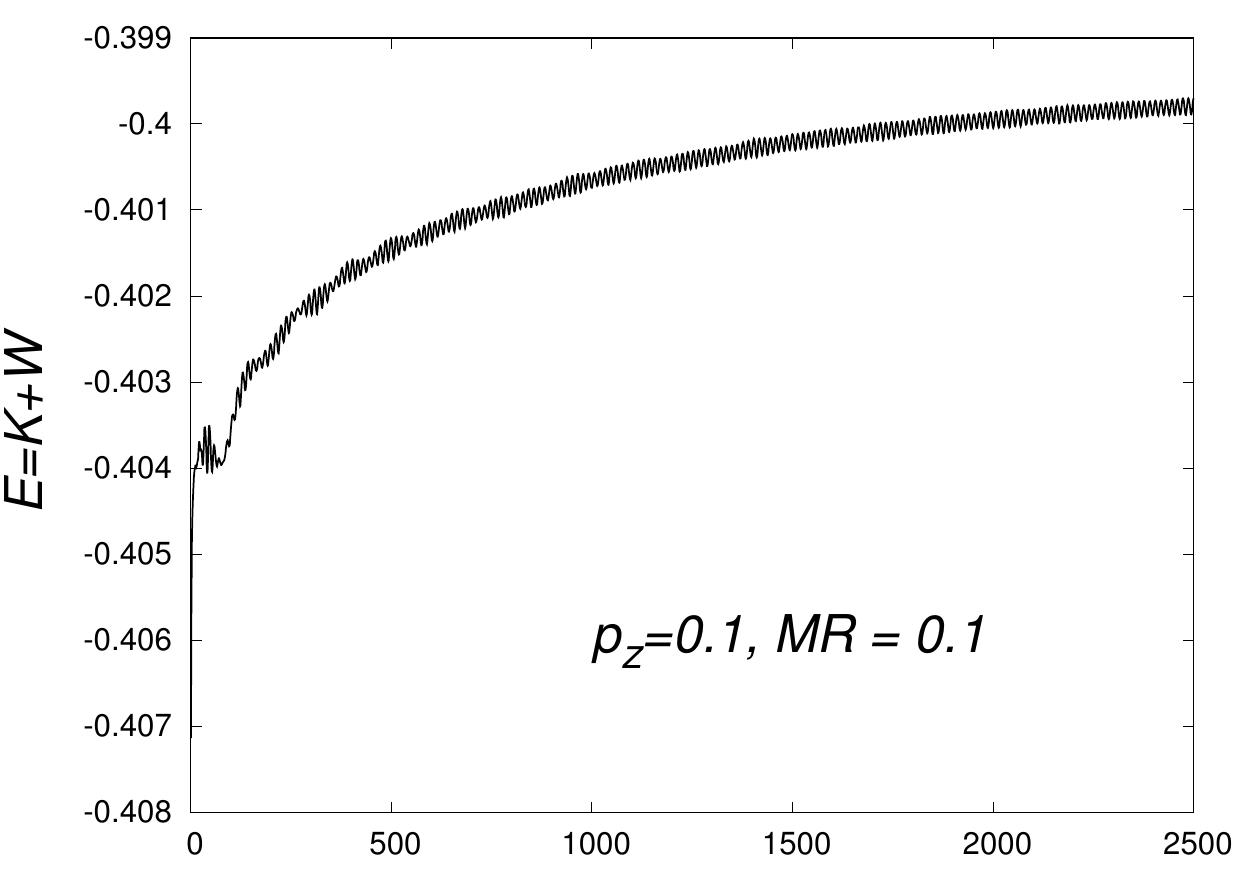}
\includegraphics[width= 3.25cm]{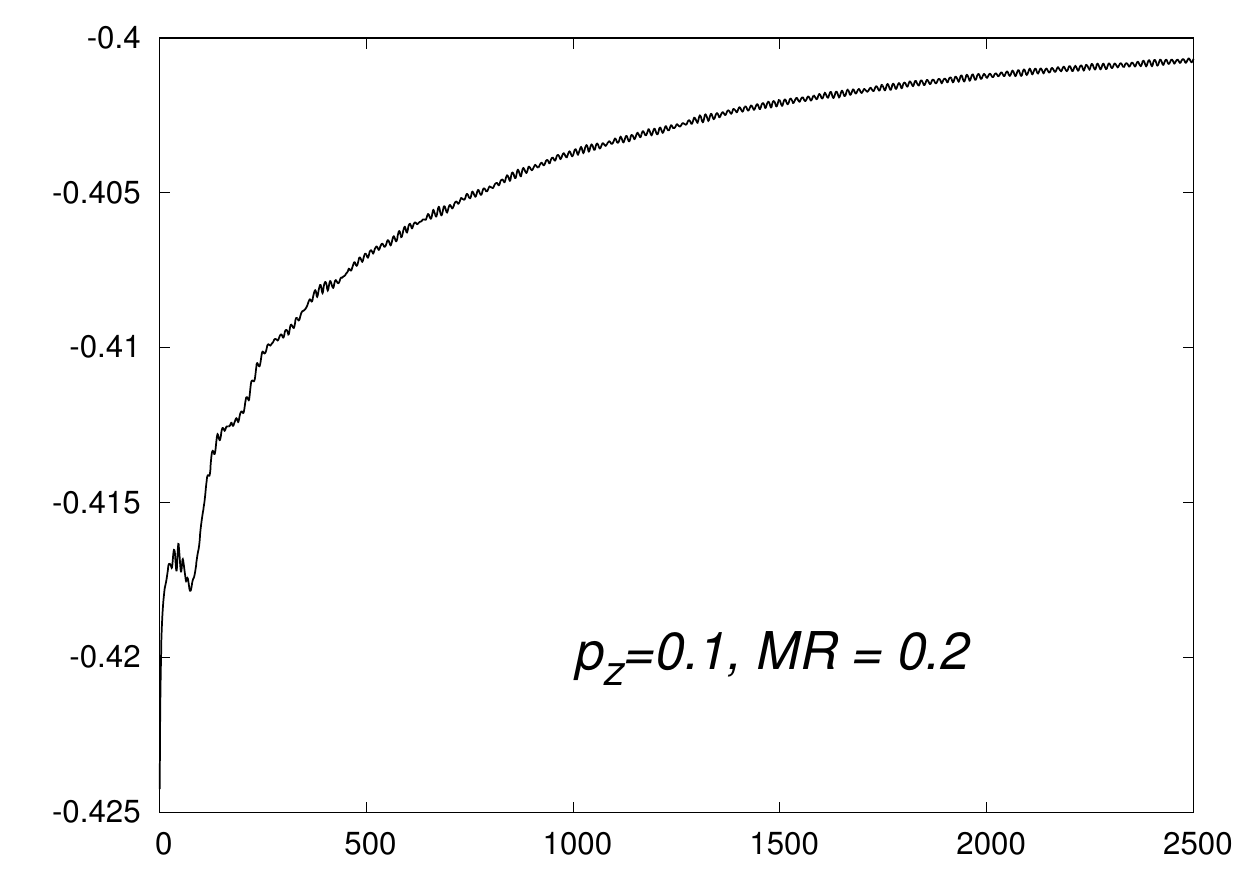}
\includegraphics[width= 3.25cm]{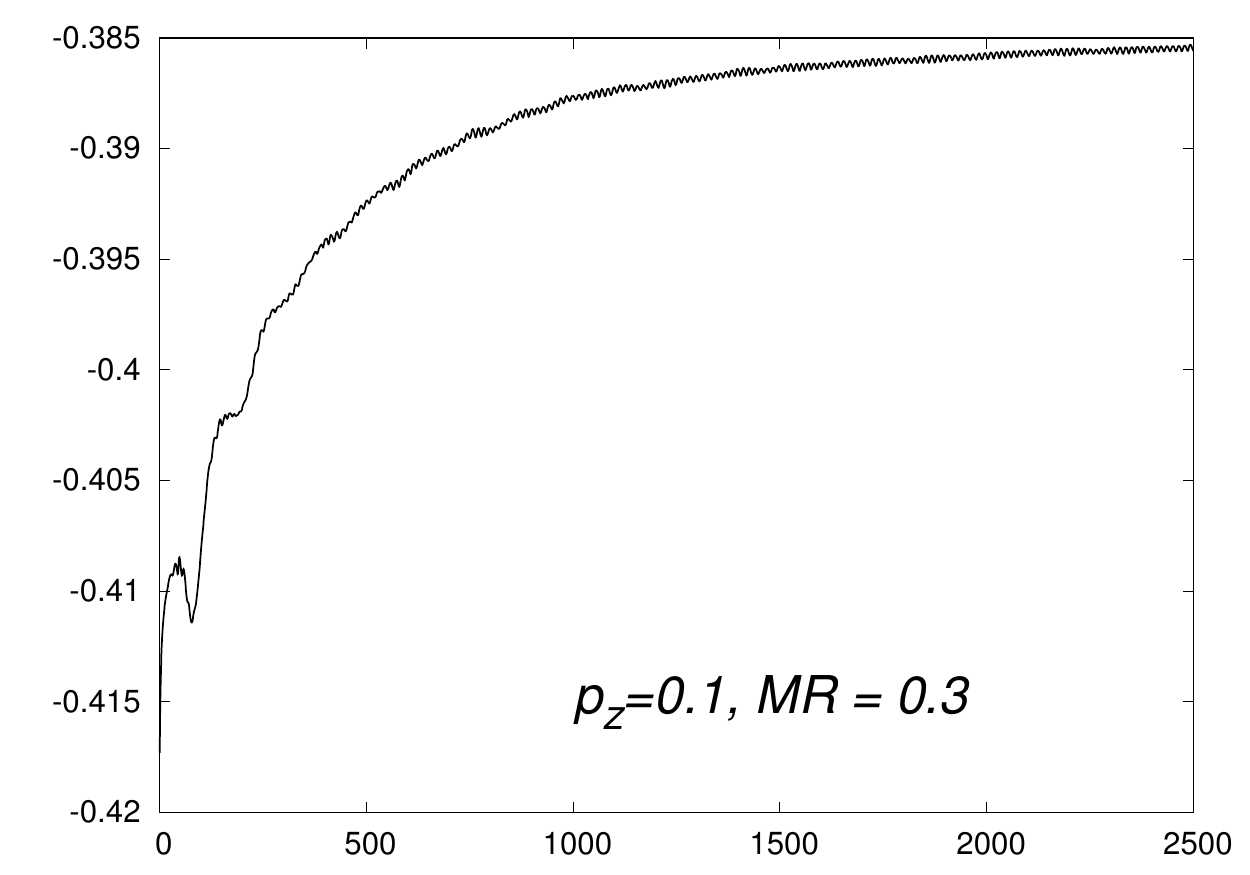}
\includegraphics[width= 3.25cm]{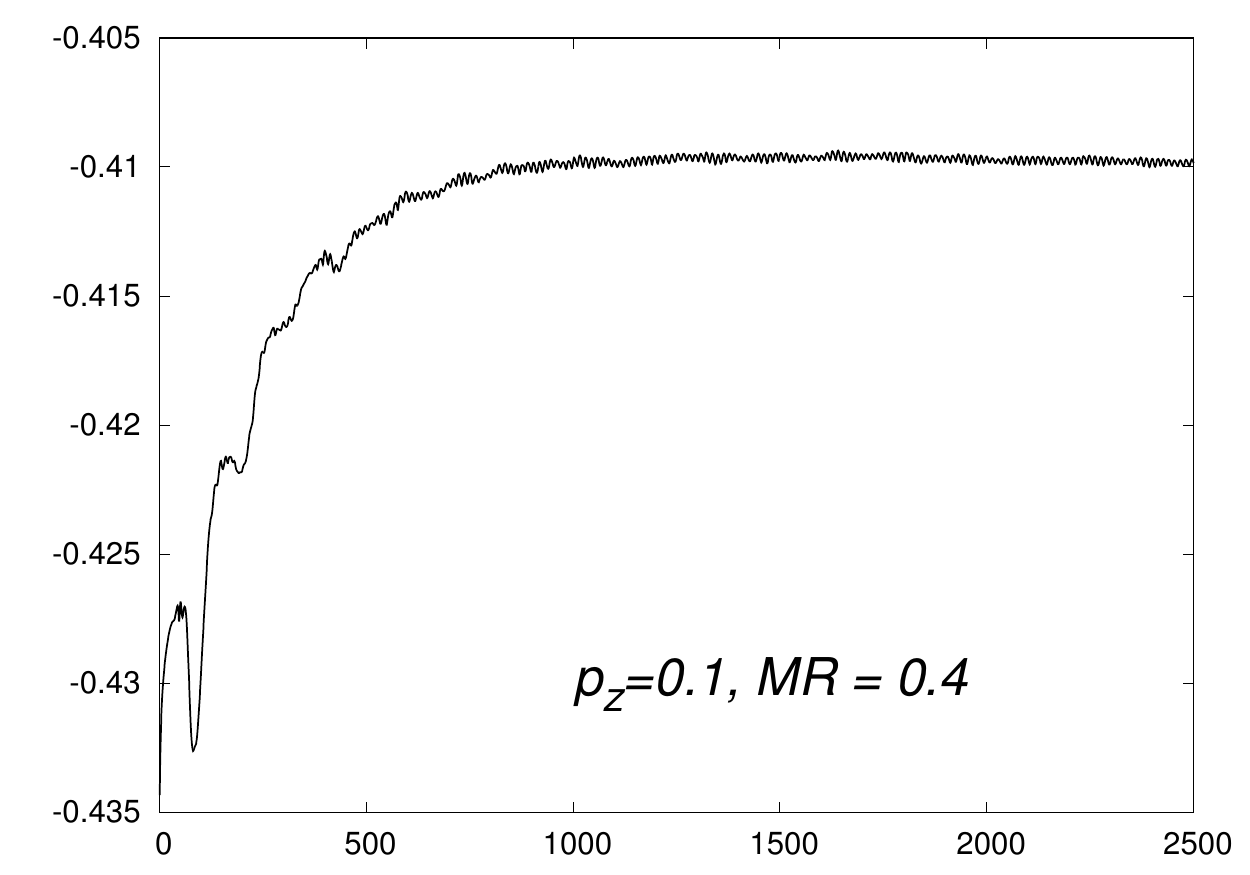}
\includegraphics[width= 3.25cm]{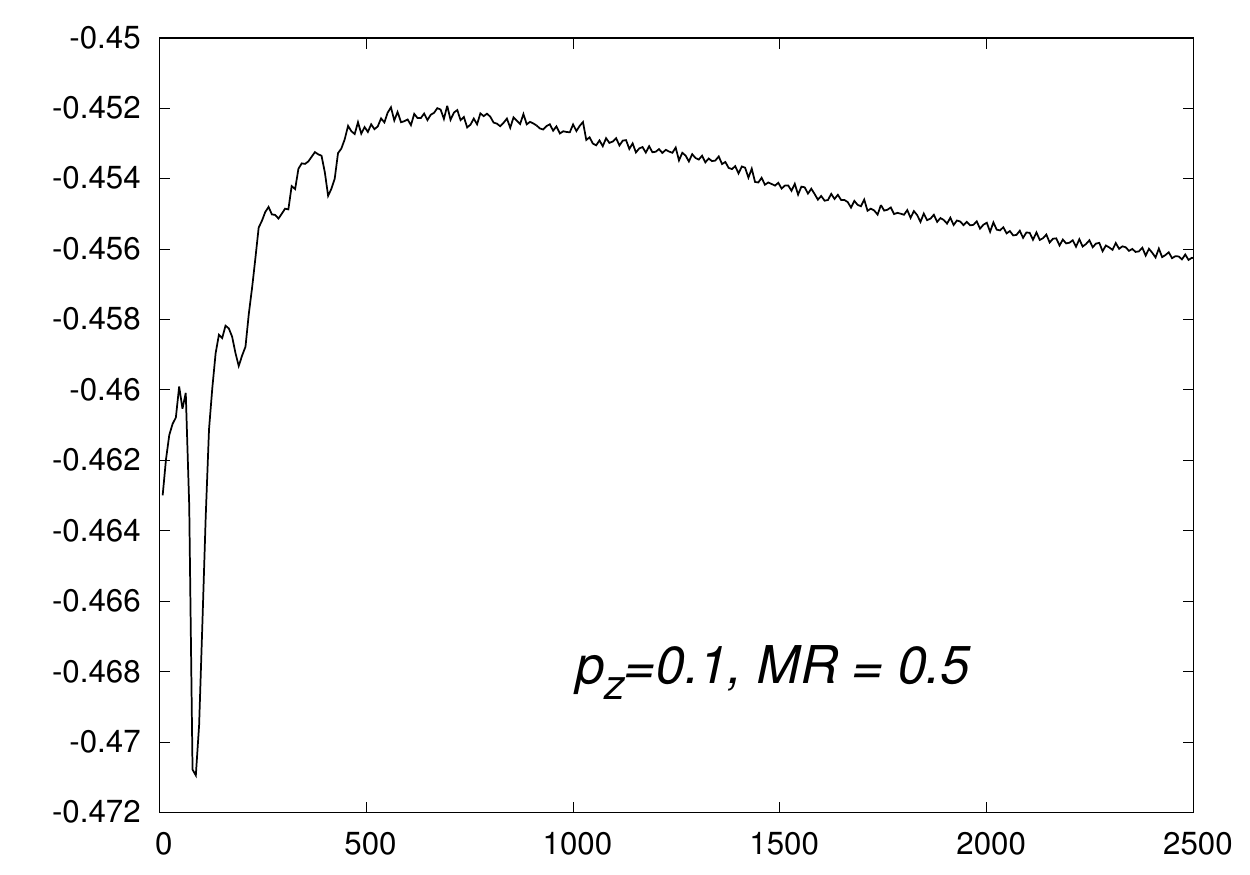}
\includegraphics[width= 3.25cm]{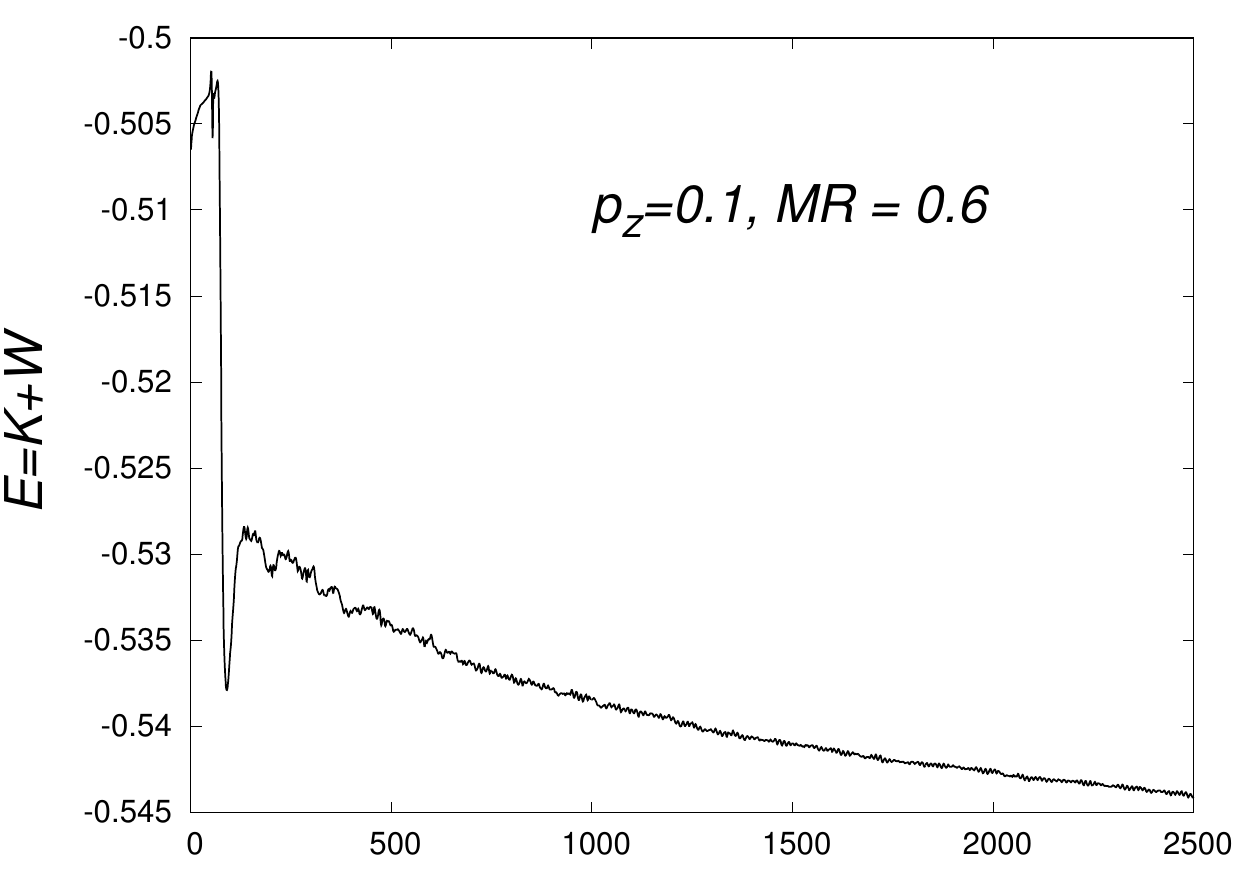}
\includegraphics[width= 3.25cm]{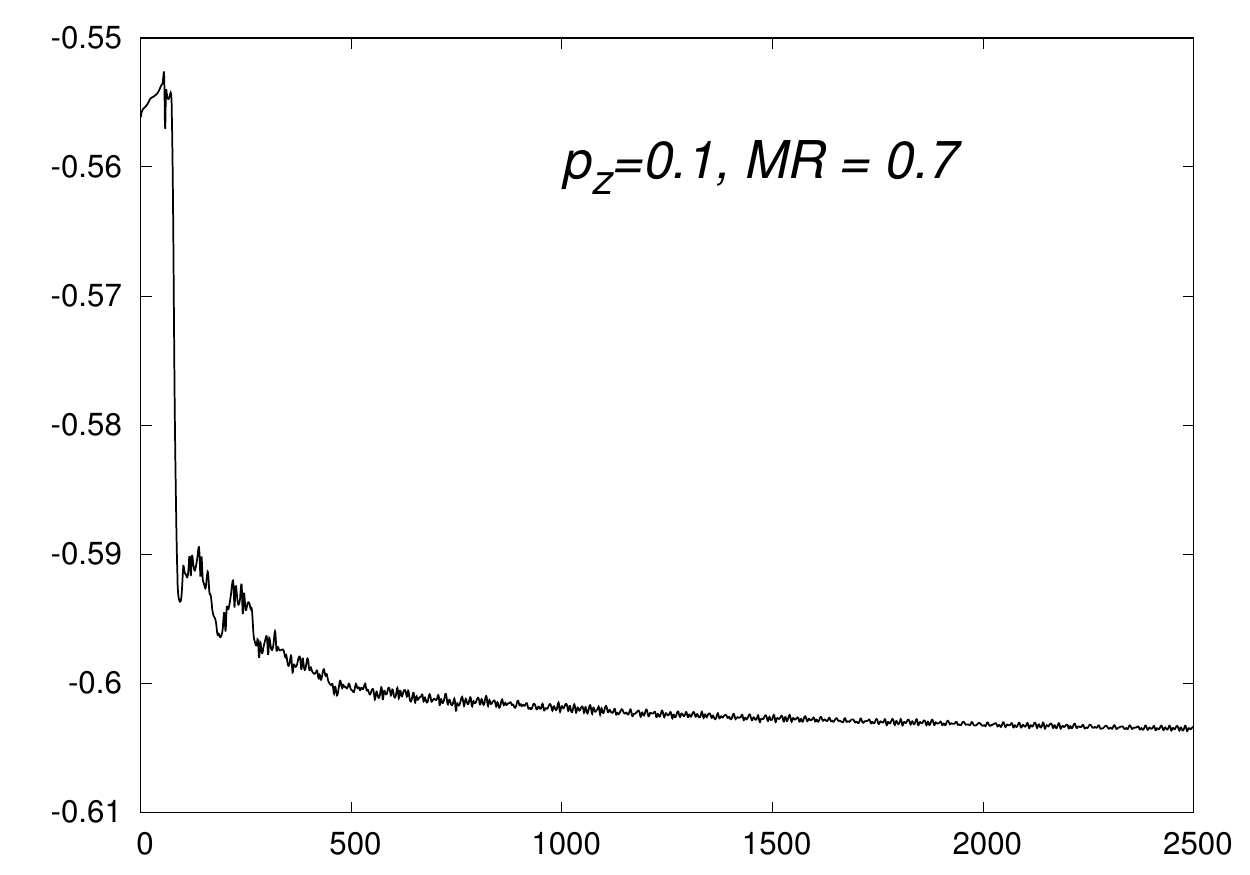}
\includegraphics[width= 3.25cm]{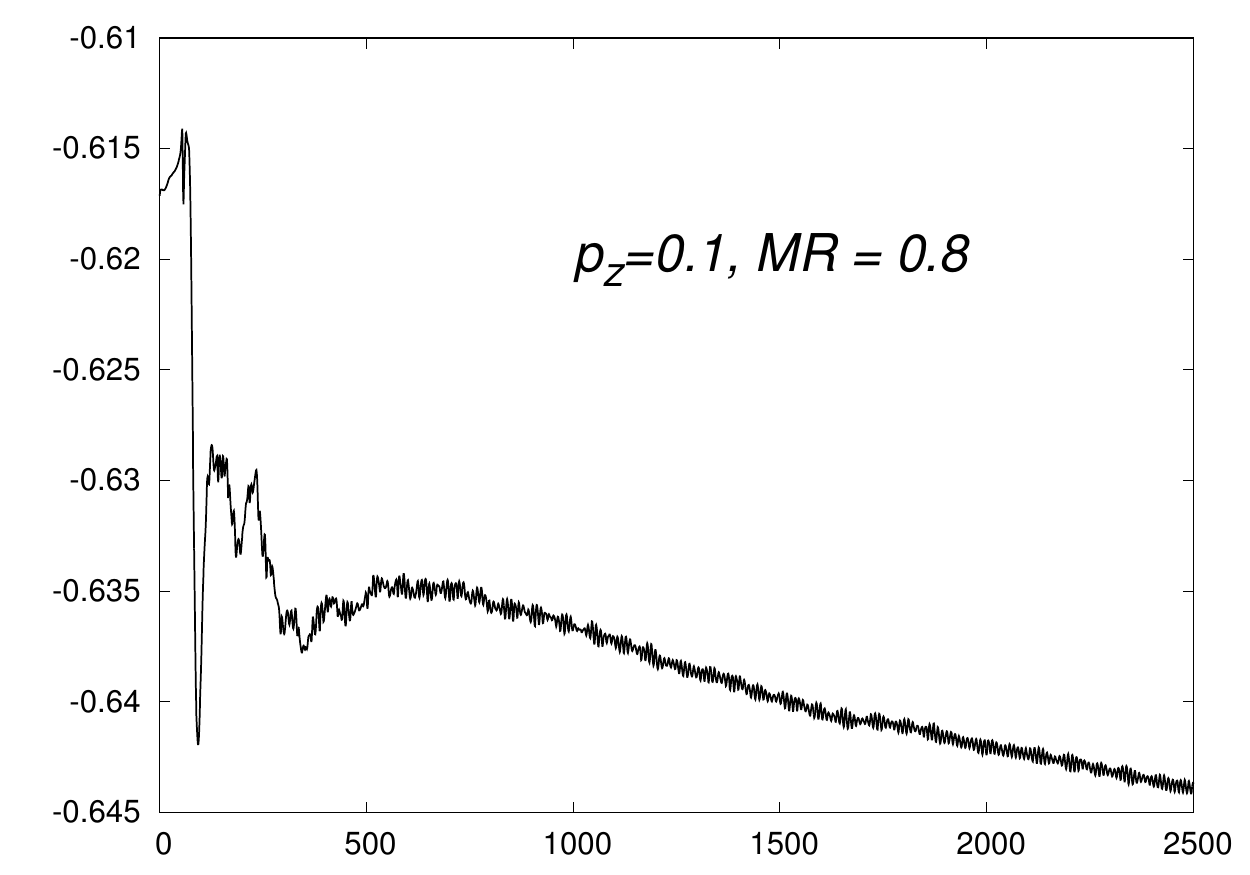}
\includegraphics[width= 3.25cm]{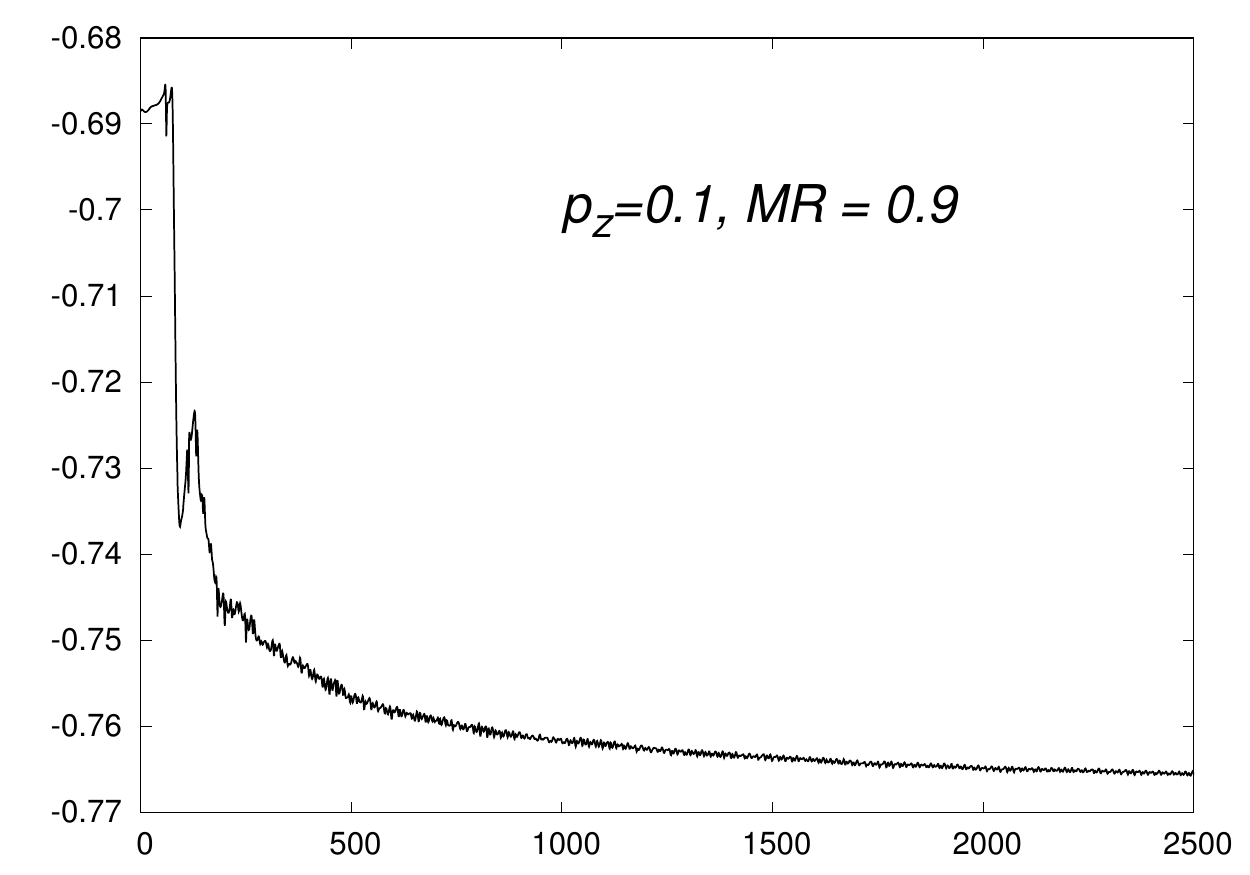}
\includegraphics[width= 3.25cm]{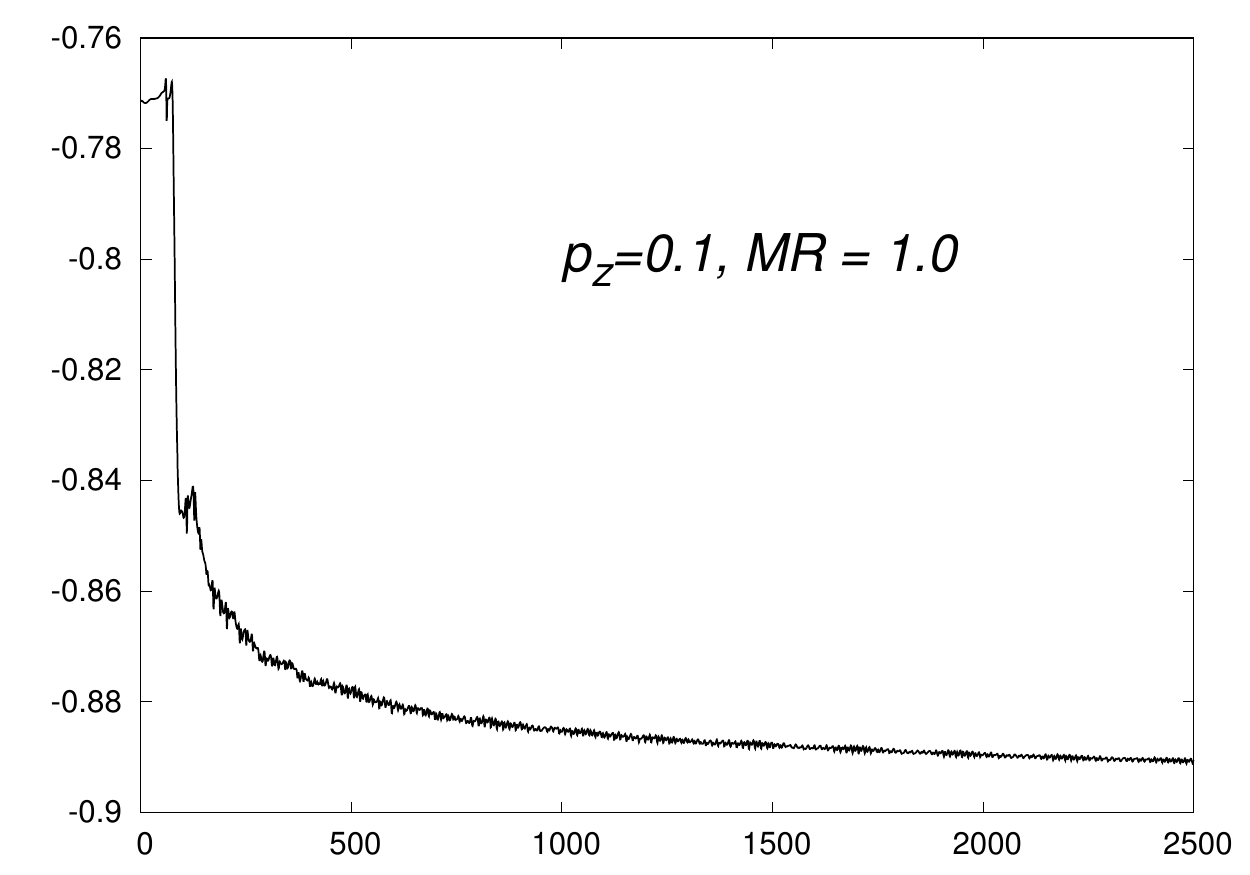}
\caption{Diagnostics of the relaxation process for the case $p_z=0.1$ and  the various mass ratios $MR=0.1,...,1.0$. We show the quantity $2K+W$, the central density where the final structure centers $\rho(0,0)$ and the total energy $E=K+W$  as function of time $t$.}
\label{fig:coolingpz0_1}
\end{figure*}

\begin{figure*}
\centering
\includegraphics[width= 3.25cm]{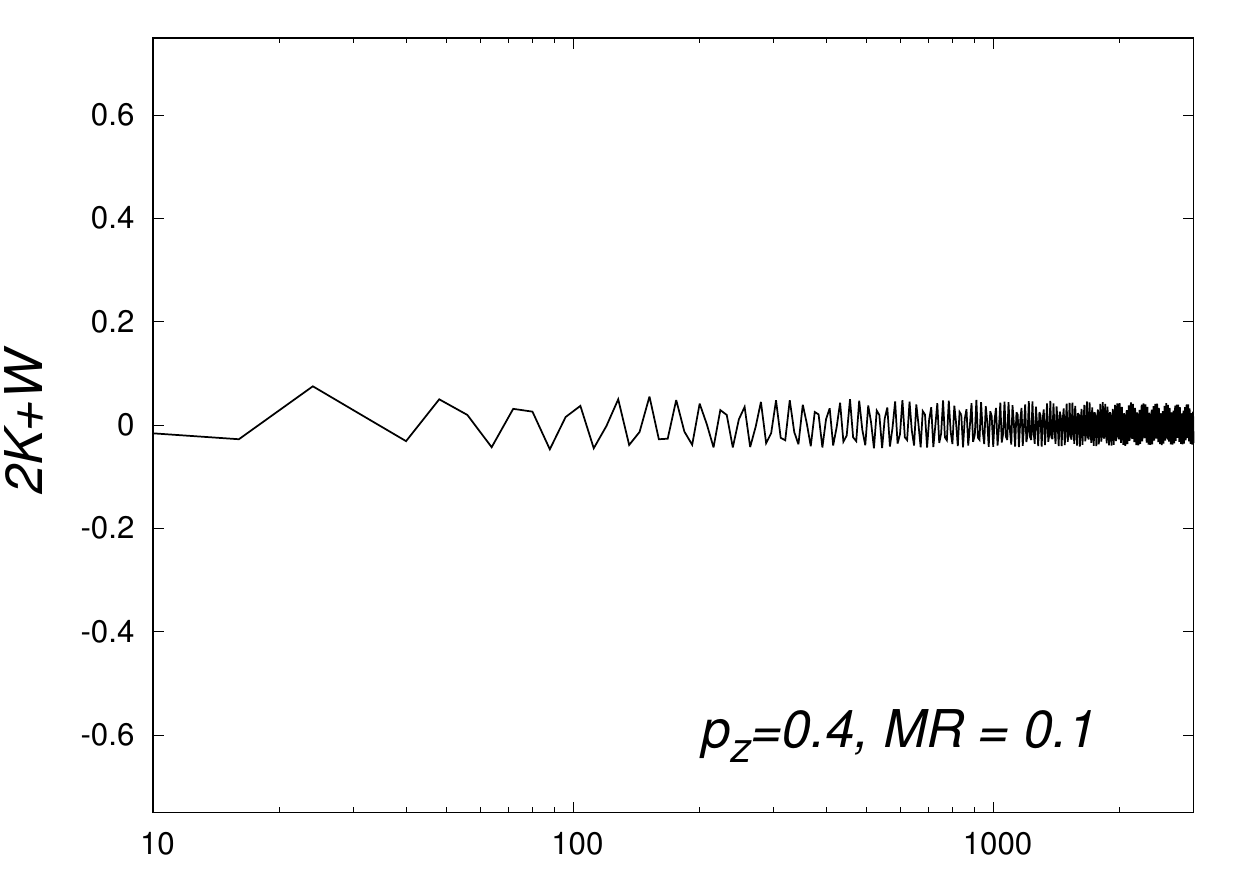}
\includegraphics[width= 3.25cm]{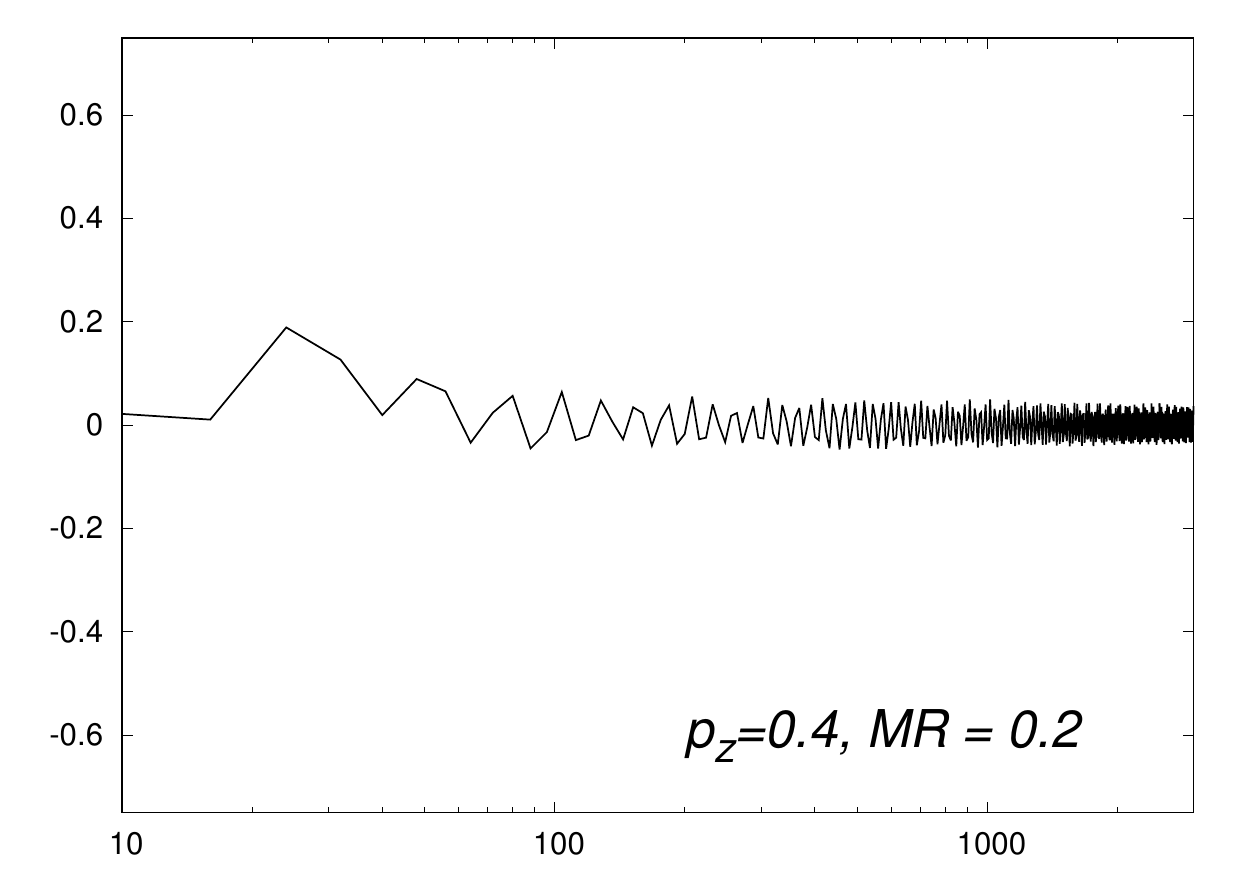}
\includegraphics[width= 3.25cm]{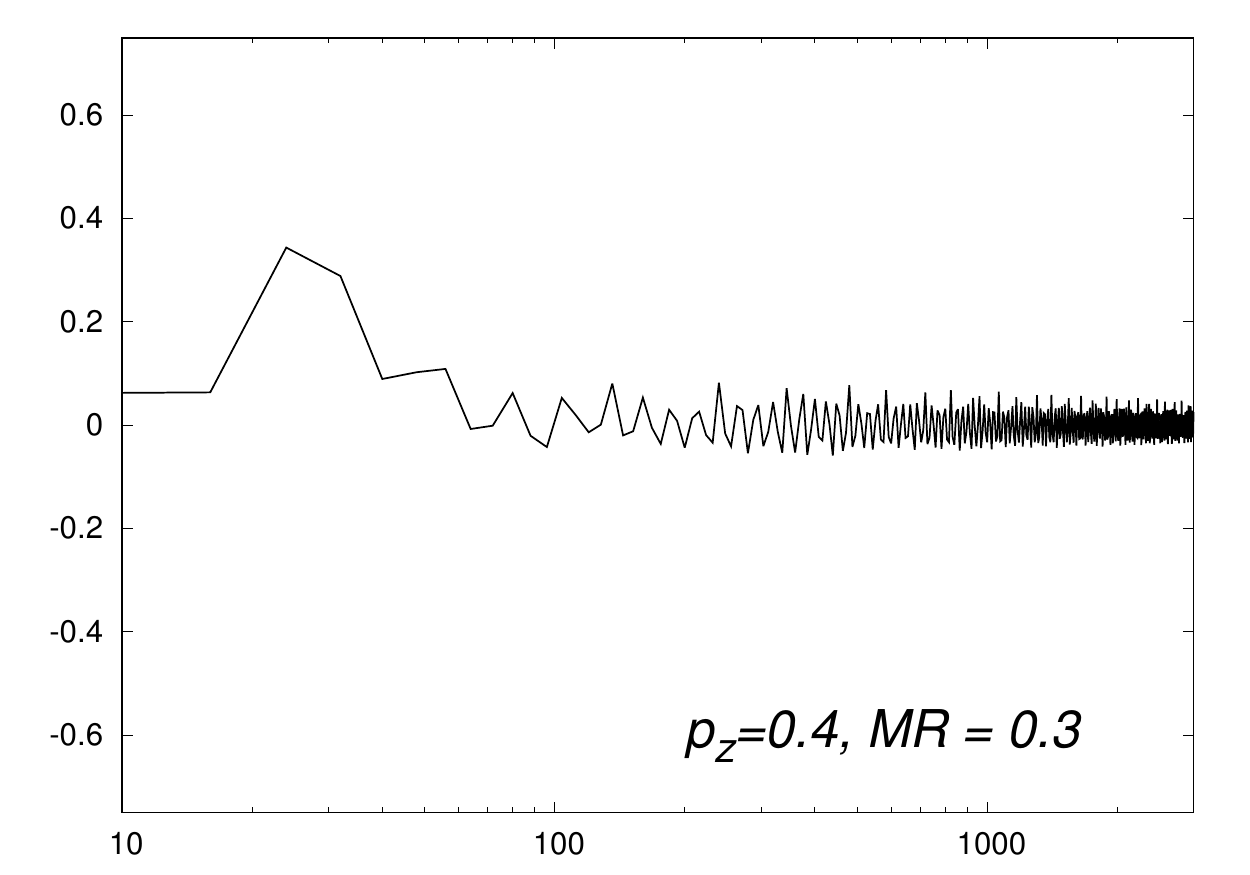}
\includegraphics[width= 3.25cm]{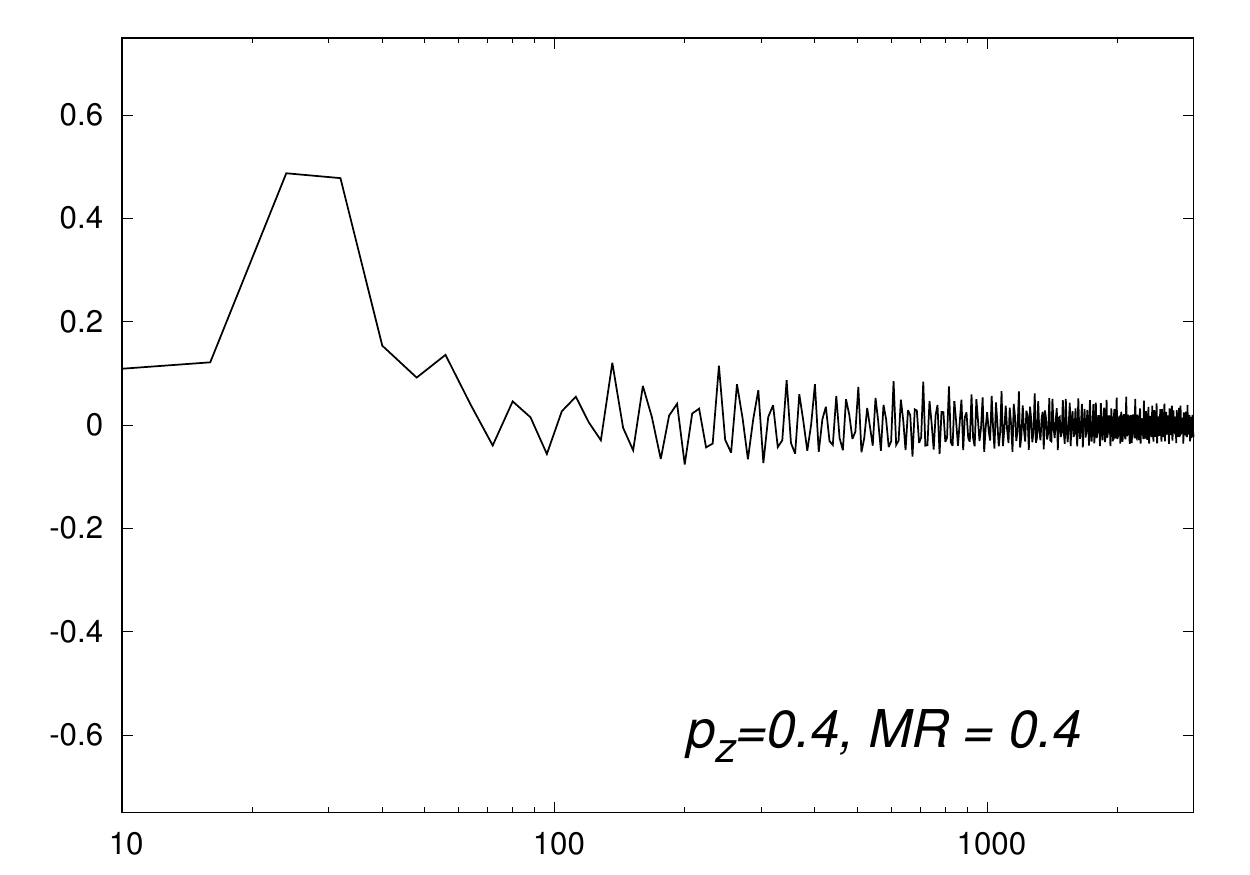}
\includegraphics[width= 3.25cm]{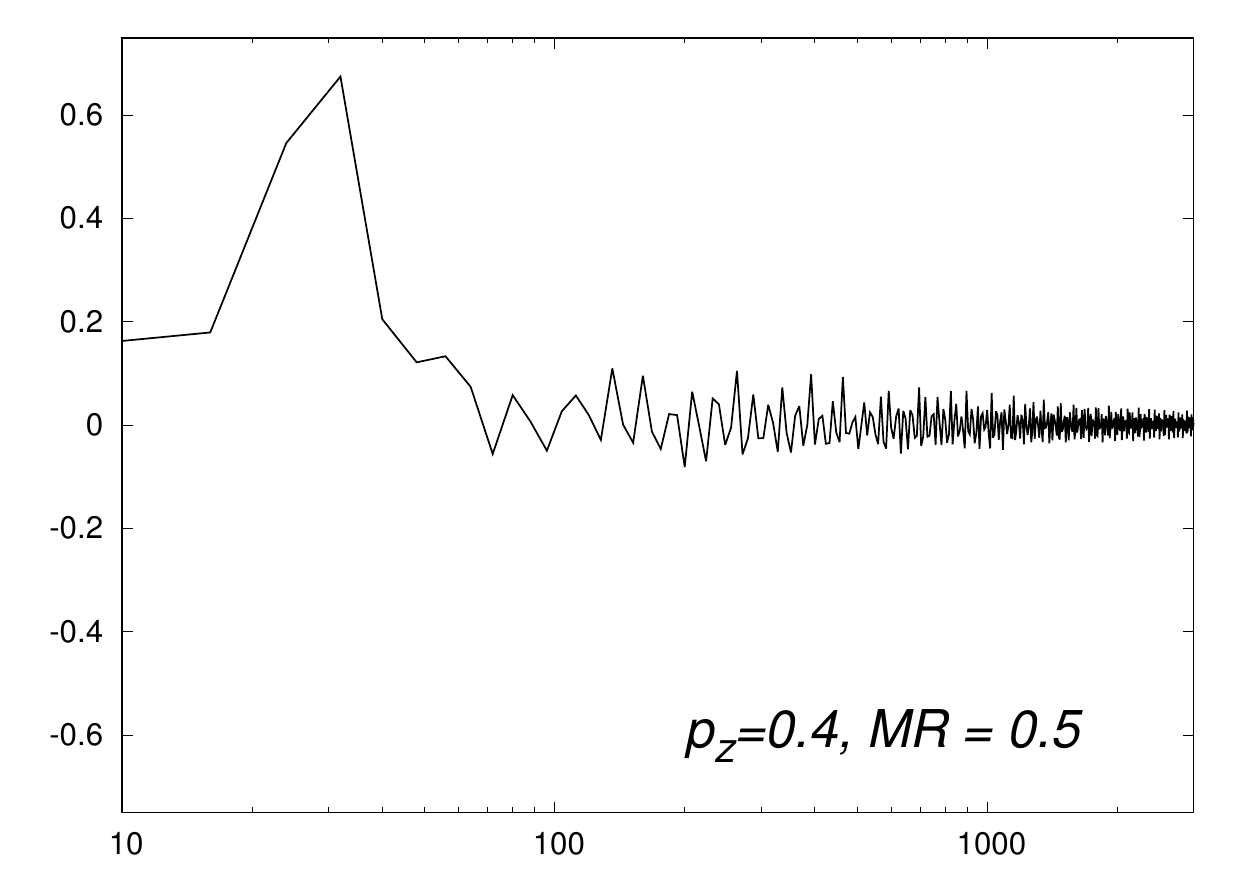}
\includegraphics[width= 3.25cm]{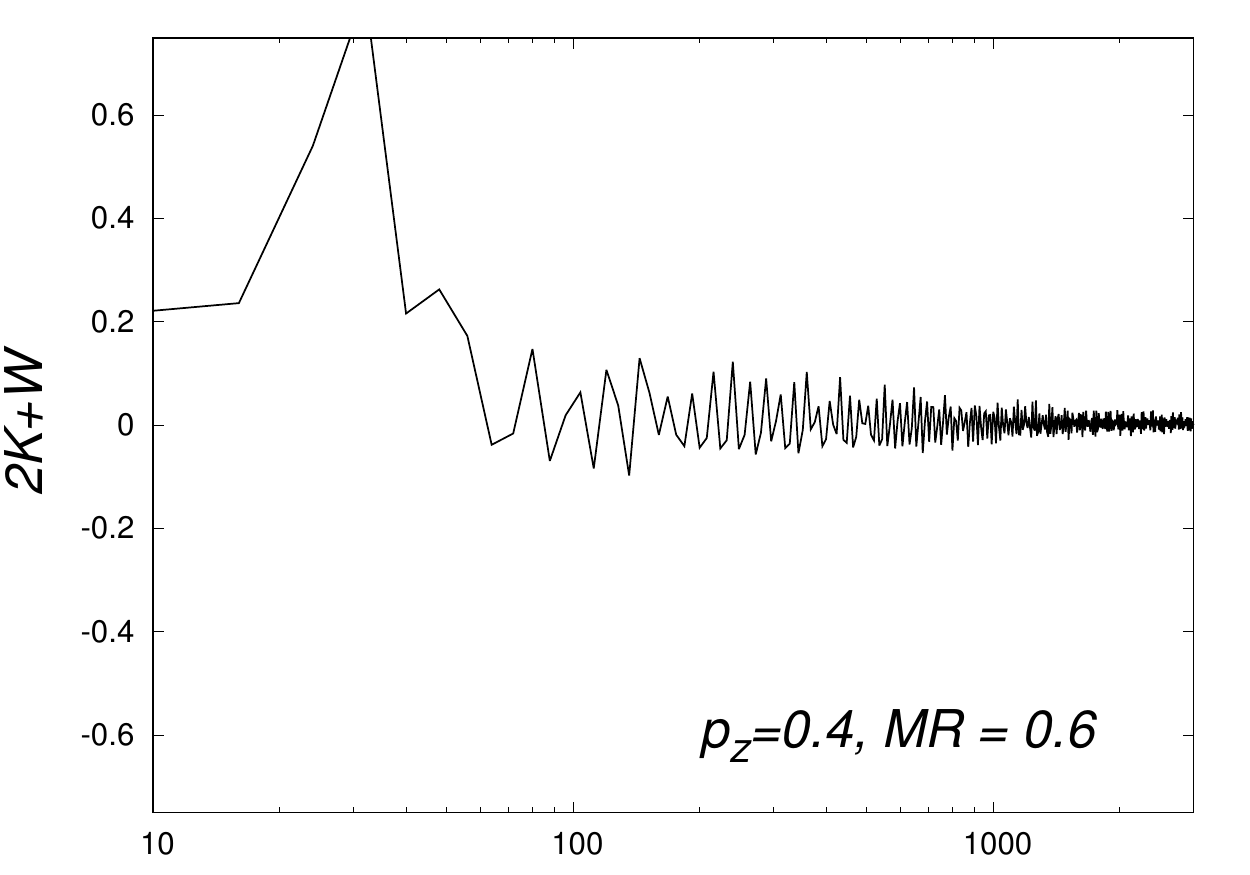}
\includegraphics[width= 3.25cm]{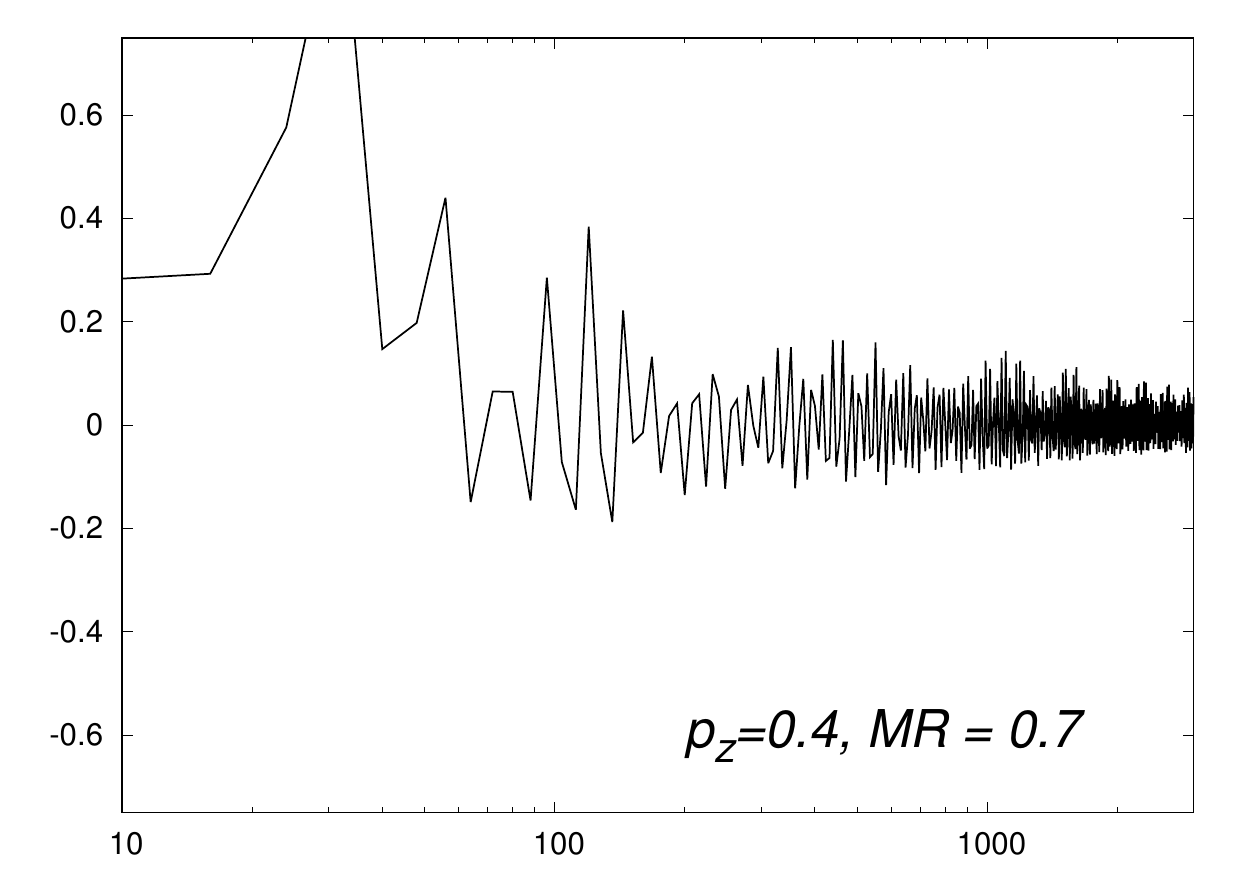}
\includegraphics[width= 3.25cm]{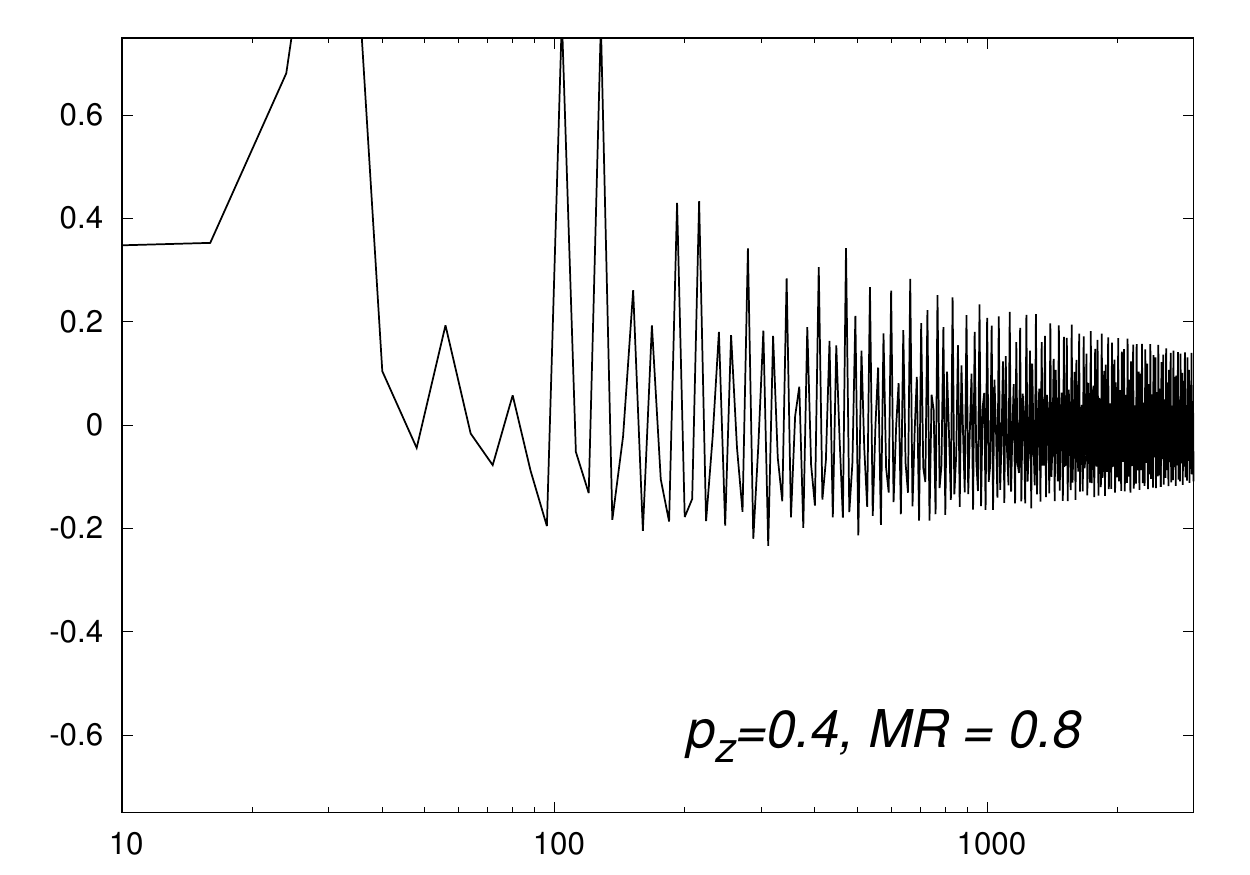}
\includegraphics[width= 3.25cm]{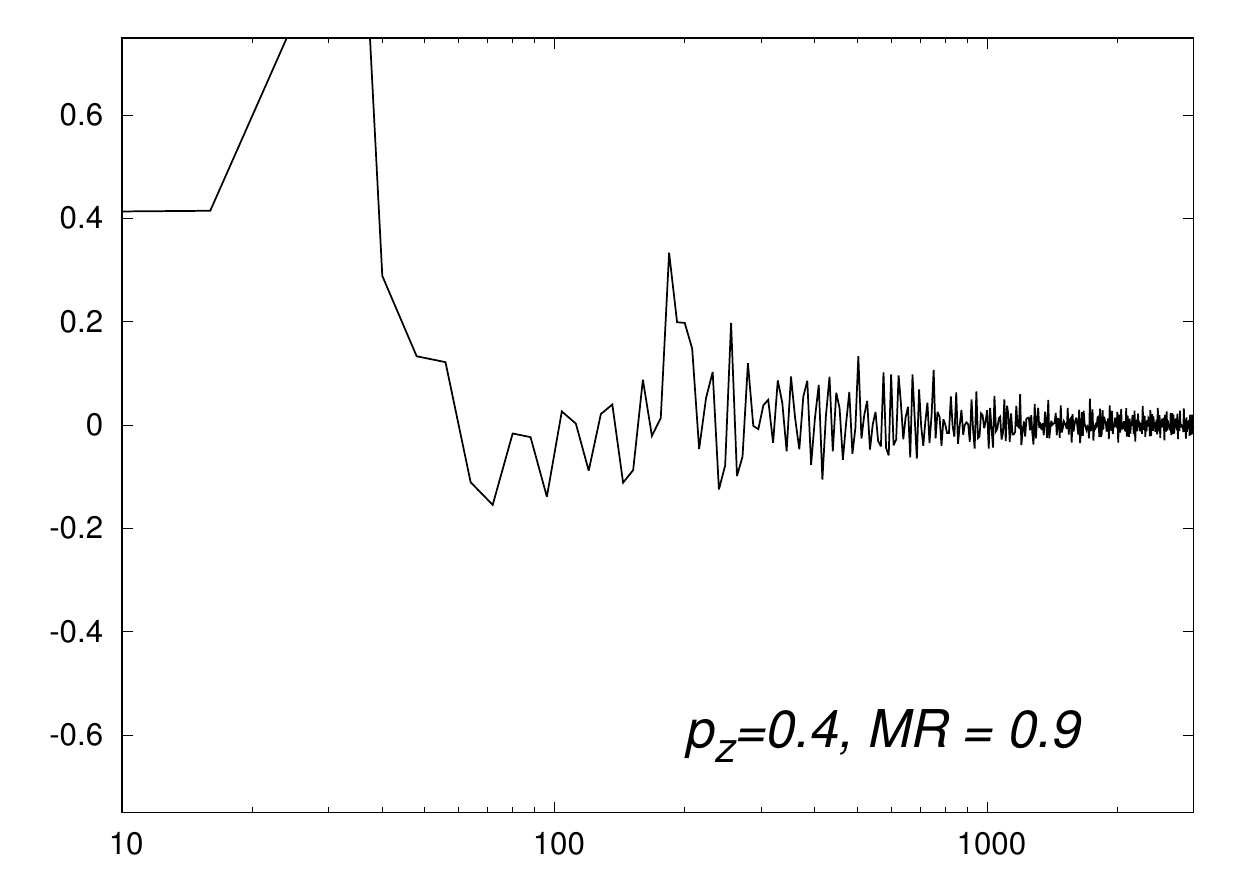}
\includegraphics[width= 3.25cm]{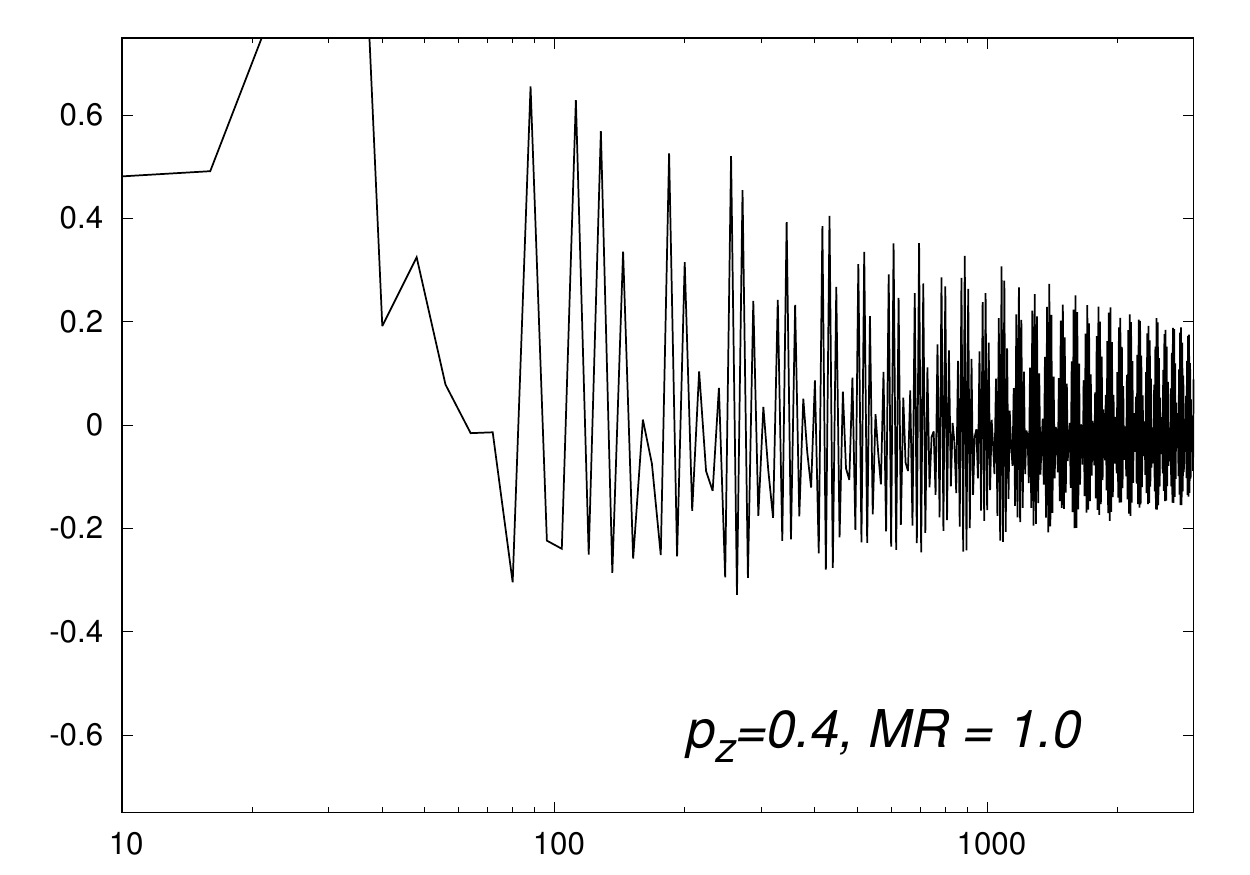}
\includegraphics[width= 3.25cm]{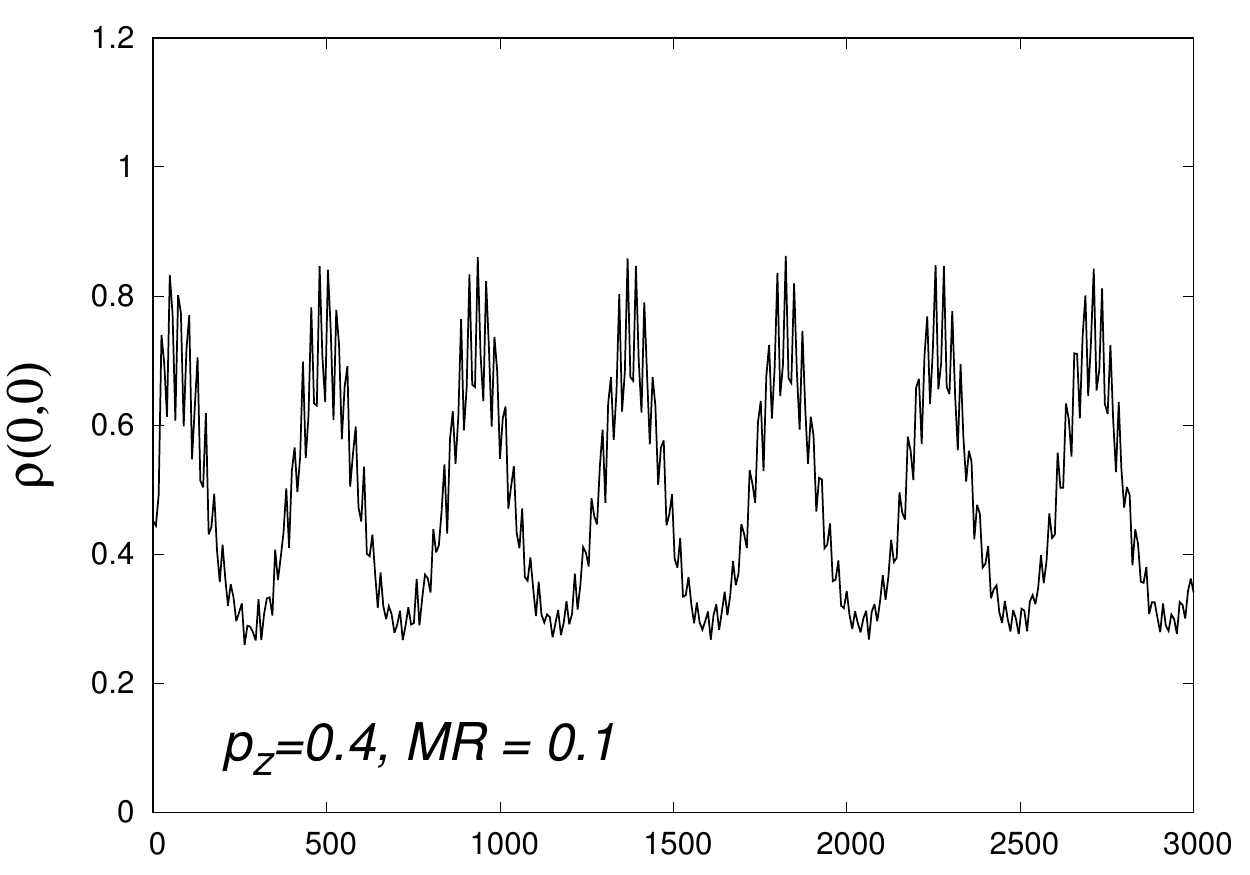}
\includegraphics[width= 3.25cm]{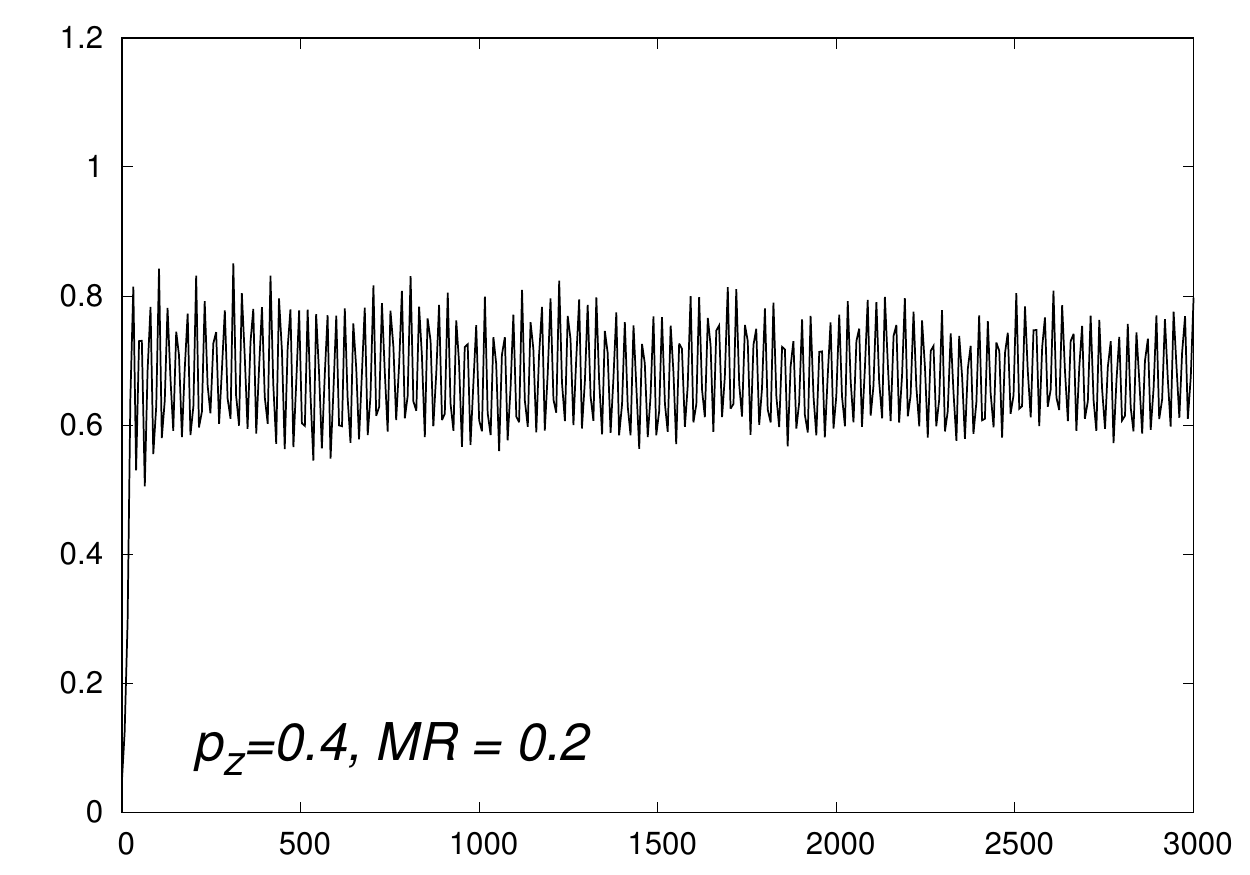}
\includegraphics[width= 3.25cm]{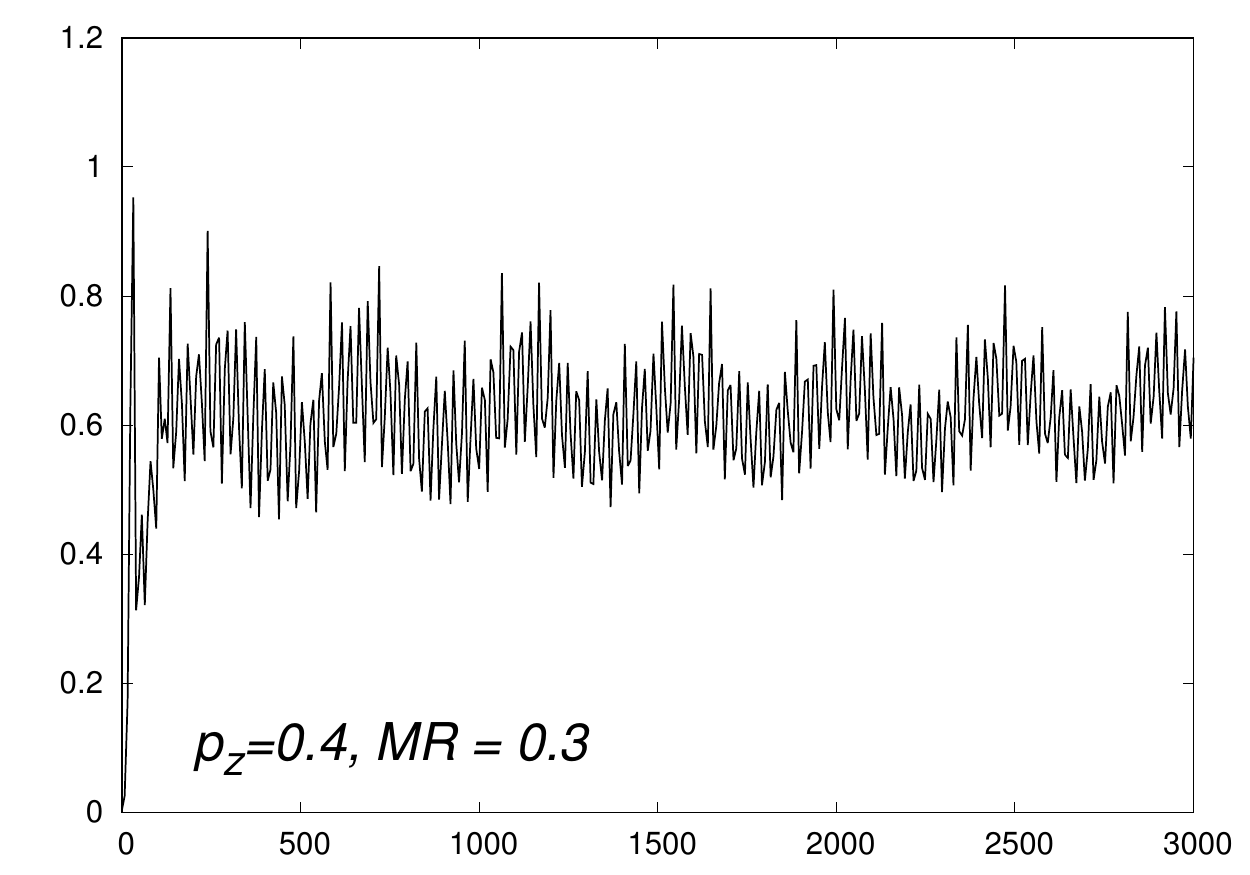}
\includegraphics[width= 3.25cm]{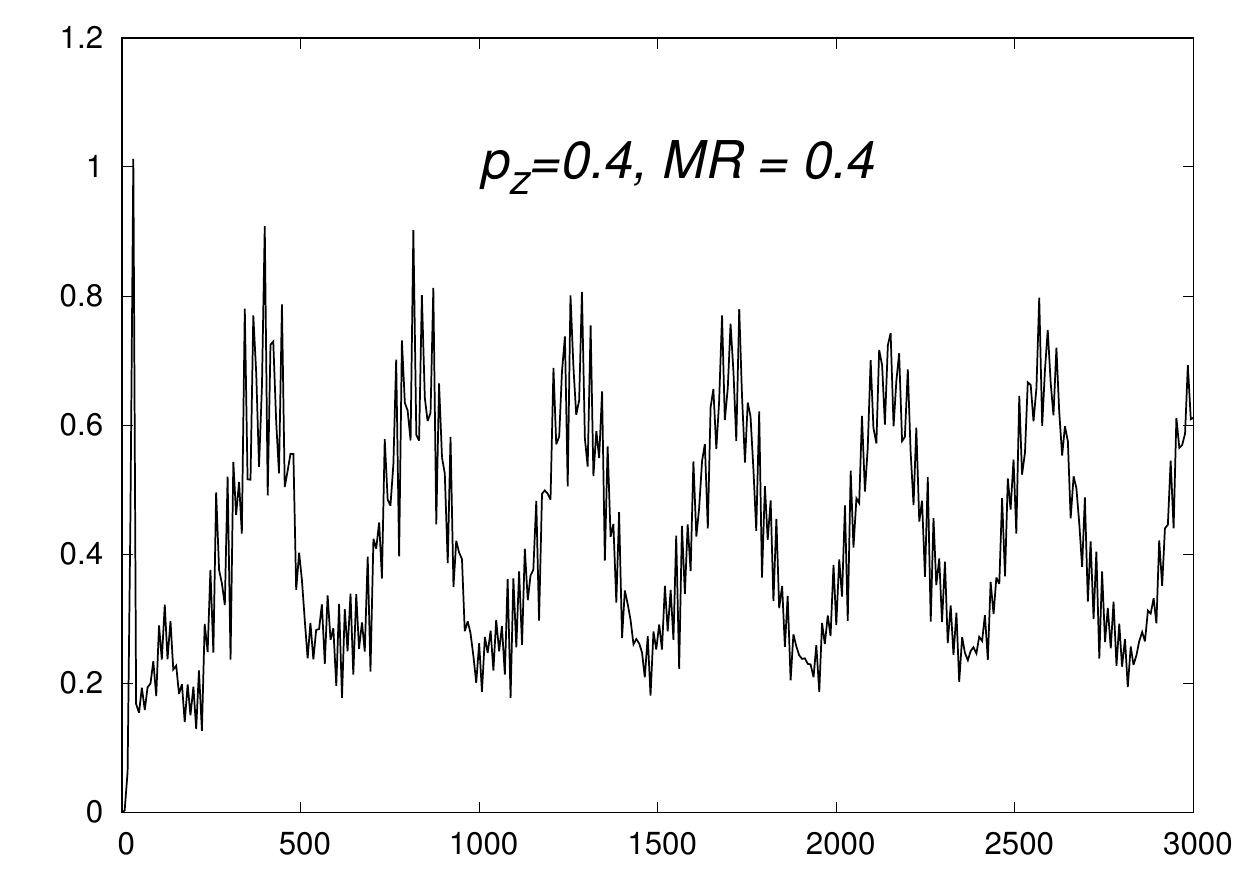}
\includegraphics[width= 3.25cm]{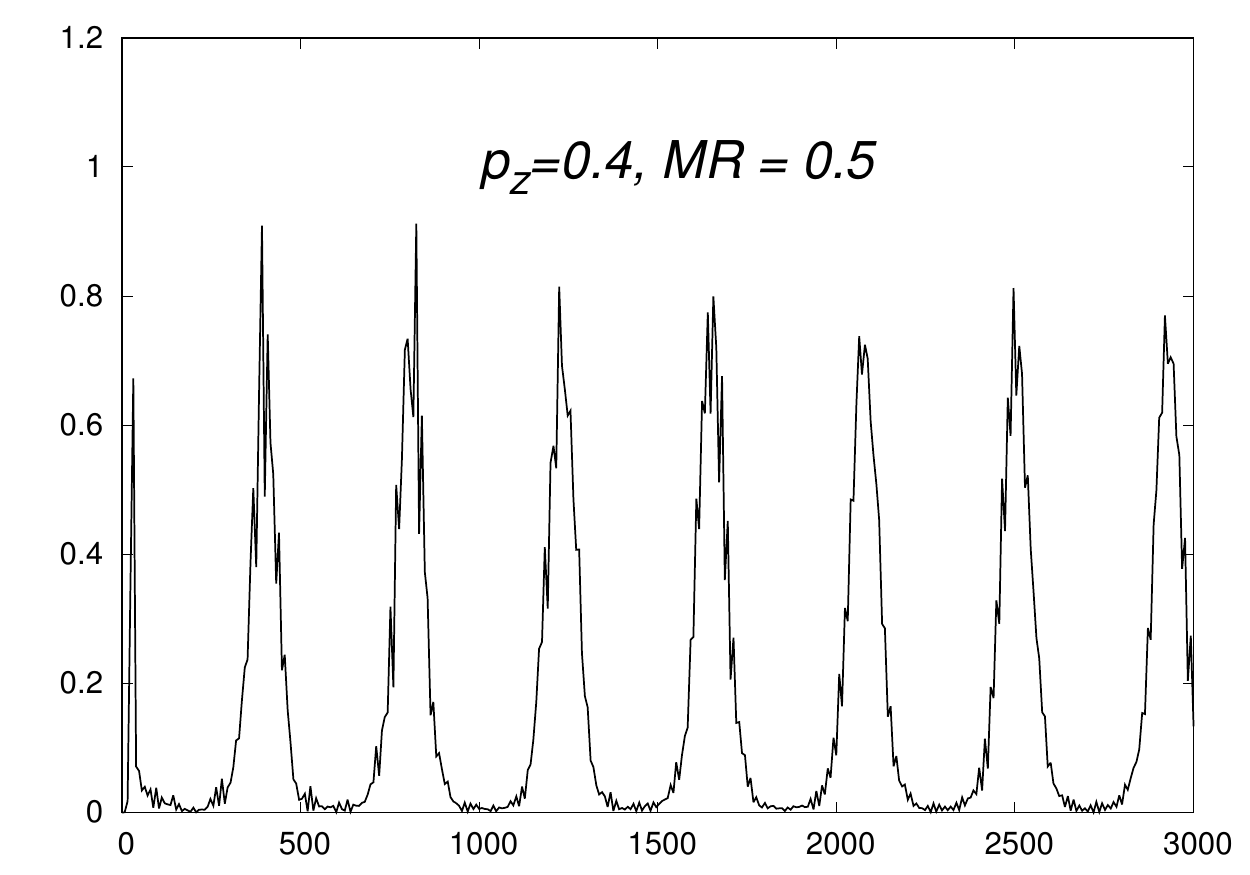}
\includegraphics[width= 3.25cm]{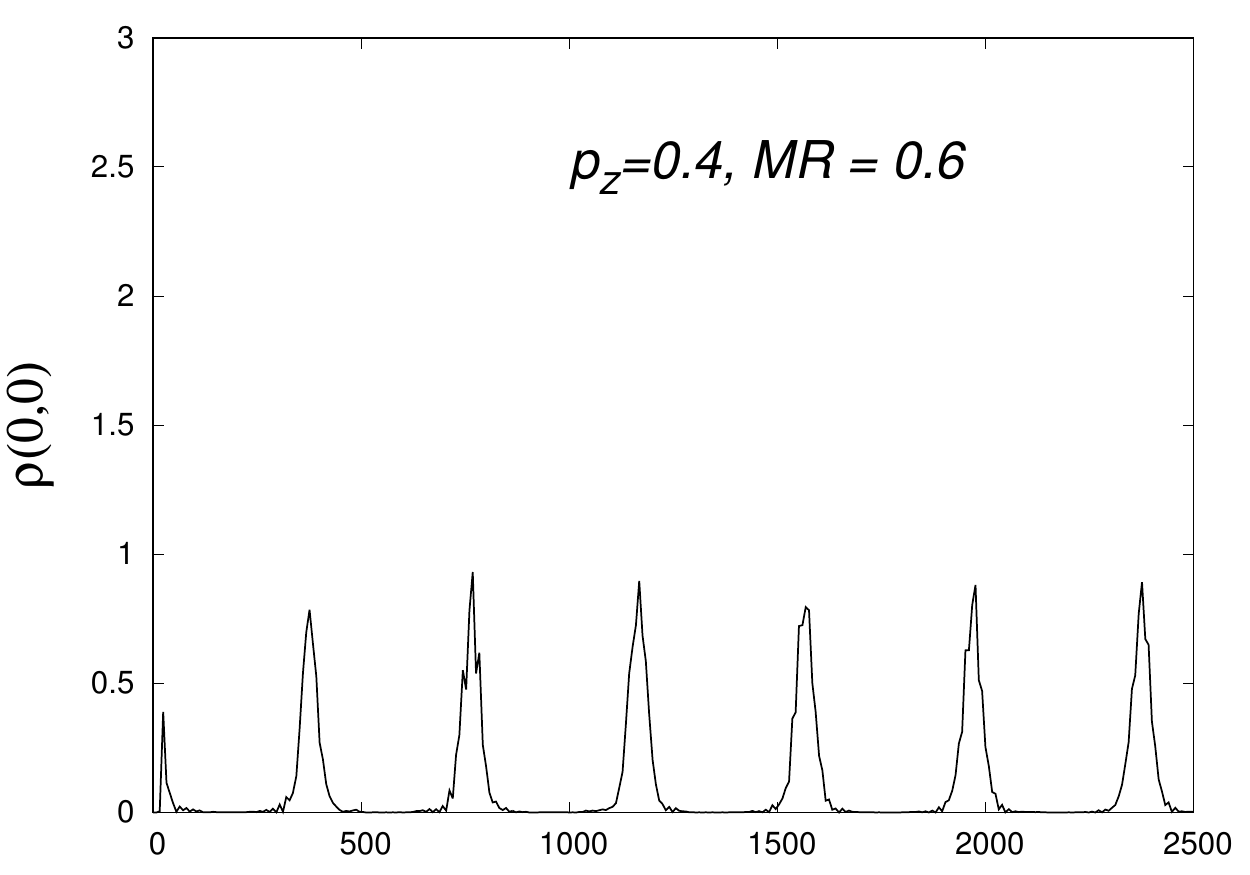}
\includegraphics[width= 3.25cm]{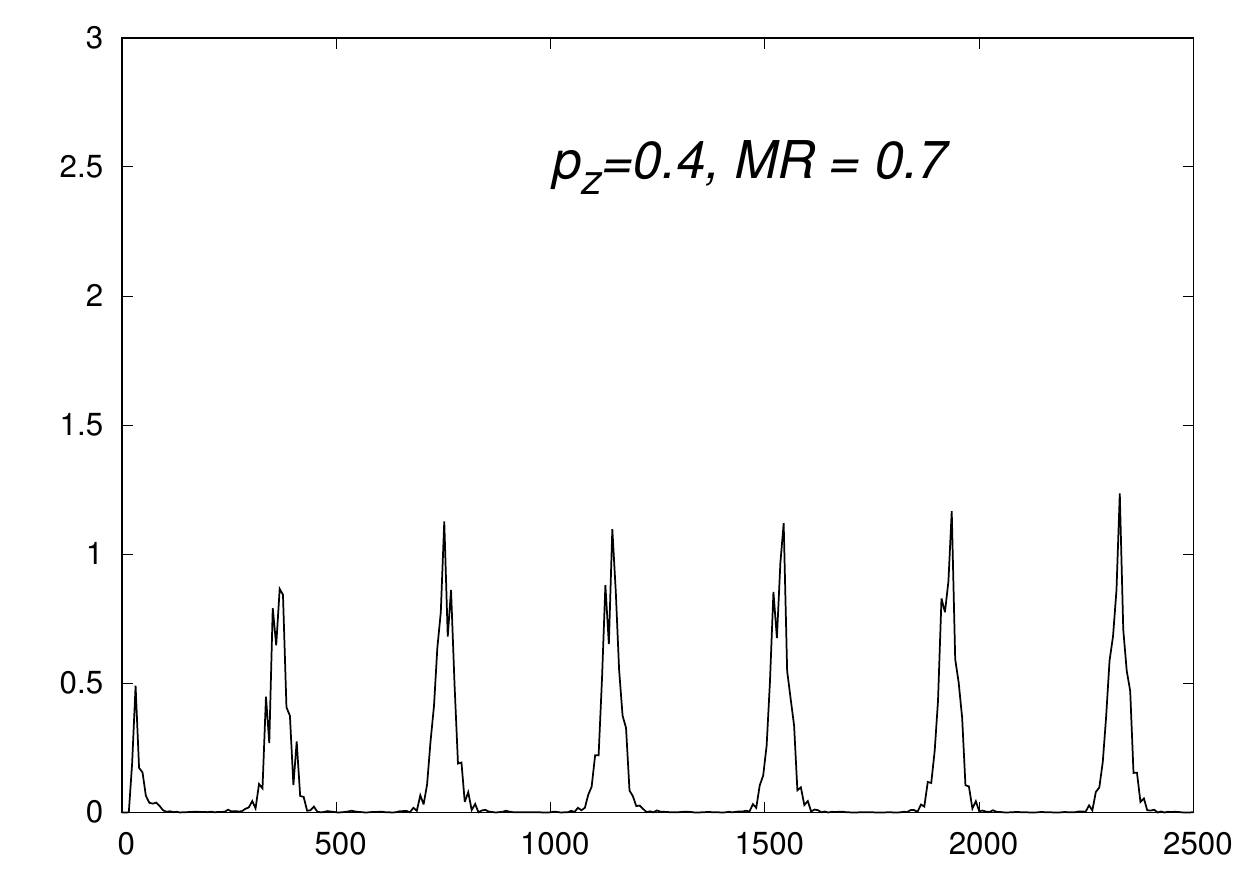}
\includegraphics[width= 3.25cm]{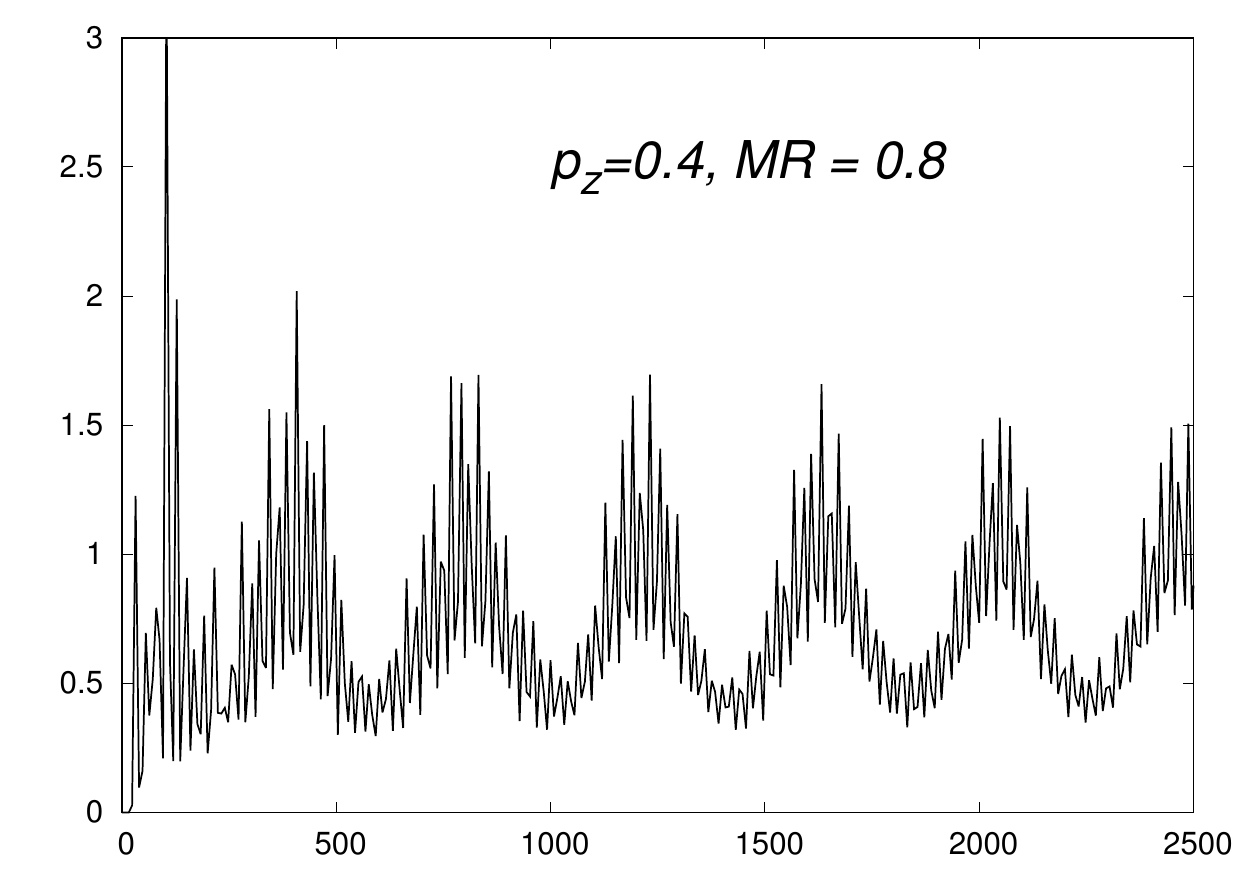}
\includegraphics[width= 3.25cm]{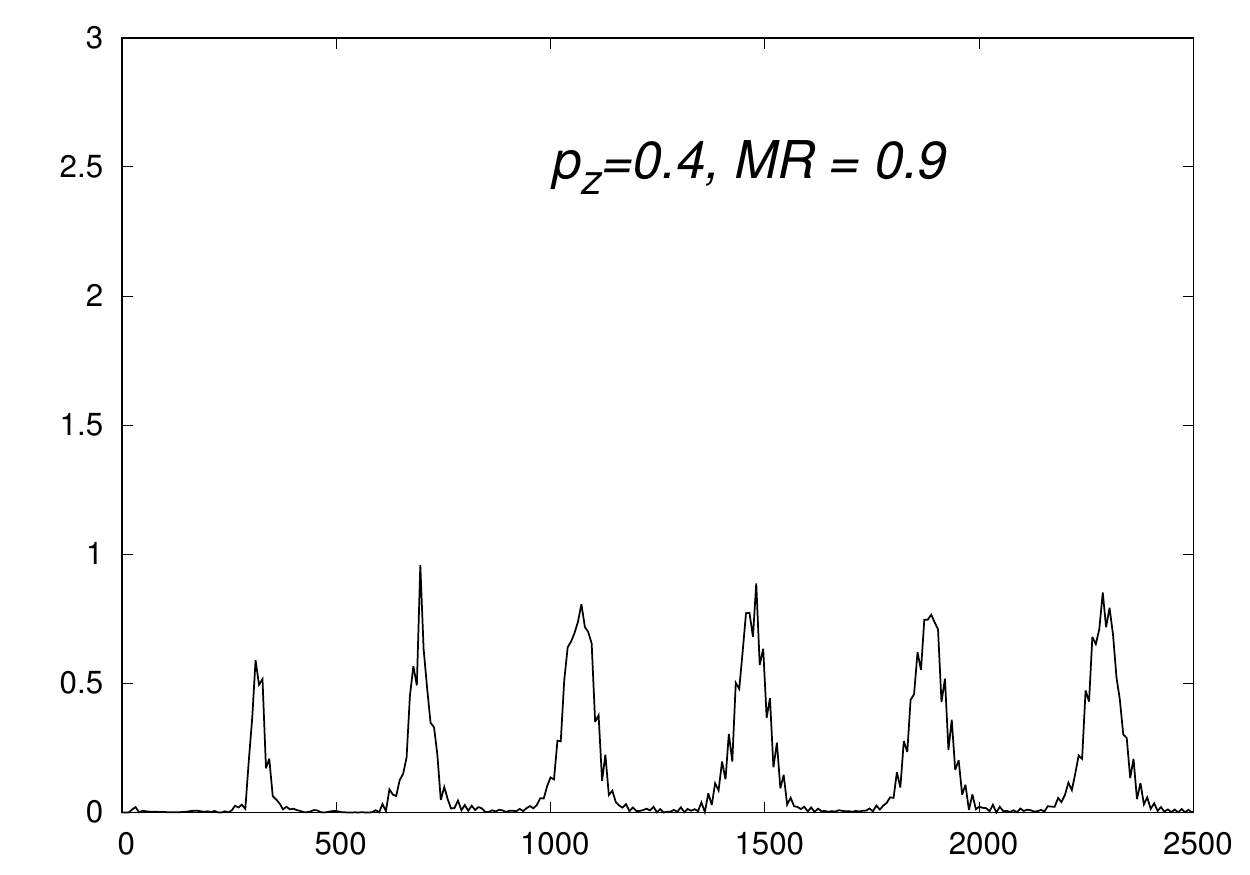}
\includegraphics[width= 3.25cm]{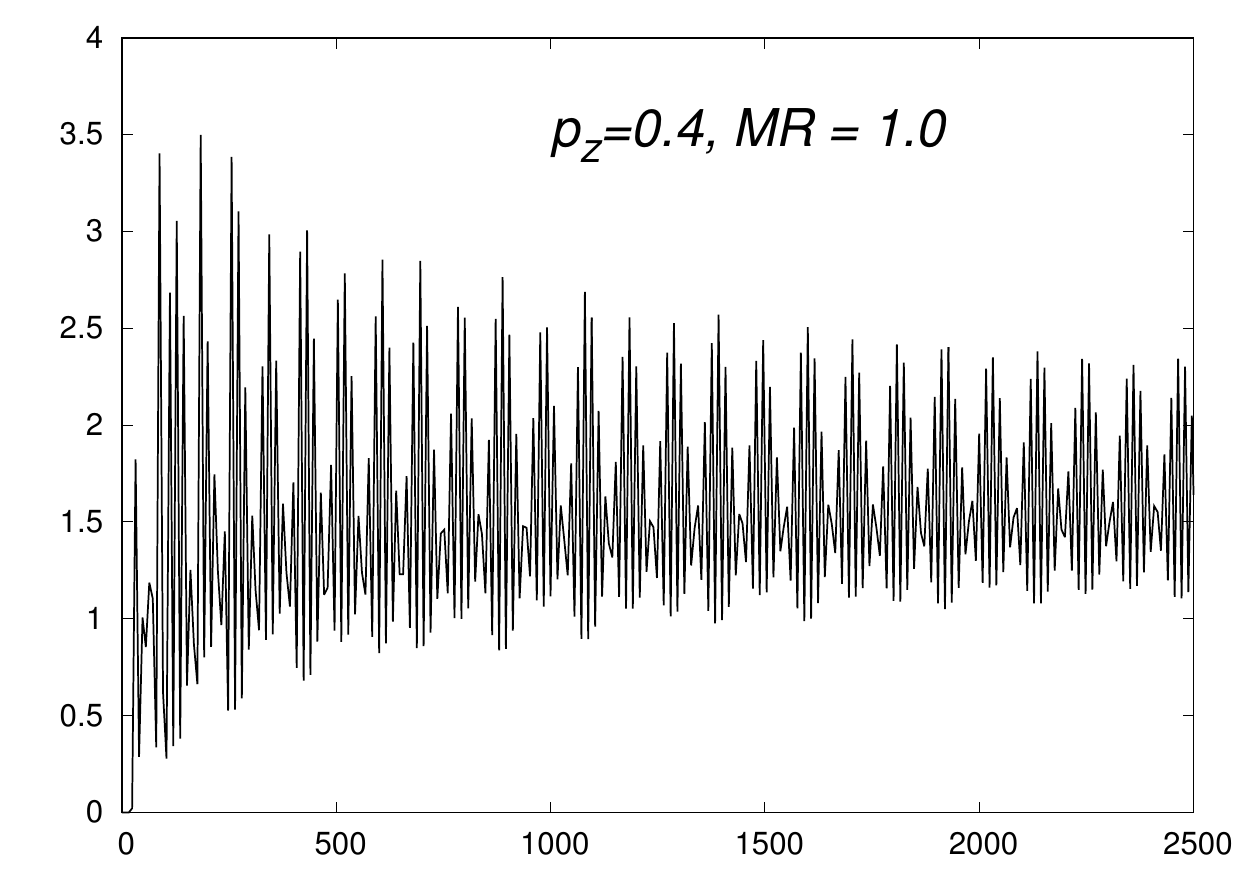}
\includegraphics[width= 3.25cm]{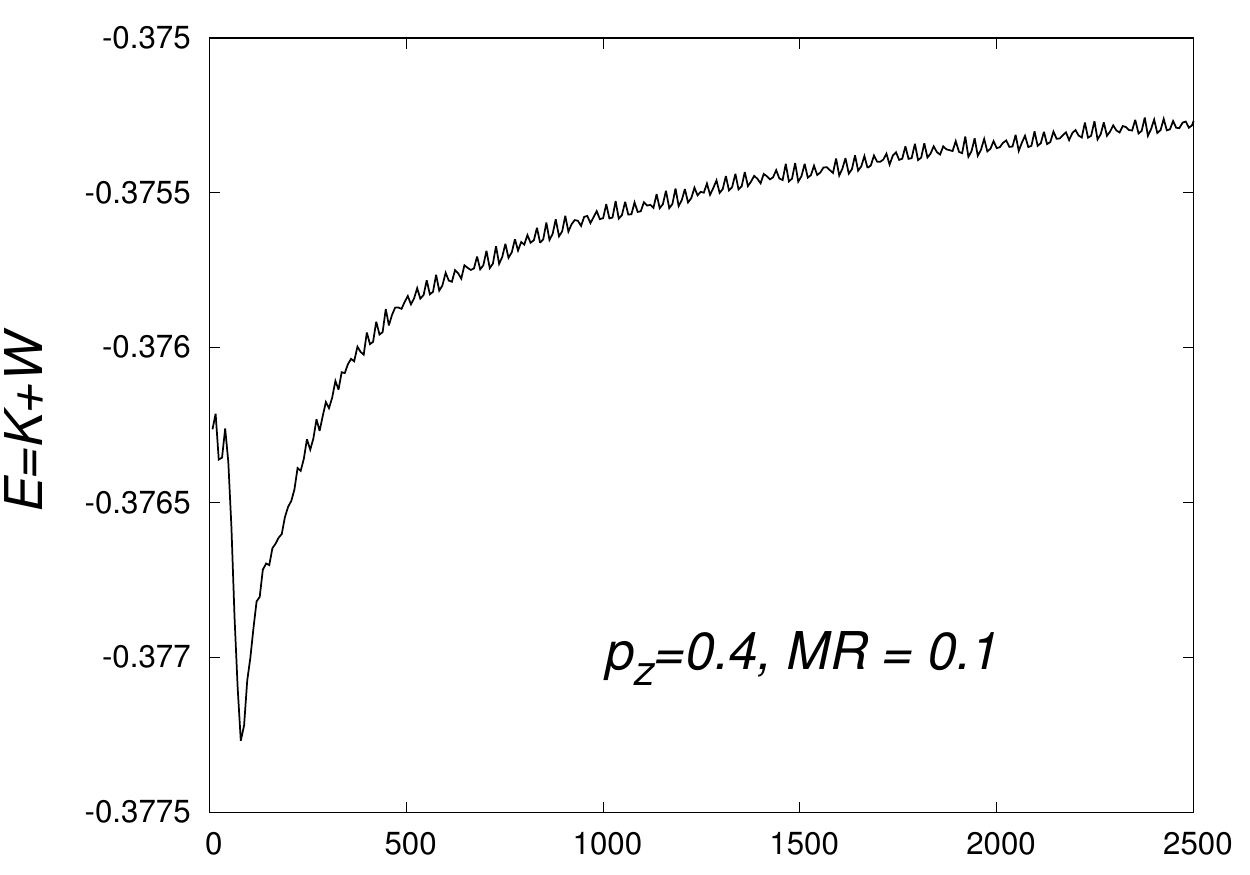}
\includegraphics[width= 3.25cm]{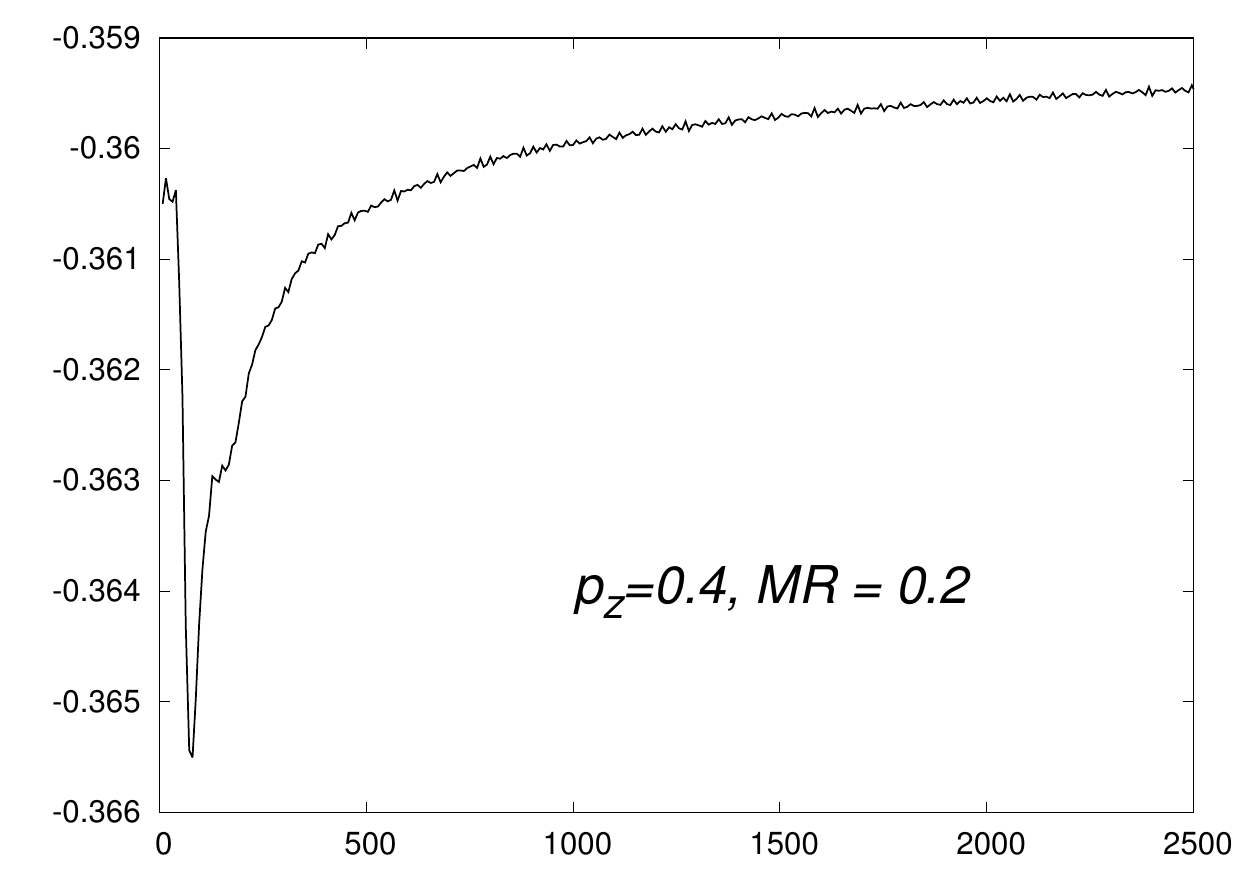}
\includegraphics[width= 3.25cm]{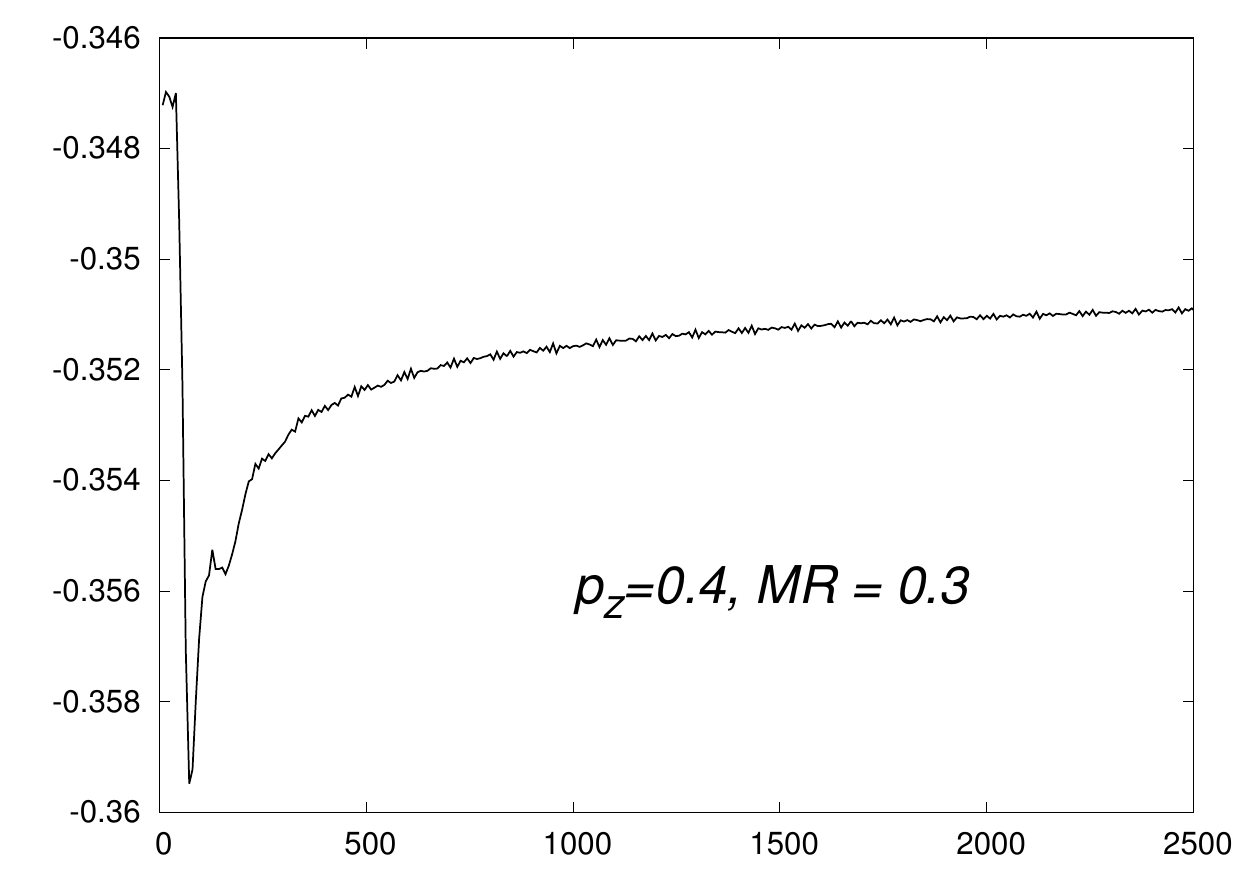}
\includegraphics[width= 3.25cm]{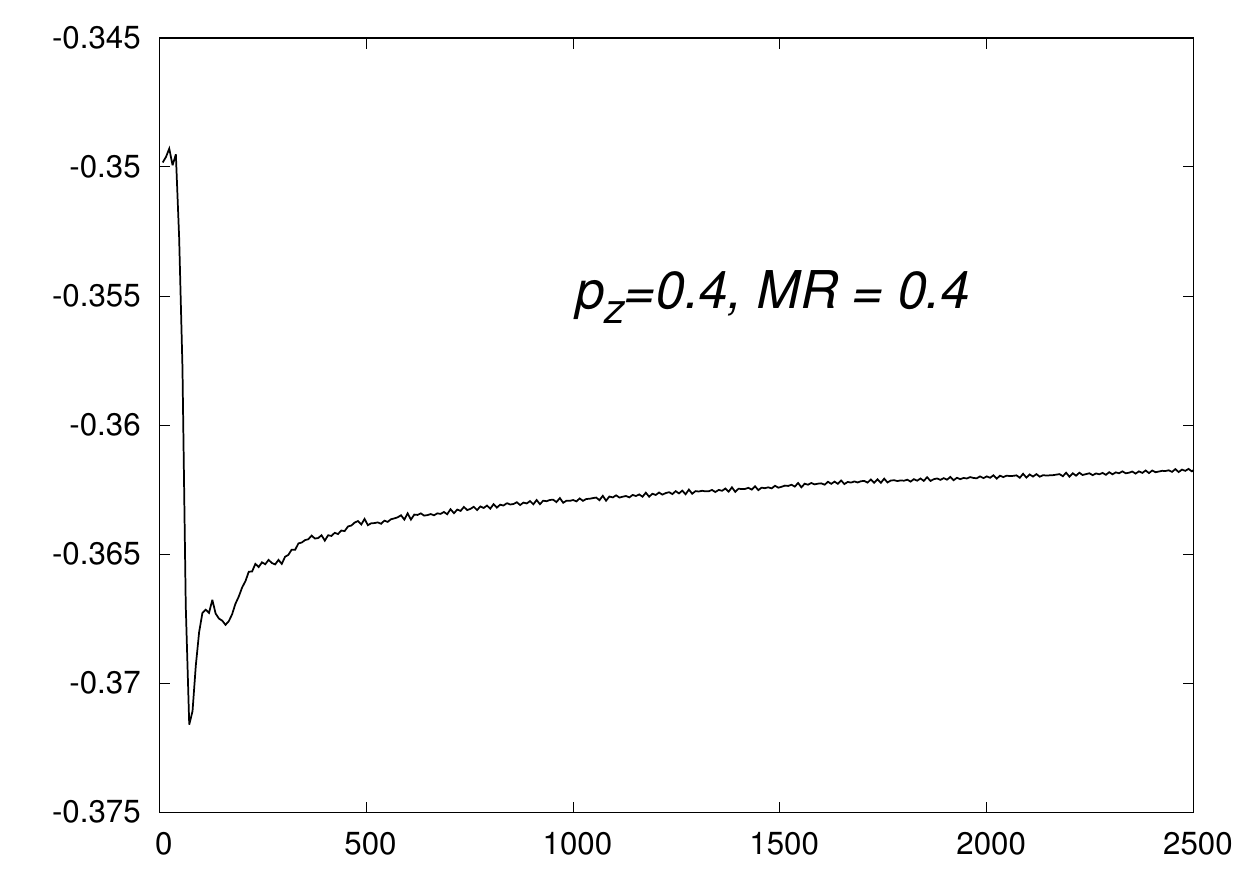}
\includegraphics[width= 3.25cm]{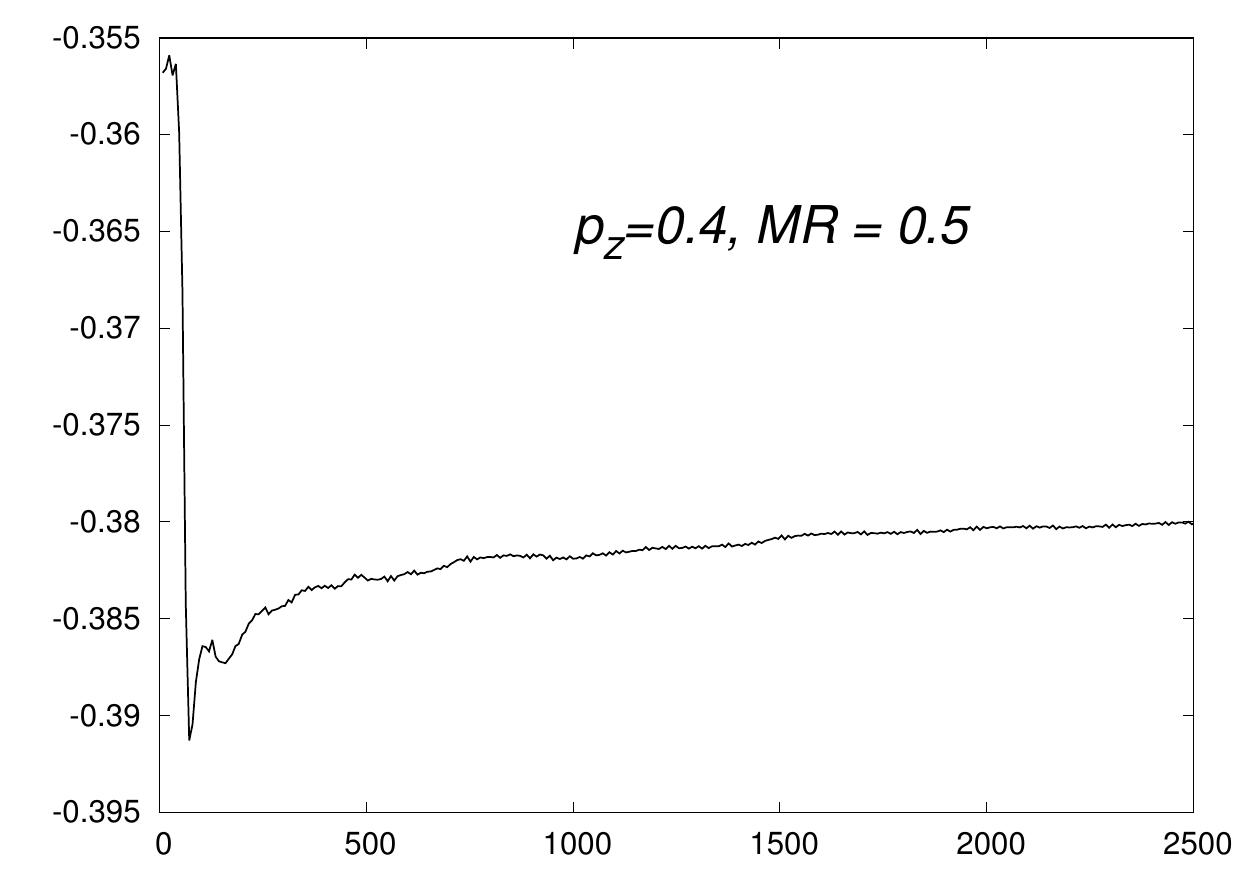}
\includegraphics[width= 3.25cm]{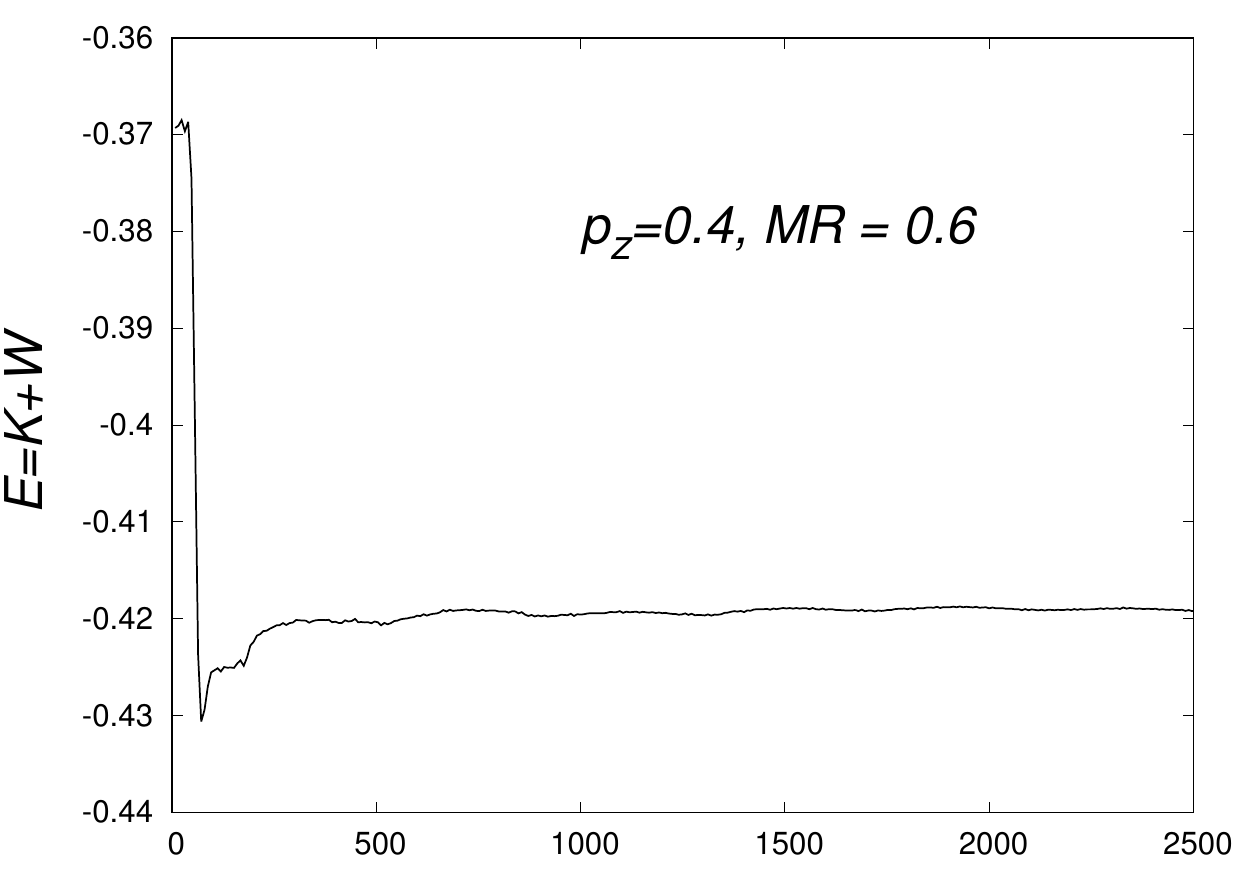}
\includegraphics[width= 3.25cm]{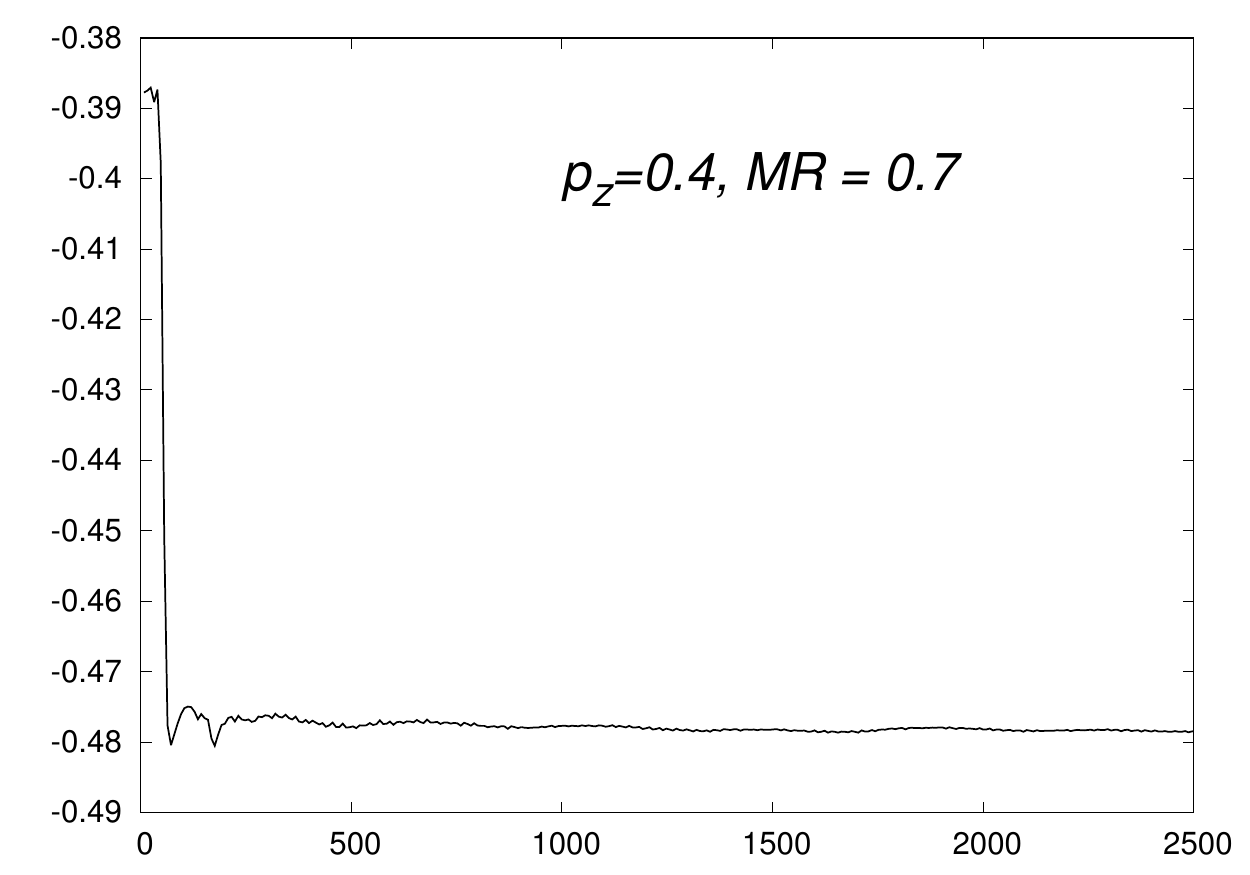}
\includegraphics[width= 3.25cm]{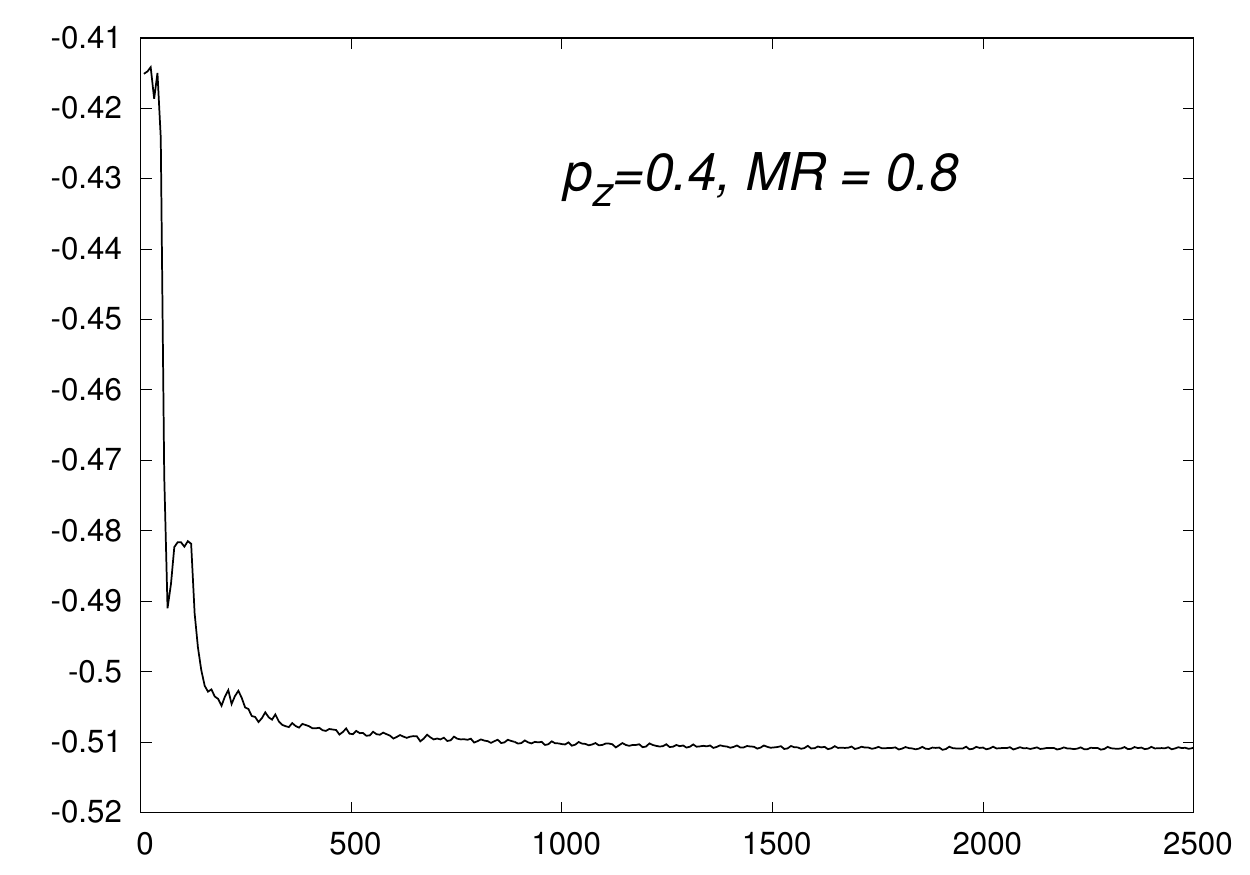}
\includegraphics[width= 3.25cm]{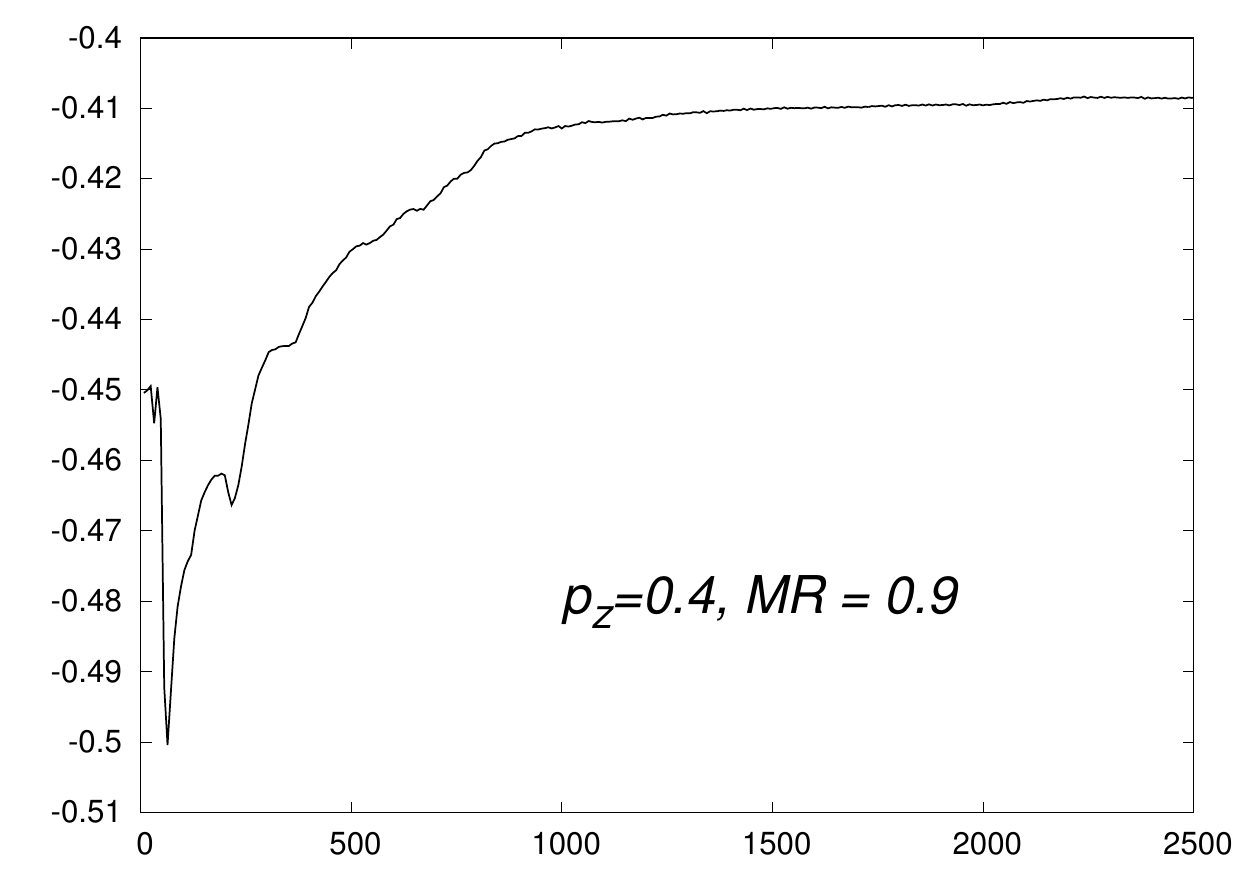}
\includegraphics[width= 3.25cm]{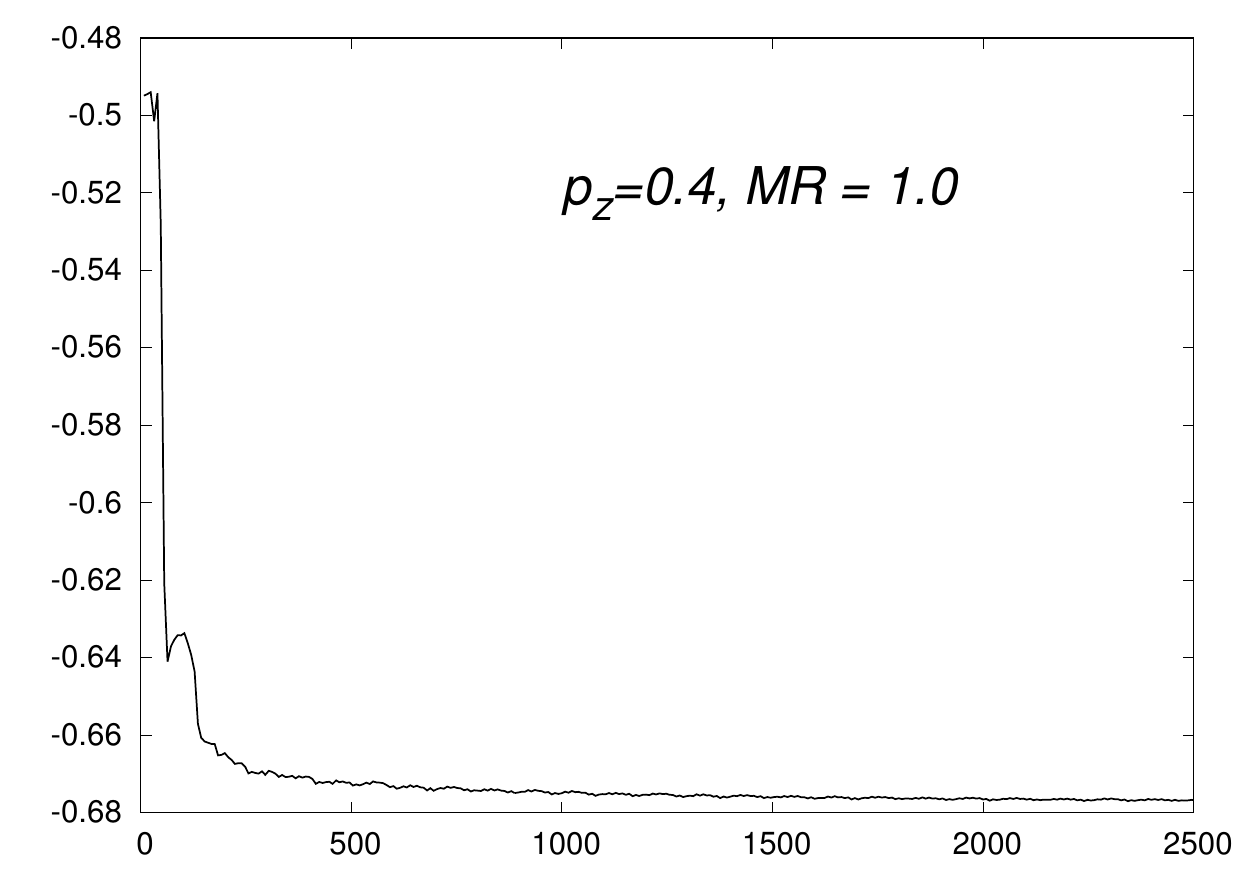}
\caption{Diagnostics of the relaxation process for the case $p_z=0.4$ and  the various mass ratios $MR=0.1,...,1.0$. We show the quantity $2K+W$, the central density where the final structure centers $\rho(0,0)$ and the total energy $E=K+W$ as function of time $t$.}
\label{fig:coolingpz0_4}
\end{figure*}

\begin{figure*}
\centering
\includegraphics[width= 3.25cm]{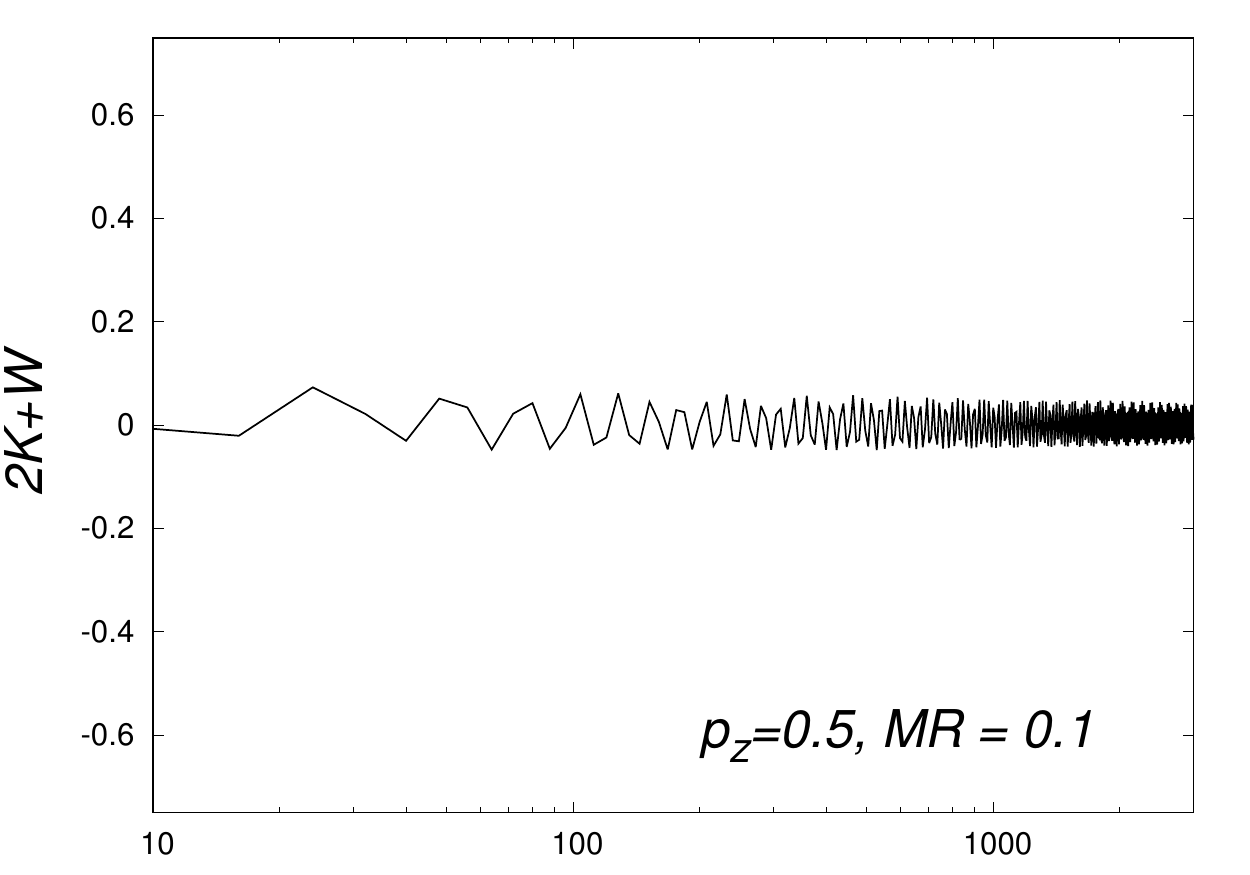}
\includegraphics[width= 3.25cm]{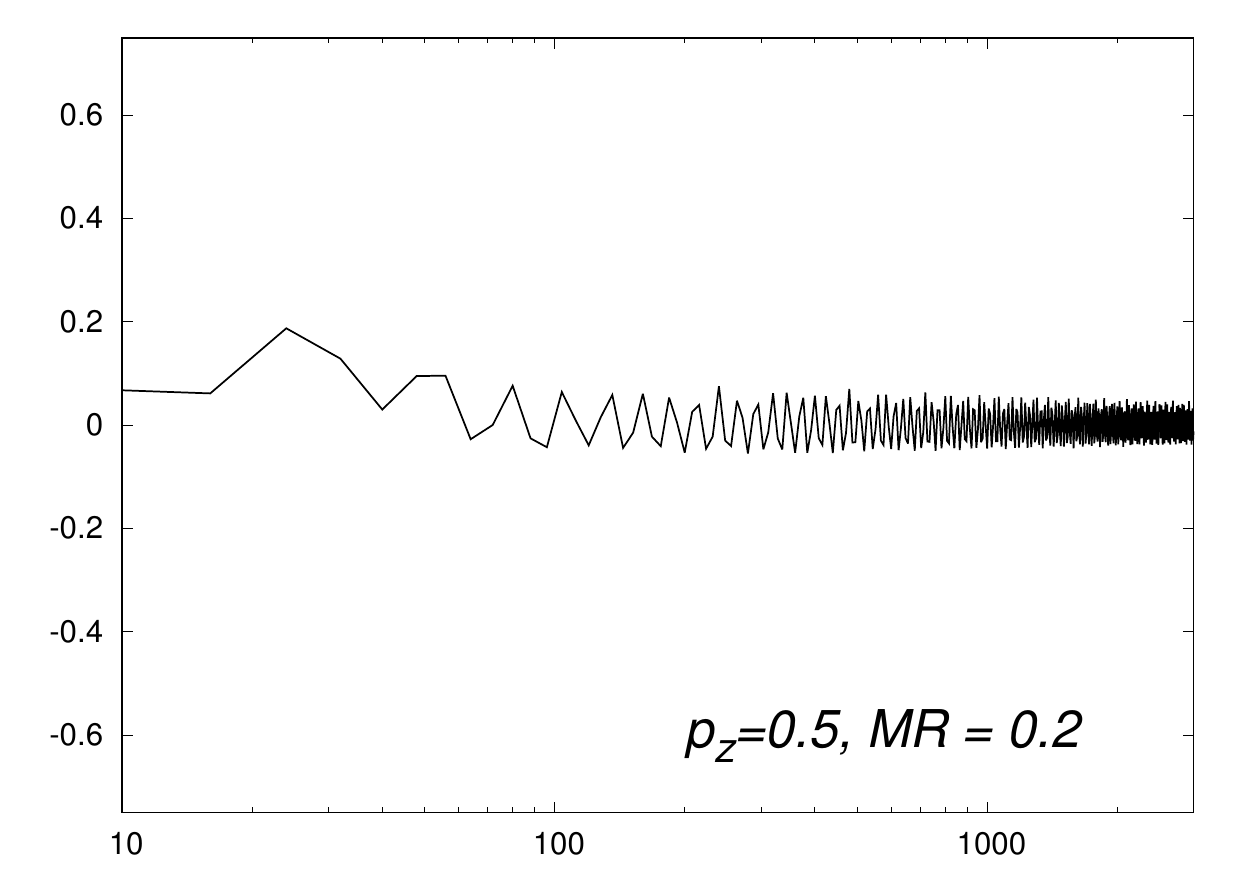}
\includegraphics[width= 3.25cm]{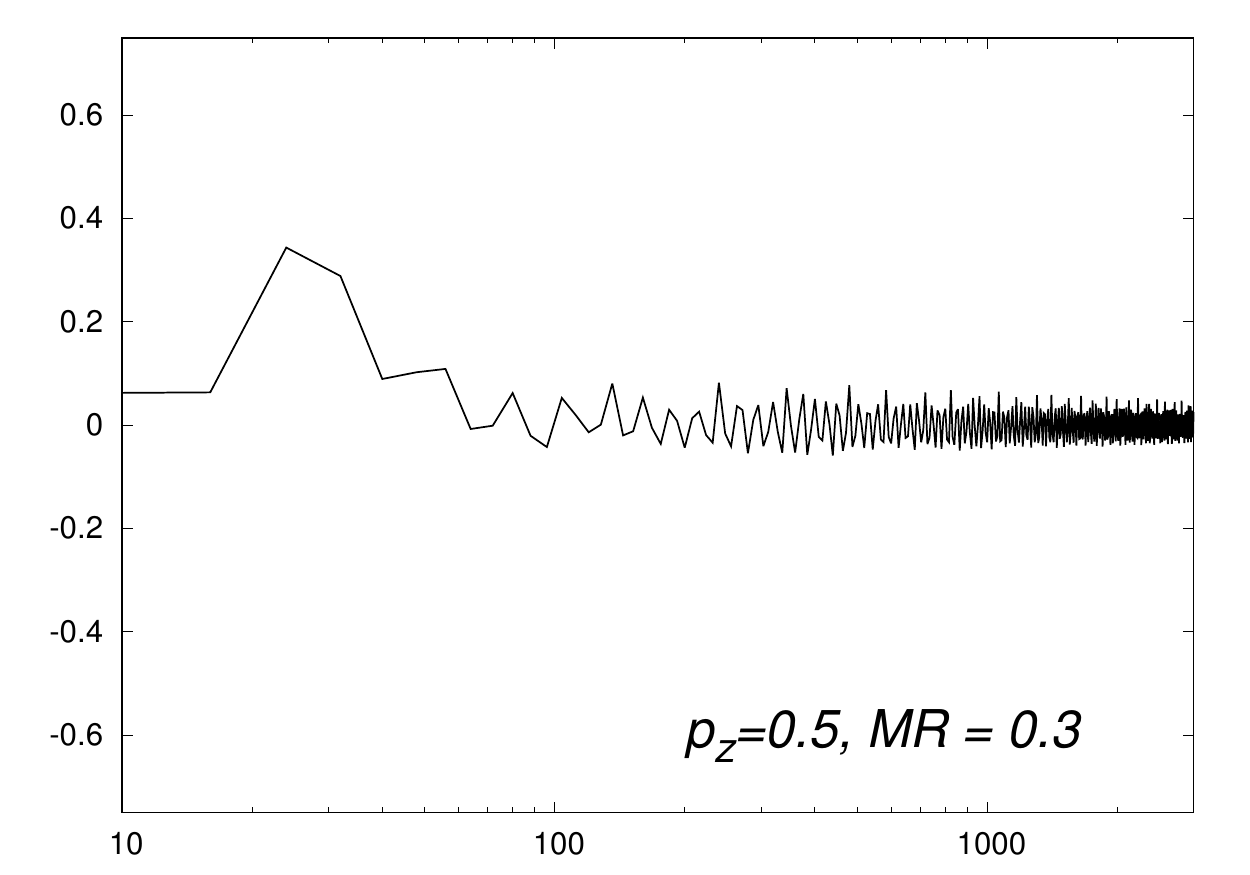}
\includegraphics[width= 3.25cm]{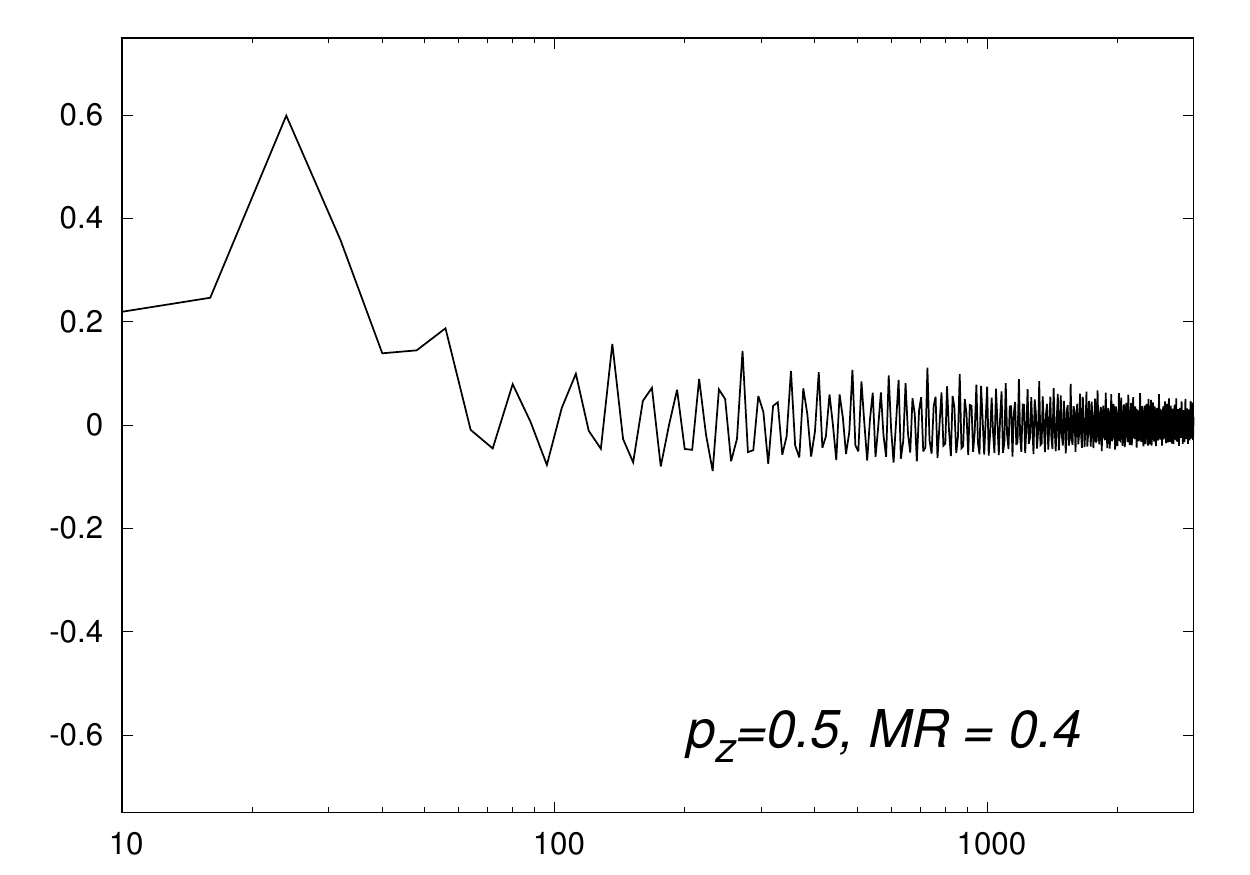}
\includegraphics[width= 3.25cm]{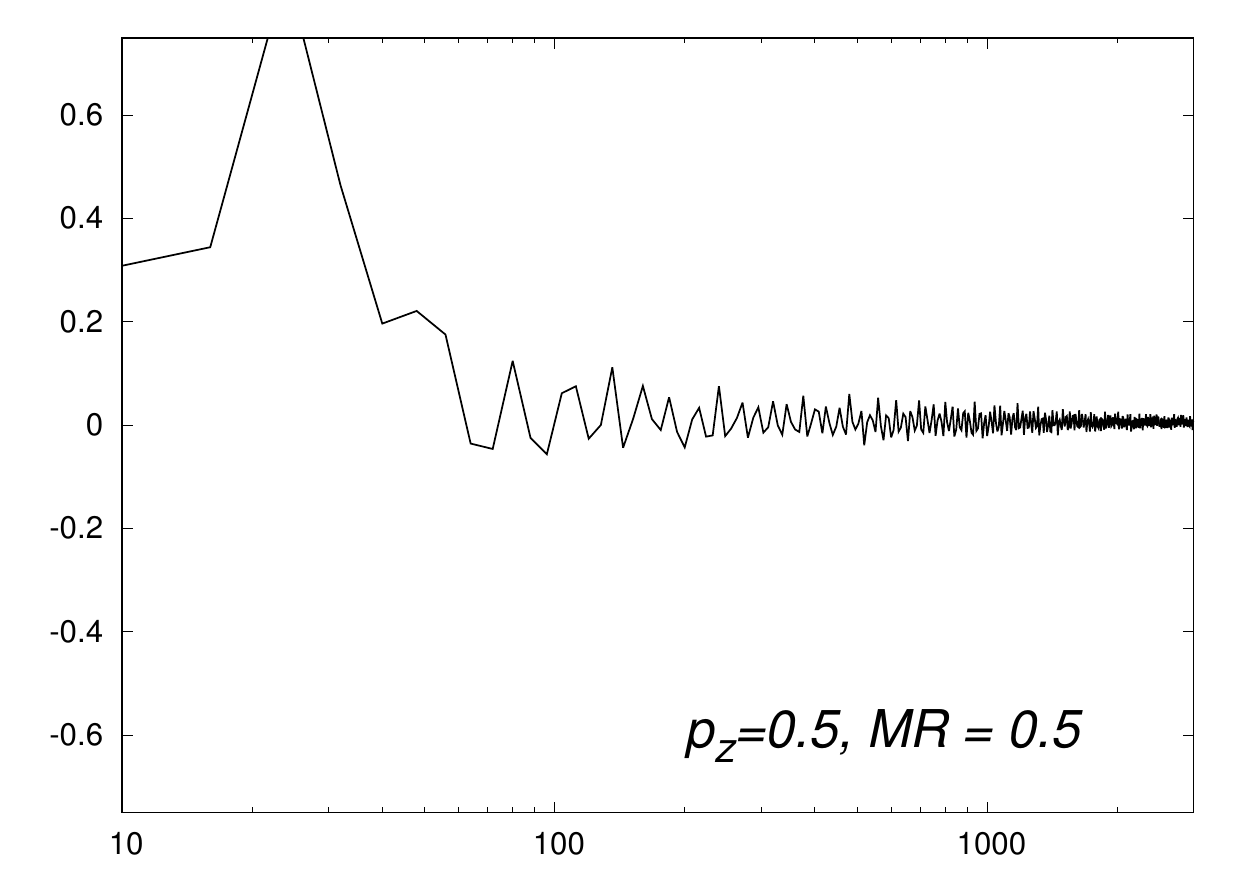}
\includegraphics[width= 3.25cm]{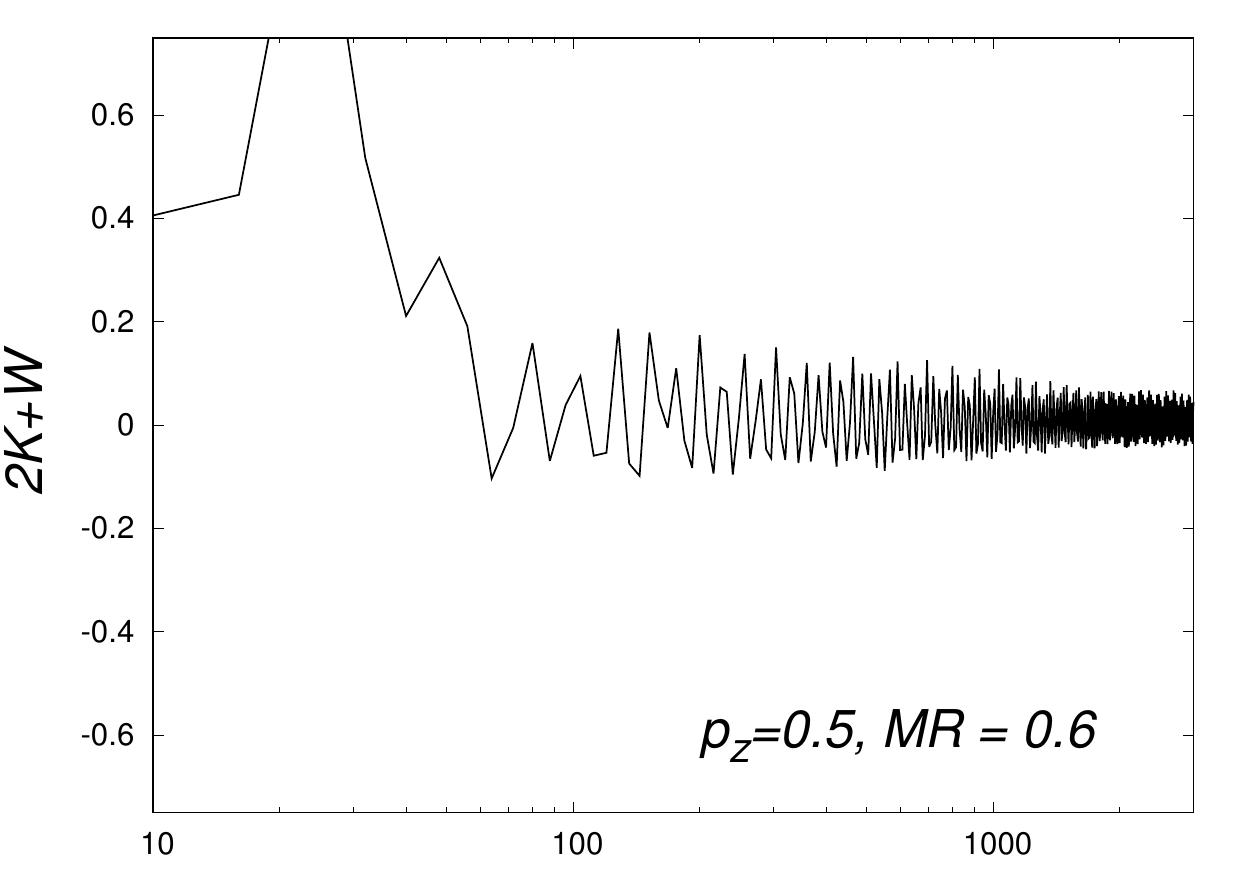}
\includegraphics[width= 3.25cm]{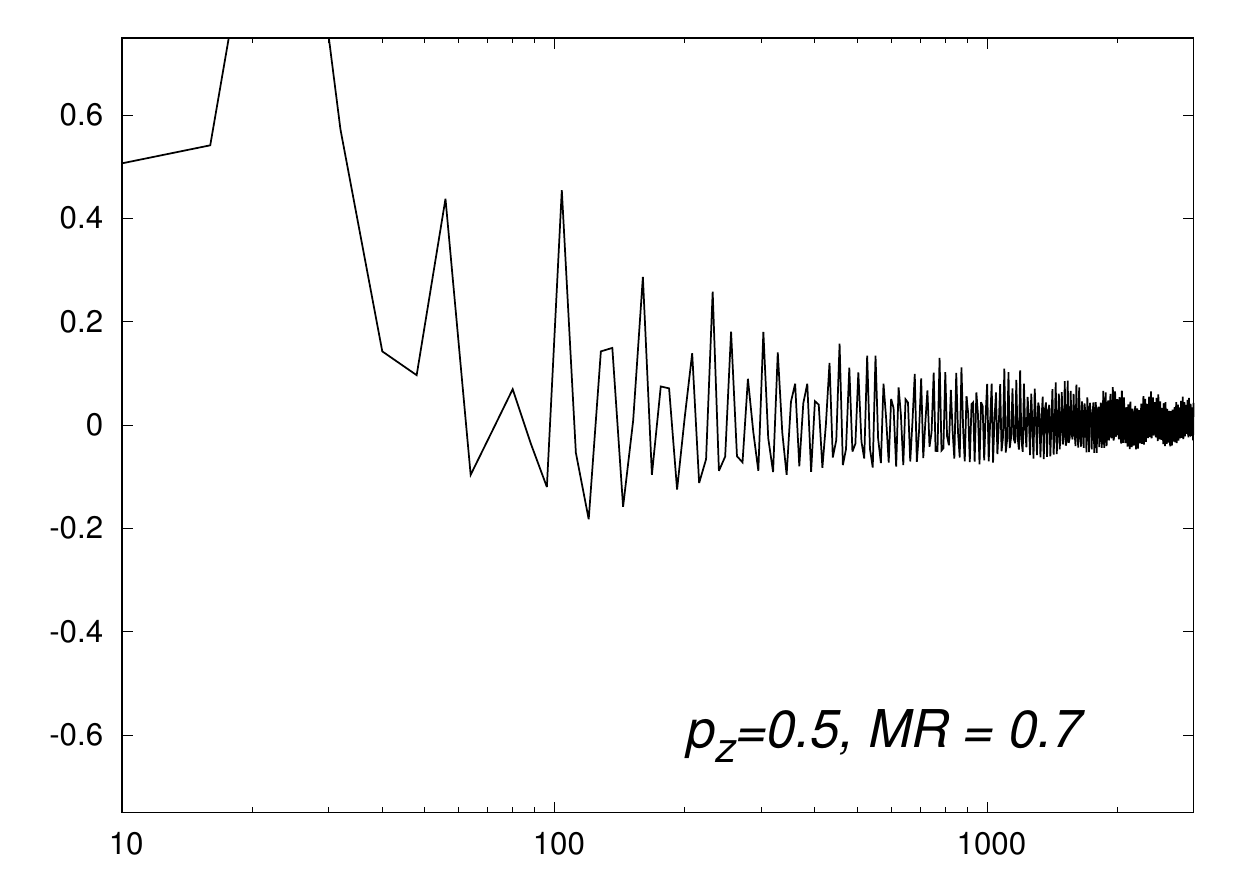}
\includegraphics[width= 3.25cm]{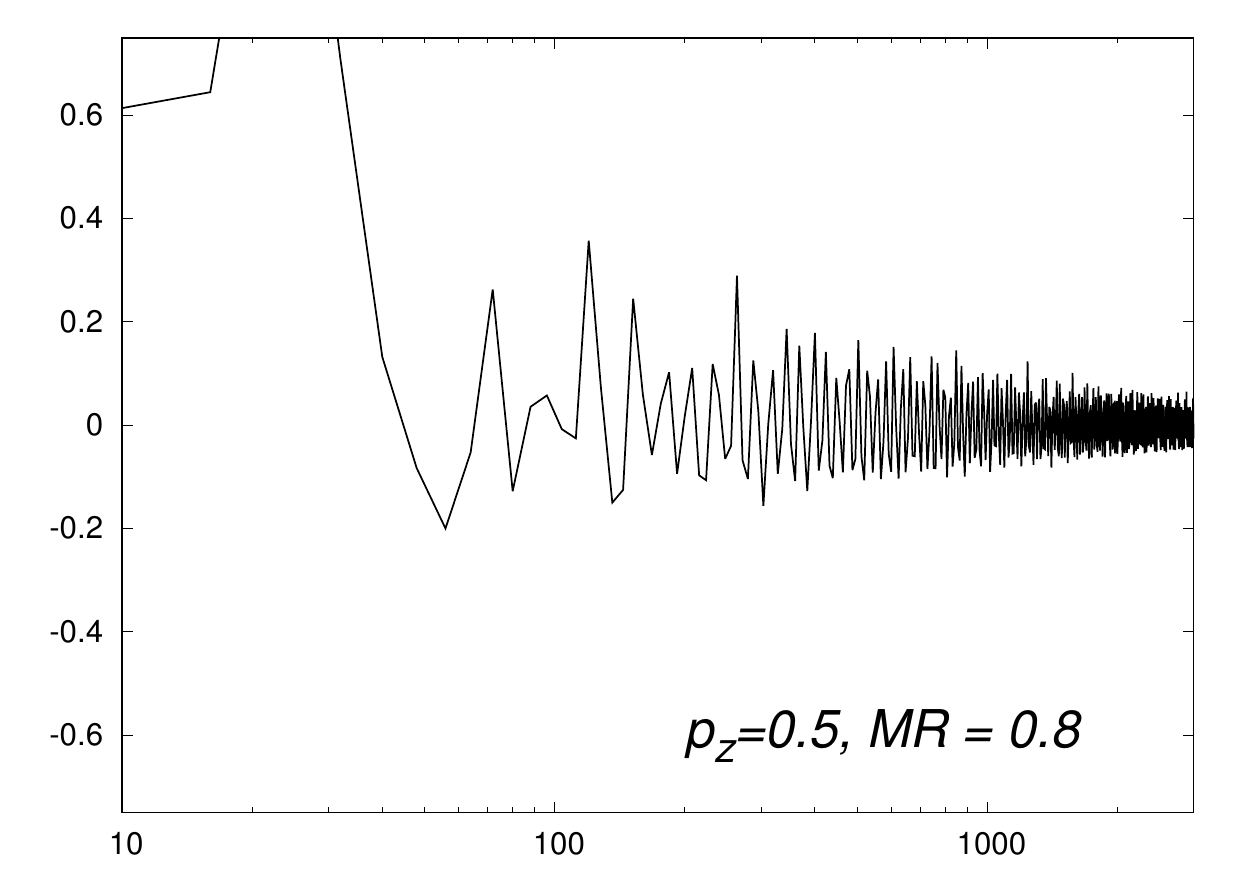}
\includegraphics[width= 3.25cm]{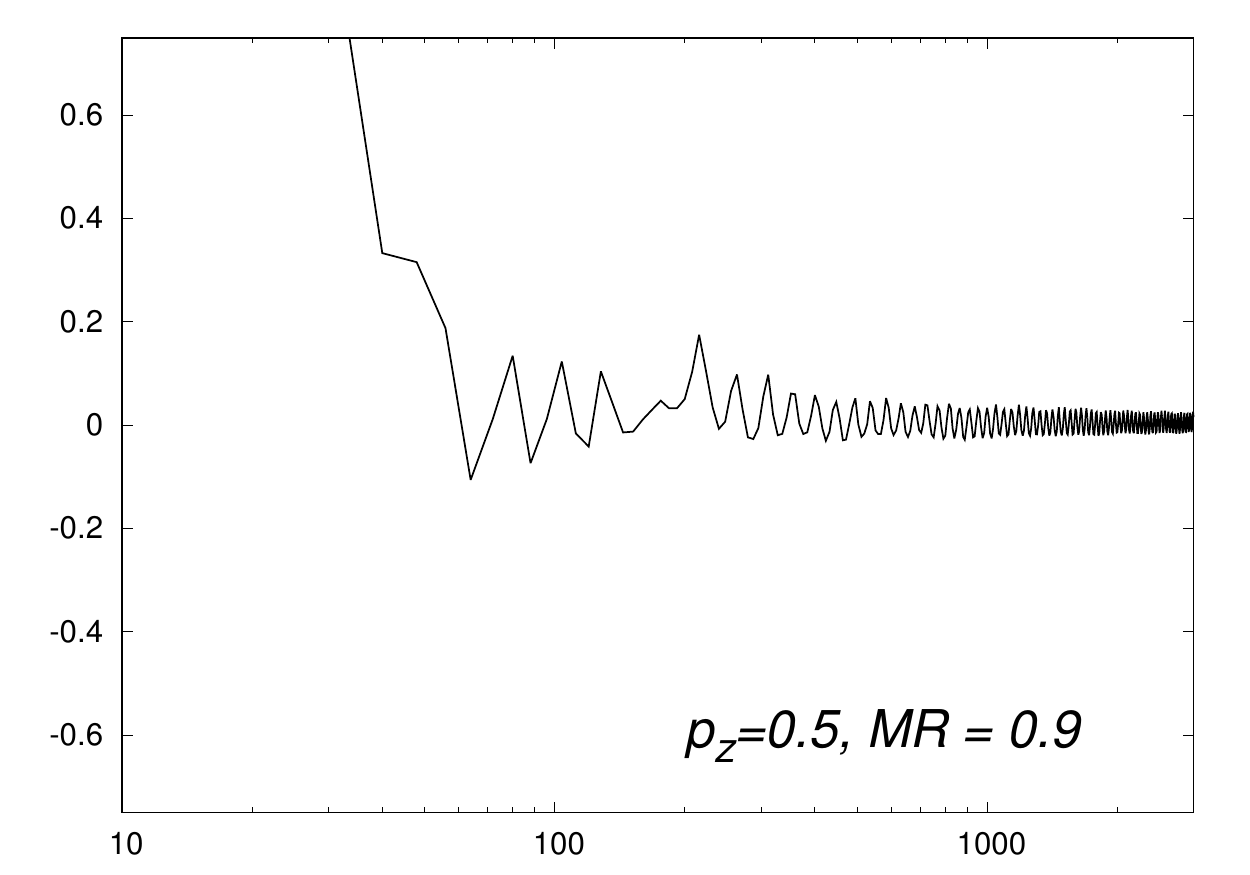}
\includegraphics[width= 3.25cm]{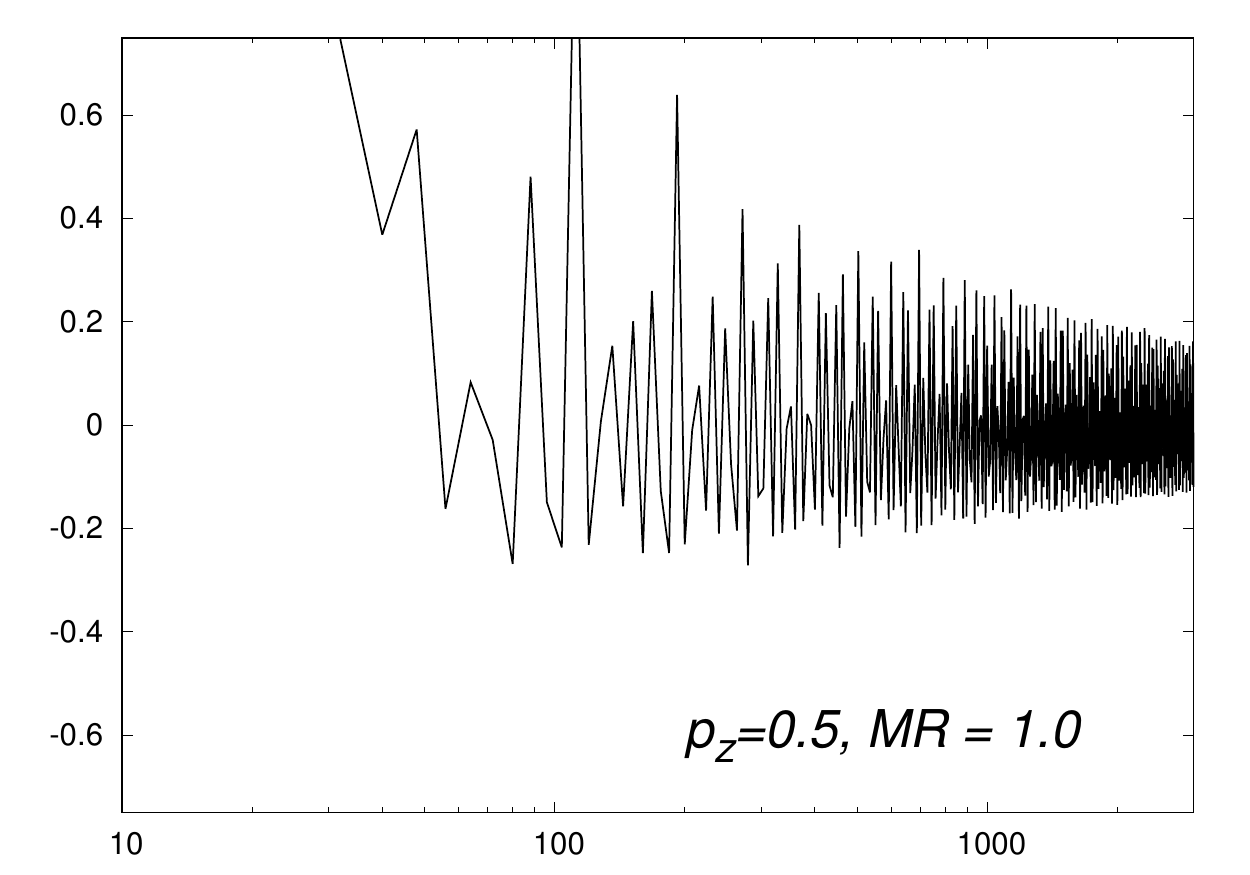}
\includegraphics[width= 3.25cm]{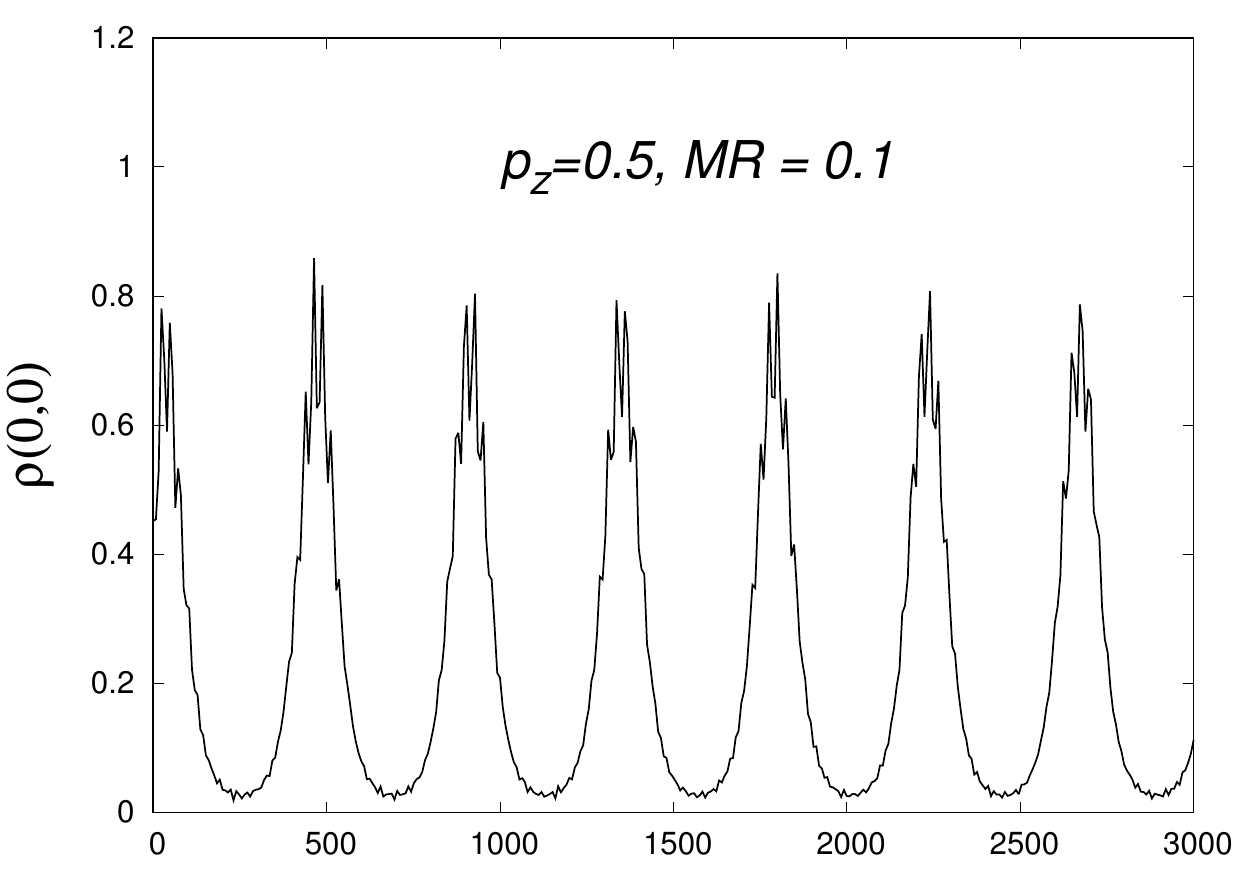}
\includegraphics[width= 3.25cm]{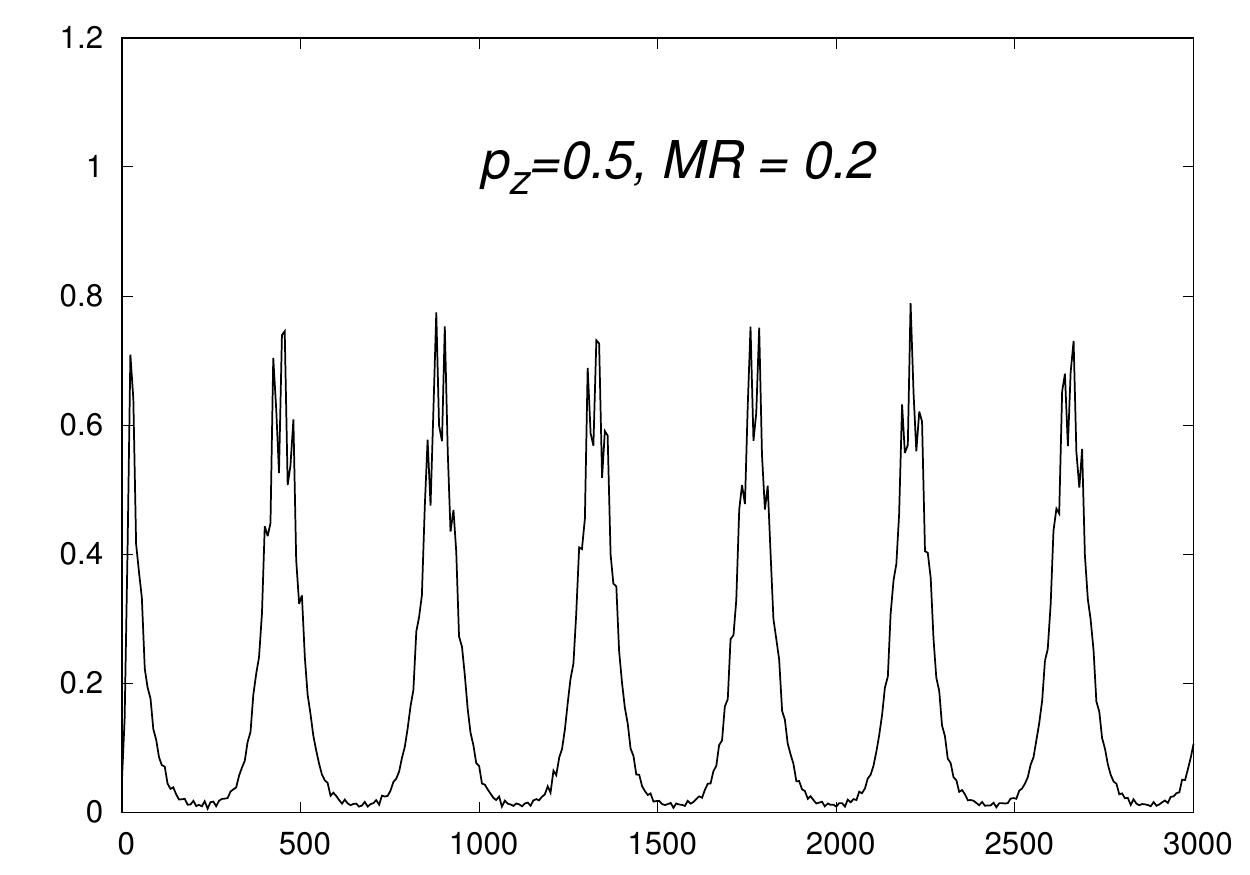}
\includegraphics[width= 3.25cm]{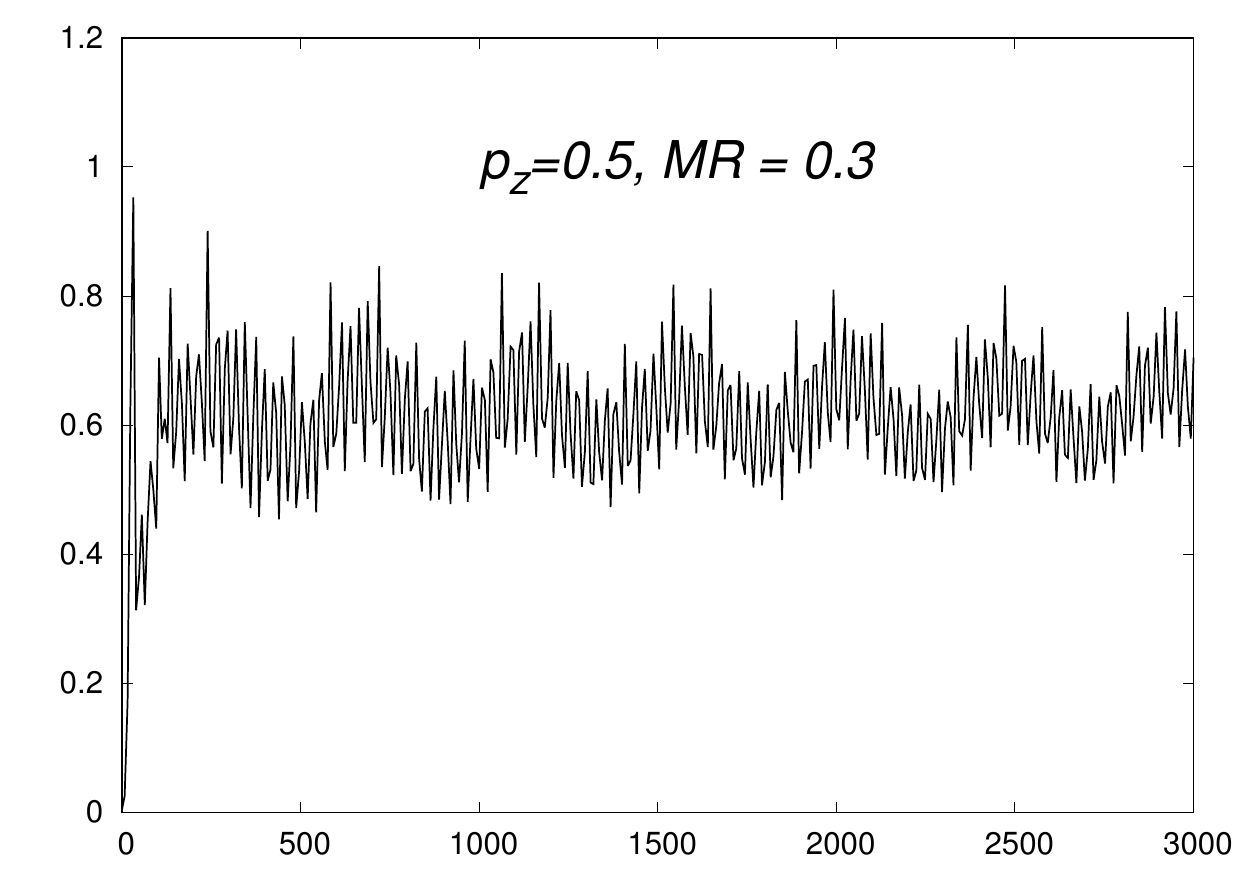}
\includegraphics[width= 3.25cm]{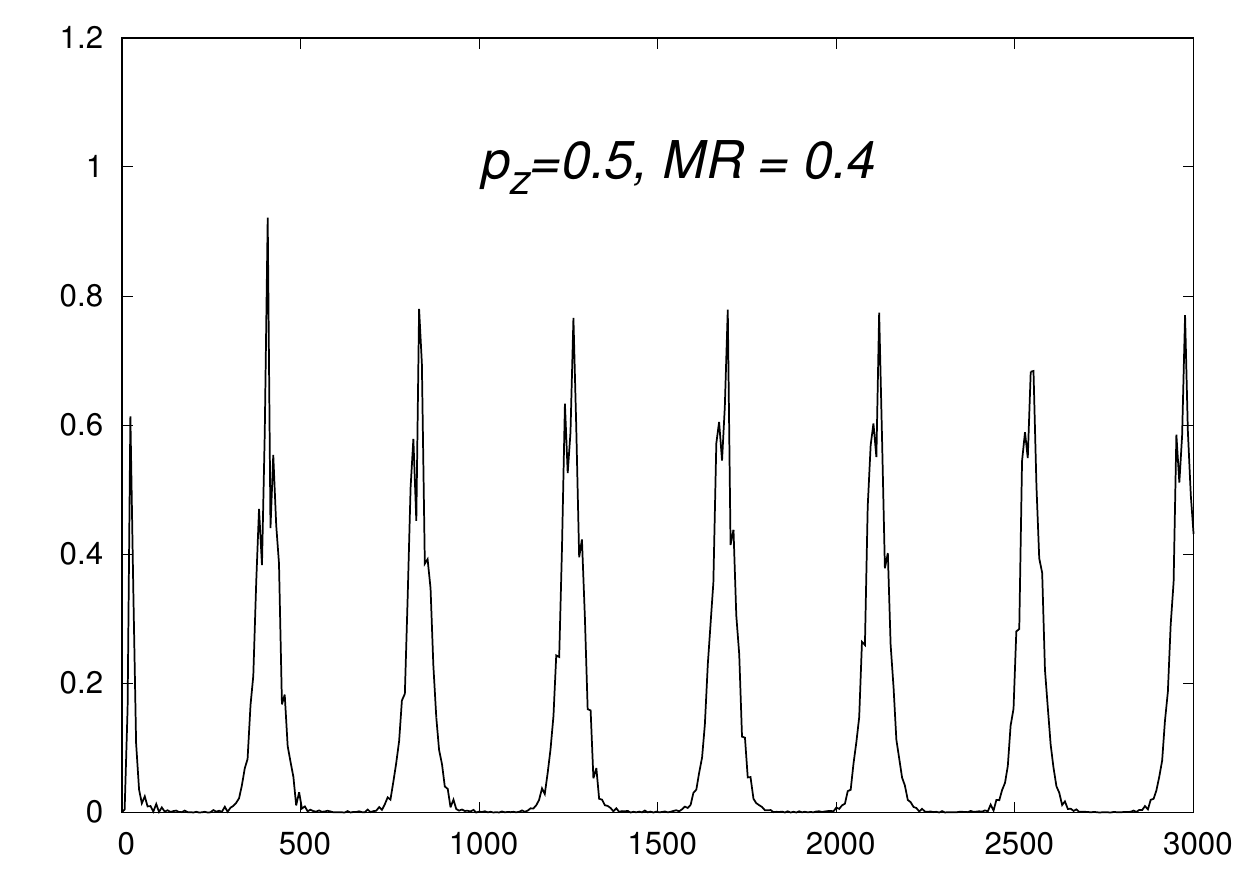}
\includegraphics[width= 3.25cm]{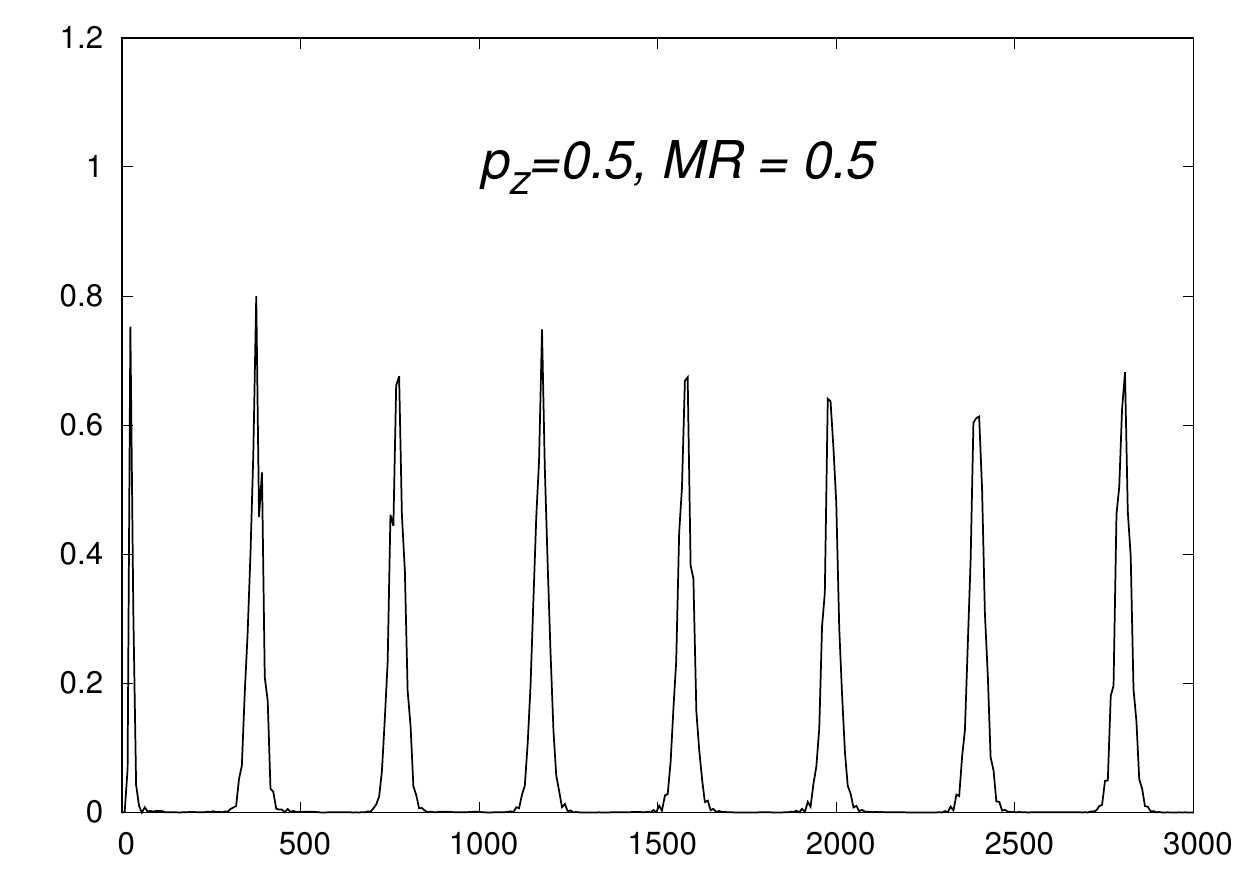}
\includegraphics[width= 3.25cm]{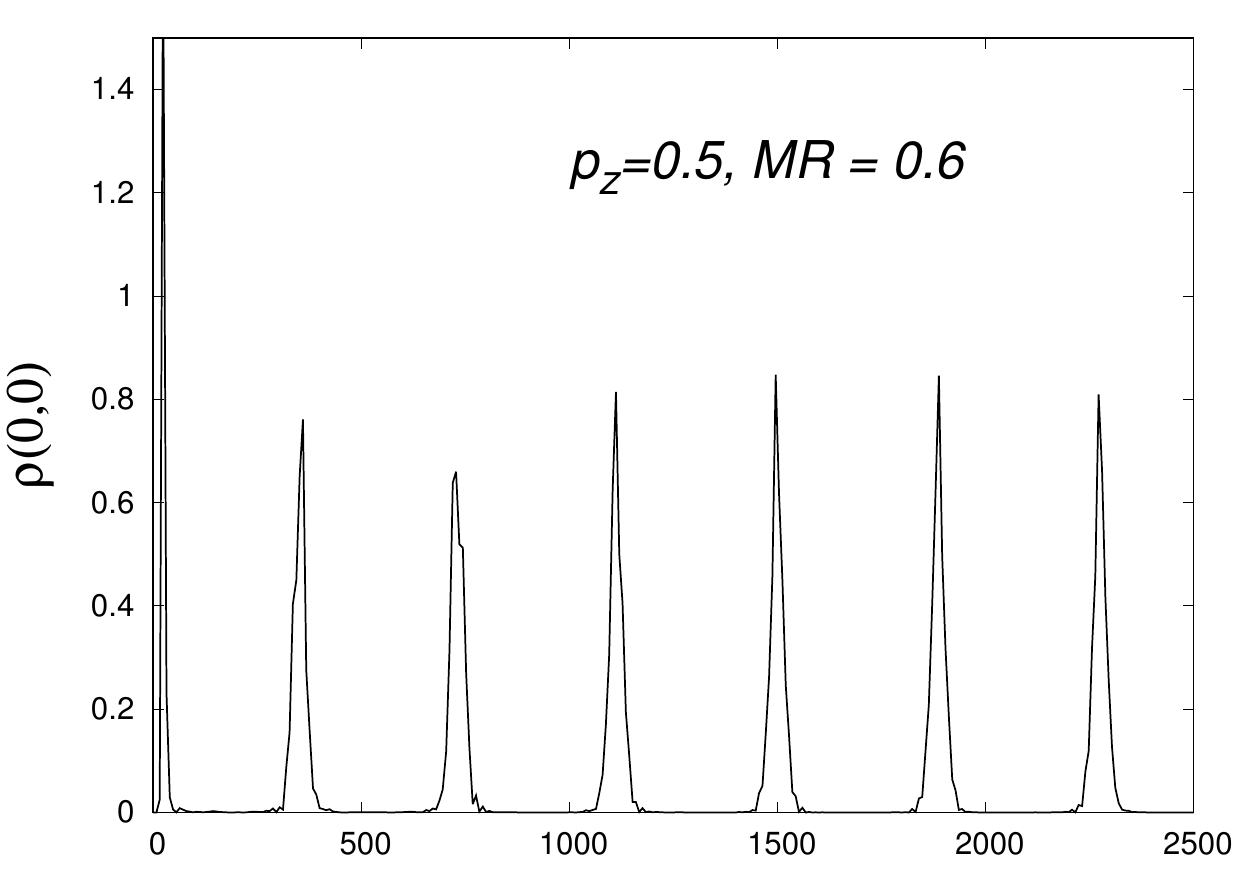}
\includegraphics[width= 3.25cm]{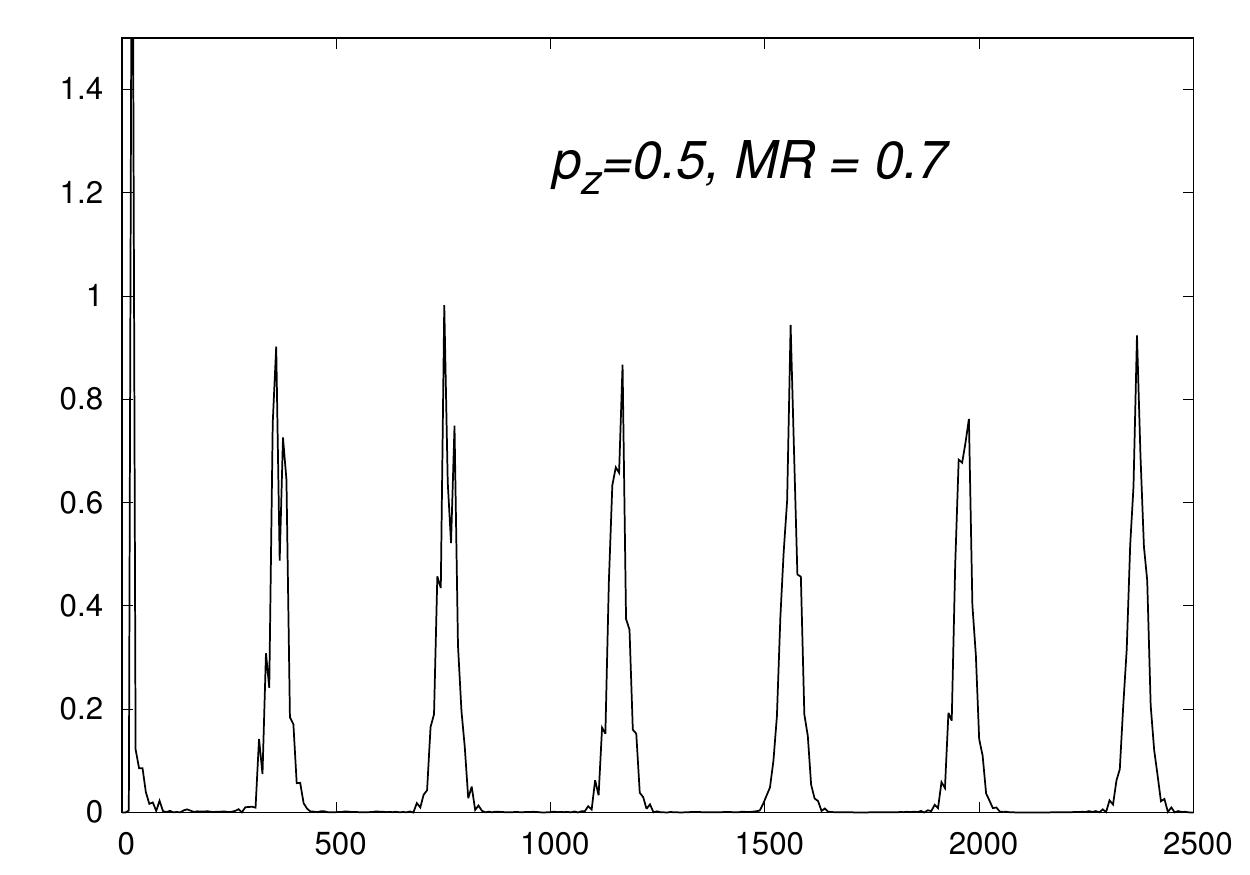}
\includegraphics[width= 3.25cm]{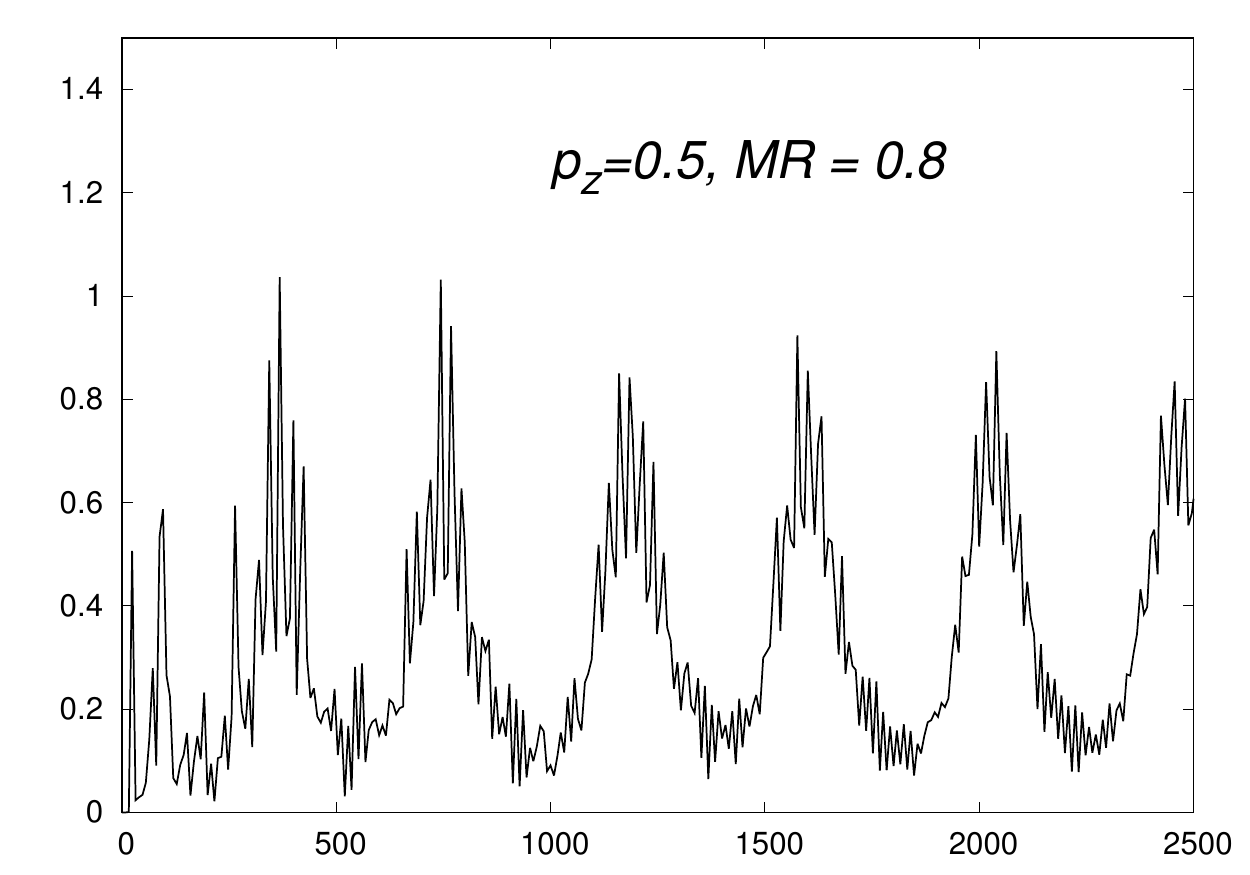}
\includegraphics[width= 3.25cm]{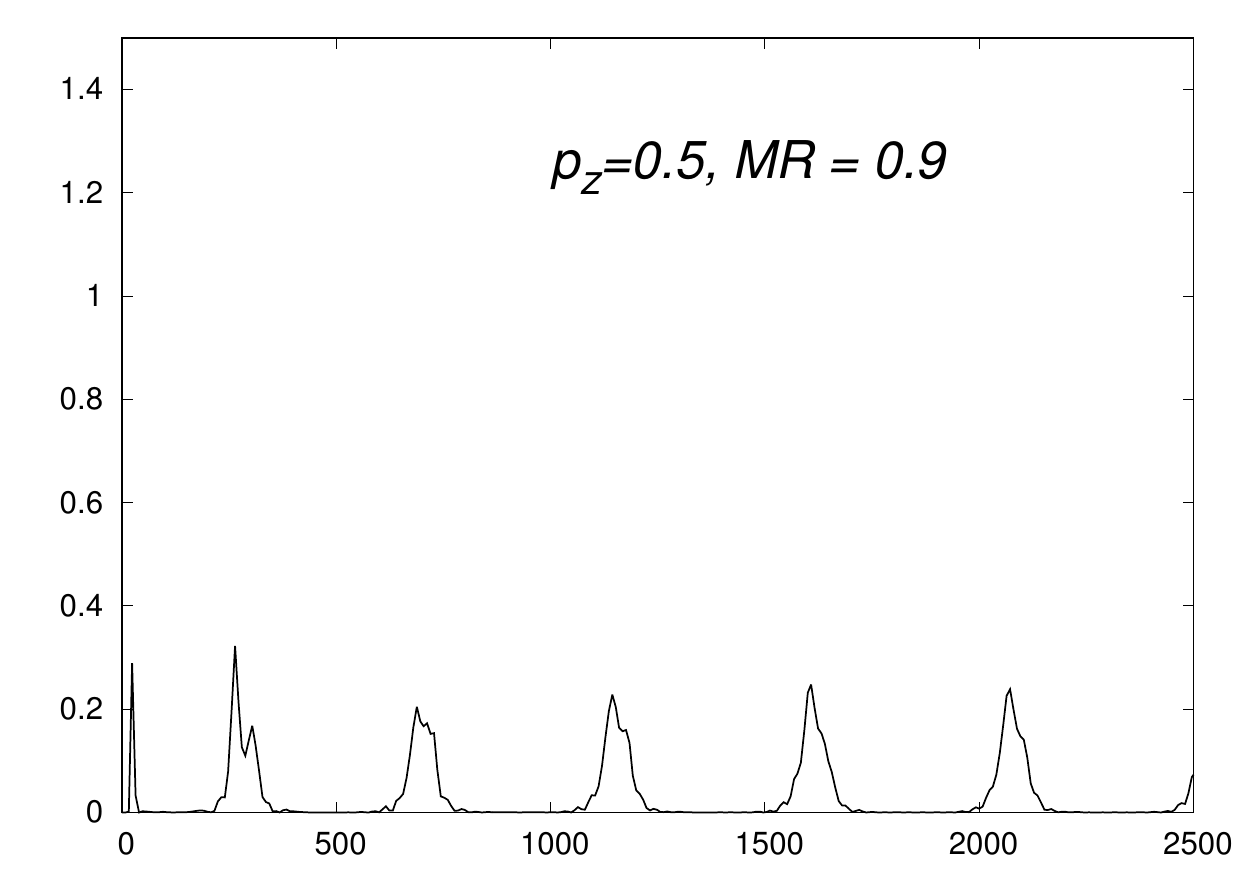}
\includegraphics[width= 3.25cm]{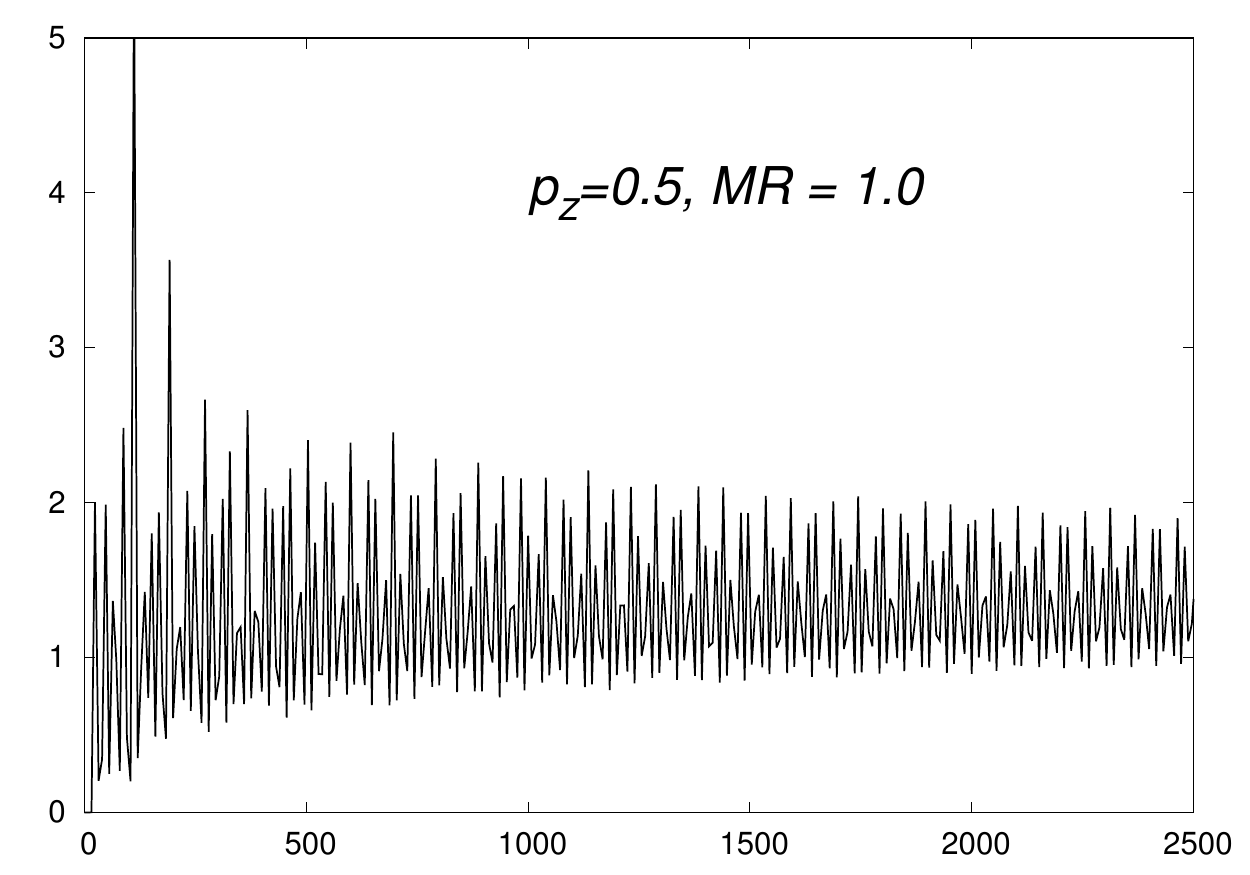}
\includegraphics[width= 3.25cm]{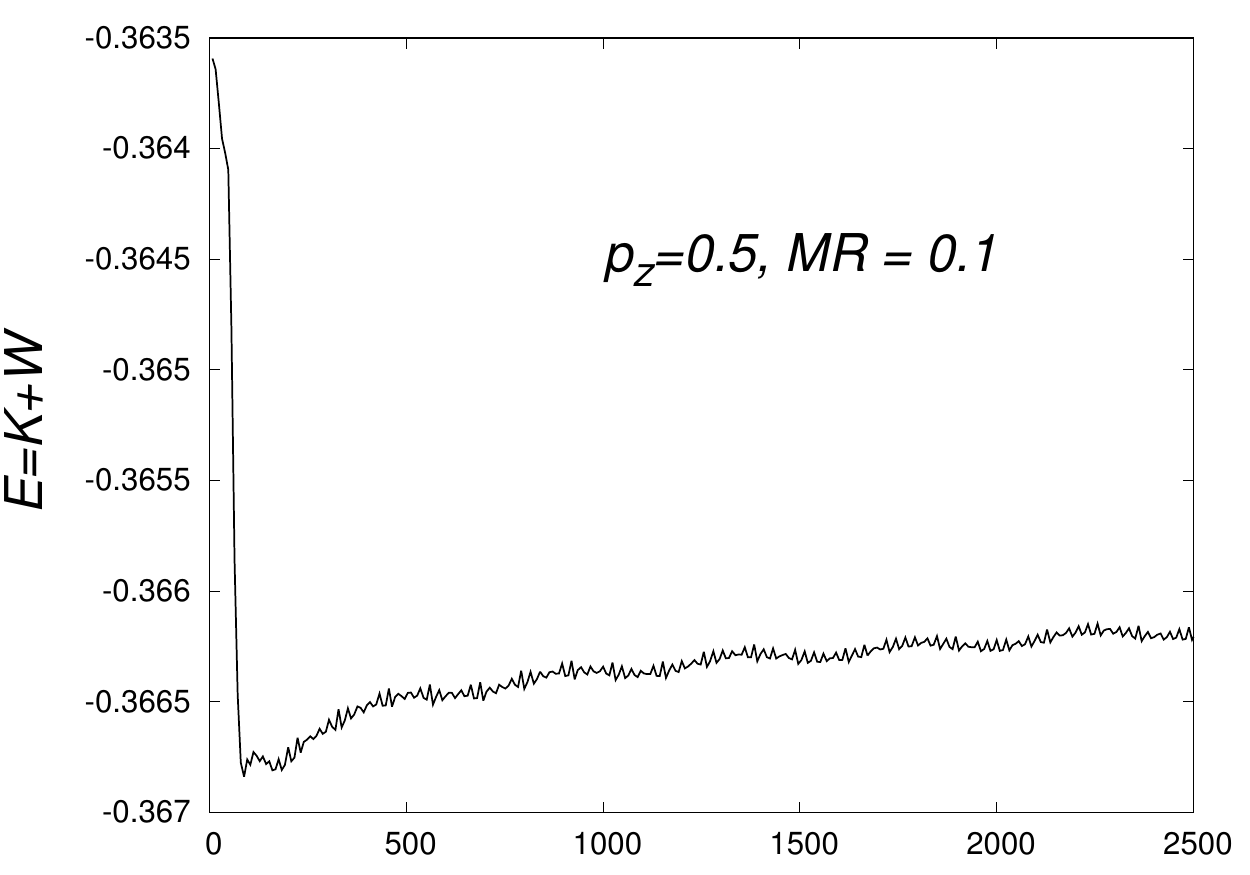}
\includegraphics[width= 3.25cm]{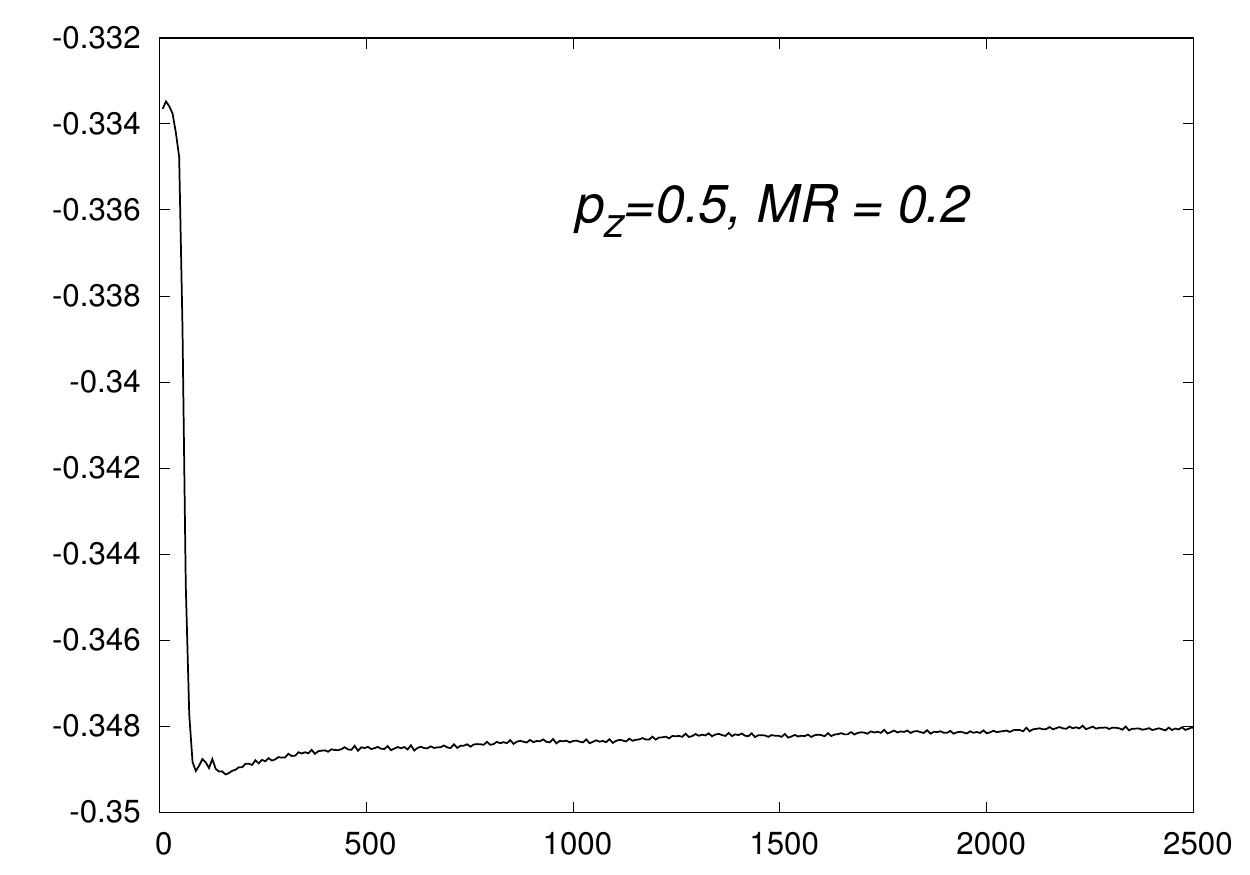}
\includegraphics[width= 3.25cm]{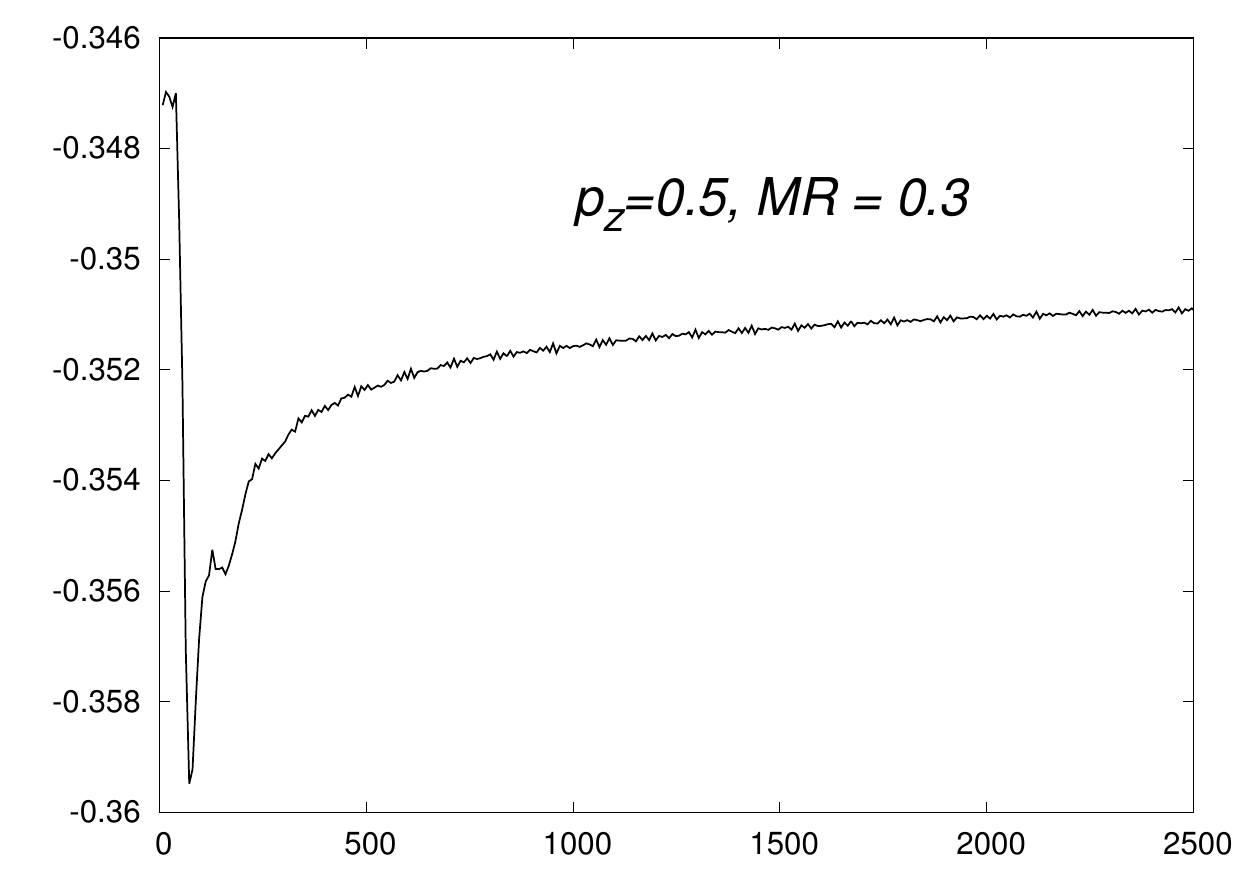}
\includegraphics[width= 3.25cm]{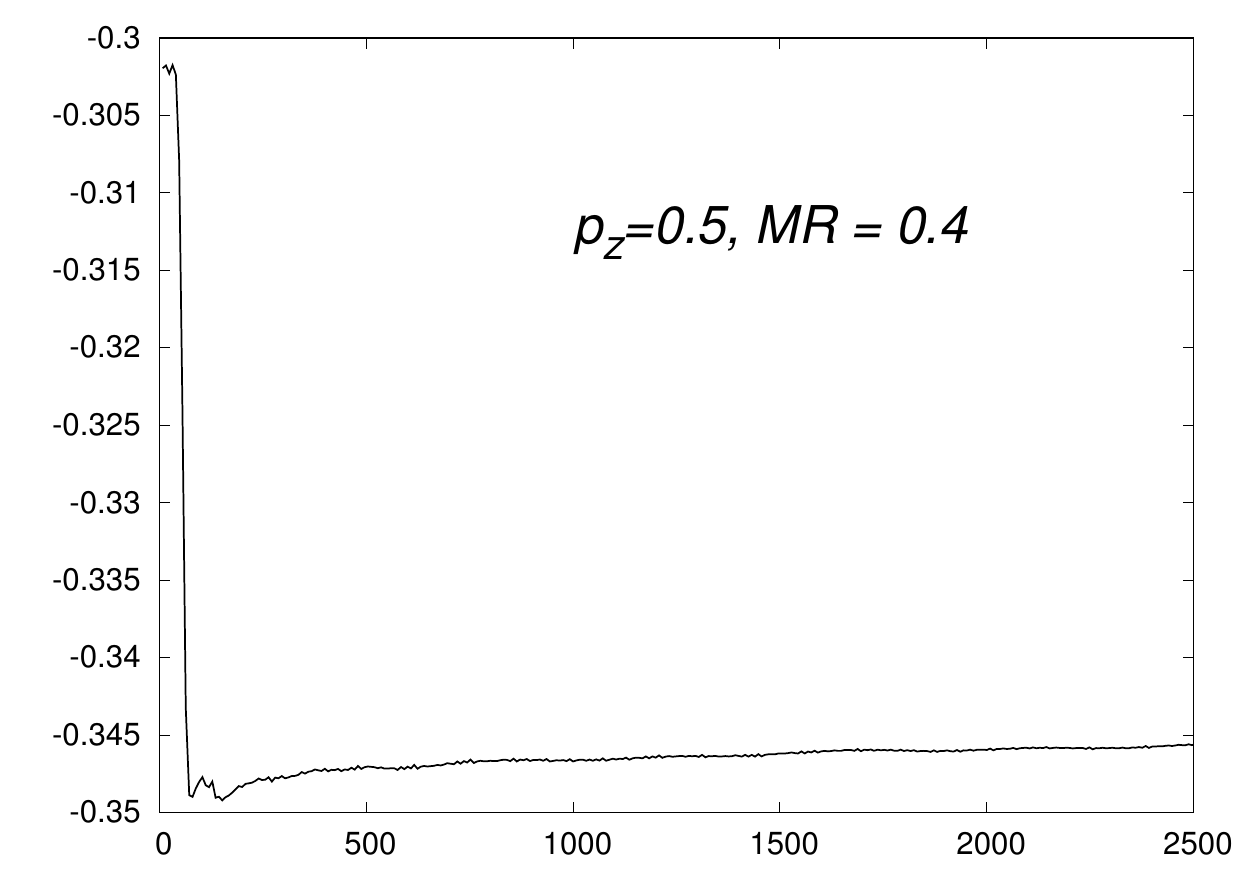}
\includegraphics[width= 3.25cm]{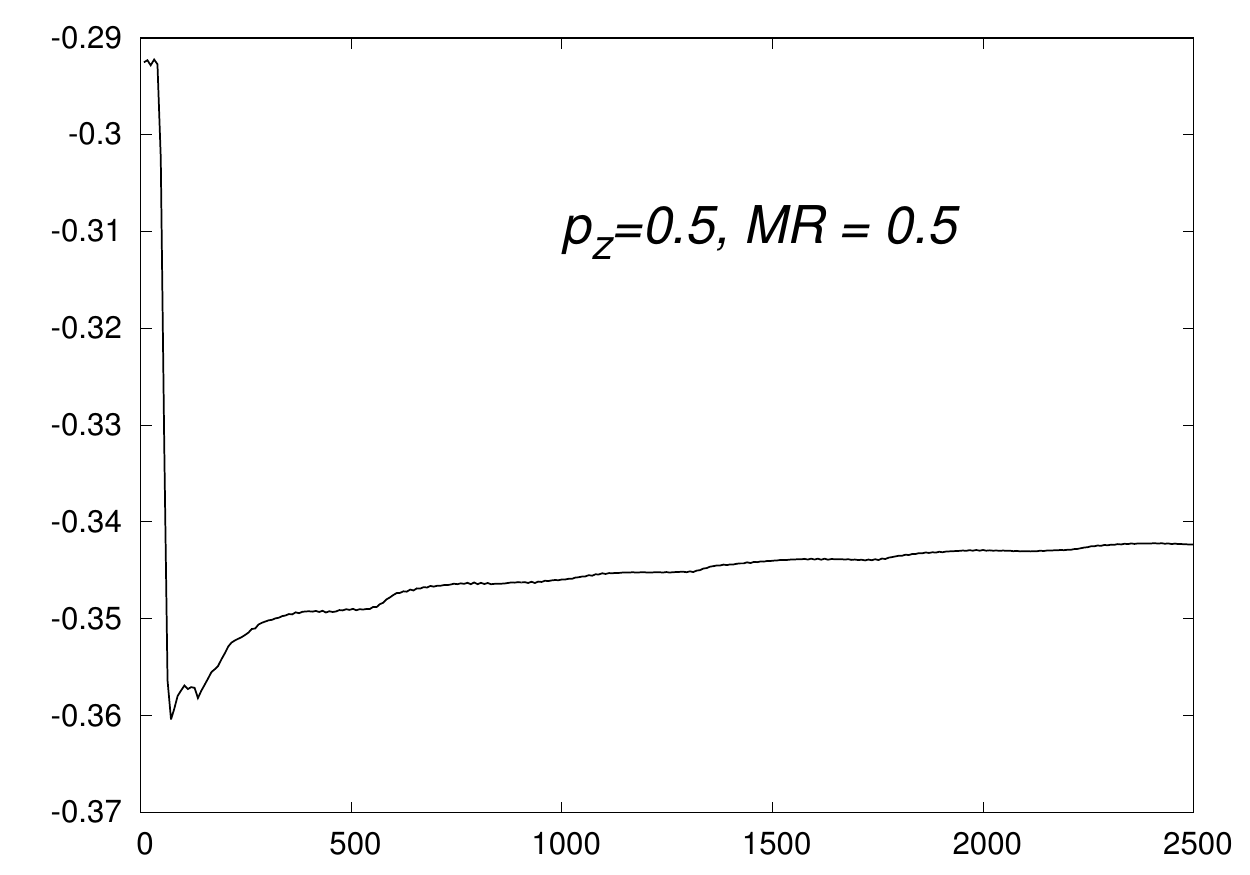}
\includegraphics[width= 3.25cm]{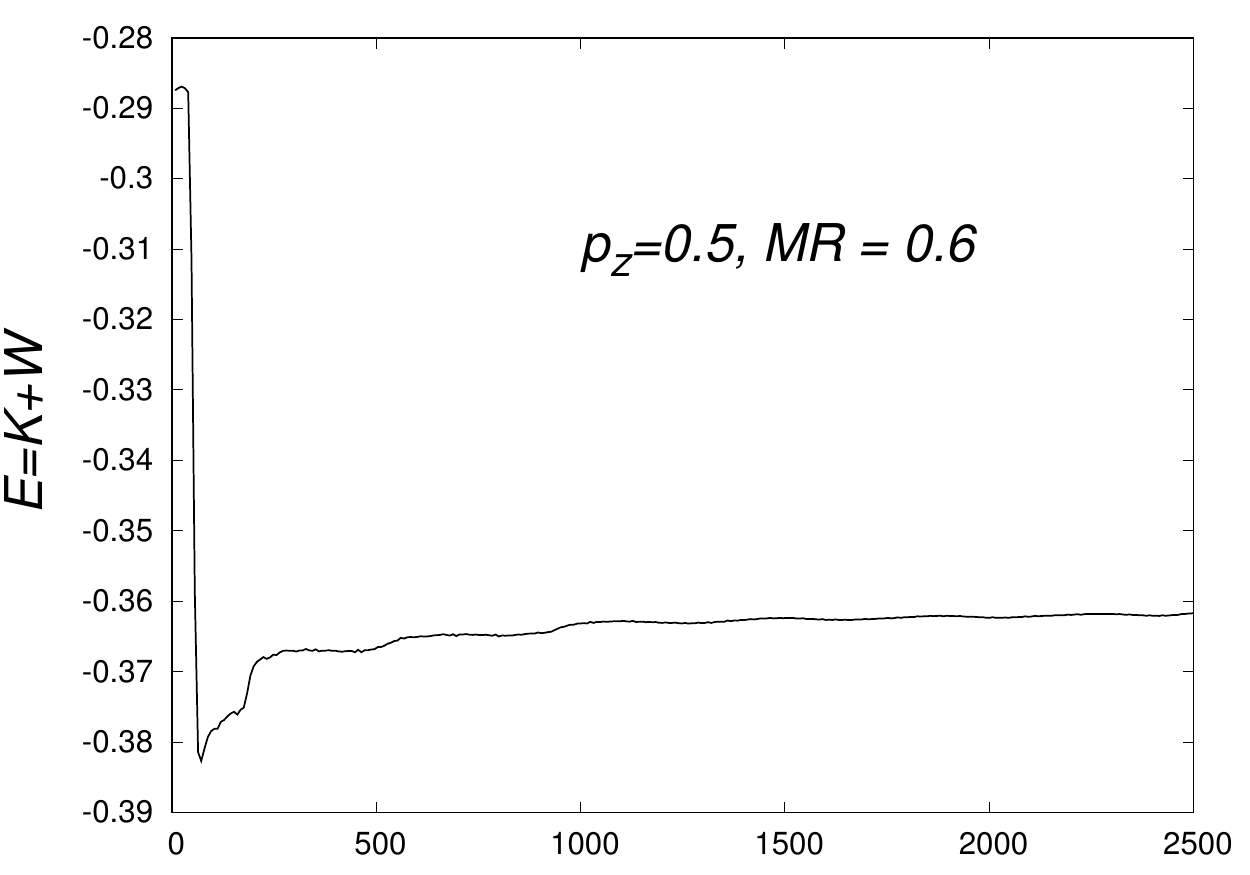}
\includegraphics[width= 3.25cm]{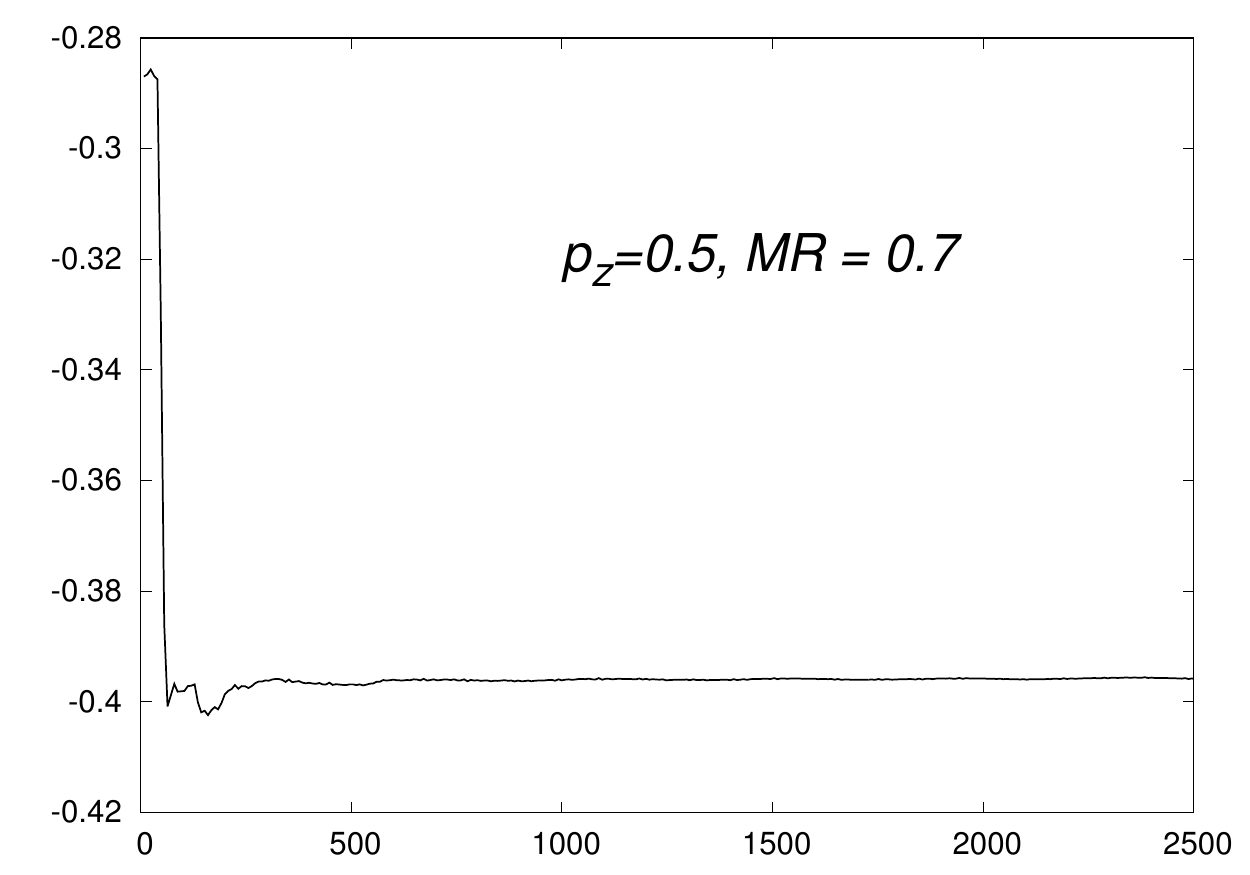}
\includegraphics[width= 3.25cm]{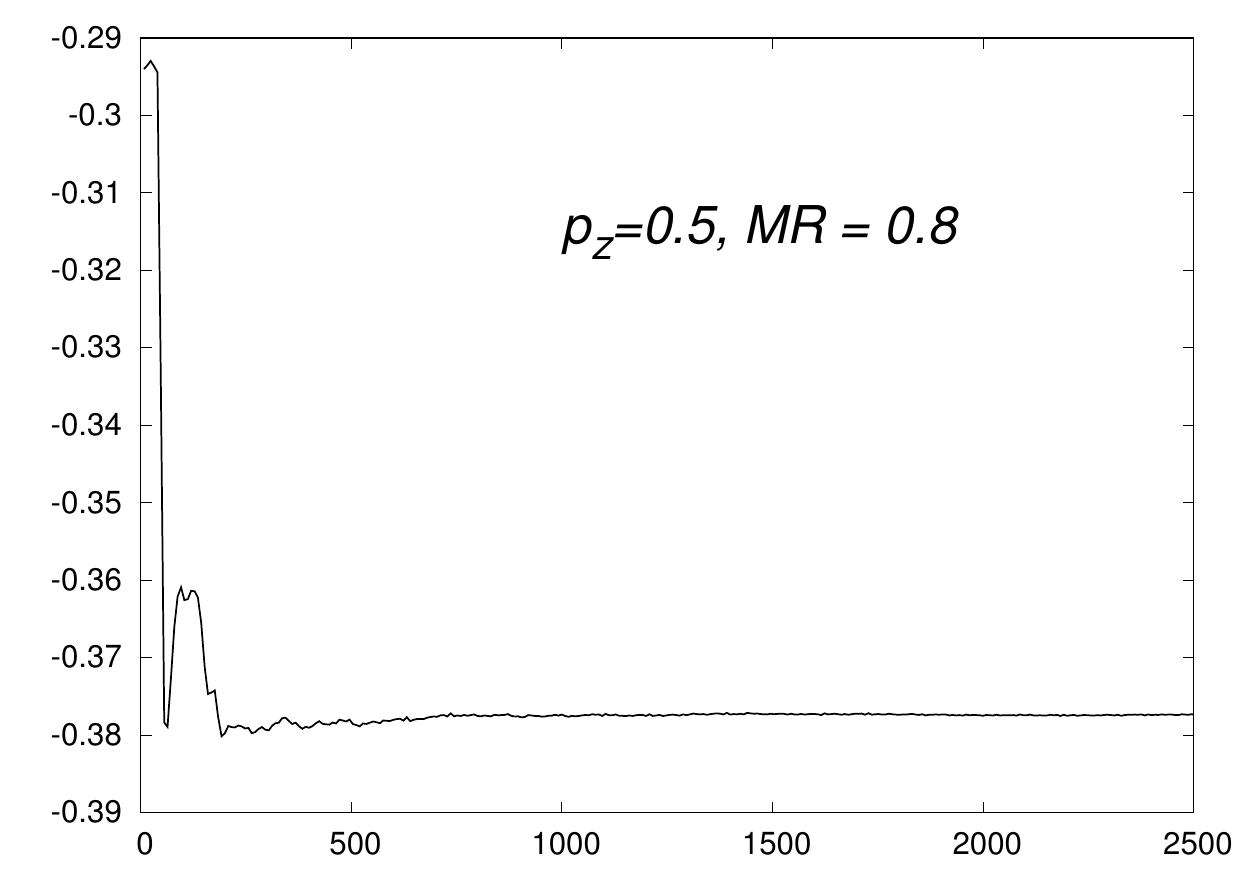}
\includegraphics[width= 3.25cm]{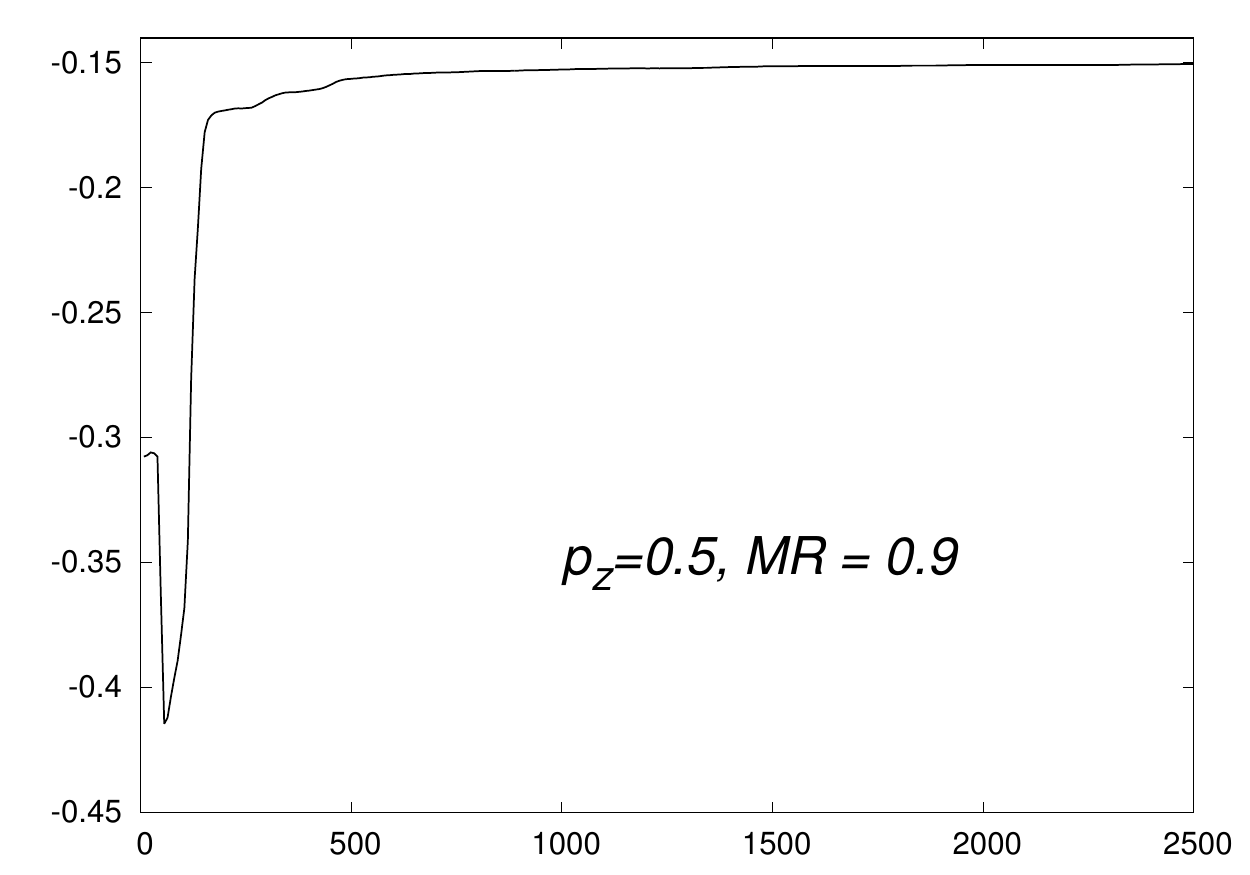}
\includegraphics[width= 3.25cm]{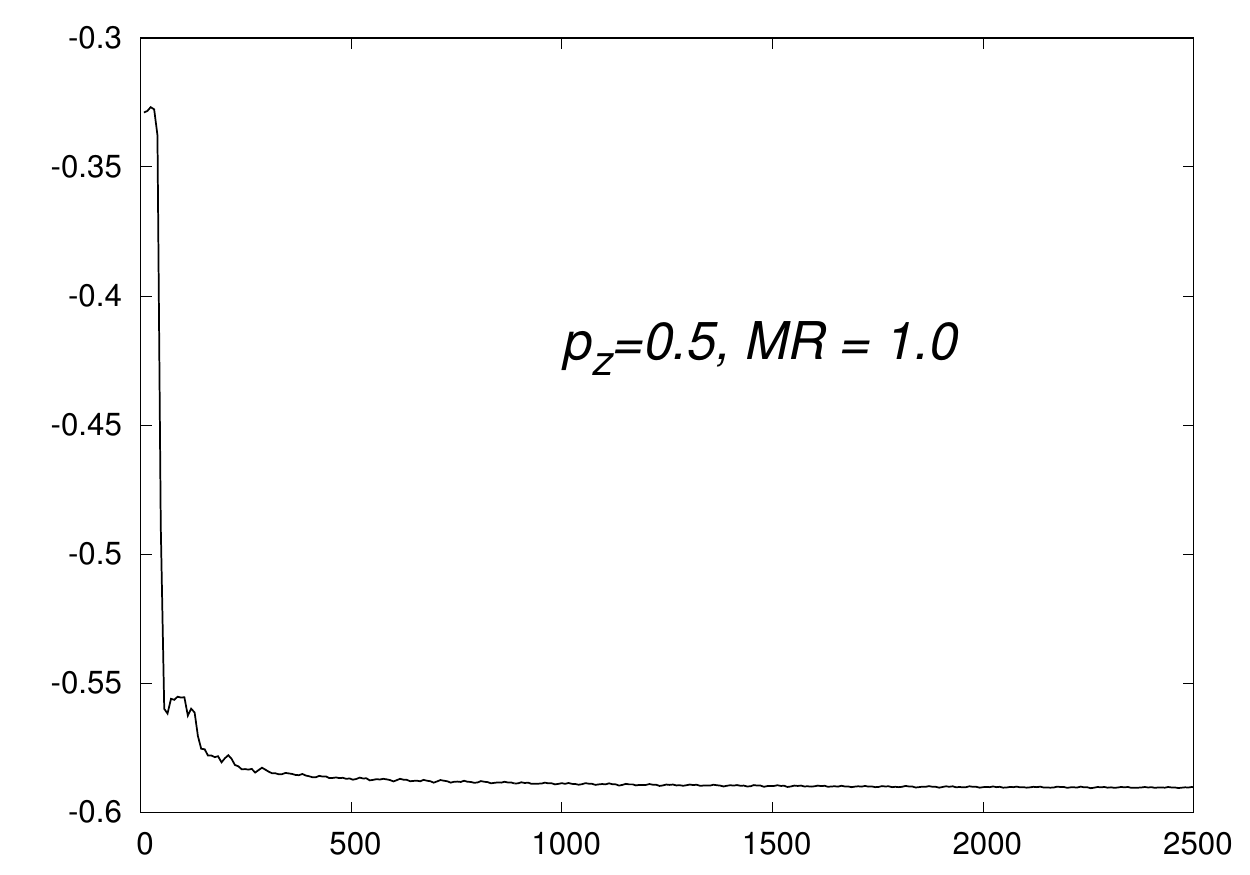}
\caption{Diagnostics of the relaxation process for the case $p_z=0.5$ and  the various mass ratios $MR=0.1,...,1.0$. We show the quantity $2K+W$, the central density where the final structure centers $\rho(0,0)$ and the total energy $E=K+W$ as function of time $t$.}
\label{fig:coolingpz0_5}
\end{figure*}

For equal mass head-on encounters, it has been found that maximum head-on momentum that guarantees the total energy $E=K+W<0$ is near to $p_z=0.7$ \cite{BernalGuzman2006a}. Then in our analysis, in order to keep the condition $E<0$ for a bounded system required for merger, and knowing that the momentum of the blob with mass $M_{\lambda}$ is $p_{z}^{R}=p_{z}^{L}/\lambda := p_z/\lambda > p_z$, we study the momentum range $0\le p_z \le 0.5$ only.
Then we divide the cases in low momentum regime that we cover with two values $p_z=0,~0.1$, and a high momentum regime that we cover with momentum values $p_z=0.4,~0.5$. 
The mass ratio parameter range is covered with runs considering from the equal mass case, to high mass ratios $MR=1,0.9,0.8,0.7,0.6,0.5,0.4,0.3,0.2,0.1$. In all the simulations we use values of $p_z$ such that the total energy $E=K+W$ is negative, so that the system is bounded in all the cases explored and results  in a merger.
Notice that, since the only configuration that changes is the one with mass $M_{\lambda}$ while $M_1$ is kept fixed, there are no repeated cases among the cases explored. 
The results of the simulations corresponding to the parameter space explored can be found in Figures \ref{fig:coolingpz0_0}, \ref{fig:coolingpz0_1}, \ref{fig:coolingpz0_4} and \ref{fig:coolingpz0_5}. 

\begin{figure}
\includegraphics[width= 0.45\textwidth]{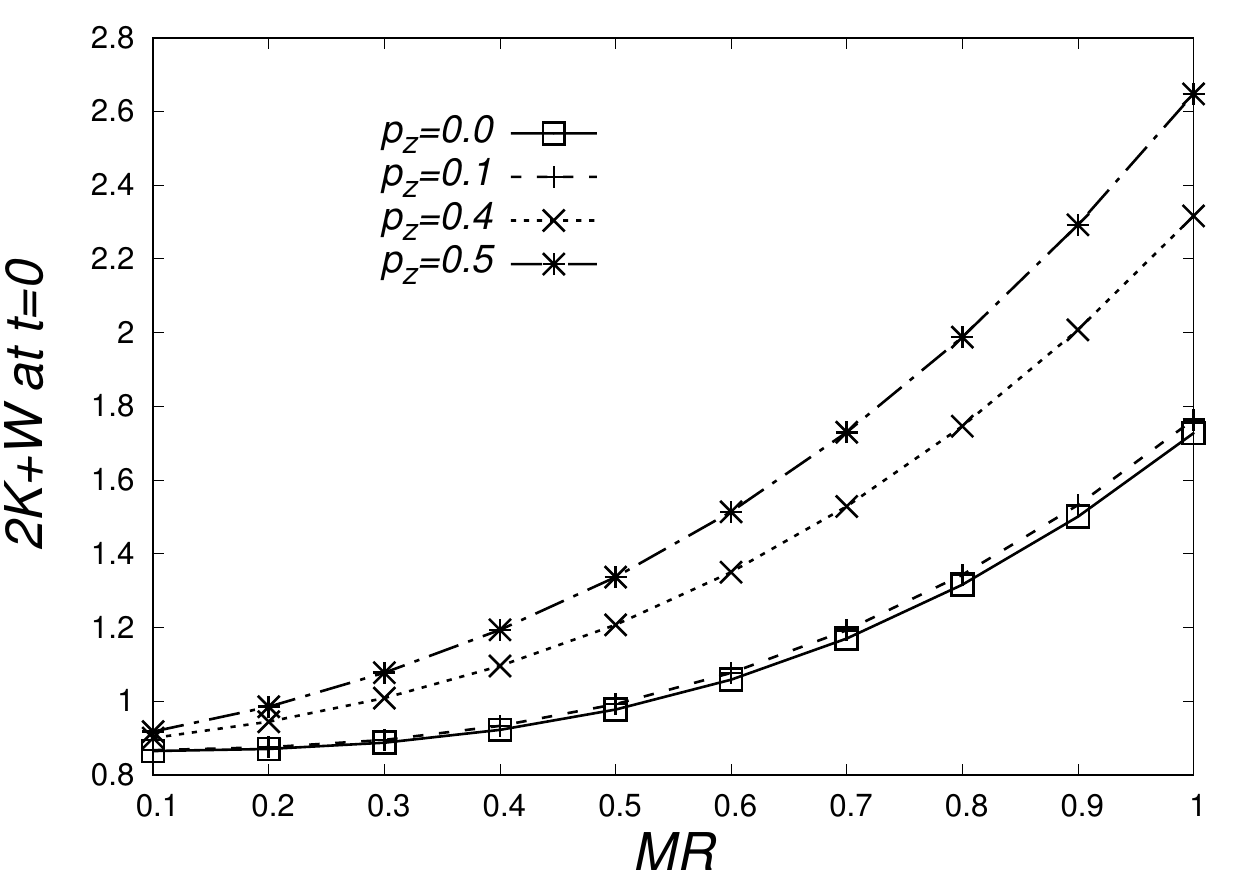}
\caption{ Initial values of $2K+W$. As expected these curves grow with $MR$ and the contribution of the kinetic energy for the different values of $p_z$.}
\label{fig:initial2KW}
\end{figure}

%++++++++++++++++++++++++++++++++++++++++++++++++++

\subsection{The Relaxation Process}
\label{sec:relax}

The relaxation process can be monitored by the quantity $C=2K+W$, which for a virialized systems is zero. For time-dependent scenarios, in which a structure is relaxing, one expects $C$ to oscillate around zero with decreasing amplitude (see for instance \cite{SeidelSuen1994,GuzmanUrena2003,GuzmanUrena2006}), 
but first let us look at its initial value. In Figure \ref{fig:initial2KW} we indicate the initial value of $C=2K+W$ for each of the configurations studied. First of all it serves to show that the state of all the different cases are different. It also illustrates how kinetic energy dominates over gravitational energy for the equal mass case over the unequal mass cases, and shows also that $C$ is bigger for higher values of $p_z$.

In our head-on mergers, for all the values of $p_z$, the quantity $C$ shows two different behaviors depending on whether $MR$ is smaller or bigger than 0.5. When the mass ratio $MR<0.5$, the amount $C=2K+W$ does show oscillations around zero, however within the time window of the simulations there is no clear sign of 
decrease. The quantity $C$ shows a global maximum around $t_{max}=100$ for the low momentum regime and around $t_{max}=50$ for the high momentum regime, where $t_{max}$ is clearly related to the initial velocity of the two configurations.  After that point, the system looses gravitational energy $W$ and $C$ starts oscillating around zero.

In the second regime, when  $MR> 0.5$, the gravitational cooling process  becomes different, $C=2W+K$ enhances and at $t_{max}$ takes its absolute maximum and the amplitude of its oscillations at larger times starts to decrease. The reason is that, for our initial set up, the initial kinetic energy is higher for bigger $MR$ due to the larger total mass $M=M_1+M_{\lambda}$ for a given $p_z$. In this case, the total energy $E$ decreases in the long term until it seems to stabilize; this means that the kinetic energy $K$ (which has positive sign) is being released in the process, and consequently the gravitational cooling is more evident.

In order to outline this process quantitatively, we estimate the relaxation time for all the simulations. For that purpose, we determine an envelope function of $C$, which after the encounter approaches zero at the same rate than the amplitude of $C$.  

\begin{figure*}
\includegraphics[width= 0.45\textwidth]{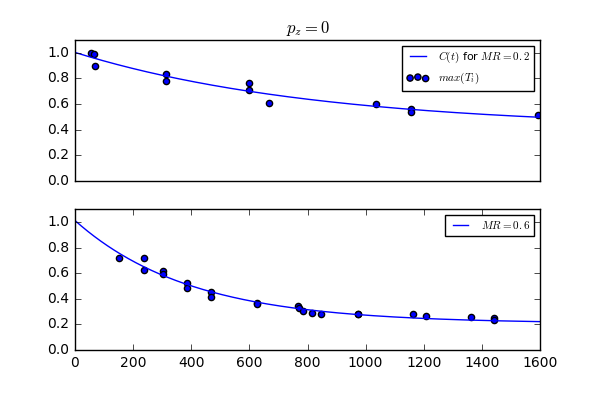}
\includegraphics[width= 0.45\textwidth]{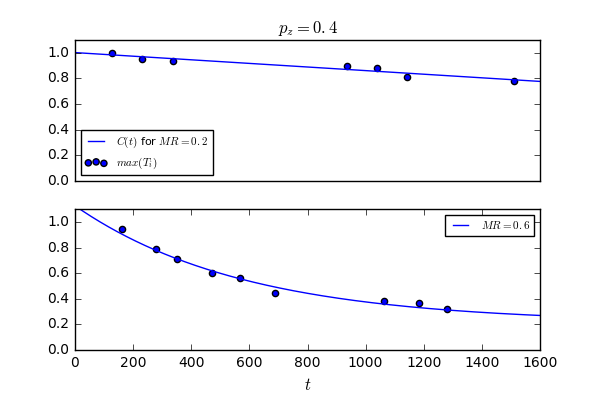}
\caption{Envelope function $\mathcal{E}(t)$ for the amplitude of $C=2K+W$ for a sample different binary systems in the low and high momentum regimes. The dots correspond to the maximum elements of the subintervals $T_i$ framing the time-series of $C$. Solid lines correspond to the best-fits of the framing-dots in each case which obey an exponential law. This quantity tracks the decay of the virializing estimator $C$ and is a key-piece to estimate the relaxation time.}
\label{fig:Env1}
\end{figure*}

Let us describe the details of our calculation. Firstly, we construct the normalized $\hat{C}(t)\equiv C(t)/C(0)$ estimator of gravitational cooling. Secondly, we pick a set of points by splitting the whole time domain in $N$ bins with uniform size given by $T/N$. Afterwards, we construct the following succession of arrays $\hat{T}=\{T_i=[iT/N,T]\}_{i=0,..,N-1}$ which are defined by cutting off a left subinterval of size equal to $iT/N$. The
desired set of points shaping the amplitude maxima are obtained as a new succession made of the maximum elements in the $\hat{T}_i$ arrays, that is $\{\mathcal{E}_i\equiv \max(\hat{T}_i)\}$ corresponding to the dots in Figure \ref{fig:Env1} for a few particular scenarios. Later, we determine a suitable minimum-square fit to $\{\mathcal{E}_i\}$, $\mathcal{E}(t)$ dubbed as the `` envelope '' of $C(t)$.  

\begin{figure*}
%\centering
\includegraphics[width= 0.7\textwidth]{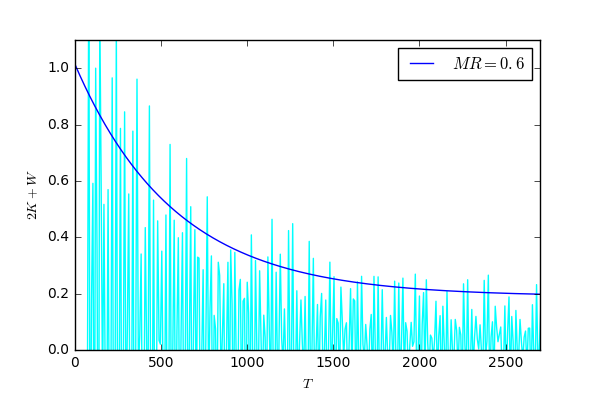}
\caption{ Envelope of the normalized time-series of $C=2K+W$ for a merger with initial $MR=0.6$ decaying exponentialy according to the model  (\ref{eq:exp_fit}) with parameters $a=0.9$, $t_0=500$ and  $c=0.2$.}
\label{fig:Env2}
\end{figure*}

Lines in Figure \ref{fig:Env1} correspond to minimum-square fits using the proper functional form (linear and exponential as we shall explain in short) for a representative set of binary systems in each momentum regime and illustrative values of $MR$. The dark-blue line in Figure \ref{fig:Env2} illustrates how the envelope $\mathcal{E}(t)$ frames the peaks of $C(t)$ for the particular case of a binary system with $p_z=0.4$ and $MR=0.6$. We  observe that the decay rate of $C(t)$ and the envelope are equal.  

In the high momentum regime, $\mathcal{E}(t)$ for the lowest mass ratios $MR=0.1$ and $MR=0.2$ corresponds to a linear fit to the resulting dots in the $T-(2K+W)$ space (see Fig. \ref{fig:Env1}). Besides, $MR>0.2$ and for all mass ratios in the low momentum regime where the gravitational cooling manifests more clearly, $\mathcal{E}(t)$ obeys a decreasing exponential law given by  

\begin{equation}\label{eq:exp_fit}
C(t)=a\,e^{-t/t_0}+c.
\end{equation}

\noindent Finally, we estimate the relaxation time $T_v$ as the time that passes since the collision until the instant when $\mathcal{E}(t)$ reduces to $1\%$. In order to accomplish that, we solve the  equation $C(T_v)-0.01=0$ numerically by using a modification of the Powell hybrid method as implemented in MINPACK \cite{LaCruz2006}. The resulting dots in $T-MR$ turn out to be clearly correlated according to the following power-law for the best-fit 

\begin{eqnarray}
T_v= T_0 (MR)^{-a}.
\label{eq:BFit_Mutimes_pz00}
\end{eqnarray}

\begin{figure*}
%\centering
 \includegraphics[width= 0.45\textwidth]{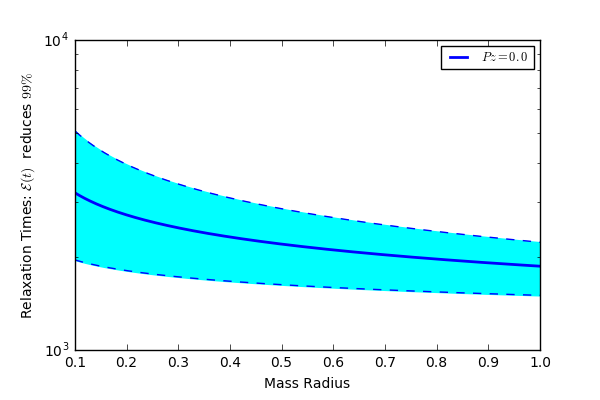}
  \includegraphics[width= 0.45\textwidth]{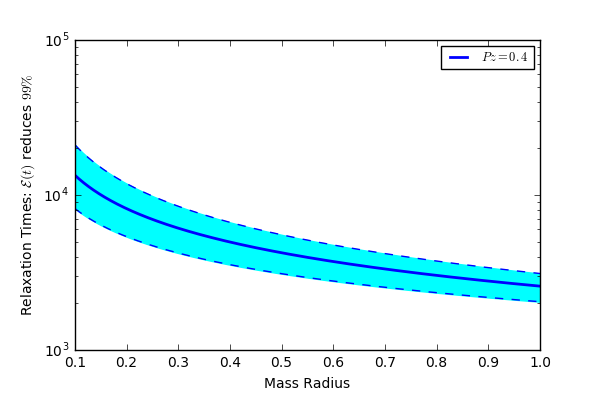}
 \caption{Correlation between relaxation times and the mass ratio parameter for representative values of total momentum $p_z=0$ and $p_z=0.4$ in both, low and high momentum regimes. In the two cases the central solid blue line corresponds to the best-fit given by equation (\ref{eq:BFit_Mutimes_pz00}). Dotted lines enclose the $1\sigma$ region of the relation.}
\label{fig:MuTime}
 \end{figure*}

\noindent Figure \ref{fig:MuTime} shows the minimum-square-root best fits for the resulting dots in the $T-MR$ space and its corresponding $1\sigma$ confidence region for representative $p_z$ values in the high and low momentum regimes. This extrapolation provides a quantitative way to estimate the relaxation time of a merger with any $MR$ in different momentum regimes. The best-fit parameters and their corresponding $1\sigma$ errors for representative $p_z=\{0,0.4\}$ are summarized in Table I.

\begin{table}
\begin{tabular}{|c|c|c|}\hline
 		& High $p_z=0.4$ 	& Low $p_z=0.0$\\\hline
 $T_0$	& $2582\pm 533$ 	& $1864 \pm 363$ \\\hline
  $a$ 	& $0.72 \pm 0.12$ 	& $0.24 \pm 0.12$\\\hline
\end{tabular}
\label{tab:ajustesfits}
\caption{Best fitting values of $a$ and $T_0$. }
\end{table}

Let us now look closely to special features of the relaxation process in different momentum regimes.

{\it Low momentum regime.} In Figures \ref{fig:coolingpz0_0}  and \ref{fig:coolingpz0_1} we show the relaxation process for the low head-on momentum cases, namely the free fall head-on encounter $p_z=0$ and the case $p_z=0.1$. As mentioned before, an interesting common feature either for low or high momentum regimes is the different behavior of $C(t)$ when $MR$ takes small and large values. However the process by which this transition happens is different in both momentum regimes. By looking at the behavior of the times-series of the  total energy $E=K+W$ for the low momentum regime  in Figures \ref{fig:coolingpz0_0} and \ref{fig:coolingpz0_1} and low mass ratios ($MR<0.5$),  $E$ is initially dominated by the gravitational energy, this is clear since right after the encounter, the gravitational well is enhanced. At later times kinetic energy increases and the derivative of the total energy tends towards a constant value. A binary system with small $MR$ is  similar to a big configuration being perturbed by a small one, where the total energy barely changes at large times and the derivative is bigger for larger $MR$. At some point $0.4<MR<0.5$ the asymptotic behavior of the derivative changes dramatically. By looking at $E=K+W$ we can clearly notice that the system starts loosing energy and the relaxation process becomes visible. In contrast to the high momentum regime which we shall look in short, the transition in this case is softer.

{\it High momentum regime.} We show the evolution process of cases $p_z=0.4$ and $p_z=0.5$ in Figures \ref{fig:coolingpz0_4} and \ref{fig:coolingpz0_5} respectively. In this high momentum regime we can spot a sudden turn-over in the  slope of the total energy. This  suggests that a low energy mode dominates and higher frequency modes somehow are suppressed. On the other hand, the total energy for this regime has a peculiar behavior since it suffers a turn over in comparison to the low momentum cases. The existence of this turn-over  is a symptom of a loss of total energy  produced by an excess of kinetic energy carried by the smaller of the two configurations. This excess of kinetic energy allows the modes carrying high energy to scatter away from the gravitational binding potential.  This turn-over of $E$ in time when $MR$ is varied does not occur in the low momentum regime.

%%%%%%%%%%%%%%.    Subsection    %%%%%%%%%%%%%%%%%%%%%%%%%%%%%%%%%%
\subsection{Further discussion of results}

{\it Gravitational cooling.} The traditional mechanism for relaxation of the Schr\"oedinger-Poisson (as originally dubbed) is the gravitational cooling \cite{SeidelSuen1994,GuzmanUrena2006}. Since there is no other interaction than gravitation and the wave function with itself, the only mechanism to relax would be the emission of matter, which changes the value of the gravitational energy of a system. Thus the gravitational cooling is understood as the mechanism that consists in the emission of matter and is more effective when the system has more kinetic energy that can help carrying matter off the gravitational potential well. The process of gravitational cooling is faster for cases with initial conditions that have a bigger excess of kinetic energy.

{\it The low frequency mode observed in all cases.} For all the values of the momentum, the central density shows a low frequency mode that calls the attention. For some combinations of $p_z$ and $MR$ the amplitude of the density is small, like for $p_z=0,~MR=0.3$  in Fig. \ref{fig:coolingpz0_0}. For other combinations, the oscillations of the central density are prominent like for example in cases $p_z=0.5$ for $MR$, except $MR=0.3,~0.8,~1$. 

\begin{figure}
\includegraphics[width= 7.5cm]{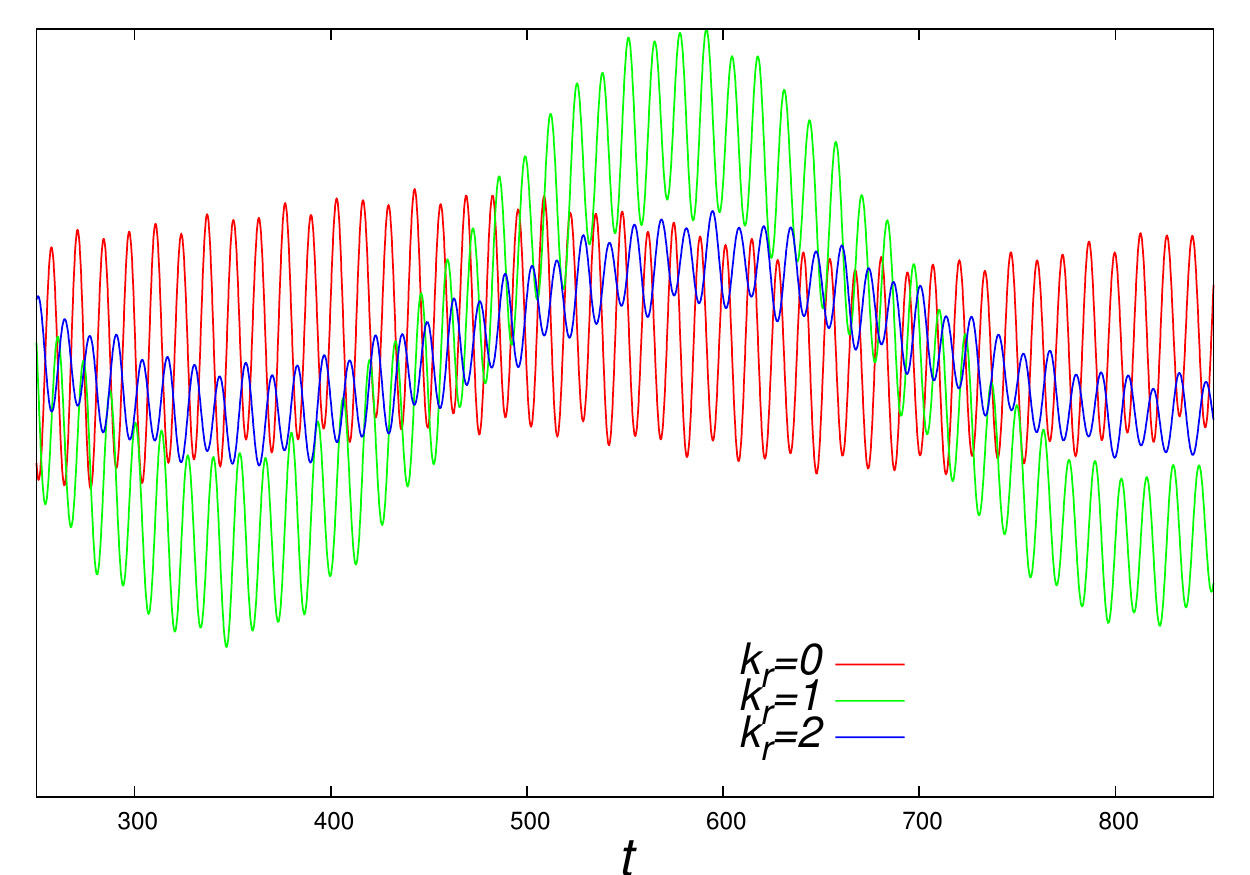}
\includegraphics[width= 7.5cm]{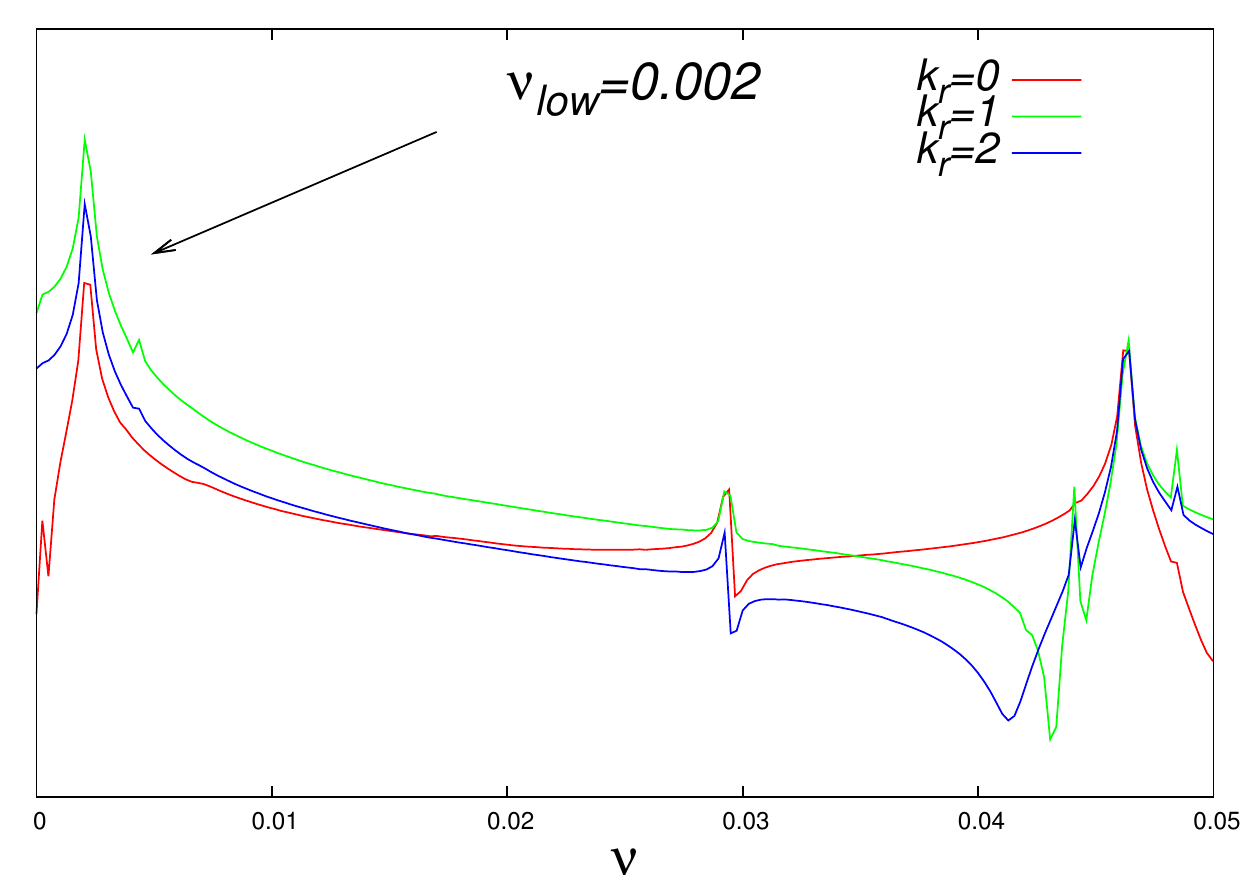}
\caption{Density measured along the $z-$axis for perturbations with $k_r=0,1,2$. Also shown is the Fourier Transform of this signal that shows the low frequency mode $\nu_{low}=0.002$. The  peak at $\nu=0.046$ corresponds to the high frequency mode observed in the time domain.}
\label{fig:perturbation}
\end{figure}

The oscillation, with such an important amplitude, can possibly be the responsible for avoiding the survival of structures when there is a merger, as discussed in \cite{GonzalezGuzman2016}. We now show this low frequency mode can be explained by a perturbation with mode $l=1,m=0$, which is consistent with the distribution of matter in a head-on encounter. In order to show this, we apply a perturbation to a single equilibrium configuration $\Psi=\Psi_e+\delta\Psi$, with $\delta \Psi = A_{lm}e^{-(\sqrt{r^2+z^2}-r_p)^2/\sigma_{p}^2} e^{-ik_r \sqrt{r^2+z^2}} Y^{0}_{1}$,  $\Psi_e$  the wave function of the equilibrium configuration, $r_p$ is the radius at which a Gaussian perturbation shell is centered. We use three different wave numbers $k_r=0,1,2$ to make sure the resulting modes do not depend on the velocity of the perturbation, and $Y_{0}^{1}$ is the spherical harmonic. The perturbation is such that the mass of the perturbation is $0.1\%$ of the mass of the equilibrium configuration. The density measured along the $z-$ axis and its Fourier Transform are shown in Fig. \ref{fig:perturbation}. The low frequency mode has peak frequency  $\nu_{low}\sim0.002$, which corresponds to a period of $T\sim 500$ similar to that of the low frequency mode of the configurations resulting from the merger.

\begin{figure}
\includegraphics[width= 8cm]{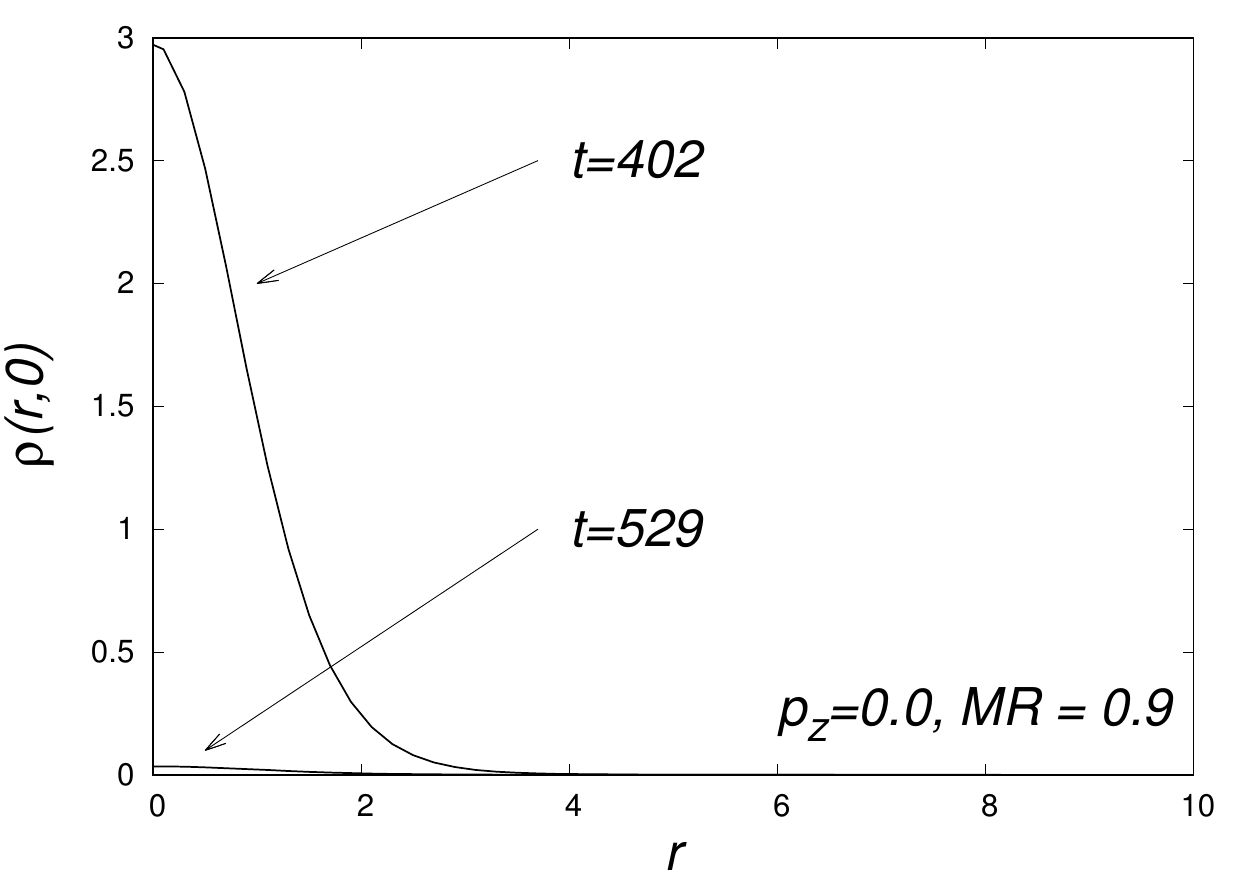}
\includegraphics[width= 8cm]{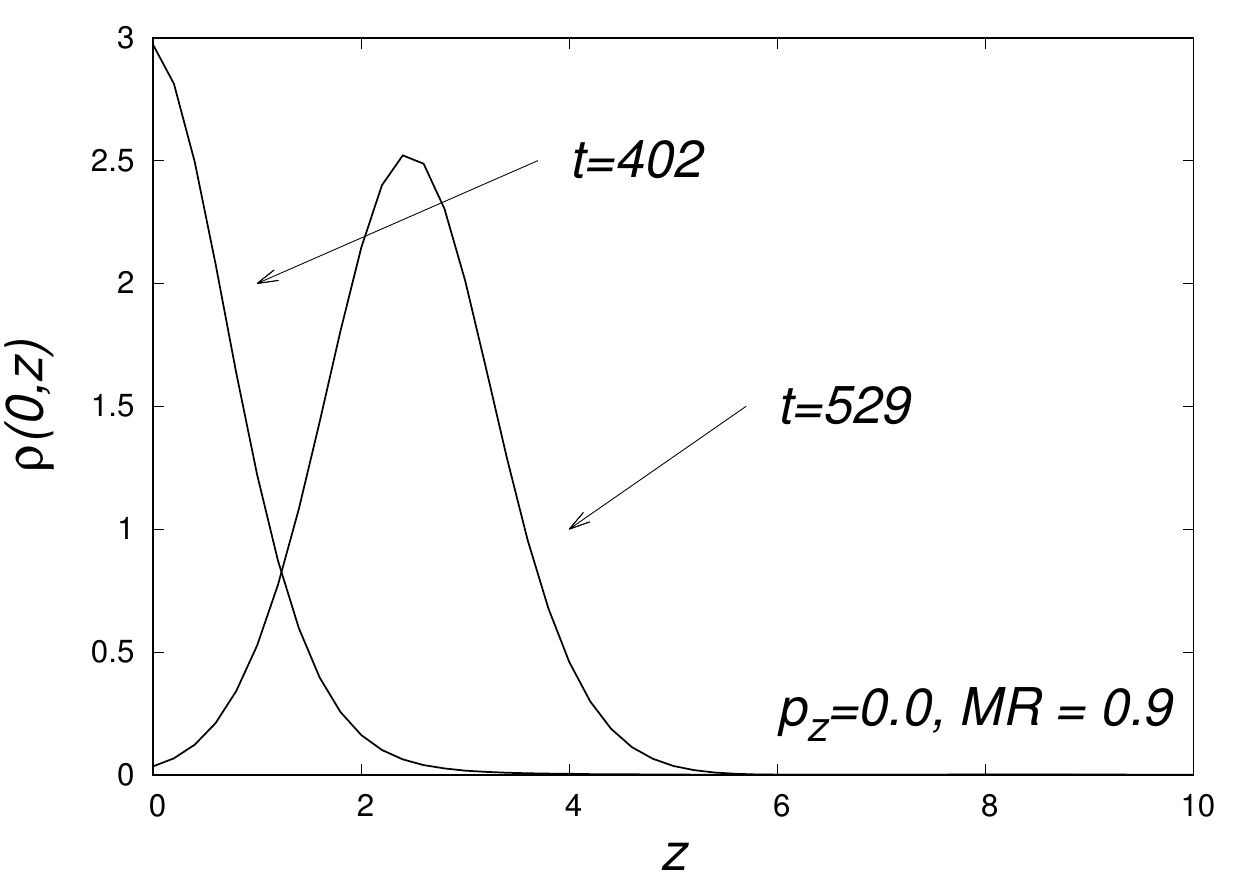}
\caption{Snapshots of the density along the radial and positive axial directions for $p_z=0$ and $MR=0.9$ at times $t=402$ and $t=529$, which correspond to a maximum and minimum of the central density for the low frequency oscillation mode shown in Fig. \ref{fig:coolingpz0_0}. This illustrates how at the maximum central  density, the profile is nearly isotropic  whereas at the minimum central density, the matter distributes in two blobs along the $z-$axis.}
\label{fig:bigoscillation}
\end{figure}

{\it High amplitude oscillations.} In order to show how dynamical these oscillations can be, in Fig. \ref{fig:bigoscillation} we show the density profile  along the radial and axial directions at times where the central density is at the maximum when $t\sim402$ and at the minimum when $t\sim 529$. This example corresponds to the free-fall case $p_z=0$ with $MR=0.9$. The cases for other values of $p_z$ and $MR$ show a similar behavior. Notice how the  density distributes in blobs along the $z$ axis when the central density is around a minimum. The process is so violent that the central density at $t=402$ is two orders of magnitude higher than at $t=529$. It calls the attention that  this example corresponds to the free fall case, in which the total initial kinetic energy of the system is the smallest one.

This highly dynamical behavior can have two important consequences. On the one hand, luminous matter might not survive a merger of two structures and would be expelled off the center of the configuration as pointed out in \cite{GonzalezGuzman2016}. On the other hand, this is a warning when calculating the mass function of structures. As far as we know, mass functions are averages of matter distributions in full 3D structure evolution within the GPP model, however it is not known whether this extremely dynamical behavior is being considered in the calculations.

{\it Time scales in physical units.} So far the results in the time domain use code units. A practical way of identifying code and physical units uses the eigenfrequency of an equilibrium configuration,  where the core radius of such equilibrium configuration and the boson mass are the required parameters for the translation \cite{Mocz2017}. For example, assuming the mass of the boson $m=2.5 \times10^{-22}eV$, then for possible  values of the core radius, say $r_c=1,~0.5,~0.25$ and $0.125$kpc, then every 1000 units of time in Figures 2 to 5, which contain the essential time-series used in our analysis, correspond to $\sim 76,~19,~4.75$ and $1.19$Gyr respectively.

Using this scaling one can also estimate the period of the low frequency oscillations discussed above. For the specific  case with $p_z=0$ and $MR=0.9$, for which the period of this mode is $T_{low}\sim 380$ in code units, for $m=2.5 \times10^{-22}eV$, assuming the possible core radius of the final configuration  $r_c=1,~0.5,~0.25$ and $0.125$kpc  the period is respectively $\sim 28.8,~7.2,~1.8$ and $0.45$Gyr.

% ----------------------------------------------------------------
% ----->     Final comments     <-----
% ----------------------------------------------------------------
\section{Conclusions and final comments}
\label{sec:final}

We carried out a systematic study of  head-on mergers between two equilibrium configurations that are solutions of the GPP system of equations. The parameter space explored includes the mass ratio between the two configurations and the head-on momentum.

We estimate the relaxation based on the decrease rate in time of the diagnostics quantity $C=2K+W$, which should be zero in an ideally virialized system. In the collision scenarios studied here, the final configuration does not show this ideal behavior, instead this quantity $C$ oscillates around zero with a decreasing amplitude that depends on the values of $p_z$ and $MR$.

There is this interesting threshold  $MR=0.5$ for all the values of the head-on momentum, above which the cooling process is more noticeable due to the amount of kinetic energy that can be released during the evolution of the system. For smaller mass ratios one of the configurations acts more like a perturbation and there is less kinetic energy to be released, then the relaxation process is slower.

There is a low frequency oscillation mode that calls the attention, because in some of the particular combinations of $p_z$ and $MR$, the central density of the final configuration oscillates violently.  It would be interesting to carry out a systematic analysis similar to that in \cite{GonzalezGuzman2016}, that could establish the consequences of the kinematics of luminous matter in such a violently oscillating environment.

In many cases of the scenarios explored, the central density of the final configuration  oscillates with this low frequency and amplitudes  that change even in two orders of magnitude. The distribution of matter in these violent cases evolve periodically  from a nearly spherical distribution to a clearly $l=0,m=0$ shaped distribution. This type of cases could be important in calculations of the mass function estimates related to the core-halo profile of structures obtained from 3D simulations. Depending on the expected core radius of the final structure, the period of this oscillation could range from a fraction of gigayear up to various gigayears.

% ----->     ACKNOWLEDGMENTS     <-----

\section*{Acknowledgments}
This research is supported by Grants No. CIC-UMSNH-4.9 and CONACyT 258726. A.A.L acknowledges financial support from CONACyT and Vicerrector\'ia de Investigaci\'on y Estudios de Posgrado BUAP. The simulations were carried out in the LNS under grants to A.A.L and F.S.G., the  computer farm funded by CONACyT 106466 and the Big Mamma cluster at the IFM.
%2017-01077

% -------------------------------------------------------
% -----     REFERENCES     ----------
% -------------------------------------------------------

%\bibliographystyle{plain}

%\end{document}
\end{document}